\documentclass[10pt,letterpaper,aps,prc,superscriptaddress,nofootinbib,floatfix,onecolumn]{revtex4-2}
\usepackage[utf8]{inputenc}

\usepackage{graphicx}
\usepackage{dcolumn}

\usepackage{xspace}
\usepackage{color}
                 
\newcommand\ie {{\it i.e. }} 
\newcommand\vs {{\it vs }} 
\newcommand\eg {{\it e.g. }}
\newcommand\etc{{\it etc. }}

\newcommand{\taa}{\mbox{$T_{\rm AA}$}\xspace}
\newcommand{\TAB}{\mbox{$T_{\rm AB}$}\xspace}
\newcommand{\Teff}{\mbox{$T_{\rm eff}$}\xspace}
\newcommand{\pt}{\mbox{$p_{\rm T}$}\xspace}
\newcommand{\kt}{\mbox{$k_{\rm T}$}\xspace}
\newcommand{\mt}{\mbox{$m_{\rm T}$}\xspace}

\newcommand{\raa}{\mbox{$R_{\rm AA}$}\xspace}
\newcommand{\rab}{\mbox{$R_{\rm AB}$}\xspace}

\newcommand{\Npart}{\mbox{$N_{\rm part}$}\xspace}
\newcommand{\Ncoll}{\mbox{$N_{\rm coll}$}\xspace}
\newcommand{\Nqp}{\mbox{$N_{\rm qp}$}\xspace}
\newcommand{\Nch}{\mbox{$N_{\rm ch}$}\xspace}

\newcommand{\Et}{\mbox{$E_{\rm T}$}\xspace}
\newcommand{\xt}{\mbox{$x_{\rm T}$}\xspace}

\newcommand{\meanet}{\mbox{$\langle E_{\rm T} \rangle$}\xspace}
\newcommand{\sqs}{\mbox{$\sqrt{s}$}\xspace}
\newcommand{\sqsn}{\mbox{$\sqrt{s_{_{NN}}}$}\xspace}
\newcommand{\snn}{\mbox{$\sqrt{s_{_{NN}}}$}\xspace}
\newcommand{\sqsntwo}{\mbox{$\sqrt{s_{_{NN}}}=200$~GeV}\xspace}
\newcommand{\NN}{\mbox{$NN$}\xspace}
\newcommand{\pp}{\mbox{$pp$}\xspace}
\newcommand{\pbarp}{\mbox{$p\bar{p}$}\xspace}
\newcommand{\pn}{\mbox{$pn$}\xspace}
\newcommand{\nn}{\mbox{$nn$}\xspace}
\renewcommand{\AA}{\mbox{A$+$A}\xspace}
\newcommand{\AB}{\mbox{A$+$B}\xspace}
\newcommand{\pA}{\mbox{p$+$A}\xspace}
\newcommand{\dau}{\mbox{$d$$+$Au}\xspace}
\newcommand{\pdau}{\mbox{$p(d)$$+$Au}\xspace}
\newcommand{\pau}{\mbox{$p$$+$Au}\xspace}
\newcommand{\auau}{\mbox{Au$+$Au}\xspace}
\newcommand{\cucu}{\mbox{Cu$+$Cu}\xspace}
\newcommand{\cuau}{\mbox{Cu$+$Au}\xspace}
\newcommand{\pbpb}{\mbox{Pb$+$Pb}\xspace}

\newcommand{\piz}{\mbox{$\pi^0$}\xspace}

\newcommand{\dnchdeta}{\mbox{$dN_{\rm ch}/d\eta$}\xspace}
\newcommand{\dngamdeta}{\mbox{$dN_{\gamma}/d\eta$}\xspace}

\newcommand{\vtwo}{\mbox{$v_2$}\xspace}
\newcommand{\vthr}{\mbox{$v_3$}\xspace}
\newcommand{\rgam}{\mbox{$R_{\gamma}$}\xspace}
\newcommand{\mgg}{\mbox{$m_{\gamma\gamma}$}\xspace}
\newcommand{\mee}{\mbox{$m_{e^{+}e^{-}}$}\xspace}

\newcommand{\gam}{\mbox{$\gamma$}\xspace}
\newcommand{\gev}{\mbox{GeV}\xspace}
\newcommand{\gevc}{\mbox{GeV/$c$}\xspace}
\newcommand{\fmc}{\mbox{fm/$c$}\xspace}

\def\be{\begin{equation}}
\def\ee{\end{equation}}

\begin{document} 

\begin{center}
{\bf Direct real photons in relativistic heavy ion collisions}
\end{center}

\vspace{0.2in}

\begin{center}
{\it Gabor David \\
Stony Brook University, Brookhaven National Laboratory}
\end{center}

\vspace{0.3in}

{\bf Abstract.}  Direct real photons are arguably the most versatile
tools to study relativistic heavy ion collisions.  They are produced,
by various mechanisms, during the entire space-time history of the
strongly interacting system.  Also, being colorless, most the time
they escape without further interaction, \ie they are penetrating
probes.  This makes them rich in information, but hard to decypher and
interpret.  This review presents the experimental and theoretical
developments related to direct real photons since the 1970s,
with a special emphasis on the recently emerged ``direct photon
puzzle'', the simultaneous presence of large yields and strong azimuthal
asymmetries of photons in heavy ion collisions, an observation that 
so far eluded full and coherent explanation.

\tableofcontents

\newpage


\section{\bf Introduction}
\label{sec:intro}

The centuries old quest to reveal the ``ultimate'' constituents of
matter and the laws of Nature governing them, and the realization from
quantum mechanics that mapping smaller and smaller objects requires
ever larger energy probes, made high energy physics one of the dominant
scientific disciplines of the XXth century.  At first we took
advantage of the Universe as an ``accelerator'' providing high energy
probes (cloud chamber and emulsion experiments), but soon we started 
to build our own accelerators, constantly increasing their energy and 
luminosity.  The most spectacular and best known results came in
elementary particle physics, but astrophysics and nuclear physics were
a close second, benefiting enormously from the progress in
experimental and theoretical tools.  

The central question in particle
physics was the substructure of hadrons and the nature of confinement,
while high energy nuclear physics' foremost concern was the behavior
of high density nuclear matter, including its collectivity and
possible phase transition into partonic matter, which in turn evoked
the early stages of the Universe.  
Albeit particle and nuclear physics came with different perspectives, 
both were concerned with the strong interaction and its underlying
theory, quantum chromodynamics (QCD), so the birth of a new
discipline at their intersection, 
relativistic heavy ion physics, was almost inevitable.  
From Princeton and Berkeley to Dubna the race for larger and larger 
ions, energies and luminosities was on.  Hadrons and nuclear fragments 
were studied extensively, and the applicability of thermo- and
hydrodynamics gradually recognized. 
An excellent review of these first facilities and early
developments -- both experimental and theoretical -- 
can be found in~\cite{Goldhaber:1978pv} by Goldhaber, while a
(literally) insightful account of the genesis and achievements of RHIC
and LHC was given by Baym in~\cite{Baym:2016wox}.

Although lepton and photon production in hadronic and nuclear collisions
was studied almost since the birth of quantum mechanics and quantum
electrodynamics~\cite{Fermi:1925fq,meitner:1933,Heisenberg:1996bt,Low:1958sn},
the golden era started in the early 1970s
when ISR and Fermilab became operational.  The available large
energies allowed almost from the beginning to explore both fundamental
facets of direct photons: at high transverse momentum (\pt) they are
probes of QCD hard scattering, while at low \pt they help to shed
light on multiparticle production.  The high \pt aspect was first
pointed out by Escobar~\cite{Escobar:1975wx}, 
Farrar and Frautschi~\cite{Farrar:1975ke} in 1975.  
At low \pt, the first one to emphasize their importance in the context 
of multiparticle production was Feinberg~\cite{Feinberg:1976ua} in
1976, famously stating that
``the thermodynamical approach naturally leads to direct $\gamma$'s
and dilepton production. ...  If the multihadron production process
contains an intermediate stage of a thermodynamical hadronic matter,
then this kind of $\gamma$ and dilepton production {\it inevitably}
exists.  The question is only on its intensity and on the detailed
behaviour of the spectra.''\footnote{Feinberg
  in~\cite{Feinberg:1976ua} even provided an estimate (based on Landau
  theory) of the $N_{\gamma}$ photon yield w.r.t. the $N_{\pi}$ pion
  yield, with the expression

\begin{equation}
N_{\gamma} = 6.4Ae^2N_{\pi}^{4/3}\Big(1-\frac{1}{N_{\pi}^{1/3}}+0.21lnN_{\pi}\Big)
\end{equation}

\noindent
which gives $N_{\gamma}/N_{\pi}$ in the 0.1-0.15 range for realistic
pion multiplicities; remarkably close to today's measurements!
}
A few years later Shuryak~\cite{Shuryak:1978ij} 
already estimated the lepton and photon production
from the ``Quark-Gluon Plasma'' (along with the hadronic production).
The idea that photons can test the most diverse manifestations of the
strong interaction took hold.  The rapid growth of the field has been
documented in several reviews in the past 
(see for
instance~\cite{Ferbel:1984ef,Alam:1996fd,Cassing:1999es,Peitzmann:2001mz,Stankus:2005eq,David:2006sr,Linnyk:2015rco}).  
By now, photon physics
became mature, but by no means a closed chapter -- just the opposite.

\vspace{0.1in}
{\it Promises.} \\
\vspace{0.1in}
One of the earliest motivations to measure direct photons in heavy ion
collisions was to gain access to the initial temperature of the system
and to study the properties of the QGP via its thermal radiation.
With time, first observations and evolving theory, expectations 
grew to include information on geometry, viscosity (and the
time-evolution thereof), effects of the conjectured large magnetic
field, gluon saturation, and the initial conditions in general.  Also,
just as in \pp, high \pt direct photons back-to-back to a jet in \AA
were expected to be the ultimate calibration tool for the energy of the
hard scattered parton, and of the possible energy loss in a colored
medium. 

\vspace{0.1in}
{\it The blessing and curse of direct photons.} \\
\vspace{0.1in}
Photons play a special role in the study of high-energy hadronic and
nuclear interactions, because they are {\it penetrating probes}.  
Being color neutral, their mean free
path is rather large not only in very dense hadronic matter, but also
in a medium of deconfined quarks and gluons, the quark-gluon plasma, or
QGP (see Sec.~\ref{sec:aahigh}).  This ensures that if they are
created at any time, they will escape the interaction region 
(mostly) unaltered and will be detectable.  This is true even if they
have to cross the hot, dense medium of QGP.  On the other hand, to our
current best knowledge at every stage (or at the very least at 
{\it most} stages) of the collision there are
physics mechanisms to produce photons, providing direct information 
both on the process itself and the environment 
(\eg the initial state including geometry, the expansion of the 
plasma or the hadron gas, and so on).  In this sense photons are the
perfect ``historians'' of the evolution of the system.  

Unfortunately,
all that information on instantaneous rates and expansion dynamics is
convolved (integrated in space-time) leaving us with just a few 
high level experimental observables: the all-inclusive
spectrum, possible azimuthal asymmetries, energy deposit or lack
thereof around the photons (isolation), rapidity distribution, 
correlations, and so on.  Therefore, disentagling the
contributions from the various processes is nearly impossible without
relying on models.  Finally, the number of photons created in the 
collision, before the final chemical and kinematic freeze-out of the
system ({\it direct photons}) is usually small compared to the photons
coming from the decay of final state hadrons, like \piz, $\eta$.
While  these decay photons are highly interesting in hadron 
physics, they pose a serious background problem for direct photon
measurements. \\

This review was written with the following goals in mind.  
It is trying to be
{\it accessible} even for physicists outside our field, but providing
plenty of references for those who want to learn the details.  It intends to
show the {\it historic context and evolution of ideas}, in the firm
belief that nothing is more instructive and promotes creativity 
and progress better,
than understanding not only {\it what} do we know now, but also the process
{\it  how} did we acquire this current knowledge.  It is aiming to be 
as {\it up-to-date} as possible, by including even very recent
preliminary results (as of early 2019).
Also, it is meant to be {\it realistic} in its claims on which issues
are really settled and which ones are not.  If in doubt, we would
rather err on the side of under-, than overconfidence.

Photons taught us a lot already,
but in the process new questions emerged, and a fully self-consistent
description of direct photons in heavy ion collisions is not available
yet, not the least because important measurements are still missing or
not sufficiently precise to discriminate between theories.
The field is very much open and if this review helps
to raise fresh interest and generate new efforts, it achieved more
than we ever dared to hope for.

In this review first we  introduce
the basic theory concepts (Sec.~\ref{sec:sources}) and experimental
techniques (Sec.~\ref{sec:experimental}),
then we discuss the results from past decades, concentrating on
the direct photon spectra in hadron-hadron and heavy ion collisions.
As mentioned above, the relevant physics (and even the experimental
techniques) of high and low \pt photon production are quite
different, so we describe them in separate sections, in their own
historic context.  On the other hand observations in heavy ion
collisions can not be understood without the corresponding
``baseline'' in \pp.  Therefore, we group together all high \pt photon results
in \pp and \AA and their impact on jet physics, collision centrality
determination, nuclear PDFs and other issues in Sec.~\ref{sec:highpt},
whereas the low \pt (or ``thermal'') production in all colliding
systems will be discussed in Sec.~\ref{sec:thermal}.  Finally, 
we review the challenging developments of the past few years, starting
with the 2011 observation of elliptic flow of ``thermal'' photons
and resulting in the  so-called ``direct photon puzzle'', its
consequences, the attempts to resolve the ``puzzle'' and the questions 
that are still open in Sec.~\ref{sec:puzzle}.

\section{{\bf Sources of photons}}
\label{sec:sources}

\subsection{Terminology}

\begin{center}
\begin{figure}[htbp]
  \includegraphics[width=0.79\linewidth]{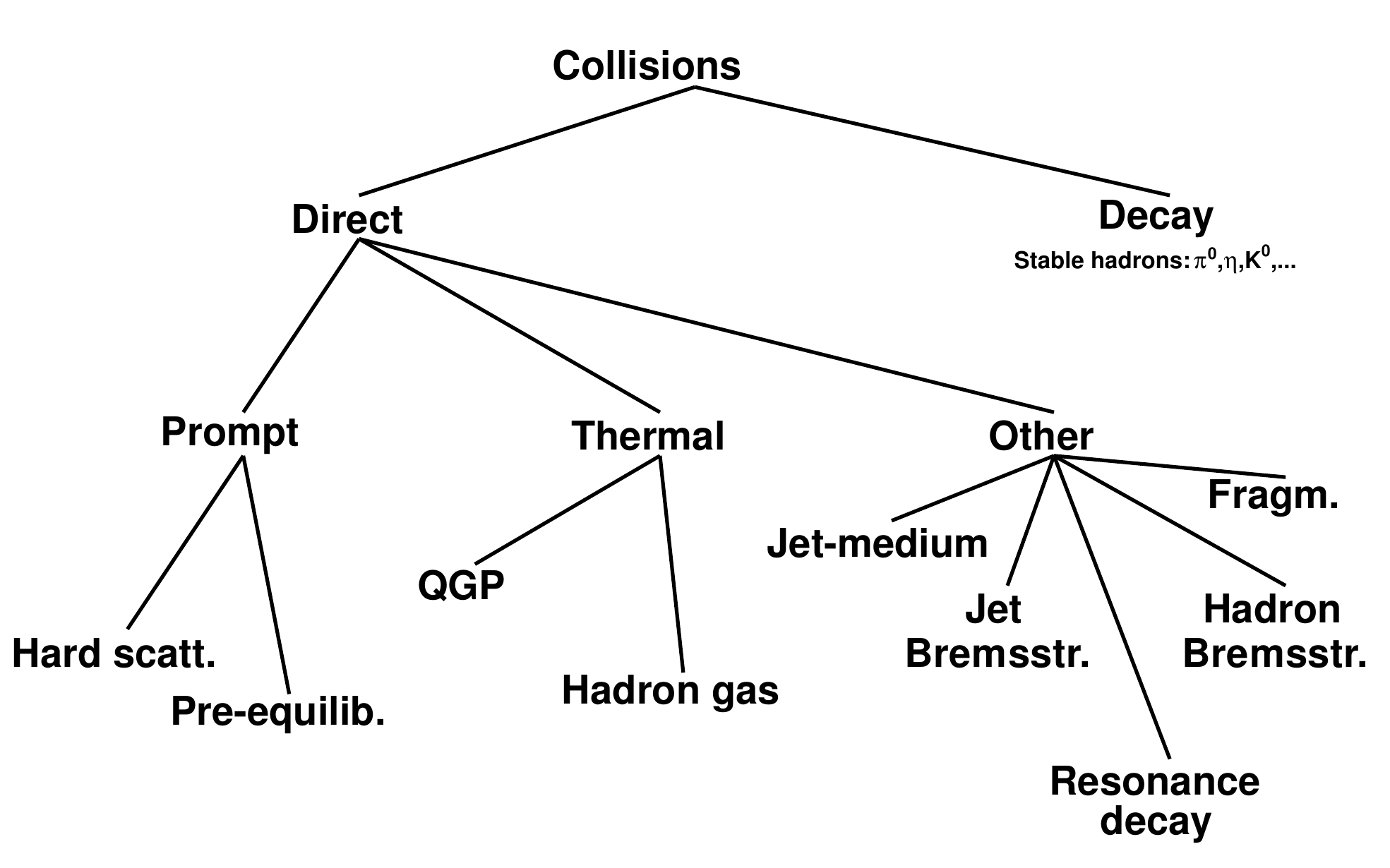}
  \caption{Prevalent terminology, referring to the sources of real
  photons in heavy ion physics 
  }
    \label{fig:nomenclature}
\end{figure}
\end{center}

The prevalent (although not completely unique) terminology of real
photons in heavy ion physics, referring to their sources, is shown 
in Fig.~\ref{fig:nomenclature}.  The two main categories are {\it direct}
and {\it decay} photons.  Decay photons come from electromagnetic
decays of long lifetime final state hadrons\footnote{Long lifetime
  means that it significantly exceeds 10-100\,fm/$c$ after which the
  colliding system is completely decoupled.
} and as such are extremely valuable -- sufficient to say that \piz's
reconstructed from $\piz \rightarrow \gamma \gamma$ provided the first
strong hint that QGP has been formed in relativistic heavy ion
collisions~\cite{Adcox:2001jp}. 
However, when measuring {\it direct} photons, \ie those that are 
produced any time during the collision proper, before the final
products completely decouple, the decay photons are a large background
and often the principal source of systematic uncertainties of the
direct photon measurement.

The subcategory {\it prompt}
as a rule includes photons from initial hard parton-parton
scattering (see for instance~\cite{Owens:1986mp}).
Usually other conjectured early sources, like photons from the 
``hot glue'', created before local thermalization~\cite{Shuryak:1992bt}
or the Glasma~\cite{McLerran:2014hza,Berges:2017eom}, photons from the
strong initial magnetic field~\cite{Tuchin:2010gx,Ayala:2017vex},
synchrotron radiation~\cite{Zakharov:2016kte}, 
 are also included here as {\it pre-equilibrium}.  

The subcategory {\it thermal} is widely used but problematic, even
theoretically, since thermalization is local at best, and the
instantaneous rates calculated with evolving temperature are in
addition subject to blue-shift due to medium expansion, so the inverse
slope parameter of \pt spectra does not directly represent any
``temperature''~\cite{Linnyk:2013hta,Shen:2013vja,Paquet:2017wji}.  
Depending on their origin (partonic or hadronic processes,
see Sec.~\ref{sec:processes}) they are called photons from the 
{\it  QGP} or the {\it hadron gas}~\cite{Kapusta:1991qp,Baier:1991em}.  
To complicate 
things, several other sources (like Bremsstrahlung) emit photons in the
thermal range, which are indistinguishable from truly
``thermal'' photons.  Therefore, in experimental papers ``thermal'' is
usually just a shorthand for photons in the few hundred MeV -- few GeV
\pt range, and we will put the word in quotation marks whenever
experimental results are discussed.

{\it Other} sources include photons from {\it jet fragmentation} in
vacuum, well known and measured in
\pp~\cite{Owens:1986mp,Gluck:1992zx,Bourhis:1997yu,Aurenche:2006vj,Klasen:2014xfa}.   
They should be distinguished
from {\it jet Bremsstrahlung} which occurs while the parton is still
traversing the (QGP) medium and losing energy in it~\cite{Owens:1986mp}.  
{\it Jet-medium} or {\it jet-photon conversion, jet-thermal} 
photons~\cite{Fries:2002kt,Renk:2013kya} are a special case
of the ultimate parton energy loss where a high \pt quark collides
with a thermal parton and transfers all its momentum to a photon
flying out in the same direction (see~\ref{sec:aahigh}).  
{\it Hadron  Bremsstrahlung} happening in the hadron
gas is yet another source of photons~\cite{Haglin:2003sh,Liu:2007zzw}).


\vspace{0.1in}
{\it Words of caution.} \\
\vspace{0.1in}
There are many sources of direct photons that are
hard or impossible to disentangle experimentally.  This often leads to
some misunderstanding when comparing data to model calculations.
For instance, in the high \pt region (above 4-5\,\gevc), dominated by
hard scattering, experiments often publish results on {\it isolated}
photons even in \AA collisions~\cite{Aad:2015lcb}, using well-defined
isolation criteria.  These results are then compared to perturbative
QCD (pQCD) calculations, but the comparison is only valid if the same
isolation criteria are applied as in the data.  In case of \pp this is
relatively straightforward but in \AA the underlying event has to be
properly simulated, too -- a very non-trivial task.
Also, in \AA ``jet-conversion'' photons (from the interaction of a
hard-scattered fast parton with the medium) are
an additional source of isolated photons in the experiment, but
seldom included in theory calculations.  

Even the distinction between direct and decay photons can become
problematic.  Short-lived {\it resonances}, like 
$\omega, \phi, a_1$ are sources of decay 
photons~\cite{Bratkovskaya:2008iq}, but rarely if ever are actually
subtracted by the experiments from the inclusive photon yields (not
the least because the parent distributions are usually not or poorly
known.  Typically only \piz and $\eta$ decays are considered and the
effect of all other hadron decays included in the systematic
uncertainties). While raising this issue may sound somewhat pedantic,
we should point out that at some point for instance the $a_1$ has been 
predicted to be a major source of photons~\cite{Xiong:1992ui}.

\subsection{The fundamental processes to produce direct photons}
\label{sec:processes}

In this section we review the fundamental sources that were believed
for a long time to be the main sources of photons in relativistic
heavy ion collisions.  More ``exotic'' mechanisms will be described in
the context of the ``direct photon puzzle'' (see Sec.~\ref{sec:puzzle}).

\subsubsection{Partonic processes in hadron-hadron collisions} 
\label{sec:partonic}

\begin{center}
\begin{figure}[htbp]
  \includegraphics[width=0.79\linewidth]{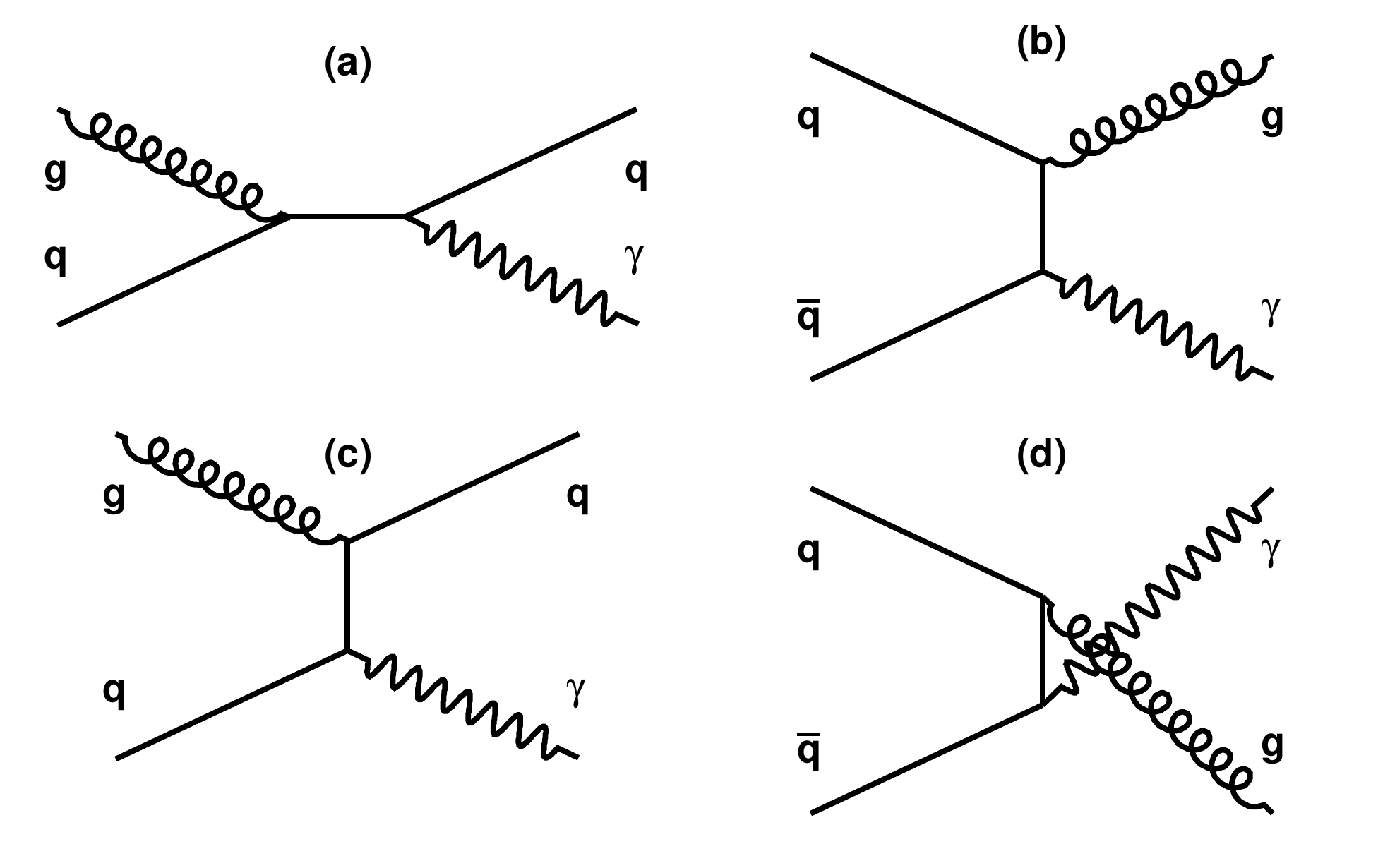}
  \caption{Leading order Feynman diagrams contributing to direct
  photon production.  (a) and (c): $s$ and $u$ channel quark-gluon Compton
  scattering.  (b) and (d): $t$ and $u$ channel quark-antiquark
  annihilation. 
  }
    \label{fig:partongraph}
\end{figure}
\end{center}

To leading order there are two types of partonic processes that
produce photons: quark-gluon Compton-scattering and quark-antiquark
annihilation (see Fig.~\ref{fig:partongraph}).  

For massive quarks and using the Mandelstam variables 
$s=(p_g + p_q)^2$, $t=(p_g - p_{\gamma})^2$ and 
$u=(p_q - p_{\gamma})^2$ the cross section for the 
$gq\rightarrow\gamma q$ Compton process is~\cite{wong:1994}

\begin{equation}
\frac{d\sigma}{dt}(gq\rightarrow\gamma q) =
\Big(\frac{e_q}{e}\Big)^2 \frac{8\pi\alpha_s\alpha_{em}}{(s-m^2)^2}
\Big\{
\Big(\frac{m^2}{s-m^2}+\frac{m^2}{u-m^2}\Big)^2 +
\Big(\frac{m^2}{s-m^2}+\frac{m^2}{u-m^2}\Big) -
\frac{1}{4}\Big(\frac{s-m^2}{u-m^2}+\frac{u-m^2}{s-m^2}\Big)
\Big\}
\label{eq:wongcompton}
\end{equation}

\noindent
and with $s=(p_q + p_{\bar{q}})^2$, $t=(p_q - p_{\gamma})^2$ and 
$u=(p_{\bar{q}} - p_{\gamma})^2$ the cross section for the 
$q\bar{q}\rightarrow\gamma g$ annihilation process is

\begin{equation}
\frac{d\sigma}{dt}(q\bar{q}\rightarrow\gamma g) =
\Big(\frac{e_q}{e}\Big)^2 \frac{8\pi\alpha_s\alpha_{em}}{s(s-4m^2)}
\Big\{
\Big(\frac{m^2}{t-m^2}+\frac{m^2}{u-m^2}\Big)^2 +
\Big(\frac{m^2}{t-m^2}+\frac{m^2}{u-m^2}\Big) -
\frac{1}{4}\Big(\frac{t-m^2}{u-m^2}+\frac{u-m^2}{t-m^2}\Big)
\Big\}
\label{eq:wongannihilation}
\end{equation}

These are the cross-sections of the lowest order photon-producing 
partonic $2\rightarrow 2$ processes (for higher order corrections in
hadron-hadron collisions see for instance~\cite{Aurenche:2006vj}).  In
order to calculate the photon rates the initial parton distribution 
functions (PDF) have to be known.  At higher order the  
fragmentation functions (FF) of the outgoing partons into photons are
also needed~\cite{Gluck:1992zx,Bourhis:1997yu,Aurenche:2006vj,Klasen:2014xfa}.
The FFs are defined as  $D^{\gamma}_{q,g}(z,Q^2)$ providing the
probability that a quark or gluon fragments into a photon carrying the
fraction $z$ of the original parton momentum.  (For a recent review 
see~\cite{Metz:2016swz}.)

For high momentum transfer $Q^2$ (hard scattering) {\it factorization}
holds~\cite{Owens:1986mp,Collins:1989gx,Brock:1993sz}, 
meaning that the cross-section of an
$AB\rightarrow\gamma A$ process is the incoherent sum of the
cross-sections $\sigma_{ij\rightarrow kl}$ of all contributing
constituent scattering processes, each convolved in the available
phase-space with the respective initial parton distributions (PDFs),
and, if relevant, with the final state parton fragmentation
functions (FFs).  Schematically, for an inclusive hadron production in \pp

\begin{equation}
\sigma (pp \rightarrow hX) = \hat{\sigma} \otimes PDF \otimes PDF
\otimes FF
\end{equation}

The PDFs and FFs cannot be calculated using
perturbation theory, but they are {\it universal} and can be obtained
from data for various types of well-controlled (\eg. $e^+e^-$) hard
processes~\cite{Owens:1986mp}.  The cross-sections 
$\sigma_{ij\rightarrow kl}$ can be calculated in pQCD.

For instance, the single inclusive photon cross-section in hadron-hadron
collisions ($h_1h_2 \rightarrow \gamma X$) to lowest
order will have the form~\cite{Aurenche:1983ws}

\begin{equation}
\frac{d\sigma}{dyd^2p_T} =
{\displaystyle\sum_{i,j}}\frac{1}{\pi}{\displaystyle \int}
dx_1F_{h_1,i}(x_1,Q^2_s)dx_2F_{h_2,j}(x_2,Q^2_s)
\frac{1}{\hat{s}}
\Big( \frac{1}{v} \frac{d\sigma_{i,j}}{dv}(\hat{s}v)
\delta(1-w) + \mathrm{corr.} \Big)
\end{equation}

\noindent
with $v=1-x_2(p_T/\sqrt{\hat{s}})e^{-y}$,
$w=(1/vx_1)(p_T/\sqrt{\hat{s}})e^{y}$, the indices $i,j$ run over the
quarks, antiquarks and gluons of the initial hadrons, and 
$F_{h_{1,i}},F_{h_{2,j}}$ are the respective parton distribution functions,
while the next-to-leading order corrections are omitted.
For fragmentation photons (an outgoing parton $k$ emits a photon) an
additional convolution is necessary with the photon fragmentation
function $D_{k,\gamma}$ 

\begin{equation}
\frac{d\sigma}{dyd^2p_T} =
{\displaystyle\sum_{i,j,k}}\frac{1}{\pi}{\displaystyle \int}
dx_1F_{h_1,i}(x_1,Q^2_s)dx_2F_{h_2,j}(x_2,Q^2_s)
\frac{dx_3}{x_3}D_{k,\gamma}(x_3,Q^2_d)
\frac{1}{\hat{s}}
\frac{1}{v} \frac{d\sigma_{i,j}}{dv}(\hat{s}v)\delta(1-w) 
\end{equation}

In lowest order only the Compton ($qg\rightarrow q\gamma$) and
annihilation ($q\bar{q} \rightarrow q\gamma$) processes contribute to
prompt photon production, the latter being suppressed in \pp due to
the lack of valence antiquarks.  This provides (at least in principle)
a way to disentangle the two processes by measuring the cross-section
differences $\sigma(p\bar{p} \rightarrow \gamma X) - 
\sigma(pp \rightarrow \gamma X)$ which provides direct access to the
valence-quark and gluon PDFs~\cite{Aurenche:1988vi}.

Some higher order processes -- like Bremsstrahlung or
fragmentation -- can contribute to the partonic rates at a strength
comparable to the fundamental Compton-scattering and annihilation.
The calculations are quite complex, and usually implemented in Monte
Carlo programs, like JETPHOX~\cite{jetphox}, but they are reasonably
well understood and are consistent with the available
data~\cite{Aurenche:2006vj}. 

\subsubsection{Radiation from  the QGP} 
\label{sec:fromqgp}

When the QGP, a medium of deconfined quarks and gluons is formed in a
heavy ion collision, it will also radiate photons.  The basic partonic 
interactions producing photons are still the same as discussed above, 
but the role of the traditional PDFs (obtained for partons 
{\it  confined} in hadrons in vacuum and fixed long time before the
collision)  is taken over by the dynamically 
evolving distribution of {\it deconfined}, interacting partons, mostly
produced in the collision itself, and collectively forming the
``medium''.  Under these circumstances there are two complementary
ways to handle the problem of having parton distributions that now 
strongly depend (including their number!) on space-time, and keep 
changing as the medium evolves. The first is to prepare the initial state
of the partons then follow their paths and interactions one-by-one
({\it microscopic transport}), circumventing the concept of the
medium.  The second is to model the medium as a statistical ensemble,
describe its properties and evolution, like that of a gas or fluid
(thermo- or hydrodynamical system)\footnote{There are also hybrid
  techniques like {\it coarse graining} introduced
  in~\cite{Huovinen:2002im}. 
}.

The very first attempt to predict
radiation from the QGP~\cite{Shuryak:1978ij} starts with an ensemble of
quarks and gluons assumed to be already in local thermal equilibrium 
with initial temperature $T_i$ ``at which the thermodynamical
description becomes reasonable'', and continues up to the final 
temperature $T_f$ ``where the system breaks into secondaries''.
The production cross-section of a (penetrating) particle ``a'' is then
given by the integral over the space-time plasma region

\begin{equation}
\sigma_a = \sigma_{in} \int_{T_i}^{T_f} W_a(T)\Phi(T)dT
\end{equation}

\noindent
where $\sigma_{in}$ is the total inelastic cross-section, $W_a$ the
production rate per unit volume of the plasma, $\Phi(T)$ is a
temperature-dependent weighting factor, estimated as
$\Phi(T)=A(s)T^{-7}$ for various expansion (Feynman scaling and Landau
hydrodynamics) scenarios.  Note that
$\Phi(T)$ strongly favors small $T$, \ie production in the later
stages.  In specific, the author~\cite{Shuryak:1978ij} finds
for the \pt distribution of direct leptons from the elementary 
$gq(\bar{q}) \rightarrow l^+l^-$ processes

\begin{equation}
d\sigma/dp_T^2 =
(\alpha^2 \sigma_{in} A(s)/\sqrt{2\pi}18p_T^4) \Gamma(7/2,M/T_i)
\end{equation}

\noindent
where $\sigma_{in}$ is the total inelastic cross-section, $A(s)$ is a slowly
varying function of the c.m.s. energy (like $\ln{s}$ or $\sqrt{s}$)
and $\Gamma(\alpha,x)$ is the incomplete gamma function.
The direct photon yield is a factor of
$20\alpha_s\ln{(p_T/m_0)}/3\alpha \sim 600$
higher~\cite{Shuryak:1978ij}.  In addition to its historic value, it is
interesting to note that this early calculation claims that photons of
quite high \pt (3-5\,\gevc) are produced predominantly not by hard
scattering but in the plasma, and late (at lower $T$).

A more formal treatment is based on the imaginary part of the photon
self-energy $\Pi_{\mu\nu}$ in the medium which is related to the 
escape rate of photons from the 
medium~\cite{Weldon:1983jn,McLerran:1984ay,Gale:1990pn}.  For small
thermal systems (compared to the photon mean free path) the emission
rate R of photons, valid to all orders in $\alpha_s$ and first order
in $\alpha_{em}$ is~\cite{Kapusta:1991qp}

\begin{equation}
E\frac{dR}{d^3p} = \frac{-2}{(2\pi)^3} \mathrm{Im} \Pi^{R,\mu}_{\mu}
\frac{1}{e^{E/T}-1}
\end{equation}

\begin{center}
\begin{figure}[htbp]
  \includegraphics[width=0.79\linewidth]{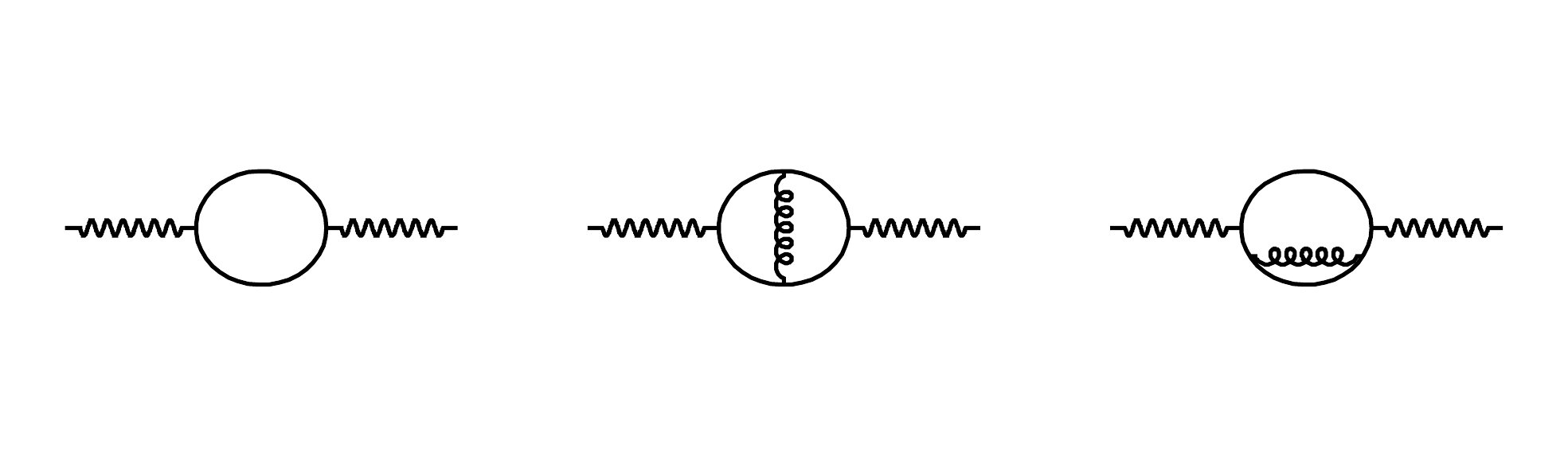}
  \caption{One- and two-loop contributions to the photon self-energy
  in QCD
  }
    \label{fig:selfenergy}
\end{figure}
\end{center}

The one- and two-loop contributions to $\Pi_{\mu,\nu}$ are shown in
Fig.~\ref{fig:selfenergy}, and $\mathrm{Im} \Pi_{\mu,\nu}$
obtained by cutting the two-loop diagrams\footnote{Cutting the
  one-loop diagram leads to $q\bar{q}\rightarrow \gamma$ prohibited
  for a photon on mass shell.
}, which then reproduces the basic graphs in
Fig.~\ref{fig:partongraph}.  For vanishing quark masses the
differential cross-sections go to infinity as $t\rightarrow 0$ and/or
$u\rightarrow 0$  (see Eqs.~\ref{eq:wongcompton},\ref{eq:wongannihilation}).
This infrared divergence can be regulated using the resummation
technique of Braaten and Pisarski~\cite{Braaten:1989mz}.  Assuming a
thermalized QGP the first compact formula for total photon radiation
rate, including hard momentum transfers (Compton scattering,
annihilation) and soft momentum transfers (regularized by the
Braaten-Pisarski method), at
$E\gg T$ was given in~\cite{Kapusta:1991qp} as

\begin{equation}
E \frac{dR}{d^3p} = \frac{5}{9} \frac{\alpha \alpha_s}{2\pi^2}
T^2 e^{-E/T} \ln\Big(\frac{2.912}{g^2}\frac{E}{T}\Big)
\label{eq:kapustaqcd}
\end{equation}

In the following decade, in part inspired by the availability of new
data from the SPS, the calculations were extended to include
Bremsstrahlung in the plasma~\cite{Aurenche:1998nw}, the effect of soft
gluons~\cite{Aurenche:1999tq},  and the Landau-Pomeranchuk-Migdal
effect~\cite{Aurenche:2000gf}.  A comprehensive set of rates
contributing in leading order to photon emission from the QGP were
given in~\cite{Arnold:2001ms}; those ``AMY rates'' are often used even
today in hydro codes (see for instance~\cite{Paquet:2015lta,Basar:2017ocn}).
It should be noted, however, that the AMY rates (and most other
explicit calculations) assume a weakly coupled ultrarelativistic
plasma~\cite{Arnold:2001ms}, calculating photon emission one by one for
various leading order processes (diagrammatic approach).  
Note that these are only the {\it rates}, which then have to be
convolved with some description of the space-time evolution of the
QGP.  The first attempt to describe the QGP evolution by longitudinal (1+1D)
hydrodynamics was made by Bjorken~\cite{Bjorken:1982qr}, who in turn
relied heavily on Landau's seminal work on the hydrodynamic theory of
multiple particle production~\cite{Belenkij:1956cd}, largely inspired
by Fermi~\cite{Fermi:1951zz}.  Deep are the roots...  We will discuss
some more recent models in Sec.~\ref{sec:puzzle}.

\begin{center}
\begin{figure}[htbp]
  \includegraphics[width=0.79\linewidth]{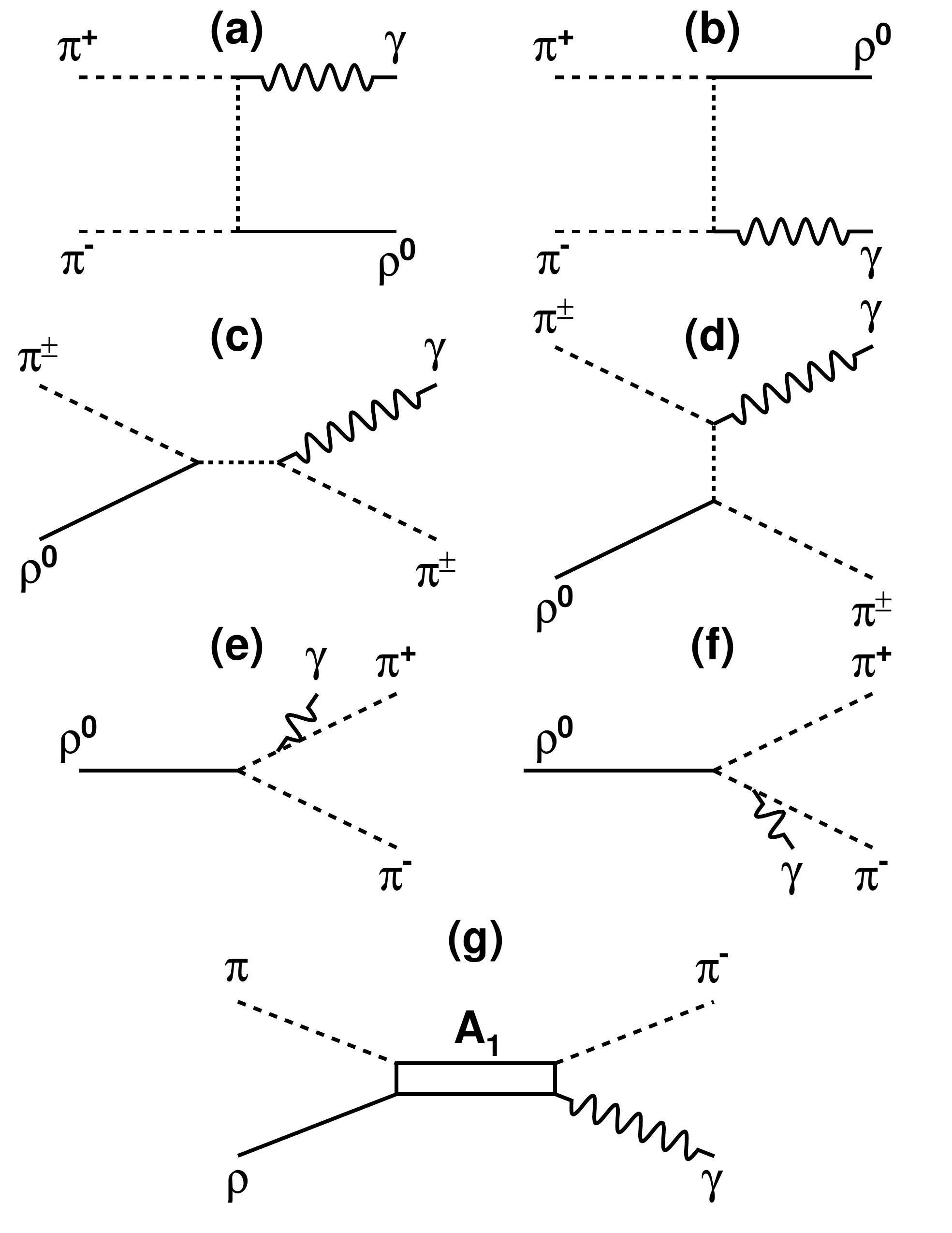}
  \caption{Panels (a)-(f): some photon producing hadronic reactions
    arising from the imaginary part of the photon self-energy
    involving charged pions and neutral $\rho$ mesons.  Above
    $E_{\gamma}>0.7$\gev the dominant process is    (c)~\cite{Kapusta:1991qp}.    
    Panel (g): contribution from the  $a_1$ resonance that according
    to~\cite{Xiong:1992ui} ``outshines'' process (c).
  }
    \label{fig:hadrongraph}
\end{figure}
\end{center}

\subsubsection{Radiation from the hadron gas}
\label{sec:fromhg}

Radiation from the hadron gas can be calculated in similar ways 
as the radiation from the QGP  -- from kinetic theory or by loop 
expansion of the photon self-energy -- but the processes are
different. Some basic diagrams are shown in Fig.~\ref{fig:hadrongraph}.   
The first estimate of the photon emission rate from hot hadronic gas
at a fixed temperature $T$  was made in~\cite{Kapusta:1991qp}
calculating processes (a)-(f) in Fig.~\ref{fig:hadrongraph} 
(all except the $a_1$ channel~\cite{Xiong:1992ui}), and the authors
concluded that {\it at the same temperature} the QGP and the hadron
gas radiate photons in the 1-3\,\gevc \pt range at the same rate,
summarized in the oft-quoted sentence: ``The hadron gas shines just as
brightly as the quark-gluon plasma.''  It was assumed that the phase
transition is first order and the system will spend a long time at the
transition temperature $T_c$ \ie the thermal photon spectrum would
allow to measure $T_c$.  

The concept of 
{\it quark-hadron  duality}~\cite{rapp:1999ej} , inspired by dilepton 
production at the SPS (CERES and NA50) and rooted in 
{\it chiral symmetry restoration}
initially led to similar conclusions\footnote{In its original form
  duality meant that the hadronic and quark-gluon degrees of freedom
  are equally well describing the emissivity of strongly interacting
  matter if the temperature is the same. 
}.  
Ironically, this has cast early
on some doubts on the usefulness of  ``thermal'' photons to diagnose
the QGP~\cite{Stankus:2005eq}.  However, initially the expansion
of the hot matter (QGP or hadronic) was often not taken into account,
despite the pioneering work by Bjorken in 1983~\cite{Bjorken:1982qr}
pointing out the applicability of hydrodynamics to the temporal
evolution of the hadronic matter in highly relativistic
nucleus-nucleus collisions\footnote{And even in~\cite{Bjorken:1982qr}
  only longitudinal flow was discussed, the transverse motion of the
  fluid, which became an all-important issue in understanding
  ``thermal'' photon production, was not addressed.
}.  
In~\cite{Song:1993ae} the
temperature-dependent emission rates for ``thermal'' photons from
hadronic matter are described using an effective chiral Lagrangian 
($\pi\rho \rightarrow \pi\gamma$,  $\pi\pi \rightarrow \rho\gamma$, 
and $\rho \rightarrow \pi\pi\gamma$, allowing intermediate $a_1$
states).  Substantial enhancement of the rates in case of finite pion
chemical potential has been pointed out in~\cite{Steele:1996su}.
Calculations in the Hadron String Dynamics framework (see also
Sec.~\ref{sec:puzzle}) have been presented in~\cite{Cassing:1999es}.
Results with another microscopic transport model (uRQMD) are shown
in~\cite{Bass:2000ib}.

The role of radial expansion ($v_0$) in modifying the photon spectra
was first emphasized in the attempts to explain the WA80/WA98
data~\cite{Dumitru:1994vc,Cleymans:1996xg}
and the effective temperature \Teff was first considered
in~\cite{Gallmeister:2000si} where $v_0=0.3$ described the low end of
the available WA98 data~\cite{Aggarwal:2000th} well.  A major
update of thermal photon production in a radially expanding
fireball~\cite{Turbide:2003si} introduced strangeness-bearing channels
but found the $\pi\rho a_1$ less important than earlier thought.  A
year later this work was followed by a calculation of the high \pt
\piz and direct photons~\cite{Turbide:2005fk} 
simultaneously\footnote{It is very important that the \piz and photon
  production will be accounted for in the same theoretical framework --
  unfortunately few publications satisfy this requirement.  A good
  early example from 1995 is the three-fluid model in~\cite{Dumitru:1994vc}.
},
and, in parallel, first predictions of azimuthal asymmetries (\vtwo, 
or {\it elliptic} flow) of direct photons have been
made~\cite{Turbide:2005bz,Turbide:2007mi}, including {\it negative}
\vtwo for high \pt photons.

\begin{center}
\begin{figure}[htbp]
  \includegraphics[width=0.45\linewidth]{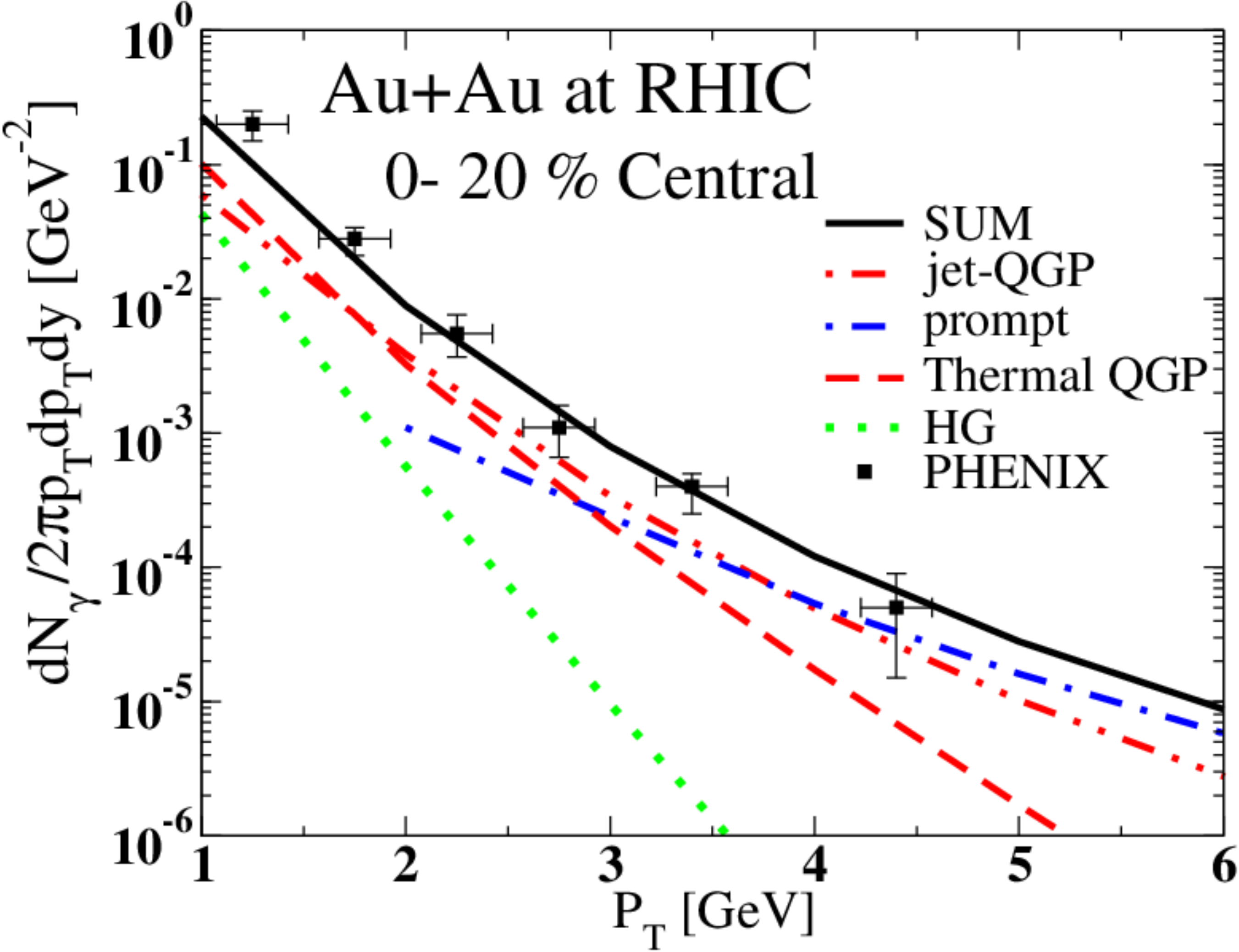}
  \includegraphics[width=0.45\linewidth]{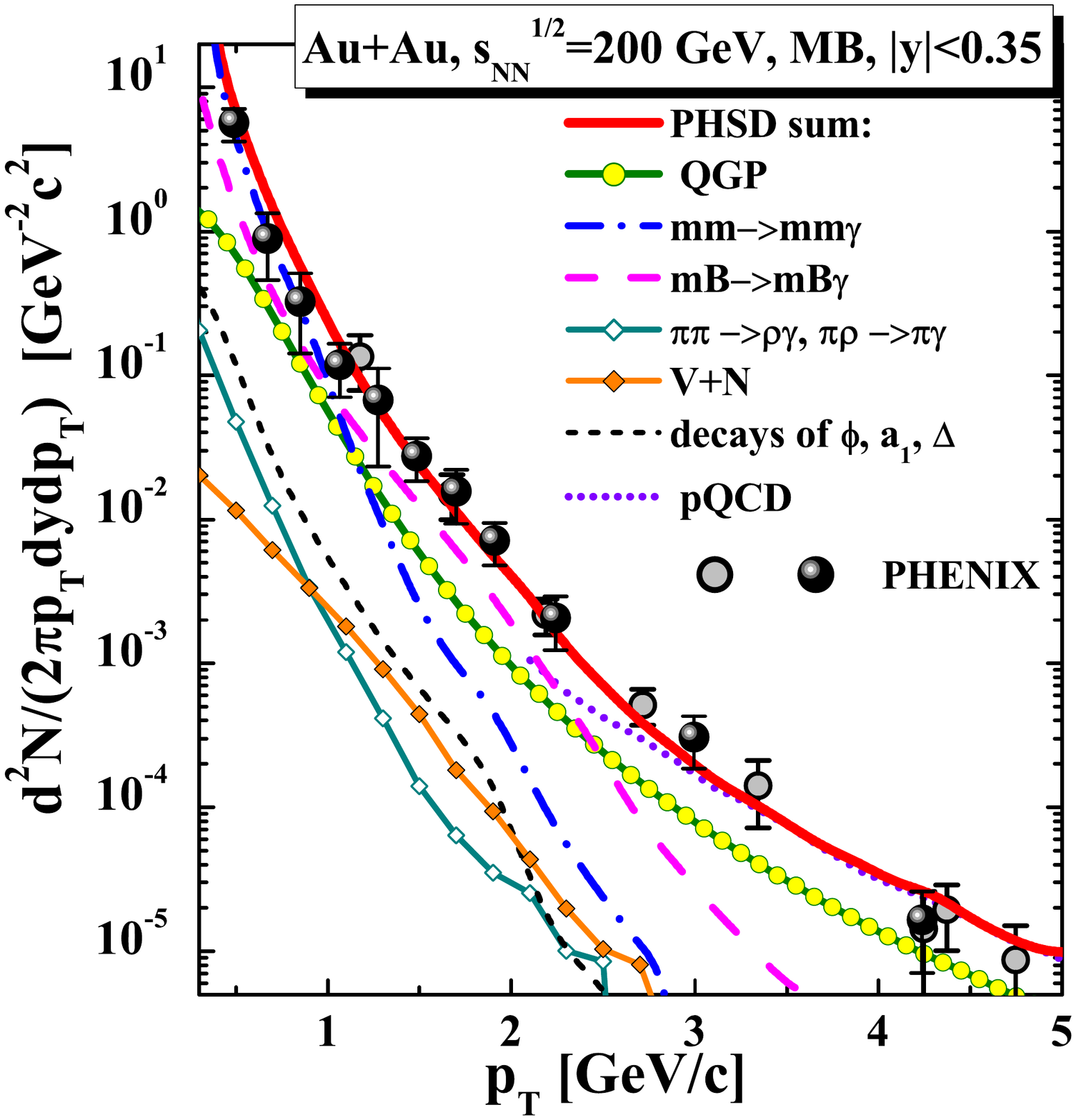}
  \caption{(Left panel) Contributions from different sources and total
    yield of photons in central \auau collisions
    at RHIC, according to a 2008 hydrodynamic calculation.
    (Figure taken from~\cite{Turbide:2007mi}.)
    (Right panel) Contributions from the QGP and various hadronic
    sources and the total yield of photons in \auau collisions at
    RHIC, according to a 2015 PHSD calculation.
    (Figure taken from~\cite{Linnyk:2015tha}.)
  }
    \label{fig:turbide_linnyk}
\end{figure}
\end{center}

Unlike Eq.~\ref{eq:kapustaqcd} there is no simple ``pocket formula''
for the radiation from the hadron gas.  Schematically the total cross
section for a particular process like $\pi\pi\rightarrow\rho\gamma$
reads~\cite{Linnyk:2015rco} 

\begin{equation}
\sigma_{\pi\pi\rightarrow\rho\gamma}(s,\rho_N) =
\int_{M_{min}}^{M_{max}}dM \sigma^0_{\pi\pi\rightarrow\rho\gamma}(s,M)A(M,\rho_N)P(s)
\label{eq:linnyk146}
\end{equation}

\noindent
\ie folding the vacuum cross section
$\sigma^0_{\pi\pi\rightarrow\rho\gamma}(s,M)$ with the in-medium
spectral function $A(M,\rho_N)$, where $\rho_N$ is the nuclear
density, and $P(s)$ accounts for the fraction
of the available part of the full $\rho$ spectral function available
in the phase space limited by $\sqrt{s}$.  Similar schemes apply for
other hadronic processes, but while the $\rho$ spectral function is
relatively well known, others, like $a_1$ are not.  Calculations of
the meson-meson and meson-baryon Bremsstrahlung are equally
involved~\cite{Liu:2007zzw}.  The complexity of calculating the direct
photon yield, particularly the hadronic part is well illustrated in
Fig.~\ref{fig:turbide_linnyk}, where the contributions from different
sources are shown from a 2008 hydrodynamic and a 2015 microscopic
transport calculation.

\subsection{Who outshines whom? From pre-equilibrium to QGP to hadron gas}
\label{sec:evolution}

Like with any new phenomenon, the hunt for the QGP started with
theorizing about the decisive signal, whose presence (or
disappearance) would unambiguously prove that this new state of matter
has been formed.  Historically the earliest suggestions were
strangeness enhancement\footnote{See a lively historic review
  in~\cite{Rafelski:2015cxa}.
}, radiation from the QGP~\cite{Shuryak:1978ij}, and $J/\Psi$
suppression~\cite{Matsui:1986dk}.  

First estimates of direct photon radiation from relativistic heavy ion
collisions~\cite{Shuryak:1978ij,Kajantie:1981wg,Halzen:1981kz} assumed
that the dominant source will be a thermally equilibrated QGP, and the
shape of the \pt spectrum would reflect its properties.  Then in 1991
a very influential paper~\cite{Kapusta:1991qp} came to the conclusion
that ``the hadron gas shines just as brightly as the quark-gluon
plasma''.  The primacy of QGP radiation has been questioned.  Note
that in this calculation the space-time evolution of the system was
not yet considered. 
Due to technical and conceptual
difficulties it took some time before the dynamical expansion of the 
radiating system became standard part of photon yield calculations.

Around 2000, based on \pbpb direct photon 
results from WA98~\cite{Aggarwal:2000th},
quark-hadron duality and the space-time evolution of the fireball
modeled in Bjorken hydrodynamics~\cite{Bjorken:1982qr} the
calculations became more sophisticated and the predictions more
differential (in \pt).  For instance, in~\cite{Steffen:2001pv} it has
been predicted that at RHIC and LHC the QGP will ``outshine'' the hot
hadron gas above \pt of 3.2 and 2.5\,\gevc, respectively, while below
that (and at the SPS in the entire spectrum) radiation from the hadron
gas will dominate.  A somewhat different conclusion (below 1\,\gevc
quark matter still dominates emissions even at SPS) has been reached 
in~\cite{Srivastava:1999ct,Srivastava:1999bt}.
The problem is quite complex\footnote{Not the least due to the
incomplete knowledge of the equation of state, governing the
space-time-temperature evolution of the system.}, and data are often
only of limited help, since 
direct photon yields are a convolution
of many terms that are hard to ``factorize'' experimentally. 

With the observation of an unexpectedly large direct photon flow
\vtwo~\cite{Adare:2011zr} (essentially the same magnitude 
as hadron \vtwo) the
situation became even more complicated: models had to simultaneously
explain large yields and \vtwo.  Large yields are usually easier to
get early (at higher temperatures, \ie from the QGP phase), but large \vtwo
usually indicates late production (hadronic phase).  However, this is not
the only possibility: in the past few years some mechanisms producing
photons with large asymmetries very early, pre-equilibrium have also
been proposed.  We will discuss the current models in
Sec.~\ref{sec:puzzle}, but it is safe to say that for the time being
we are not sure which stage of the evolution dominates the emission of
low \pt photons, but most likely it is not the QGP.  

\section{\bf Experimental techniques}
\label{sec:experimental}

Direct photon measurements are ``notoriously difficult'' due to a
combination of low rates, large physics background (photons from
hadron decays), occasional instrumental background (photons from
secondary interactions with detector components) and issues with
photon identification (contamination from misidentified hadrons
at low \pt and hadron decay photon pairs merged and misidentified as a
single photon at high \pt).  

There are two fundamentally different techniques to measure real
photons: {\it electromagnetic calorimetry} where the energy of the
photon itself is measured directly, and {\it conversion}, where
the photon converts into an $e^{\pm}$ pair\footnote{Preferably, but
  not necessarily at a known place in the detector.
}; the original photon direction and energy are then reconstructed
from the measured $e^{\pm}$ pair momenta.

\subsection{Real photons (calorimetry)}
\label{sec:expreal}

Electromagnetic calorimeters are detectors in which the impinging
photons and electrons lose (ideally) all their energy, part of which
is converted into some detectable signal.  The energy loss mechanism
is the production of secondary particles in an alternating sequence 
of pair creation ($\gamma\rightarrow e^{+}e^{-}$) and Bremsstrahlung
($e^{\pm}\rightarrow e^{\pm}\gamma$), until the energy of the
secondaries fall below the critical energy~\cite{Fabjan:1982ev}.  The 
set of all secondaries is the (electromagnetic) {\it shower}, a
statistical object.  The average depth of the shower is proportional
to $\ln(E)$, and is usually measured in units of 
{\it radiation  length} $X_0$ (related to the mean free path of the
shower particles in the material), while the transverse size is
characterized by the {\it Moliere-radius} $\rho_M$ (about 95\% of the
electromagnetic shower is contained laterally in a cylinder with radius
$2\rho_M$~\cite{Fabjan:1982ev}).  Part of the
deposited energy is converted into some detectable signal, like
Cerenkov or scintillation light, captured by some photosensitive
device, or the number of electrons at various depths, typically
measured by silicon detectors.  The transverse size of the detector
modules read out individually ({\it granularity}) is of the order of
$\rho_M$, optimizing the contradictory requirements of energy and
position resolution\footnote{In heavy ion physics calorimeters are
  usually designed such that a shower from a single impinging
  particle deposits energy in several neighboring modules
  (``cluster'').  Energy measurement is usually better, if the cluster
  consists of fewer modules, but impact position measurement improves
  with   the number of modules.
}.  
To prevent {\it shower leakage} at the far end
the depth of the modules should be at least 18-20$X_0$.

Calorimeters can be {\it homogeneous}, like scintillating crystals or
strong Cerenkov emitters; in these transparent detectors the entire
volume is {\it active}, because light is produced everywhere and most
of it can be collected and observed. In {\it sampling calorimeters}
highly absorbing {\it passive} regions (usually high $Z$ material)
absorb most of the energy without producing detectable signals.  These 
alternate with regions of active material, where the remaining energy
of the shower produces observable light or charge 
(``visible energy''). The ratio of visible to total energy is called
the {\it sampling fraction}.  Due to the statistical nature of shower
development the resolution of sampling calorimeters is inferior to
homogenous ones, but they can be much more compact and
cost-effective.  For good calorimeters the resolution is dominated by
the {\it statistical term}
$\sigma_E /E \propto A/\sqrt{E(GeV)}$, with $A$ in the few percent
range for homogenous and $\approx$8-15\% for sampling calorimeters.
Excellent reviews of calorimeter technologies, performance,
applications (and pitfalls!) can be found 
in~\cite{Fabjan:2003aq,Wigmans:2017vgs,Wigmans:2018fua}.

Photon reconstruction in electromagnetic calorimeters involves
identifying showers (clusters in the raw data) that are likely coming
from photons, rejecting hadrons, reconstructing the photon energy and
its impact point on the calorimeter surface.  An additional, extremely
important step is at higher energies to determine whether a single,
photon-like cluster comes indeed from a single photon, or two nearby
photons, \eg from a \piz decay (merging).  Hadron rejection is sometimes
aided by a thin, charge-sensitive device in front of the
calorimeter (charge veto, see for instance~\cite{Janssen:2000ac}), 
while the position measurement and the
resolution of two nearby photons can be enhanced by a high granularity
``pre-shower'' detector (see for instance~\cite{Gallinaro:2004tm}.  
The principal tool for photon identification
is the analysis of the shower shape (size, compactness, dispersion,
ellipticity, comparison to the predictions of a shower model \etc),
where the different characteristics are often combined
stochastically~\cite{Afanasiev:2012dg}.  -- Yet another way to distinguish
between single photons and merged decay photons of the same total
energy is  to use a longitudinally segmented calorimeter (see for
instance the  UA1 detector at CERN~\cite{Albajar:1988im}): the
penetration of the two smaller energy photons is shallower, so the
ratio of energy deposit in the first and second segment discriminates
between a single high energy and two lower energy, but merged
photons. 

As mentioned before, the direct photon signal has a large background
from hadron decay photons.  In a low multiplicity environment, like
\pp collisions, such decay photons can be {\it tagged} in each event
with  a reasonable efficiency by checking if it has an invariant mass
\mgg consistent with \piz if combined with {\it any} other photon in the
event.  In high multiplicity events, like \AA, such tagging isn't
possible, because the {\it combinatorial background} -- two, in
reality uncorrelated, photons having by accident \mgg consistent with
the \piz mass -- is too large.  In \AA the direct photon yield is 
usually obtained {\it statistically}, by subtracting the {\it estimated}
decay photon yield from the observed {\it inclusive} yield.  
The decay kinematics is known, but it is hard to overemphasize how
much in these type of measurements the accuracy of the final direct 
photon result depends on the knowledge of hadron yields, particularly
those of \piz and $\eta$.
Ideally those are measured in the same experiment, with the same
setup, to minimize systematic uncertainties from acceptance,
absolute calibration, and so on.  Note that the photon contribution from
other meson decays is usually small compared to other uncertainties
of the measurement.

Once the \piz and $\eta$ spectra are known, their decay photon
contribution in the detector has to be simulated (including the
acceptance and analysis cuts), then this simulated decay spectrum is
subtracted from the inclusive photon spectrum.  Finally, the
difference (inclusive - decay), \ie the direct photon spectrum has to be 
unfolded for detector resolution and other effects~\cite{Afanasiev:2012dg}.

\subsection{External and internal conversion}
\label{sec:expconv}

\begin{center}
\begin{figure}[htbp]
  \includegraphics[width=0.43\linewidth]{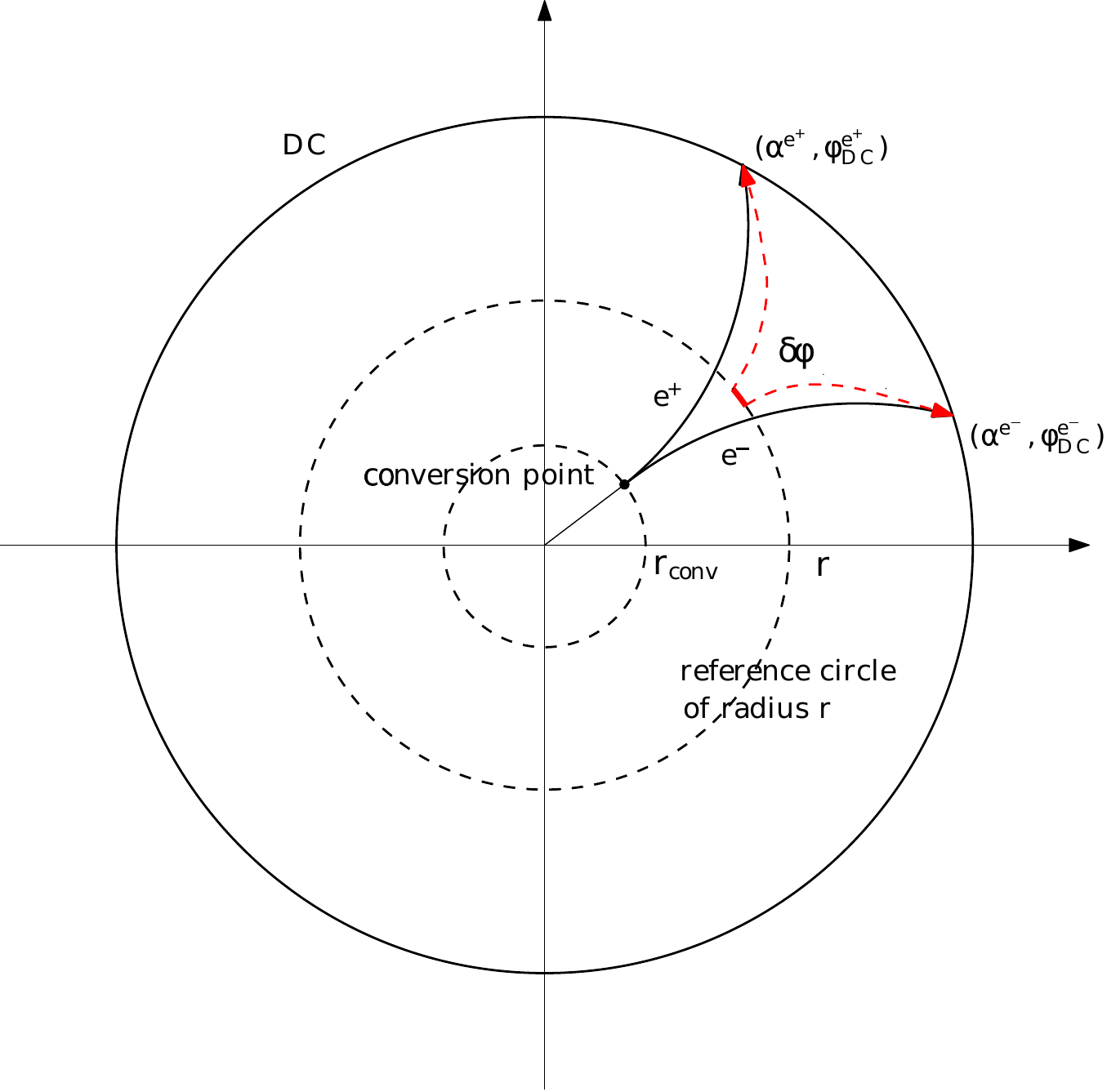}
  \includegraphics[width=0.48\linewidth]{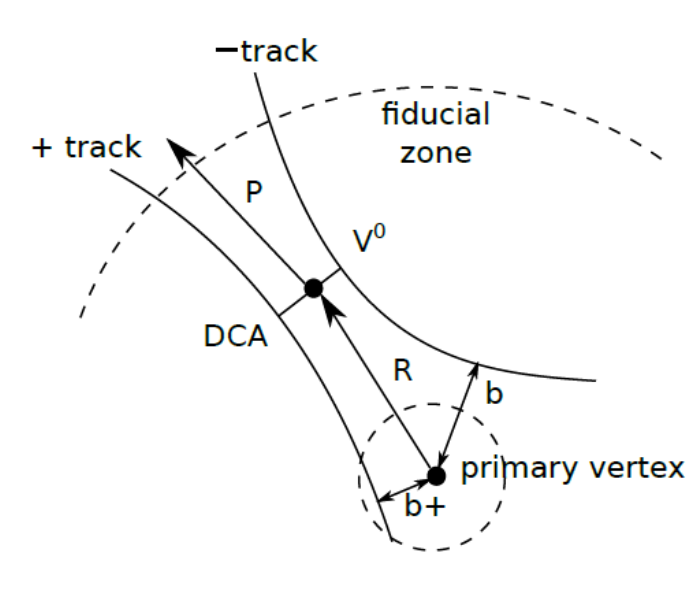}
  \caption{(Left)  $e^+e^-$ conversion pair reconstruction in PHENIX,
    based on the fact that the tracks cross at a point with known
    detector material and at that point their opening angle is zero.
    Electrons are identified via Cherenkov-radiation, energy deposit
    pattern and $E/p$ ratio in the calorimeter.
    (Figure courtesy of Wenqing Fan.)
    (Right) Conversion pair reconstruction in ALICE based on the
    distance of closest approach (DCA) far from the collision
    vertex.  Electrons are identified via $dE/dx$ in the TPC.
    (Figure courtesy of Friederike Bock.)
  }
    \label{fig:extconv}
\end{figure}
\end{center}

In high multiplicity heavy ion collisions calorimetry is not ideal
for photon measurements in the low \pt (less than 3-4\,\gevc)
range\footnote{The principal reasons are rapidly deteriorating resolution
  ($\propto 1/\sqrt{E_{\gamma}}$), increasing contamination from
  hadrons, including neutral ones like $n,\bar{n}$, and lack of
  direction which hinders rejection of instrumental background photons.
}.
Instead, photons that converted into an $e^{+}e^{-}$ pair in some (external)
detector material are reconstructed from the invariant mass \mee of
dielectrons.  Momentum resolution in tracking is usually much superior
to energy resolution in calorimeters, so the photon energy is measured
more precisely and the direction of the photon is reconstructed, too.
The method is described for the case where the conversion point
(radius) is assumed in~\cite{Adare:2014fwh}, or explicitely known from
secondary vertex reconstruction in~\cite{Adam:2015lda}, illustrated in
Fig.~\ref{fig:extconv}, right panel.  Hadron
contamination is typically very small (less than 1-5\% of the sample);
the main background is due to dielectrons from \piz Dalitz-decay.  
The drawback is small statistics, since detectors, for obvious
reasons, tend to put as little material as possible in the region
where high precision tracking is done, so the probability for real
photons to convert on such material is small.  This limits the high \pt
reach of the measurement typically to a few \gevc.  --  

Recently, another technique has been introduced that does not require 
{\it a priori} knowledge of the conversion point
(radius)~\cite{Fan:2017fuj}.  It relies on the fact that if a pair of
well-identified $e^+e^-$ tracks cross at a certain point, which is at
macroscopic distance from the vertex with some detector material
nearby {\it and} the opening angle of the tracks at this point is
zero, it is virtually certain that they come from a converted photon
(see Fig.~\ref{fig:extconv}, left panel).

\begin{center}
\begin{figure}[htbp]
  \includegraphics[width=0.6\linewidth]{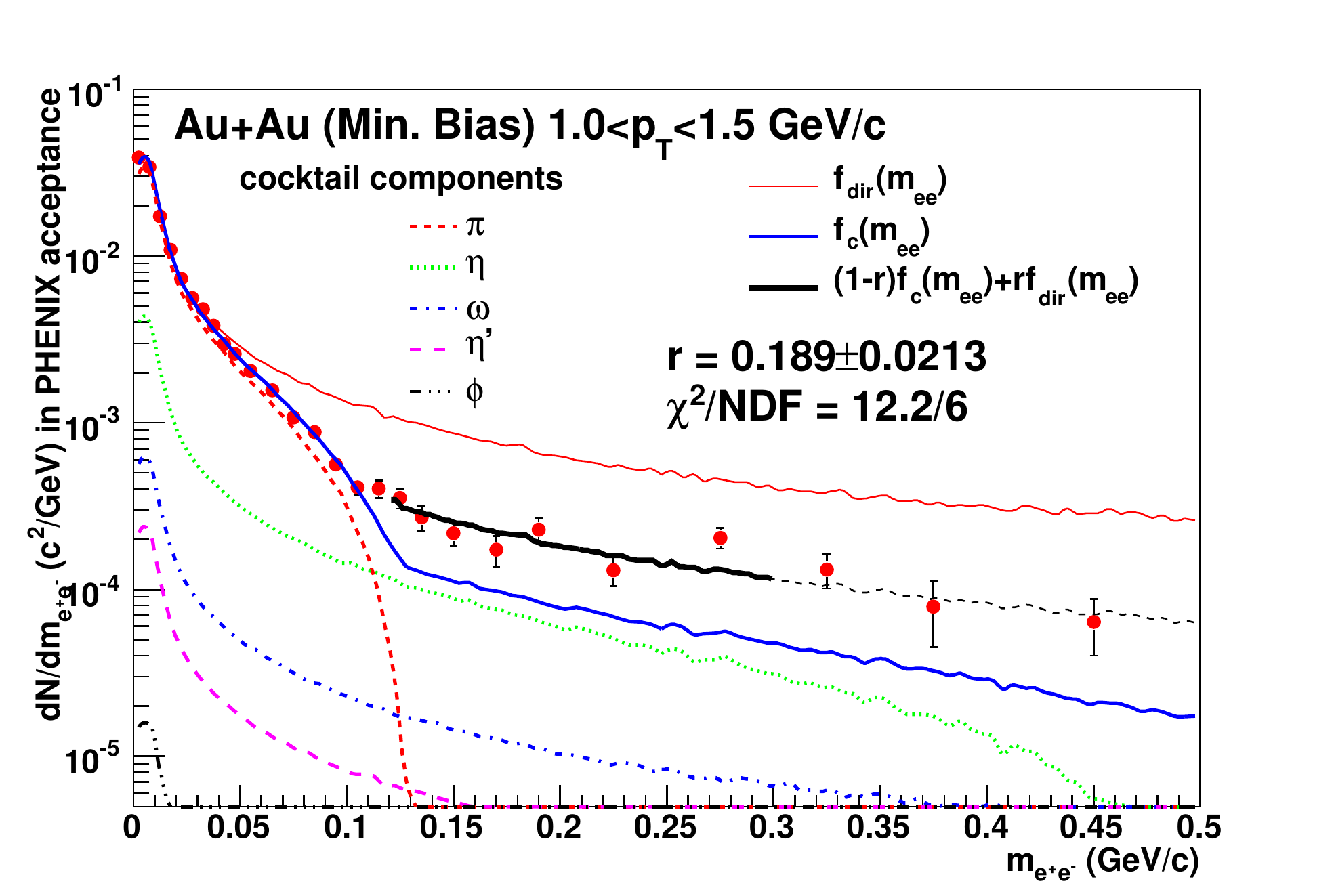}
  \caption{Electron pair mass distribution for minimum bias \auau
  collisions, $1.0<p_T<1.5$\gevc.  The red curve shows the expected
  shape from direct (virtual) photons ($f_{dir}(m)$), 
  the blue curve is the expected  shape from the hadron cocktail
  ($f_c(m)$), $r$ is the direct photon fraction of the total
  dielectron yield, the free parameter of the two-component fit (black
  curve). (Figure taken from~\cite{Adare:2009qk}.)
  }
    \label{fig:intconv}
\end{figure}
\end{center}

Yet another method (``internal conversion'') takes advantage of the
fact that any process emitting real photons can also produce off-shell
virtual photons ($\gamma^{*}$), which then emerge as low invariant
mass $e^{+}e^{-}$ pairs.  The relation between real photon production
$dN_{\gamma}/dp_T$ and associated $e^{+}e^{-}$ production can be
written as~\cite{Kroll:1955zu,Lichard:1994yx,Adare:2009qk}

\begin{equation}
\frac{d^2N_{ee}}{dm_{ee}dp_T} =
\frac{2\alpha}{3\pi} \frac{1}{m_{ee}}
\sqrt{1-\frac{4m^2_{e}}{m^2_{ee}}}
  \Big(1+\frac{2m^2_{e}}{m^2_{ee}}\Big)
S(m_{ee},p_T) \frac{dN_{\gamma}}{dp_T}
\label{eq:krollwada}
\end{equation}

\noindent
where $\alpha$ is the fine structure constant, $m_e$ is the electron
mass, $m_{ee}$ the mass of the dielectron pair, $S(m_{ee},p_T)$ is a
process-dependent factor encoding the differences between real and
virtual photon production.  The terms containing $m^2_e/m^2_{ee}$ go to
unity for $m_{ee}>>m_e$.  The factor $S(m_{ee},p_T)$ is 
{\it assumed}\footnote{There are some arguments about the validity of
  this assumption, for instance in~\cite{Dusling:2009ej}.
}
to become unity as $m_{ee}\rightarrow 0$ or $m_{ee}/p_T\rightarrow 0$, 
\ie at sufficiently high \pt~\cite{Adare:2008ab,Adare:2009qk,STAR:2016use}.  This way
Eq.~\ref{eq:krollwada} simplifies to

\begin{equation}
\frac{d^2N_{ee}}{dm_{ee}dp_T} \approx
\frac{2\alpha}{3\pi} \frac{1}{m_{ee}}
\frac{dN_{\gamma}}{dp_T}
\label{eq:intconv}
\end{equation}

\noindent
The measurement then proceeds as illustrated in
Fig.~\ref{fig:intconv}, providing the direct photon excess ratio $r$.
At any given \pt the \mee distribution is fitted with a two-component
function 

\begin{equation}
f(m_{ee},r)=(1-r)f_c(m_{ee}+rf_{dir}(m_{ee}))
\end{equation}

\noindent
where $f_c(m_{ee})$ is the expected shape of the background mass
distribution, 
(sum of expected $e^+e^-$ pairs from hadron decays, or ``cocktail'', as
shown in Fig.~\ref{fig:intconv}), $f_{dir}(m_{ee})$ is the expected
shape of the virtual direct photon internal conversion mass
distribution (for \pt $>$1\,\gevc its shape is $1/m_{ee}$, see
Eq.~\ref{eq:intconv}), both separately normalized to the data for
$m_{ee}<30$\,MeV/$c^2$. The direct photon excess ratio $r$ is the only
free parameter.  If the inclusive photon yield is known, the
direct photon yield can be calculated as 
$dN^{dir}_{\gamma}(p_T) = r dN^{incl}_{\gamma}(p_T)$.

\subsection{Other tools and techniques}
\label{sec:othertools}

{\it Pre-shower / photon multiplicity detectors} \\

\vspace{0.1in}
In a high multiplicity environment electromagnetic calorimeters often 
cannot measure individual photons (and their energy) due to the
inherent limitation given by the transverse size of the electromagnetic 
showers\footnote{They still can measure total neutral energy in a
  solid angle, a quantity relevant for instance in jet physics.
}.  
However, sometimes the sheer number of photons -- irrespective of
their energy -- can be an important observable 
(see Sec.~\ref{sec:spsags}).  Photon multiplicity detectors are
simple devices, with a thin (2-3$X_0$) converter followed by a high
granularity sensitive layer, like small scintillator 
pads~\cite{Aggarwal:1996sy,Aggarwal:1998py}, read out separately.  Most
hadrons crossing the scintillator deposit only minimum ionization
energy, so counting the pads with energy above this threshold is a
good proxy of the number of photons.  The readout device can be as
simple as a CCD camera~\cite{Aggarwal:1998bd}.  --  
{\it Pre-shower detectors} are usually installed in front of
electromagnetic calorimeters meant to solve the problem of resolving
single photons from two close-by photons from the decay of a high
momentum \piz and to provide a very precise measurement of the impact
point.   They usually consist of layers of thin (0.5-1 $X_0$)
converters and high granularity, position sensitive detectors, like
fine pitched Si pads or wire chambers~\cite{Balanda:2004gh}.  
The electromagnetic showers
start {\it before} entering the calorimeter proper, their transverse
size is still very small, so even close-by particles can be well
distinguished.   A variation on the idea is the
{\it shower maximum detector}, a similar, charge-sensitive, high
granularity device placed at 5-6$X_0$ depth in the calorimeter, where
the showers are already well developed~\cite{Beddo:2002zx}.

\vspace{0.1in}
{\it Isolation cuts} \\

\vspace{0.1in}
Isolated high \pt prompt photons are a precious tool to investigate
pQCD, the gluon distribution
functions~\cite{Aurenche:1988vi,Ichou:2010wc,Arleo:2011gc} and, in
back-to-back correlation measurements, setting the parton energy scale
($\gamma$-jet) and measuring fragmentation functions of partons into
final state hadrons ($\gamma$-hadron)\footnote{In
  Sec.~\ref{sec:sources} we discussed fragmentation of a quark or
  gluon into a photon, \ie the photon is non-isolated, part of a jet,
  carrying a fraction of the original parton energy.
  Here we discuss back-to-back correlation of an isolated photon, that
  emerges unchanged from the hard scattering, and a jet on the opposite
  side.  The isolated photon has the same energy as the quark or gluon
  originating the opposing jet, setting the scale to the fragmentation
  function of the original parton into the final state hadrons
  observed in the jet.
}.  
Here {\it isolated} means little or no activity --
at least no {\it correlated} activity -- in the vicinity of the
photon.  Isolation cuts are not uniquely defined, they vary depending
on the experiment, colliding system and energy,
but typically they impose an upper limit on
the energy observed in a certain radius around the photon, derived
from the average ``underlying event''.  As an example, for
\sqsn=2.76\,TeV \pbpb collisions and photons of 20\,\gev energy or
higher CMS requires less than 5\,\gev energy in an isolation cone of
the size $\Delta R = \sqrt{(\Delta\eta)^2 + (\Delta\varphi)^2}<0.4$
aound the photon~\cite{Chatrchyan:2012vq}.  For \sqs=200\,\gev \pp
collisions PHENIX requires that the sum of the momenta of charged
tracks in a cone $\Delta R < 0.3$ should be less than 10\% of the
photon energy~\cite{Adare:2010yw}.  Lack of a single, unique
definition makes the comparison of isolated photon data from
different experiments difficult, and of course theory calculations
should always try to implement the same cuts as the experiment they
are compared to.

\begin{figure}[htbp]
  \includegraphics[width=0.49\linewidth]{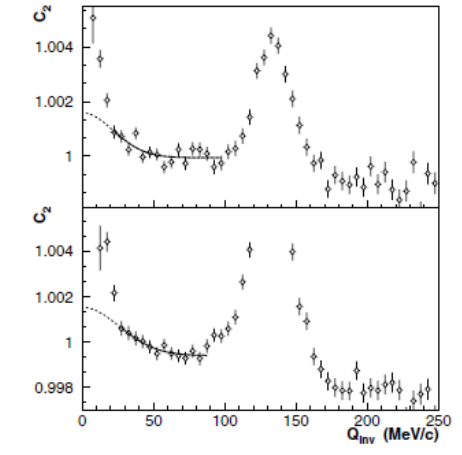}
  \caption{Two-photon correlation function for average photon momenta
  $100<p_T<200$\,MeV (top) and $200<p_T<300$\,MeV (bottom).  The
  dotted line is the extrapolation to low $Q_{inv}$ assuming chaotic
  (Gaussian) distribution to get the correlation strength $\lambda$.
  (Figure taken from~\cite{Aggarwal:2003zy}.)
  }
    \label{fig:wa98_hbt_2004}
\end{figure}

\noindent

{\it Hanbury Brown -- Twiss correlations.} \\

\vspace{0.1in}
Hanbury Brown -- Twiss (HBT) interferometry~\cite{Brown:1956zza} 
has been extensively used for hadrons~\cite{Boal:1990yh} earlier to explore
the  space-time extent of the emitting source at kinetic freeze-out.
Applying it for momentum differences of photon pairs from heavy ion
collisions sounds eminently plausible: after all, the method has been
first introduced to measure the size of distant stars using two-photon
correlations.  For a chaotic source and photons of similar momenta 
($p_1\approx p_2$) the correlation of the invariant relative momenta
$Q_{inv}=\sqrt{-(p_1-p_2)^2}$ is Gaussian and it characterizes both
the source size and its relative strength with respect to other
sources~\cite{Peressounko:2003cf}. The $\Delta Q_{inv}$ region where the
correlation is enhanced (width of the Gaussian) is inversely
proportional to the source size $R$ ($\Delta Q_{inv} \sim \hbar/R$).  
That means that HBT at very low $\Delta Q_{inv}$ can in principle 
differentiate between photons from the collision itself (length scale
is fm, $\Delta Q_{inv} \sim \hbar/fm \sim$100\,MeV) and final state
hadron decay (length scale is nm, 
$\Delta Q_{inv} \sim \hbar/nm\sim$100\,eV)\footnote{In~\cite{Bass:2004de} 
it has been suggested that HBT of photons at $k_T$=2\,\gevc could be
used to measure the system size at the time of hard scattering, 
{\it before} thermalization.   Even more detailed information on
quark-gluon dynamics using high \pt photon HBT is proposed
in~\cite{Srivastava:1993js}. 
}.
Furthermore, if $f$ is the fraction of photons from a particular
source, the $\lambda$ strength parameter of the respective correlation 
is proportional to $f^2$~\cite{Peressounko:2003cf}.  So if one can
measure the inclusive yield of photons at momentum $p$ and the 
correlation strengths of photon pairs with the same average
momentum and $\Delta Q_{inv} \sim$10-100\,MeV, one can assert the
direct photon yield at $p$.  As seen in Fig.~\ref{fig:wa98_hbt_2004}
the method is quite delicate: around $Q_{inv}\sim 0$ the background is
overwhelming, and the Gaussian has to be reconstructed and
extrapolated to $Q_{inv}=0$ in a narrow $Q_{inv}$ window.
Furthermore, to measure small differences of large momenta with high
precision is a task ill-suited for the average calorimeters.
Conversion techniques with their superior energy resolution are more
promising, but due to their low rate still very challenging and never
tried so far.  
A very clear and pedagogical description of the method is given
in~\cite{Stankus:2005eq}.

\subsection{Ways to present direct photon data}
\label{sec:waystopresent}

\begin{center}
\begin{figure}[htbp]
  \includegraphics[width=0.44\linewidth]{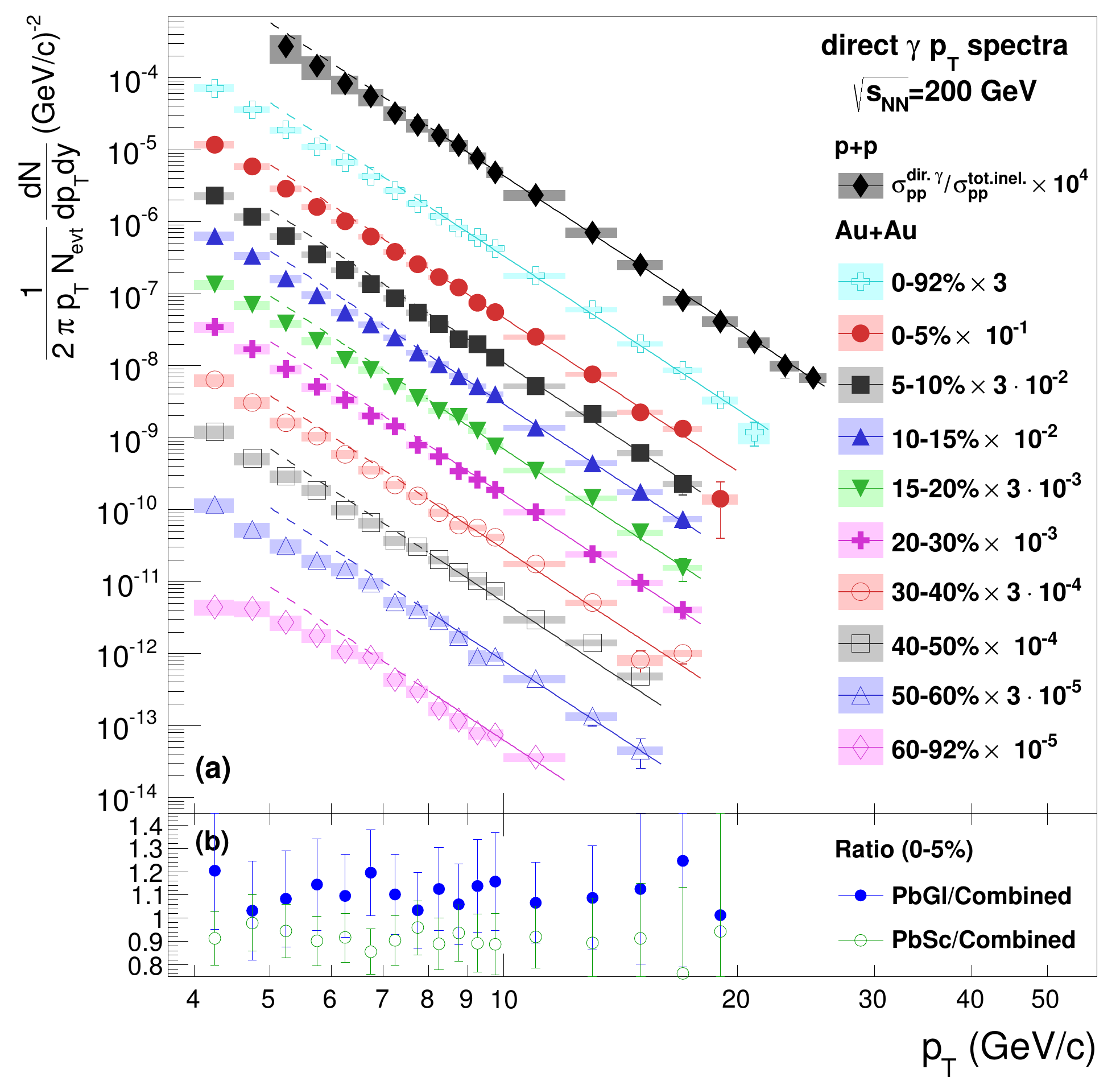}
  \includegraphics[width=0.54\linewidth]{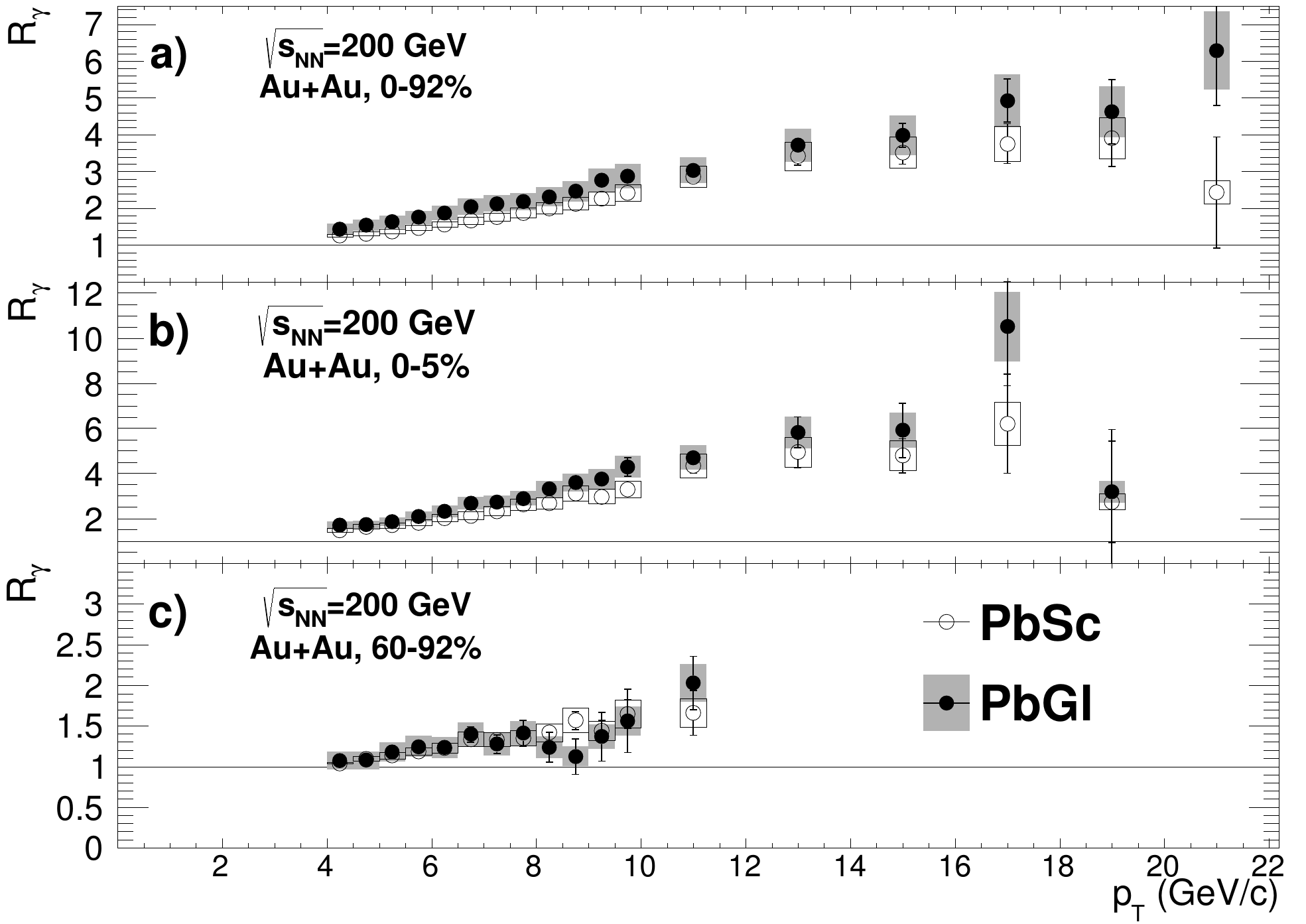}
  \caption{Different ways to present direct photon data.
    Left: direct photon invariant yields in \sqsn = 200\,GeV \auau 
    collisions at
    different centralities (figure taken from~\cite{Afanasiev:2012dg}.
  Right: direct photon excess ratio, \ie the
  inclusive/decay photon ratio $R_{\gamma}$ for select centralities in
    \sqsn = 200\,GeV \auau collisions (figure taken from~\cite{Afanasiev:2012dg}). 
    See text for details.
  }
    \label{fig:spectrargamma}
\end{figure}
\end{center}

There are several ways to present the direct photon data.  
The most transparent
one is the cross section (for \pp) or {\it invariant yield} (for \AA)

\begin{equation}
\sigma(p) = E\frac{d^3\sigma}{dp^3} \qquad {\rm and} \qquad
N_{inv}(p_T) = \frac{1}{2\pi p_T} \frac{1}{N_{evt}} \frac{dN}{dp_T dy}
\end{equation}
  
\noindent
where $y$ is the rapidity (equivalent to pseudorapidity $\eta$ for
photons).  This quantity is easy to compare to pQCD or other
calculations (see Fig.~\ref{fig:spectrargamma}, left panel).  
However, the uncertainties are usually quite
large\footnote{There are many issues, among them the irreducible
  problem that the direct photon
  spectrum is the {\it small} difference of two large numbers: the
  inclusive minus the hadron decay photons.  Uncertainty of the
  absolute energy  scale is another issue: 1\% error on $E_{\gamma}$
  translates to $\sim$6-11\% error on the yield, depending on \sqsn.
}.
Another useful observable is the {\it excess photon ratio} 

\begin{equation}
R_{\gamma}(p_T) = \frac{\gamma^{inclusive}(p_T)}{\gamma^{decay}(p_T)}
\end{equation}

\noindent
where $\gamma^{decay}(p_T)$ is the (calculated) number of hadron decay
photons in the total inclusive photon spectrum.  While information on
the absolute yield is lost, in $R_{\gamma}$ systematic uncertainties
related to particle identification and energy scale are substantially
reduced (see Fig.~\ref{fig:spectrargamma}, right panel). 
On the other hand it is not a good quantity to make comparisons
between different colliding systems, energies or centralities,
because it depends on the yield of neutral mesons, too.  If the direct
photon yields are unchanged, but some reason mesons are suppressed or 
enhanced, $R_{\gamma}$ becomes artificially high or low.

$R_{\gamma}$ is not to be confused with the ratio of direct over
inclusive photons

\begin{equation}
r_{\gamma}(p_T) = \frac{\gamma^{direct}(p_T)}{\gamma^{inclusive}(p_T)}
\end{equation}

\noindent
also called ``direct photon fraction'' and preferred in internal
conversion photon analyses~\cite{Adare:2008ab,STAR:2016use}
(also see the left panel in Fig.~\ref{fig:ppg086_fig4}).

A third way to present the results is the $\gamma/\pi^0$
ratio.  Unfortunately it is used two different ways, and it is not
always immediately clear which definition is meant in a particular
instance.  In one case its precise definition is
$N_{\gamma}^{dir}(p_T)/N_{\pi^0}(p_T)$, the ratio of direct photon and
\piz yields at the same \pt.  This ratio was very popular since the
earliest days of perturbative QCD calculations, many quantitative
predictions have been made for $\gamma/\pi^0$ in hadron-hadron
collisions  (see for 
instance~\cite{Escobar:1975wx,Farrar:1975ke,Ferbel:1984ef}).
The high \pt direct photons come primarily from quark-gluon
Compton scattering, and preserve their original \pt, while the \piz
comes from the fragmentation of the scattered parton and carries only
a fraction of the original parton \pt.  This way the ratio
$N_{\gamma}^{dir}(p_T)/N_{\pi^0}(p_T)$ {\it at the same \pt} can
become quite large.  If the fragmentation functions are known
the process provides information on the gluon PDF in the colliding
hadron~\cite{Baier:1980qy,Aurenche:1988vi}.  
Experimentally, finding high \pt (isolated) photons in hadron-hadron
collisions is moderately difficult (the multiplicity is low).

\begin{center}
\begin{figure}[htbp]
  \includegraphics[width=0.3\linewidth]{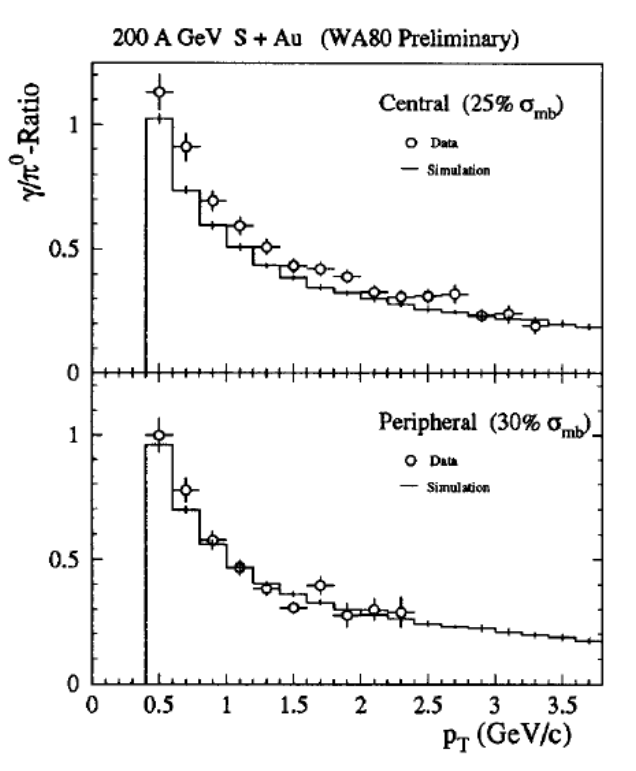}
  \includegraphics[width=0.3\linewidth]{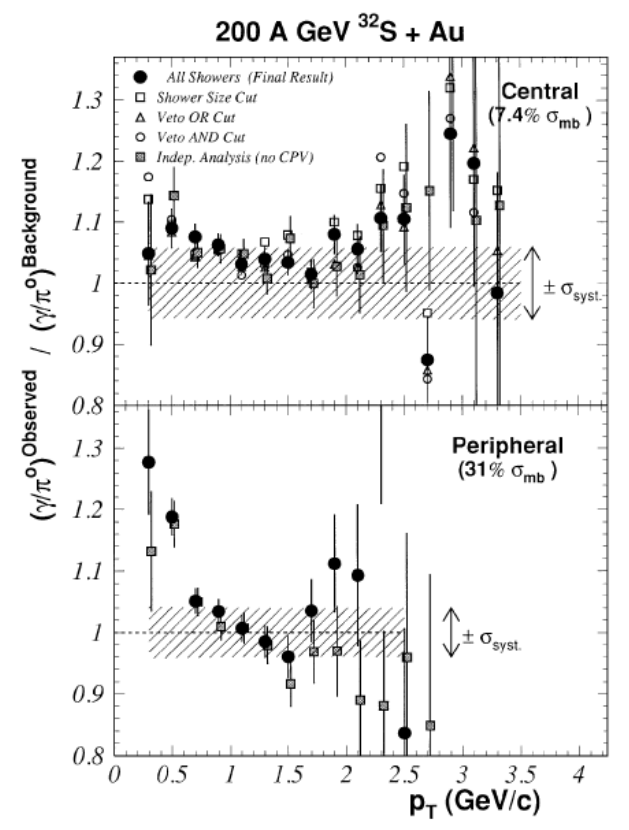}
  \includegraphics[width=0.3\linewidth]{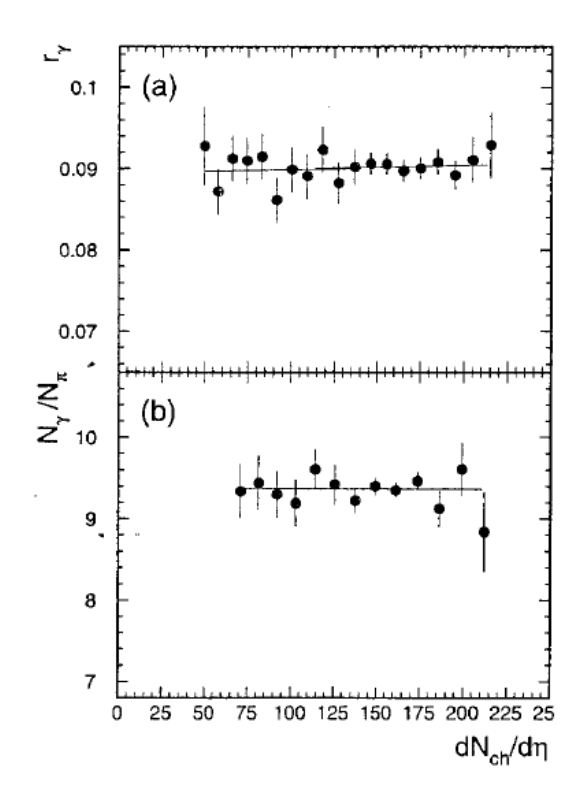}
  \caption{Different ways to present direct photon data.
    Left: $\gamma/\pi^0$ ratio~\cite{Awes:1995cj}.
    Middle: double ratio~\cite{Albrecht:1995fs} of the measured
    $\gamma/\pi^0$ and the one simulated purely from hadron decays.
    Right: $N_{\gamma}/N_{ch}$ from~\cite{Baur:1995gt}, not requiring
    reconstruction of \piz.
    See text for details.
  }
    \label{fig:gammapinch}
\end{figure}
\end{center}

The second definition of  $\gamma/\pi^0$
is the ratio $N_{\gamma}^{inc}(p_T)/N_{\pi^0}(p_T)$ of
the {\it inclusive photon} and \piz yields taken at the same \pt.
This is a very robust quantity, since inclusive (but not necessarily
direct) photons can be measured even in very high multiplicity
environments.  On the other hand,  it carries only
limited information content (see Fig.~\ref{fig:gammapinch}, left panel).  
It can indicate the presence of direct photons in addition to the
numerous decay photons, but is rarely used to extract actual yields or
cross-sections. Its usefulness is rooted in Sternheimer's
formula~\cite{Sternheimer:1955zz} stating that at sufficiently high
energies ($E>500$\,MeV) the energy spectrum of decay photons is
related to the \piz spectrum (in the same solid angle) as

\begin{equation}
N_{\gamma}(E_{\gamma}) = \int_{E_{\gamma}}^{\infty}
\frac{2}{E_{\pi^0}} N_{\pi^0}(E_{\pi^0})dE_{\pi^0}
\end{equation}

At mid-rapidity and high energies $E_{\gamma}$ and $E_{\pi^0}$ can be
replaced by the respective \pt.  Since high \pt particle spectra are
power-law ($\sim p_T^{-n}$), the decay photon spectra are related to
the \piz spectra as $N_{\gamma}(p_T) = (2/n)N_{\pi^0}(p_T)$, and the 
$\gamma/\pi^0$ ratio converges to a constant $2/n$ at higher \pt, if
and only if the sole source of photons is \piz decay.  It was
frequently used in the early days of photon physics, and even today it
is an important sanity check in any photon analysis.  It also inspired
the introduction of the {\it double ratio} of the measured inclusive
$\gamma/\pi^0$ and the simulated, purely decay $\gamma/\pi^0$

\begin{equation}
R(p_T) = \frac{\gamma^{inc}_{meas}(p_T) / \pi^0_{meas}(p_T)}
         {\gamma^{dec}_{sim}(p_T) / \pi^0_{sim}(p_T)}
\end{equation}

\noindent
a quantity very similar, but not identical to $R_{\gamma}$ discussed above (see
Fig.~\ref{fig:gammapinch}, middle panel). Finally, we should mention
the $N_{\gamma}/N_{ch}$ ratio of photons to charged particles (see 
Fig.~\ref{fig:gammapinch}, right panel), frequently used in the early
days of direct photon physics, because it didn't require
reconstruction of the \piz spectrum (see Sec.~\ref{sec:spsags}).

\section{\bf The high \pt region.}
\label{sec:highpt}

\subsection{Hard photons in $pp$ and $p\bar{p}$ collisions}
\label{sec:pphigh}

While the actual transverse momentum above which direct photons are
considered ``high \pt'' is ill-defined, the term usually refers to
photons originating from scattering of 
hard (large $x$) partons and calculable with
pQCD, with the dominant process being quark-gluon Compton scattering
($qg \rightarrow q\gamma$).  Typically \pt larger than 3-5\,\gevc is
considered high \pt (depending on \sqs), 
and photon measurements are relatively easy there.
Below that the uncertainties on calculations are large
(reaching an order of magnitude~\cite{Acharya:2018dqe}), 
not the least because the fraction of fragmentation photons increases,
but the actual values are poorly known.  As we will see later, at low
\pt  the experimentally observed yields are usually very
small~\cite{Adare:2008ab} or just upper limits~\cite{Acharya:2018dqe},
so there is little input to meaningfully test the calculations.

\subsubsection{Spectra}
\label{sec:ppspectra}

High \pt direct photons in \pp collisions
in the 30$<$\sqs$<$62\,\gev range have first been studied at the
CERN ISR\footnote{For a review of the early experiments at CERN ISR
  and Fermilab see~\cite{Ferbel:1984ef}}
up to \pt=7\gevc~\cite{Diakonou:1979sv}, by measuring the ratio
of single photons to \piz at the same \pt (\gam/\piz) for
\sqs=31, 53 and 63\,\gev.
For photons coming from \piz decays this ratio is easy to calculate
(see Sternheimer's formula, Sec.~\ref{sec:waystopresent});  in presence
of direct photons the ratio will increase. 
Within uncertainties \gam/\piz was consistent with no direct photons
 at \pt=3\,\gevc  (all measured photons were accounted for from hadron
 decays), then 
started to rise slowly, reaching about 20\% excess above 5\,\gevc.
Remarkably, \gam/\piz didn't seem to depend on \sqs.
Using the same setup and measuring the differences in same-side and 
away-side charged multiplicity  in events triggered by \piz and single 
photons R807 found evidence that the dominant process might indeed be
$qg \rightarrow q\gamma$~\cite{Diakonou:1980vz}\footnote{A photon
  produced by this process has no additional multiplicity associated
  with it.}.  
One interesting
consequence is that direct photons offer access to gluon PDFs,
another one is that if those ``isolated'' photons are
back-to-back to a jet, they provide a very good estimate of the jet
(\ie the original parton) energy. 

By 1982 inclusive cross-sections for single-\gam and \piz up to
\pt=12\,\gevc in 30$<$\sqs$<$63\,\gev \pp collisions were
published~\cite{Anassontzis:1982gm}.  The last important attempt at
the ISR was the comparison of \gam/\piz in $pp$ and $p\bar{p}$ by the
AFS collaboration~\cite{Akesson:1985za}.  This was promising, because
due  to the presence of large $x$ antiquarks in $p\bar{p}$ collisions the 
$q\bar{q}\rightarrow g\gamma$ annihilation process was expected to 
contribute to the high \pt direct photon production, as predicted by
QCD, and its amplitude could in principle be determined when photon
production in \pp and \pbarp is compared.  
Unfortunately, due to the short
running time AFS didn't find a statistically significant difference
between the \gam/\piz in \pp and $p\bar{p}$.

Shortly thereafter the situation changed when at
the CERN S$p\bar{p}$S 
the available energy in \pbarp collisions increased an order of
magnitude, up to \sqs=630\,\gev.  Thanks to improvements in detector 
technology and analysis techniques {\it isolated} direct photons 
could be measured\footnote{In fact, being isolated, \ie a single
  large neutral energy deposit with no other activity around it became
  the main   identification criterion of direct photons, to
  distinguish them from two very close decay photons from a \piz,
  which in turn was expected to be part of a jet, with plenty of
  activity near the photon~\cite{Appel:1986ix}.
}
up to 100\,\gevc at mid-rapidity~\cite{Appel:1986ix,Albajar:1988im}.  
Due to the presence of valence antiquarks \pbarp collisions made
it possible to study for the first time the 
$q\bar{q}\rightarrow\gamma\gamma$ 
process\footnote{A real {\it tour de force} with altogether 6 events
  found.
},
by observing back-to-back, isolated, high \pt
``double photons''~\cite{Albajar:1988im}.   This rare process in principle
provides information on the intrinsic \kt of the partons, 
by the transverse momentum imbalance of the two photons\footnote{For
  more detailed discussion see Sec.~\ref{sec:ppcorr}.
}, 
although perturbative corrections may destroy the
significance~\cite{Aurenche:1985yk}.  Somewhat later, using an internal
hydrogen gas jet target, \pp and \pbarp data were also taken at
\sqs=24.3\,\gev by UA6 and the difference
$\sigma(\bar{p}p\rightarrow\gamma X) - \sigma(pp\rightarrow\gamma X) $
was measured for the first time~\cite{Sozzi:1993sm,Ballocchi:1998au}.  This
difference isolates  the leading order $q\bar{q}$ annihilation term.
Once the quark distributions are known (from deep inelastic scattering),
one could determine the gluon distributions from the other leading order
process, the $qg \rightarrow \gamma q$ Compton scattering, and even
$\alpha_s$ can be measured~\cite{Ballocchi:1993sq}

In the early 1990s
Fermilab experiments CDF and D0 measured prompt photon cross sections
in \sqs=1.8\,TeV $p\bar{p}$ collisions up to
\pt=120\,\gevc.  CDF measured at central
rapidities~\cite{Abe:1994rra}, while D0 also published results for
forward rapidities~\cite{Abachi:1996qz}, providing constraints on the
low-$x$ gluon distributions.  All these data provided input for
incremental improvement of NLO, then NNLO
calculations~\cite{Baer:1990ra,Gordon:1993qc,Kidonakis:1999hq} 
without any major surprises.  However, one particular fixed target
experiment (FNAL E706~\cite{Apanasevich:1997hm,Apanasevich:2004dr})
strongly disagreed with the calculations, triggering speculations that
the effect of intrinsic \kt is much larger than previously assumed,
and, to lesser extent, the results from WA70~\cite{Bonesini:1987bv}
also deviated from the general trend, shown below.  Due to this
discrepancy, photon data in \pp and \pbarp were omitted from
global-fit analyses of proton PDFs for about a
decade~\cite{Ichou:2010wc}. 

The relatively wide \sqs gap between the CERN and FNAL fixed target and
collider data has been filled by RHIC when PHENIX published direct
photon cross-sections in \pp at 
\sqs=200\,\gev~\cite{Adler:2006yt,Adare:2012yt} and STAR 
in~\cite{Abelev:2009hx}.  Although \pp data have been taken at
\sqs=510\,\gev as well, direct photon spectra have not been published
yet.  The impact of the FNAL and RHIC data on the gluon distribution
in the proton is discussed in~\cite{Ichou:2010wc}.  At central
rapidities various PDF parametrizations provide cross sections 
within 15\%, while at $y=4$ (low $x$) the differences are 
within 30\%.  The uncertainties are largest at low $E_T^{\gamma}$
which has some impact on the ``thermal'' photon measurements in heavy
ion collisions, too.

Beyond FNAL energies at the LHC both ATLAS and CMS measured
isolated prompt photon cross-sections at 
\sqs=7\,TeV~\cite{Aad:2010sp,Khachatryan:2010fm}, 
CMS and ALICE provided data at
2.76\,TeV~\cite{Chatrchyan:2012vq,Acharya:2018dqe}, 
ATLAS and ALICE published
results for 8~\cite{Aad:2016xcr,Acharya:2018dqe}
and ATLAS for 13\,Tev, too~\cite{Aaboud:2017cbm}
(see Table~\ref{tab:ppdata}).

A convenient and physics driven way to compare \pp photon data taken at very
different \sqs and 
covering orders of magnitude both in \pt and cross-section is to present
them as a function of the {\it scaling} variable \xt=2\pt/\sqs.
For the hard scattering region~\cite{Cahalan:1974tp}

\begin{equation}
E\frac{d^3\sigma}{dp^3} = \frac{1}{\sqrt{s}^{n(x_T,\sqrt{s})}}G(x_T)
\end{equation}

\noindent
where $n(x_T,\sqrt{s})=4$ for leading order QCD without evolution of
$\alpha_s$, and all effects from
the structure function and the fragmentation function into photons 
are encoded in $G(x_T)$.
Higher order effects usually increase the value of $n(x_T,\sqrt{s})$.

An excellent compilation of the data on prompt photon production in 
hadron-hadron interactions, available until 1997, 
and comparisons in terms of \xt to contemporary NLO calculations can 
be found in~\cite{Vogelsang:1997cq}.  
The comparisons were moderately successful, meaning that in addition
to differences in absolute magnitude often the {\it shapes} of the
spectra in data and theory differed significantly.  This can be
explained in part by lack of proper tools to implement the precise 
experimental cuts (like isolation cuts) in the calculations.

\begin{figure}[htbp]
  \includegraphics[width=0.54\linewidth]{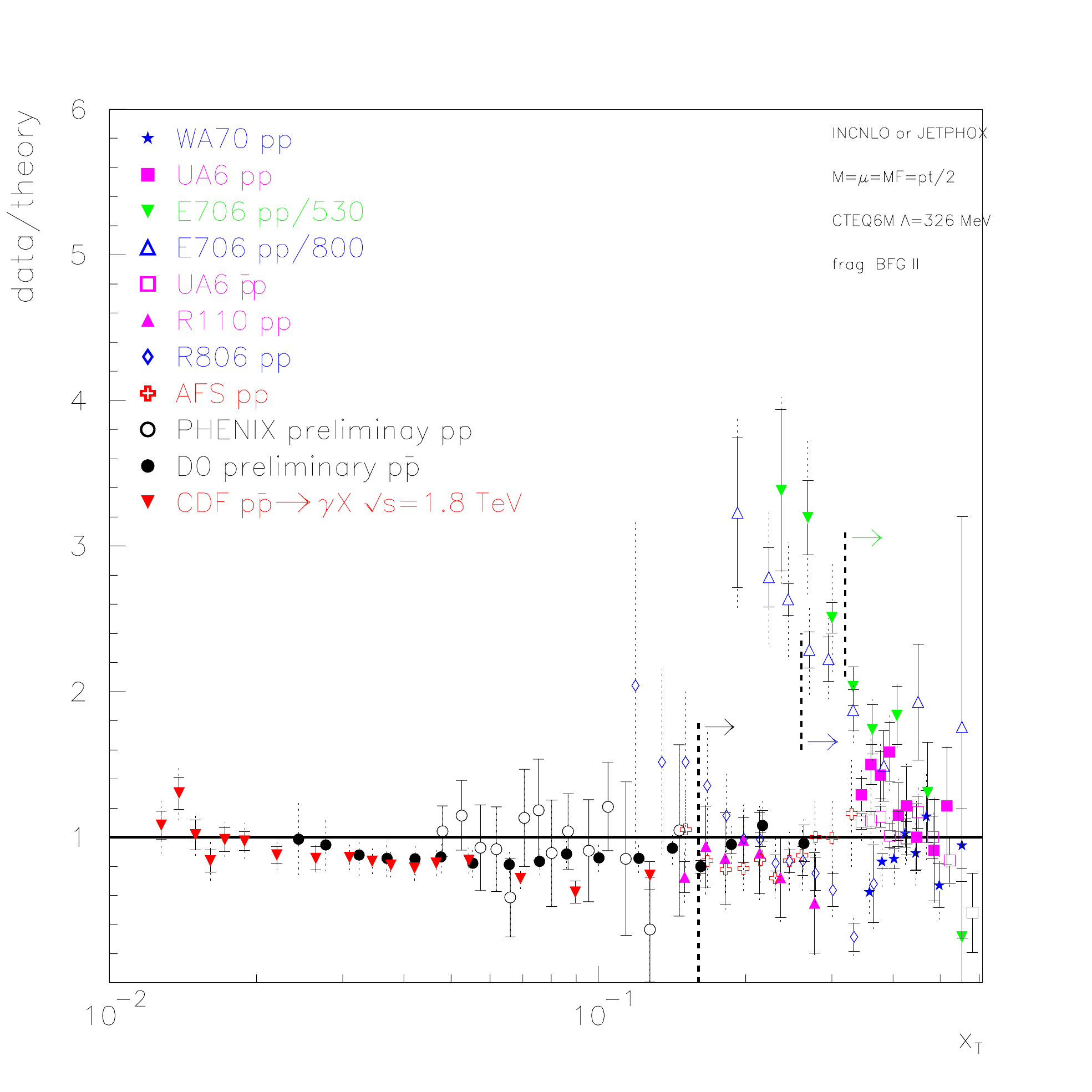}
  \includegraphics[width=0.44\linewidth]{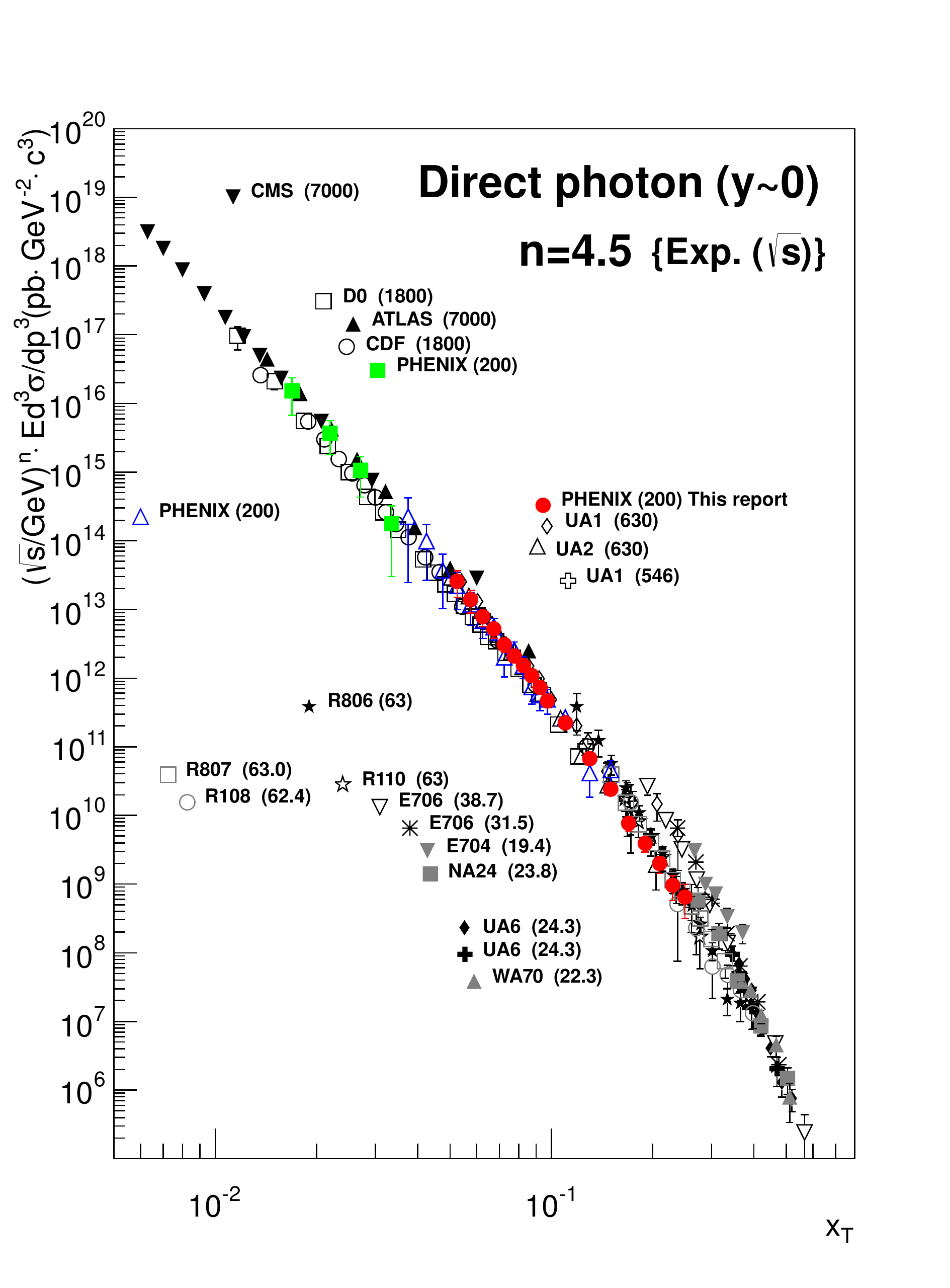}
  \caption{Left: direct photon data compared to NLO calculations
    for select $pp$ and $p\bar{p}$ results as of 2006 (figure courtesy 
    of P. Aurenche and M. Werlen,
    see also~\cite{Aurenche:2006vj}).  
    Sources of data:
    WA70~\cite{Bonesini:1987bv}, UA6~\cite{Ballocchi:1998au},
    E706~\cite{Apanasevich:2004dr}, R110~\cite{Angelis:1989zv},
    R806~\cite{Anassontzis:1982gm}, AFS~\cite{Akesson:1989hp},
    D0~\cite{Abazov:2001af}, CDF~\cite{Acosta:2002ya,Acosta:2004bg}.
    Right:  \xt-scaling of the world data
    on direct photons in $pp$ and $p\bar{p}$ as of 2012 (figure taken
    from~\cite{Adare:2012yt}).  Sources of additional data included here:
    CMS~\cite{Khachatryan:2010fm}, ATLAS~\cite{Aad:2011tw},
    D0~\cite{Abazov:2005wc},  CDF~\cite{Abe:1994rra,Aaltonen:2009ty},
    PHENIX~\cite{Adare:2009qk,Adare:2012yt,Adler:2006yt},
    UA1~\cite{Albajar:1988im}, UA2~\cite{Alitti:1992hn},
    R807~\cite{Akesson:1989hp}, R108~\cite{Angelis:1980yc},
    E704~\cite{Adams:1995gg}, NA24~\cite{DeMarzo:1986vi},
    UA6~\cite{Sozzi:1993sm}.
  }
    \label{fig:ppworlddata}
\end{figure}

A decade later a landmark survey of photon production in hadronic
collisions~\cite{Aurenche:2006vj}, using NLO pQCD calculations
implemented in the JETPHOX Monte Carlo code, found much better
agreement between data and theory (see Fig.~\ref{fig:ppworlddata},
left panel).
Apart of two (controversial) datasets from Fermilab
E706~\cite{Apanasevich:2004dr} and to some lesser extent the D0
results~\cite{Abazov:2001af}  the data are well described from
\sqsn = 23\,GeV to 1.96\,TeV, covering 9 orders of magnitude in cross
section.

In Fig.~\ref{fig:ppworlddata}, right panel, the \pp and $p\bar{p}$
direct photon data available in
2012 are shown.  The cross sections are multiplied by $(\sqrt{s}$)$^{4.5}$ and
plotted vs \xt. The data covering the range of 19.4 - 7000\,GeV in
$\sqrt{s}$ line up on a single curve, and the effective exponent
$n_{eff}$=4.5 indicates that the role of scaling violations from PDF
and running of $\alpha_s$ is small~\cite{dEnterria:2012kvo}.
The fact that hard photon production in \pp is well understood 
at least down to \xt $=10^{-2}$ is crucial when interpreting certain 
observations in heavy ion collisions\footnote{Recent
  measurements by ALICE~\cite{Acharya:2018dqe} extended the \xt range down to
  $10^{-4}$, although often prividing only upper limits.  The new data
  don't line up with the trend seen previously at higher \xt (see
  Bock~\cite{ect:2018}). 
}.
Direct photon cross-section measurements in \pp are 
summarized in Table~\ref{tab:ppdata}.

\begin{table}[h]
  \begin{tabular}{|l|c|c|c|c|c|c|} \hline
   Experiment   &  $\sqrt{s}$  &  method  &   $\eta$   &   $p_T$   &
   Publications  & Comment \\ \hline \hline
   R412 / CERN  ISR &  $pp$ 45, 53\,GeV  &  calor.  &
   $|\eta|\sim 0$   &   1.6-3.8\,GeV/$c$   &   
   \cite{Darriulat:1976fb}  (1976) & $\gamma/\pi^0$ \\ 
   R107 / CERN  ISR &  $pp$ 53\,GeV  &  calor.  &
   $|\eta|\sim 0$   &   2.3-3.7\,GeV/$c$   &   
   \cite{Amaldi:1978hs}  (1978) & $\gamma/\pi^0$ \\ 
   AFS / CERN  ISR &  $pp$ 31, 53, 63\,GeV  &  calor.  &
   $|\eta|\sim 0$   &   3 - 7(9)\,GeV/$c$   &   
   \cite{Diakonou:1979sv,Diakonou:1980vz}  (1979/80) & $\gamma/\pi^0$ \\ 
   CCOR / CERN  ISR &  $pp$ 62.4\,GeV  &  calor.  &
   $|\eta|<1.1$   &   5 - 13\,GeV/$c$   &   \cite{Angelis:1980yc} (1980) &  \\ 
   AFS / CERN  ISR &  $pp$ 31, 45, 53, 63\,GeV  &  calor.  &
   $|\eta|<\sim 0$   &   3 - 12\,GeV/$c$   &
   \cite{Anassontzis:1982gm}  (1982) &   \\ 
   AFS R807 / CERN  ISR &  $pp$ 63\,GeV  &  calor.  &
   $2.0<\eta <2.75$   &   1.5-4.25\,GeV/$c$   &
   \cite{Akesson:1983xr}  (1983) &  $\gamma/\pi^0$ \\ 
   AFS R808 / CERN  ISR &  $pp,p\bar{p}$ 53\,GeV  &  calor.  &
   $|\eta| <0.4$   &   2-6\,GeV/$c$   &
   \cite{Akesson:1985za}  (1985) &  $\gamma/\pi^0$ \\ 
   UA2 / CERN  S$p\bar{p}$S &  $p\bar{p}$ 630\,GeV  &  calor.
   &   $|\eta|<1.8$   &   15-43\,GeV/$c$   &   \cite{Appel:1986ix} (1986) &   \\ 
   NA24 / CERN  SPS &  $pp$ 23.7\,GeV  &  calor.  &   
   $|\eta|<0.8$  &   3-6\,GeV/$c$   &   \cite{DeMarzo:1986vi} (1987)  &  fixed tgt \\ 
   WA70 / CERN  SPS &  $pp$ 22.3\,GeV  &  calor.  &   
   &   4-6.5\,GeV/$c$   &   \cite{Bonesini:1987bv}  (1988) &  fixed tgt \\ 
   UA1 / CERN  S$p\bar{p}$S &  $p\bar{p}$ 546, 630\,GeV  &  calor.
   &   $|\eta|<3.0$   &   16-100\,GeV/$c$   
   &   \cite{Albajar:1988im} (1988) &   \\ 
   CMOR / CERN  ISR &  $pp$ 63\,GeV  &  calor.  &   $|\eta|<1.1$
   &   4.5 - 10\,GeV/$c$   &   \cite{Angelis:1989zv}  (1989) &  \\ 
   AFS / CERN  ISR &  $pp$ 63\,GeV  &  calor.  &   $|\eta|<1$
   &   4.5 - 11\,GeV/$c$   &   \cite{Akesson:1989hp}  (1990) &  \\ 
   CDF / FNAL  &  $p\bar{p}$ 1.8TeV  &  calor,conv  &   $|\eta|<0.9$
   &   10-60\,\gevc   & \cite{Abe:1993qb}  (1993) & iso. \\ 
   UA6 / CERN  S$p\bar{p}$S &  $pp,p\bar{p}$ 24.3\,GeV  &  calor.
   &   $-0.2<\eta<1$   &   4.1 - 5.7\,GeV/$c$   &
   \cite{Sozzi:1993sm} (1993)  & fixed tgt  \\ 
   E704 / FNAL  &  $pp$ 19.4\,GeV  &  calor.  &   $|\eta|<0.15$
   &   2.5-3.8\,GeV/$c$   &   \cite{Adams:1995gg} (1995)  & fixed tgt \\ 
   UA6 / CERN  S$p\bar{p}$S &  $pp,p\bar{p}$ 24.3\,GeV  &  calor.
   &   $-0.1<\eta<0.9$   &   4.1 - 7.7\,GeV/$c$   &
   \cite{Ballocchi:1998au} (1998)  & fixed tgt  \\ 
   D0 / FNAL  &  $p\bar{p}$ 1.8TeV  &  calor.  &   $|\eta|<2.5$
   &   10-110\,GeV/$c$   &   \cite{Abbott:1999kd}  (2000) & iso. \\ 
   D0 / FNAL  &  $p\bar{p}$ 0.63 TeV  &  calor.  &   $|\eta|<2.5$
   &   10-30\,GeV/$c$   &   \cite{Abazov:2001af}  (2001) & iso. \\ 
   CDF / FNAL  &  $p\bar{p}$ 0.63, 1.8TeV  &  calor.  &   $|\eta|<0.9$
   &   10-30(110)\,GeV/$c$   &   \cite{Acosta:2002ya}  (2002) & iso. \\ 
   CDF / FNAL  &  $p\bar{p}$ 1.8TeV  &  conv.  &   $|\eta|<0.9$   &   10-60\,\gevc   &
   \cite{Acosta:2004bg}  (2004) & iso. \\ 
   E706 / FNAL &  $pp$ 31.8, 38.7\,GeV  &  calor.  &   $|\eta|<0.75$
   &   3.5 - 12\,GeV/$c$   &   \cite{Apanasevich:2004dr}  (2004) &  fixed tgt \\ 
   PHENIX / BNL RHIC  &  $pp$ 200\,GeV  &  calor.  &   $|\eta|<0.35$
   &   5.5-7\,GeV/$c$   &   \cite{Adler:2005qk}  (2005) &   \\ 
   PHENIX / BNL RHIC  &  $pp$ 200\,GeV  &  calor.  &   $|\eta|<0.35$
   &   3-16\,GeV/$c$   &   \cite{Adler:2006yt}  (2007) &  iso. \\ 
   PHENIX / BNL RHIC  &  $pp$ 200\,GeV  &  int. conv.  &   $|\eta|<0.35$
   &   1-4.5\,GeV/$c$   &   \cite{Adare:2008ab,Adare:2009qk}  (2010) &  \\ 
   STAR / BNL RHIC  &  $pp$ 200\,GeV  &  calor.  &   $|\eta|<1$
   &   6-14\,GeV/$c$   &   \cite{Abelev:2009hx}  (2010) &  \\ 
   CMS / CERN LHC  &  $pp$ 7\,TeV  &  calor.  &   $|\eta|<1.45$
   &   21-300\,GeV   &   \cite{Khachatryan:2010fm}  (2011) & iso. \\ 
   ATLAS / CERN LHC  &  $pp$ 7\,TeV  &  calor.  &   $|\eta|<1.81$
   &   15-100\,GeV   &   \cite{Aad:2010sp}  (2011) & iso. \\ 
   ATLAS / CERN LHC  &  $pp$ 7\,TeV  &  calor.  &   $|\eta|<2.37$
   &   45-400\,GeV   &   \cite{Aad:2011tw}  (2011) & iso. \\ 
   PHENIX / BNL RHIC  &  $pp$ 200\,GeV  &  calor.  &   $|\eta|<0.35$
   &   5.5-25\,GeV/$c$   &   \cite{Adare:2012yt}  (2012) &  iso. \\ 
   CMS / CERN LHC  &  $pp$ 2.76\,TeV  &  calor.  &   $|\eta|<1.44$
   &   20-80\,GeV   &   \cite{Chatrchyan:2012vq}  (2012) & iso. \\ 
   ATLAS / CERN LHC  &  $pp$ 7\,TeV  &  calor.  &   $|\eta|<2.37$
   &   100-1000\,GeV   &   \cite{Aad:2013zba} (2014) & iso. \\ 
   ATLAS / CERN LHC  &  $pp$ 8\,TeV  &  calor.  &   $|\eta|<2.37$
   &   25-1500\,GeV   &   \cite{Aad:2016xcr}  (2016) & iso. \\ 
   ATLAS / CERN LHC  &  $pp$ 13\,TeV  &  calor.  &   $|\eta|<2.37$
   &   125-1000\,GeV   &   \cite{Aaboud:2017cbm} (2017)  & iso. \\ 
   ALICE / CERN LHC  &  $pp$ 2.76 and 8\,TeV  &  comb.  &   $|\eta|<0.9$
   &   0.3-16\,GeV   &   \cite{Acharya:2018dqe} (2018)  & iso. \\
   ATLAS / CERN LHC  &  $pp$ 13 and 8\,TeV  &  calor.  &   $|\eta|<2.37$
   &   125-1500\,GeV   &   \cite{Aaboud:2019vpz} (2019)  & 
   iso. $\sigma_{13}/\sigma_{8}$ \\ \hline
  \end{tabular}
  \vspace{0.3cm}
  \caption{Summary of $pp$ and $p\bar{p}$ direct photon data (spectra,
    cross-sections).
}
  \label{tab:ppdata}
 \end{table}

\subsubsection{Photon-jet, photon-hadron and photon-photon correlations}
\label{sec:ppcorr}

With the caveat mentioned earlier (see also~\cite{Aurenche:1985yk}),
high \pt isolated photons back-to-back in azimuth with a high \pt
hadron or jet can set the energy scale of the original hard scattered
parton\footnote{This is particularly important in heavy ion collisions
  where the partons lose energy while traversing the QGP.
}.
It has been pointed out already in
1980~\cite{Baier:1980qy,Cormell:1980ja,Halzen:1980iw} 
that back-to-back isolated photon-jet correlation measurements in \pp
can provide direct infomation about the gluon distribution (PDF) in
the proton,  furthermore, if there is good particle identification on
the jet side, also provide  information on the parton fragmentation
into hadrons, primarily 
of $u$ quarks\footnote{The dominance of $u$-quarks in back-to-back 
  correlations is
  demonstrated by the observed charge asymmetry of hadrons opposite to
  the photon~\cite{Angelis:1989zv,Adare:2010yw}.
}.

These measurements are more complicated than the
inclusive photon  technique by UA6 discussed above, but they are also
richer in information~\cite{Klasen:2002xb}. 
PHENIX estimated the average parton transverse momentum \kt at RHIC 
energies in \pp~\cite{Adare:2010yw}, and measured the ratio
$z_T=p_T^{h}/p_T^{\gamma}$, a proxy for the fragmentation
function~\cite{Adare:2012qi}.   A similar measurement has been
published by STAR in~\cite{STAR:2016jdz}.
Various proton PDF sets have been tested for instance by back-to-back
isolated  photon-jet measurements in \pp at \sqs=7\,TeV by
ATLAS~\cite{Aad:2013gaa}.  Also, the distribution of the rapidity
difference  $\Delta y$ between the photon and the jet reveals that
$t$-channel quark exchange (Compton-scattering) is the dominant source
of photons, rather than gluon exchange
(fragmentation)~\cite{Aad:2013gaa}.  A recently published analysis of
  the 13\,TeV data reaches similar conclusions~\cite{Aaboud:2017kff}. 
The measurements are quite precise tests of pQCD -- the experimental
uncertainties are smaller than those of the theory calculations. 
Triple differential cross sections
$d^3\sigma/(dp_T^{\gamma}d\eta^{\gamma}d\eta^{jet})$ in \pp at 7\,TeV 
have been published by CMS and compared to LO (SHERPA) and NLO
(JETPHOX) predictions: the LO calculation underpredicts the data by
about 10-15\%, while NLO describes the data within uncertainties.
Recently \pbpb $\gamma$-jet results at 5.02\,TeV have also been
published by CMS~\cite{McGinn:2017opi}.  

While the most important properties of the free proton are encoded in
the PDFs, its multi-dimensional structure has long been conjectured
(see for instance~\cite{Collins:1989gx,Aidala:2012mv}), and a breakdown
of QCD factorization has been predicted when the nonperturbative
transverse momentum of partons are explicitely considered (transverse
momentum dependent, or TMD framework).  Factorization breaking can be
studied comparing the acoplanarity~\cite{DellaNegra:1977cyz}  in 
back-to-back dihadron (dijet) and photon-hadron (photon-jet) angular 
correlation and its dependence on the hard scale (\ie the trigger \pt).  
Measurements of \piz-h and \gam-h correlations in \pp at 
\sqs=200 and 510\,GeV~\cite{Adare:2016bug,Osborn:2018wxd} have so far 
not confirmed factorization breaking in these processes.

As mentioned earlier, the $q\bar{q}\rightarrow \gamma\gamma$
and the $gg\rightarrow \gamma\gamma$ channels give access to the 
primordial (or intrinsic) \kt of partons, the driving factor behind the
Cronin-effect~\cite{Cronin:1974zm}.  In order to get high \pt photons
valence antiquarks are needed.  The WA70 experiment at CERN studied
the $\pi^{-}p\rightarrow \gamma\gamma$ process at 280\,\gevc beam
momentum looking for two high \pt photons ($>$2.75\,\gevc) and
estimated the effective intrinsic \kt using three different
observables ($\Delta\Phi, p_{out}$ and \pt(\gam\gam)), which all
provided consistent $\langle\kt\rangle$ values in the 0.91-0.98\,\gevc
range~\cite{Bonvin:1990kx}.  At FNAL the CDF experiment measured
isolated prompt diphoton $\langle\kt\rangle$ and cross-section in 
\pbarp (\sqs=1.8\,TeV) collisions~\cite{Abe:1992cy} primarily with
the goal to estimate a possible background to the Higgs-search in the
\gam\gam channel.  Similar studies of diphoton angular and momentum
correlations have been done at the LHC 
(see for instance~\cite{Aad:2011mh}), not only to facilitate the Higgs
search, but also because they are potent probes of QCD in some
kinematic regions.

\subsection{Hard photons in heavy ion collisions}
\label{sec:aahigh}

As mentioned before, the definition of ``hard'' (high \pt) photons is
somewhat arbitrary, but at RHIC and LHC energies usually photons above 
3-5\,\gevc are considered hard.  The key issue is their origin: most 
hard photons come from low-order scattering of incoming partons with 
relatively high $x$ momentum fraction, and are calculable in pQCD 
using the free  hadronic or nuclear PDFs.  This is to be contrasted
with ``soft'' photons, that come from non-perturbative sources,
ranging from ``thermal'' production to hadron Bremsstrahlung.  Also
note, that -- somewhat misleadingly -- hard photons are sometimes
called ``prompt'' (first interaction) photons, although not all prompt
photons are necessarily high \pt and not
all high \pt photons are prompt (\eg fragmentation or jet-photon
conversion photons).

Contrary to high \pt hadrons,
invariant yields of hard photons in relativistic 
heavy ion collisions so far didn't
provide major surprises, drastically new physics insights.
Paradoxically, {\it this is the most favorable outcome possible}.
High \pt photons are indeed penetrating probes, as advertised, well
calculable in pQCD\footnote{This means that the contribution of hard, 
  but non-prompt  photons to the total yield is relatively small.
}, 
unmodified by the QGP, thus they are important reference and solid
calibration tools for hadronic processes, even if a medium is formed
in \AA with a complicated space-time evolution.

\subsubsection{Invariant yields}
\label{sec:aayields}

\begin{center}
\begin{figure}[htbp]
  \includegraphics[width=0.6\linewidth]{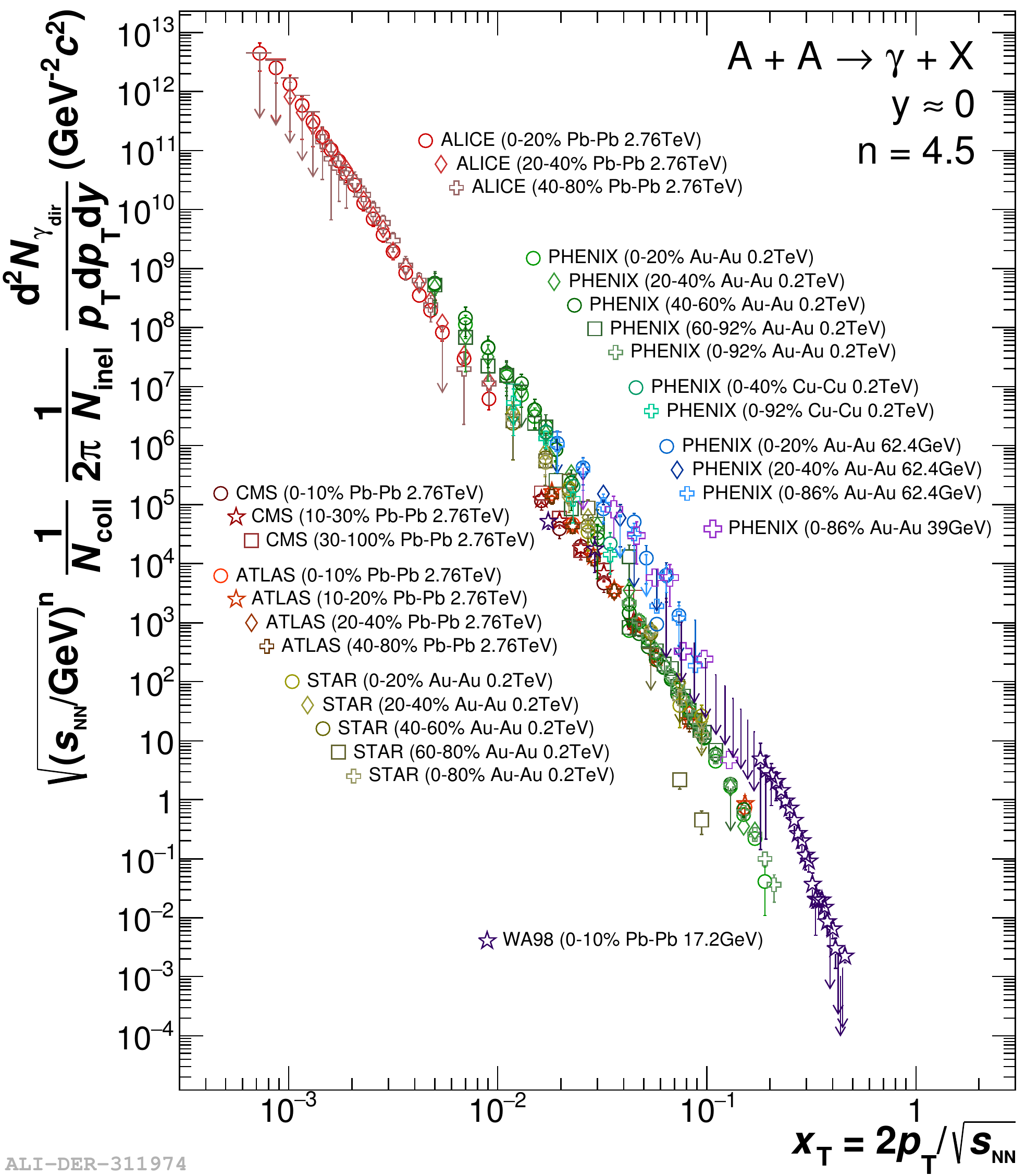}
  \caption{\xt scaling of direct photon invariant yields at
    mid-rapidity  in \AA collisions at
    different \sqsn and centralities.
    (Figure courtesy of Friederike Bock.)
  }
    \label{fig:bock_AA_yield_xT}
\end{figure}
\end{center}

The first measurement of direct photon production at moderately high
\pt\footnote{
For a summary of the published invariant yield measurements see
Table~\ref{tab:aadata}. 
} in 
collisions involving nuclei ($p$+Be at 19.4 and 23.8\,\gev)
was published by Fermilab E95 in 1979~\cite{Baltrusaitis:1979ns} and
used calorimetry to establish the (inclusive) $\gamma$/\piz ratio.  
The authors found an excess of single photons above that was
expected from \piz and $\eta$ decays.  The excess was consistent with
the predictions in~\cite{Ruckl:1978mx}, and the signal increased both with
increasing \pt and $x_F$.  The next two decades were
dominated by the CERN SPS program, with some activity at Fermilab and
BNL AGS, and restricted mostly to fixed target $pA$ 
collisions.

The first direct
photon signal in \AA collisions ($Pb+Pb$ at 17.4\,\gev) that
marginally
reaches into the hard scattering domain was published
by WA98~\cite{Aggarwal:2000th} in 2000\footnote{Photons were measured
  up to 4\,\gevc which can be considered high \pt at that \sqsn.
}, and compared to the scaled \pp yield, 
The authors found that the shape of the \pp
and \pbpb distributions is similar, but the yield in \pbpb is somewhat
enhanced.

The first measurement of direct photon production at truly high \pt 
in collisions involving nuclei ($p$+Be at 530 and 800\,\gevc) was 
published by FNAL E706~\cite{Apanasevich:1997hm,Apanasevich:2004dr} 
and has shown a large
discrepancy between the data and the NLO pQCD calculations,
irrespective of the scales applied.  A similar discrepancy has been
observed in \piz production, too, leading the authors to suggest
adding a supplemental Gaussian transverse momentum smearing to the incoming
partons with $\langle k_T \rangle$ in the 1-1.7\,\gevc range to
describe the data.  Although never officially withdrawn, the validity
of these data has been questioned~\cite{Aurenche:2006vj}.

\begin{table}[h]
  \begin{tabular}{|l|c|c|c|c|c|c|} \hline
   Experiment   &  $\sqrt{s_{NN}}$  &  method  &   $\eta$   &   $p_T$   &
   Publications  & Comment \\ \hline \hline
   E95 / FNAL &  $p+Be$ 19.4, 23.8\,GeV  &  calor.  &   $-1.74<\eta<0$
   &   1.5-4\,GeV/$c$   &   \cite{Baltrusaitis:1979ns} (1979) & $\gamma/\pi^0$  \\ 
   E629 / FNAL &  $p+C$ 19.4\,GeV  &  calor.  &   $-0.75<\eta<0.2$
   &   2.1-5\,GeV/$c$   &   \cite{McLaughlin:1983ba}  (1983) & $\gamma/\pi^0$  \\ 
   NA3 / CERN SPS &  $p+C$ 19.4\,GeV  &  calor.  &   $-0.4<\eta<1.2$
   &   3-5\,GeV/$c$   &   \cite{Badier:1985wg}  (1986) &  spectrum \\ 
   NA34 / CERN SPS &  $p+Be,p+Al$ 29\,GeV  &  calor.  &   $-0.1<\eta<2.9$
   &   0.0-0.1\,GeV/$c$   &   \cite{Schukraft:1988zg} (1989) &  spectrum \\ 
   NA34 / CERN SPS &  $p,O,S+W,Pt$ 19.4\,GeV  &  conv.  &   $1.0<\eta<1.9$
   &   0.1-1.4\,GeV/$c$   &   \cite{Akesson:1989tk} (1990) &  spectrum \\ 
   WA80 / CERN SPS &  $(p,O)+,(C,Au)$ 19\,GeV  &  calor.  &   $1.5<\eta<2.1$
   &   0.4-2.8\,GeV/$c$   &   \cite{Albrecht:1990jq}  (1991) &  $\gamma/\pi^0$ \\ 
   WA80 / CERN SPS &  $S+Au$ 19\,GeV  &  calor.  &   $2.1<\eta<2.9$
   &   0.5-2.5\,GeV/$c$   &   \cite{Albrecht:1995fs}  (1996) &  upp. lim. \\ 
   NA45 / CERN SPS &  $S+Au$ 19\,GeV  &  conv.  &   $2.1<\eta<2.65$
   &   0.4-2\,GeV/$c$   &   \cite{Baur:1995gt} (1996) &  upp. lim. \\ 
   E855 / BNL AGS &  $p+Be,W$ 18\,GeV  &  calor.  &   $-2.4<\eta<0.5$
   &   0.0-1\,GeV/$c$   &   \cite{Tincknell:1996ks} (1996) &  had. Brems. \\ 
   E706 / FNAL &  $p+Be$ 31.8, 38.7\,GeV  &  calor.  &   $|\eta|<0.75$
   &   3.5 - 12\,GeV/$c$   &   \cite{Apanasevich:1997hm} (1998) &  spectra \\ 
   WA98 / CERN SPS &  $Pb+Pb$ 17.4\,GeV  &  calor.  &   $2.35<\eta<2.95$
   &   0.5-4\,GeV/$c$   &   \cite{Aggarwal:2000th} (2000) &  spectra \\ 
   STAR / BNL RHIC  &  $Au+Au$ 130\,GeV  &  conv.  &   $|\eta|<0.5$
   &   1.65-2.4\,GeV/$c$   &   \cite{Adams:2004fm} (2004) & $R_{\gamma}^{-1}$  \\ 
   E706 / FNAL &  $p+Be$ 31.8, 38.7\,GeV  &  calor.  &   $|\eta|<0.75$
   &   3.5 - 12\,GeV/$c$   &   \cite{Apanasevich:2004dr} (2004) &  spectra \\ 
   WA98 / CERN SPS &  $Pb+Pb$ 17.4\,GeV  &  calor.  &   $2.35<\eta<2.95$
   &   0.1-0.3\,GeV/$c$   &   \cite{Aggarwal:2003zy} (2004) &  $\gamma$ HBT \\ 
   PHENIX / BNL RHIC  &  $Au+Au$ 200\,GeV  &  calor.  &   $|\eta|<0.35$
   &   1-14\,GeV/$c$   &   \cite{Adler:2005ig} (2005) &  spectra \\ 
   PHENIX / BNL RHIC  &  $Au+Au$ 200\,GeV  &  int. conv.  &   $|\eta|<0.35$
   &   1-4.5\,GeV/$c$   &   \cite{Adare:2008ab,Adare:2009qk} (2010) & spectra \\ 
   STAR / BNL RHIC  &  $d+Au$ 200\,GeV  &  calor.  &   $|\eta|<1$
   &   6-14\,GeV/$c$   &   \cite{Abelev:2009hx} (2010)  &  \\ 
   CMS / CERN LHC  &  $Pb+Pb$ 2.76TeV  &  calor.  &   $|\eta|<1.44$
   &   20-80\,GeV   &   \cite{Chatrchyan:2012vq} (2012) & iso. \\ 
   PHENIX / BNL RHIC  &  $d+Au$ 200\,GeV  &  mixed  &   $|\eta|<0.35$
   &   1-17\,GeV/$c$   &   \cite{Adare:2012vn} (2013) &  \\ 
   WA98 / CERN SPS &  $p+(C,Pb)$ 17.4\,GeV  &  calor.  &   $2.3<\eta<3.0$
   &   0.7-3\,GeV/$c$   &   \cite{Aggarwal:2011ns} (2013) &
   upp. lim. \\ 
   PHENIX / BNL RHIC  &  $Au+Au$ 200\,GeV  &  conv.  &   $|\eta|<0.35$
   &   0.6-4\,GeV/$c$   &   \cite{Adare:2014fwh} (2015) &  \\ 
   ALICE / CERN LHC  &  $Pb+Pb$ 2.76\,TeV  &  mixed  &   $|\eta|<0.9$
   &   1-14\,GeV/$c$   &   \cite{Adam:2015lda} (2016) &  \\ 
   ATLAS / CERN LHC  &  $Pb+Pb$ 2.76\,TeV  &  calor.  &   $|\eta|<2.37$
   &   22-280\,GeV/$c$   &   \cite{Aad:2015lcb} (2016) &  \\ 
   STAR / BNL RHIC  &  $Au+Au$ 200\,GeV  &  int. conv.  &   $|\eta|<1$
   &   1-10\,GeV/$c$   &   \cite{STAR:2016use} (2017) &  \\ 
   ATLAS / CERN LHC  &  $p+Pb$ 8.16\,TeV  &  calor.  &   $-2.83<\eta<1.91$
   &   25-500\,GeV/$c$   &   \cite{ATLAS:2017ojy} (2017) & iso.  \\ 
   PHENIX / BNL RHIC  &  $Cu+Cu$ 200\,GeV  &  int. conv.  &   $|\eta|<0.35$
   &   1-4\,GeV/$c$   &   \cite{Adare:2018jsz} (2018) &  \\ 
   \hline
   \end{tabular}
  \vspace{0.3cm}
  \caption{Summary of $pA$ and $AA,AB$ direct photon data (invariant yields).
}
  \label{tab:aadata}
 \end{table}

RHIC enabled for the first time the study of photon production with heavy
ions at \pt that is unquestionably in the pQCD range.  PHENIX measured
direct photons in 200\,\gev \auau up to 14\,\gevc~\cite{Adler:2005ig},
followed by a $d$+Au measurement by STAR~\cite{Abelev:2009hx}.  Apart
of the ``isospin effect'' (see Sec.~\ref{sec:isospin}) the results
were consistent with NLO pQCD
calculations~\cite{Gordon:1993qc} scaled by 
the expected number of binary nucleon-nucleon collisions (see also
Sec.~\ref{sec:rab}).  The calculation used the 
CTEQ6~\cite{Pumplin:2002vw} set of parton distribution functions and
the GRV~\cite{Gluck:1992zx} set of fragmentation functions.
With the start of the LHC heavy ion program both \sqsn and the \pt
range increased considerably, but the measurements at
CMS~\cite{Chatrchyan:2012vq} (\pbpb at 2.76\,TeV up to 80\gevc \pt),
ALICE~\cite{Adam:2015lda} (up to 14\,\gevc) and
ATLAS~\cite{Aad:2015lcb} (up to 280\,\gevc) also agreed well with the
respective NLO pQCD calculations, properly scaled, at least at central
rapidity.  These observations are strong evidence that the dominant
sources of high \pt photons are the same in \AA as is \pp, namely hard
scattering, and those photons are unaffected by any later state or
evolution of the colliding system.  The nuclear modification factor of
high \pt photons is about unity.\footnote{Some small deviations are
  expected and well understood -- see Sec.~\ref{sec:isospin}.
}

\subsubsection{Nuclear modification factor}
\label{sec:rab}

When nuclei $A$ and $B$ collide,
the nuclear modification factor \rab quantifies those 
effects on particle production in hard scattering processes that arise
because entire nuclei collided rather than individual hadrons (like \pp).
Such effects can be the formation of the QGP, in which the hard
scattered parton loses energy (final state or FS effects), but also
the modification of the parton distribution functions in the nuclei
with respect to those of free hadrons (nPDF, an initial state or IS
effect).  For an observable $a$ (like jets, hadrons, photons) the
yield measured in \AB is compared to the yield expected from the
properly scaled number of independent nucleon-nucleon collisions.
The scaling factor is the {\it nuclear overlap function} \TAB,
originated in~\cite{Bialas:1976ed}, or, more often, its equivalent
$\langle \Ncoll \rangle$, the number of binary nucleon-nucleon 
collisions, where
the two quantities are connected by the total inelastic cross section

\begin{equation}
\langle \TAB \rangle = \langle \Ncoll \rangle/\sigma^{inel}_{pp}
\end{equation}

\noindent
\Ncoll can be derived for instance from a Monte-Carlo implementation 
of the Glauber model~\cite{Glauber:1959,Glauber:2006gd,Miller:2007ri}.
The nuclear modification factor is then defined as

\begin{equation}
R^{a}_{AB}(p_T) = 
\frac{(1/N^{evt}_{AB})d^2N^{a}_{AB}/dp_Tdy}
{(\langle N_{coll} \rangle/\sigma^{inel}_{pp})d^2\sigma^{a}_{pp}/dp_Tdy}
\end{equation}

\noindent
where $d^2\sigma^{a}_{pp}/dp_Tdy$ is the measured \pp cross section
for observable $a$ (jets, leading hadrons, photons) and 
$\sigma^{inel}_{pp}$ is the total inelastic \pp cross section.
If the yield of an observable $a$ in \AA collisions is unaffected by
the environment, \ie it is a simple incoherent superposition of the yields from
\Ncoll elementary \pp collisions, \raa would be unity\footnote{While
  the reverse is not necessarily true, it would be a remarkable
  coincidence if mechanisms enhancing and depleting \raa would exactly
  cancel each other's effect over a wide \pt range.
}.  This is exactly the case for high \pt photons.

Contrary to that, already from the very first data taken at RHIC it was found
that \raa is much smaller than unity for high \pt \piz -- their
production at any given \pt is {\it suppressed} in central \auau 
collisions as compared to the \Ncoll-scaled \pp expectation~\cite{Adcox:2001jp}.  
The phenomenon, dubbed as ``jet quenching'', and confirmed by numerous
other measurements of hadrons and jets 
at RHIC and LHC, became a crucial evidence of QGP
formation in heavy ion collisions.  The interpretation was that while
in \AA collisions high \pt partons are still produced at the expected
rate (\Ncoll-scaled \pp), they suffer radiative and collisional
energy loss when moving through the colored QGP medium formed around
them, and fragment into final state particles as a smaller energy
parton would in a \pp collision.\footnote{
This explanation was supported by the observation that the
exponents $n$ at high-\pt of the power-law spectra $p_T^{-n}$ were 
very similar in \pp and \auau, therefore, the difference between the 
\AA and \Ncoll-scaled \pp yields could be interpreted as a
$\delta p_T$ shift in the \pt-scale~\cite{Adare:2015cua}.
}

\begin{figure}[htbp]
  \includegraphics[width=0.45\linewidth]{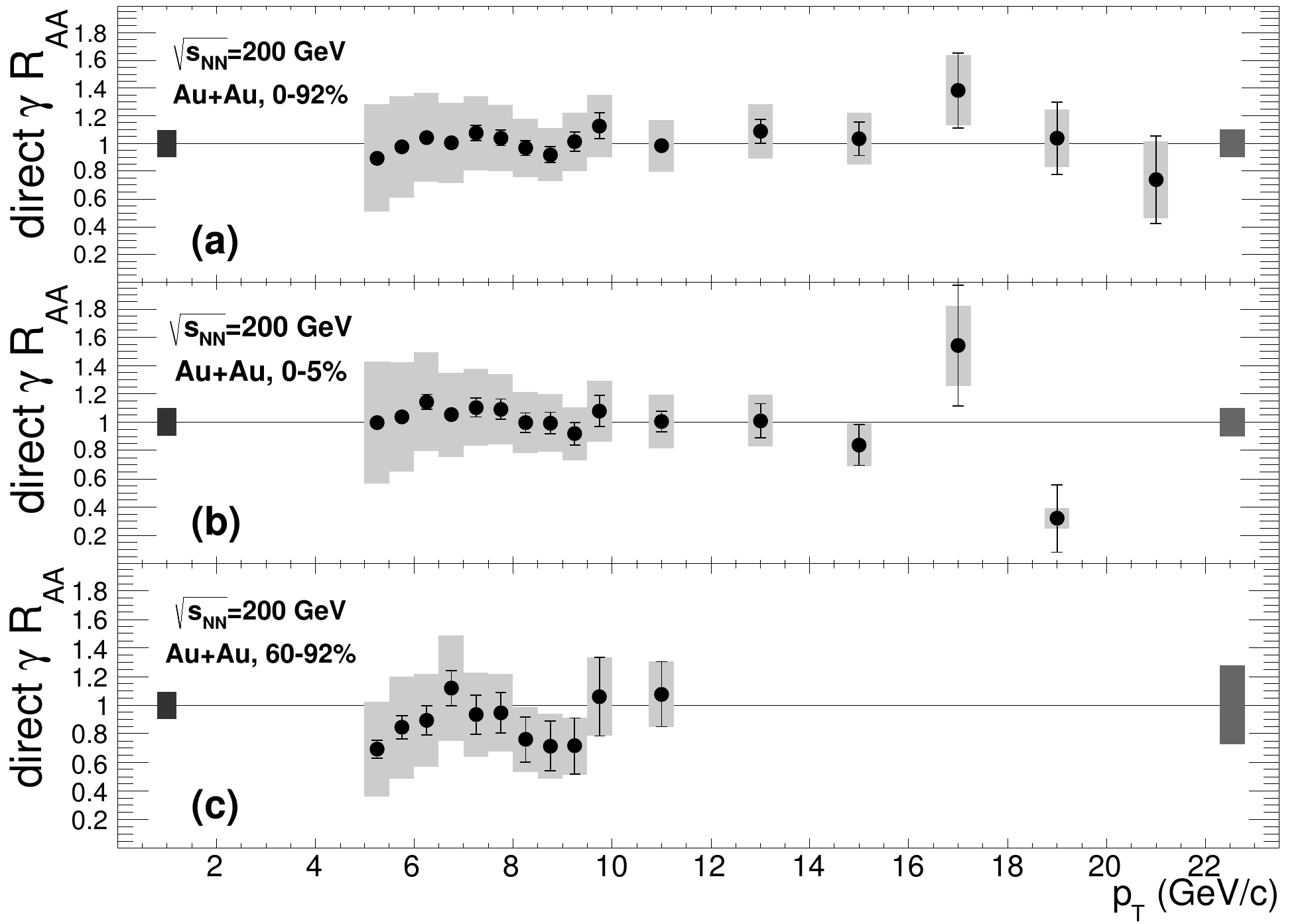}
  \includegraphics[width=0.54\linewidth]{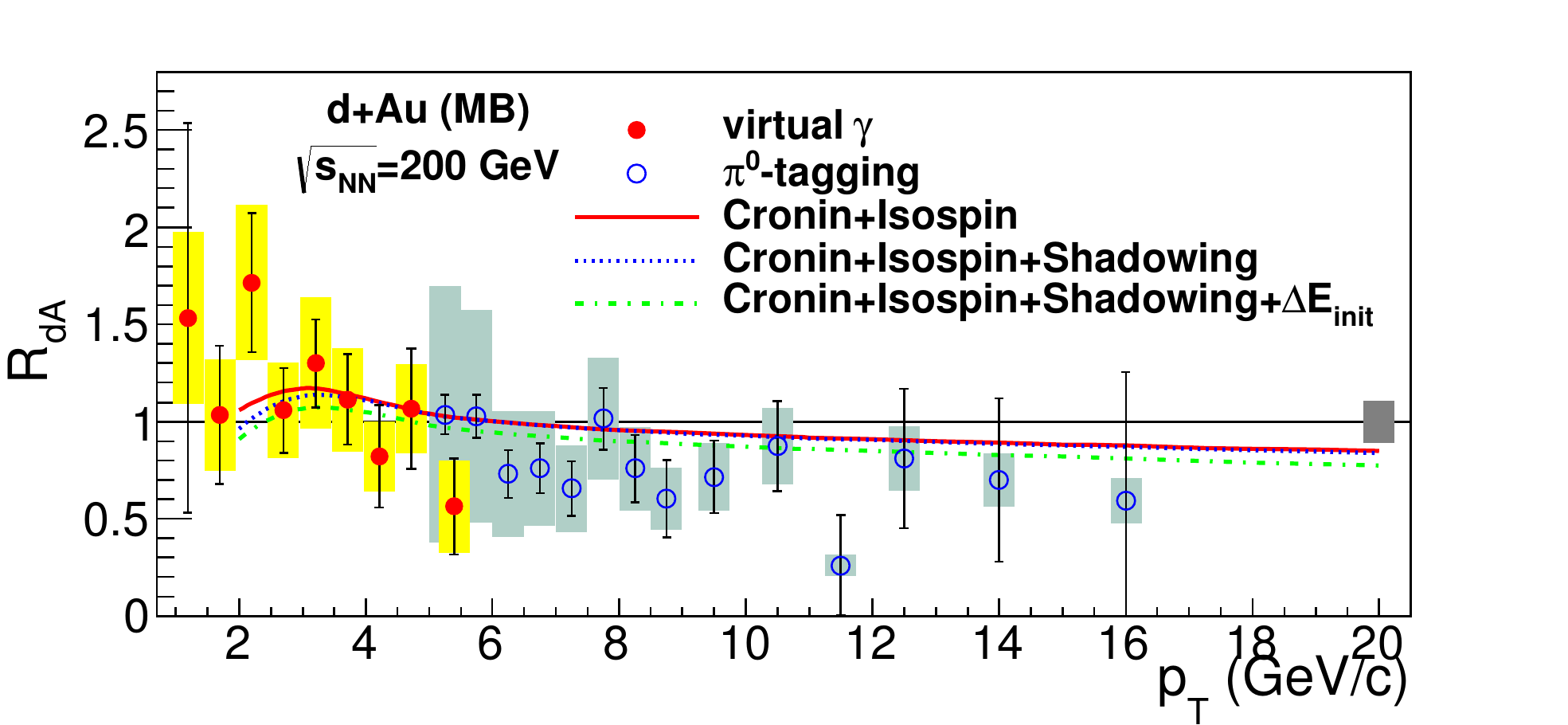}
  \caption{Left: Direct photon nuclear modification factor \raa for three
  different centrality selections in \sqsn=200\,GeV \auau collisions.
  Figure taken from~\cite{Afanasiev:2012dg}.
  Right: direct photon nuclear modification factor for minimum bias
  $d$+Au collisions at \sqsn=200\,GeV.
  Figure taken from~\cite{Adare:2012vn}, the calculations are
  from~\cite{Vitev:2008vk}. 
  }
    \label{fig:ppg139_140}
\end{figure}

\begin{figure}[htbp]
  \includegraphics[width=0.90\linewidth]{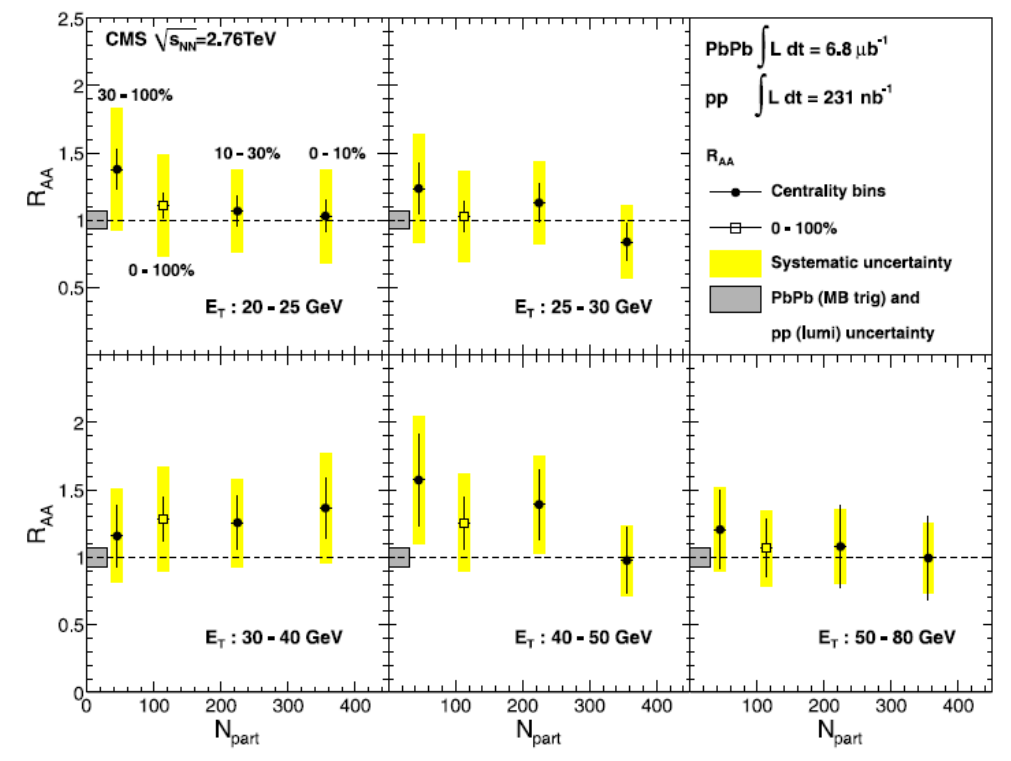}
  \caption{The measured nuclear modification factor \raa for photons 
    as a function
    of \Npart in \snn=2.76\.TeV \pbpb collisions.  Figure taken from
    the CMS publication~\cite{Chatrchyan:2012vq}.
  }
    \label{fig:cms_R_PbPb}
\end{figure}

In Fig.~\ref{fig:ppg139_140} the direct
photon \raa is shown for various centralities in \sqsn=200\,GeV \auau
collisions~\cite{Afanasiev:2012dg}. All \raa-s are consistent with unity within 
uncertainties, as expected, proving that the concept of \Ncoll and the
way it is calculated are sane, at least in the case when two large
ions collide.  Similar results were obtained at the LHC by
CMS~\cite{Chatrchyan:2012vq} (see Fig.~\ref{fig:cms_R_PbPb}), 
ATLAS at mid-rapidity~\cite{Aad:2015lcb} and
ALICE~\cite{Adam:2015lda}: within uncertainties \raa is unity.  To first
order neither enhancement, nor suppression can be observed for
photons, \ie \Ncoll, as calculated in \AA from the Glauber model, is a
meaningful quantity.  Therefore, the observed large suppression of
hadrons and jets is not a methodological artifact.

\begin{figure}[htbp]
  \includegraphics[width=0.90\linewidth]{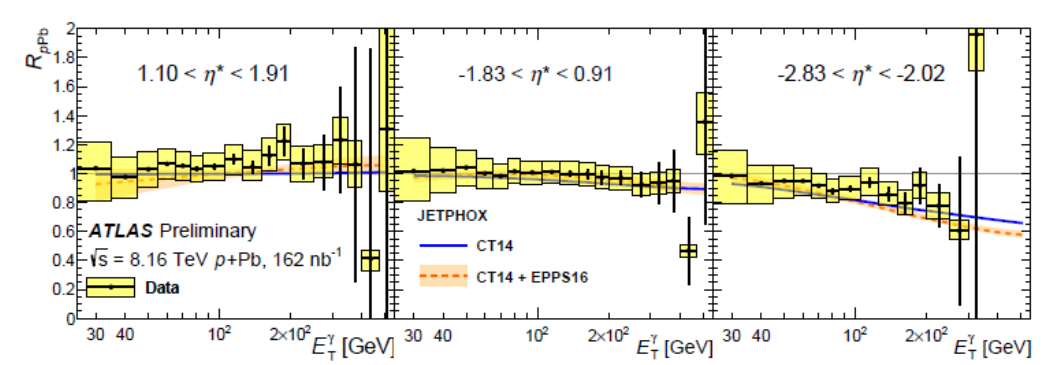}
  \caption{Direct photon $R_{pPb}$ in minimum bias \snn=8.16\,TeV
    $p+$Pb collisions at various pseudorapidities ($\eta<0$ means the
    lead-going direction).  Figure taken from~\cite{ATLAS:2017ojy}
  }
    \label{fig:atlas_R_pPb}
\end{figure}

Moreover, the nuclear modification factor for photons 
{\it at mid-rapidity} is around unity
even in very asymmetric (small-on-large nuclei) collisions, at least
in minimum bias collisions, as seen on the right panel of 
Fig.~\ref{fig:ppg139_140} for \dau collisions at RHIC and
Fig.~\ref{fig:atlas_R_pPb} for $p+$Pb at LHC\footnote{The deviation
  from unity at large rapidities may reflect a nuclear modification of
  the parton densities (nPDF vs free proton PDF), but there are not
  enough data yet to settle this issue, which quite possibly will only
  be resolved at a future electron-ion collider (EIC).
}.
Note that -- in contrast to the \pbpb case -- dependence on collision
centrality is not shown.  In fact, determination of the collision
centrality in small-on-large systems became controversial based on
results shown at the Quark Matter 2012 
conference\footnote{Now often replaced by the (purely experimental) 
  ``event activity'' or some other non-geometric quantity, see for 
  instance~\cite{Adam:2014qja}.  Lately centrality even in the most 
  peripheral \pbpb  collisions is revised in~\cite{Acharya:2018njl}.
}, for an early calculation see Fig. 6 in~\cite{Alvioli:2013vk}.  
We will discuss in more detail this and the role of photons in
eliminating possible centrality biases in Sec.~\ref{sec:standardcandle}.

\subsubsection{Collision centrality -- photons as the standard candle}
\label{sec:standardcandle}

The concept of {\it centrality} is paramount in heavy ion collisions.
Originally, in theoretical calculations, it was defined by the impact
parameter $b$, which in case of large, spherical and identical nuclei is a
well-defined quantity.  Using $b$ and the density
distribution of the nucleons in the nucleus,
the nuclear overlap function and the
average number of participating nucleons (\Npart), binary
nucleon-nucleon collisions (\Ncoll) -- or any other geometric quantity
like eccentricity -- can also be calculated in models~\cite{Miller:2007ri}. 

Experimentally, however, $b$, \Npart, \Ncoll, or any other geometric
quantity, and centrality in general, can not be directly measured.
Instead, events are classified based upon the
distribution of some bulk observable, like transverse energy \Et or
(charged) multiplicity \Nch, occasionally energy at zero degrees 
(attributed to ``spectators'') or
some combination thereof.  Centrality is defined by percentiles
of the total (minimum bias) distribution.\footnote{Counterintuitively,
  the most central - smallest $b$ - 10\% is called 0-10\%, the most
  peripheral 90-100\%.
}
These observables then are linked to the
geometric quantities ($b$, \Npart, \Ncoll, \etc) --  most often by
using some Monte Carlo
implementation of the Glauber-model (for a comprehensive review 
see~\cite{Miller:2007ri}).  The typical Glauber Monte Carlo samples the
distribution of nucleons in the colliding nuclei A and B, and based upon 
the impact parameter $b$ calculates the overlap area/volume ($T_{AB}$).
Next it propagates the nucleons of A and B in this volume on a
straight path   ({\it optical limit}), and using the nucleon-nucleon 
cross section $\sigma_{NN}$ calculates the number of nucleons that
participated in any interaction (\Npart), as well as the total number
of binary nucleon-nucleon interactions (\Ncoll), since a nucleon from 
nucleus A can interact with more than one nucleon of nucleus B and
vice versa.  

There are two crucial and non-trivial assumptions here:
the straight path and the incoherence of the nucleon-nucleon (\NN)
collisions.  Even if a nucleon collides $n$ times, each interaction
happens with the same $\sigma_{NN}$\footnote{In the last few years,
  in light of high precision RHIC and LHC data from very asymmetric
  collisions, like $p$+A, $d$+A, these assumptions have been
  questioned, see below.
}.
This is in line with the original Glauber-model (small momentum
exchange in each interaction)\footnote{Recently the original
  assumptions have been relaxed in order to explain unexpected results
  in small-on-large collisions~\cite{Alvioli:2013vk}.
}.

The connection between the observed \Et or \Nch and the theoretical
\Npart is then made by finding a kernel distribution (usually negative
binomial), which, if convolved \Npart times for a specific $b$, and
integrated over $b$ reproduces the observed \Et or \Nch distribution.
This is reasonable because
the bulk observables \Et and \Nch are dominated by contributions
from soft particles and correlated mostly with 
\Npart\footnote{Based on simultaneous studies of \pp, \dau and \auau
  collisions it has been suggested recently that the proper degree of 
  freedom is the number of constituent-quark participants
  $N_{qp}$~\cite{Adler:2013aqf}. 
}.  
On the other hand, rare hard probes (high \pt particles, jets,
originating in initial, high $Q^2$ parton-parton scattering) are
expected to scale with \Ncoll, the effective \NN
luminosity in heavy ion collisions.  This expectation is justified
as long as the individual \NN collisions are incoherent and the cross
section of the hard process in question is
$\sigma_h << \sigma_{NN}$, in other words almost all \NN
collisions are soft, low momentum exchange, and despite the increased
\NN luminosity (\Ncoll) it takes hundreds or thousands of \AA
collisions to have one single hard \NN scattering, still accompanied
by many soft \NN collisions in the same event.  
At RHIC energies and below this condition is usually 
satisfied\footnote{Maybe with the exception of specially
  selected ``extreme'' event classes, like the top 0-1\% centrality
  in~\cite{Aggarwal:2007gw}.
}, 
at least for photon-producing processes\footnote{Remember, their cross
  section os suppressed by a factor of $\alpha_{em}/\alpha_s$.
}, but at higher energies it may be violated in
multiparton interactions (MPI)~\cite{Blok:2017alw}.  

While not directly observable, \Ncoll plays a central role in the
diagnostics of the hot and dense QGP formed in heavy ion 
collisions, as discussed above (Sec.~\ref{sec:rab}).  But
straightforward as it is, the concept of \Ncoll, and its calculated
value at different centralities, hinges upon the assumptions and
actual Monte Carlo implementation of the Glauber-model.  Fortunately,
high \pt direct photons offer a purely experimental sanity check.  
{\it To leading order}, all high \pt isolated photons are produced in initial
hard scattering, like in \pp, where the yields are well understood.
The mean free path $\lambda_{\gamma}$ of the photon in the QGP can be
estimated from the equilibration time of photons in the 
plasma~\cite{wong:1994}

\begin{equation}
\tau_{\gamma} = \frac{9}{10\pi\alpha_{em}\alpha_{s}}
\frac{E_{\gamma}}{T^2}
\frac{e^{E_{\gamma}/T+1}}{e^{E_{\gamma}/T-1}}
\frac{1}{ln(3.7388E_{\gamma}/4\pi\alpha_sT)}.
\end{equation}

With $\alpha_s$=0.4 and $T$=200\,MeV $\tau_{\gamma}$=481\,\fmc
for 2\,\gevc photons and rapidly increasing with $E_{\gamma}$.
This is to be compared with the $O(10)$\,\fmc lifetime of the plasma. 
Therefore, photons from hard scattering leave the collision volume
unaltered.  The rate of hard scattering photons in \AA is \Ncoll
times the rate in \pp, so the nuclear modification factor \raa of
photons will be unity if and only if the Glauber calculation provides
the proper \Ncoll for the given (experimental) event centrality
class. High \pt direct photons should then be the 
{\it ``standard  candles''} when deriving \Ncoll, and in general,
quantities related to collision geometry.

High \pt direct photons in \AA collisions prove that the mapping
between collision geometry and bulk experimental observables via the
Glauber-model works without any obvious problems.  The reason is that 
``in heavy ion collisions, we manipulate the fact that the
majority of the initial-state nucleon-nucleon collisions will be
analogous to MB p+p collisions, with a small perturbation from much
rarer hard interactions''~\cite{Miller:2007ri}.  
The cautious phrasing is warranted, because, to quote Glauber's
original lecture ``{\it ...the approximate wave function (74) is only
  adequate for the treatment of small-angle scattering.  It does not
  contain, in general, a correct estimate of the Fourier amplitudes
  corresponding to large momentum transfer.}''~\cite{Glauber:1959}.
In other words in \AA the original Glauber model works, because in any
particular collision, even if a few nucleons suffer hard scattering
(violating the basic assumptions above), there are many more nucleons
in both nuclei that collide softly and behave like the average minimum
bias \pp.  These ``normal'' collisions then produce sufficient number of
soft particles to make the centrality determination 
{\it essentially correct}, the more so, because the {\it fluctuations}
of soft production in individual nucleon-nucleon
collisions\footnote{Negative binomial distributions, or NBDs.
} are quite large.  The only case when this logic breaks down are
extremely peripheral collisions in which only a few nucleons from both
nuclei interact at all\footnote{Extremely peripheral {\it nuclear}
  collisions should not be confused with ultraperipheral collisions
  (UPCs) discussed in Sec.~\ref{sec:collateral}.   
}.

When very asymmetric systems collide (like \pau) {\it and} a hard
scattering happens, the sole projectile nucleon is necessarily part of it,
suffering large momentum transfer, degrading its ability to produce
soft particles,  and there are
not dozens of other projectile nucleons around that would 
``make up'' with their average collisions for the missing multiplicity
(or any other global observable used to determine centrality).  The
basic conditions of applicability of the Glauber-model in its original
form are violated for the (only) projectile nucleon.  The
result is a potentially serious bias in the experimental determination
of centrality in \pdau (or other very small on large) collisions when
a hard scattering occured in the
event~\cite{David:2014zya,Kordell:2016njg}.  
The reduced multiplicity means that the event tends to be classified
as less central than it should be based on collision geometry,
potentially leading to mistaken claims of suppression of high \pt
hadrons in ``central'' and/or their enhancement in ``peripheral''
\pdau collisions. Moreover, if there is indeed such a bias, it can 
increase with the \pt of the most energetic particle or jet observed:
\rab in ``peripheral'' events will increase with \pt monotonically,
and continuously decrease with \pt in ``central'' events.  Such trends
were clearly seen in early results on leading hadron or jet \rab when
the experiments applied the traditional Glauber-model to determine
centrality in very asymmetric collisions.

One way to circumvent the bias is to categorize events according to the
experimentally measured ``event activity'' rather than the
model-calculated ``centrality'' (this path has been adopted
by the ALICE experiment, see for 
instance~\cite{Adam:2014qja,Acharya:2017okq}).  
This is not just a question of semantics.  ``Centrality'' implies that
all events in the class are in the same impact parameter range, similar
in collision geometry, governed by similar physics and therefore
directly comparable.   ``Event activity'', on the other hand, is a
neutral, purely empirical classifier, allowing the members of the
class to occasionally reflect quite different physics processes.

Another way to eliminate the bias is to assume that the direct photon
\rab at sufficiently high \pt is 
{\it always unity}\footnote{Apart of the well understood effects
  discussed in   Sec.~\ref{sec:isospin}. 
}, 
and any deviation should be treated as a (\pt dependent) bias on how
\Ncoll is calculated, \ie on the centrality
determination, and the centrality redefined
accordingly~\cite{David:2017vvn}.  While feasible, this is a
non-trivial task.  A
practical workaround that leaves the centrality itself biased, but
allows to decouple and study purely final state effects (like jet or
leading hadron suppression) is to use the double ratio
$R_{AB}^{jet,hadron}/R_{AB}^{photon}$, or, in general, the double ratio
formed with electroweak bosons~\cite{Citron:2019txn}.

In any case, we would like to emphasize that -- particularly in very
asymmetric collisions -- high \pt direct photon production should
always be the standard candle with which other high \pt observables 
are calibrated to avoid premature or outright false physics
conclusions.

\subsubsection{Expected deviations of the photon \raa from unity}
\label{sec:isospin}

The argument that the photon \raa proves the validity of the
Glauber-type calculations in heavy ion collisions
hinges upon the assumption that all high \pt
isolated photons are produced in initial hard scattering.
One should also keep in mind that when deriving direct photon \raa, 
all experiments (and many calculations) use the direct photon spectra 
in \pp.  Strictly speaking this is not correct: there are
some minor issues, second order effects to consider.  First, in \AA 
the binary (nucleon-nucleon) collisions are not only \pp, but \pn and
\nn as well.  While irrelevant for strong interactions, this is
important in electromagnetic processes\footnote{Colliding deuteron on
  deuteron at RHIC would offer clearly tagged and separated $pp$, $pn$
  and, most interesting, high energy $nn$ collisions -- 
  a unique opportunity to study isospin effects (and
  lack thereof in strong interactions) at high \sqs.
  While no new physics is expected, it would serve as a powerful
  confirmation of some QCD and QED fundamentals, and if some
  unexpected results were found, they would obviously be of utmost
  interest.  Unfortunately $dd$ collisions didn't happen so far,
  despite repeated suggestions since 2005, including those by the
  author. 
}, 
since the cross sections are
proportional to the squared sum of quark charges 
($\sum e^2_q$, see Eq.~\ref{eq:wongcompton}), and the fraction of $u$
quarks is higher in \pp than in any heavy ion
collision\footnote{Remember, $p$ is $uud$, $n$ is $udd$, and the
  electric charge of $u$ is 2/3, while $d$ is -1/3.
}.  
Therefore,
if hard scattering is indeed the only source of high \pt photons, the
difference between \pp, \pn and \nn collisions decreases the direct
photon \raa ({\it isospin effect}~\cite{Arleo:2006xb}, see also
Fig.~\ref{fig:ppg139_140}, right panel with calculations
from~\cite{Vitev:2008vk}).

Another source of high \pt (nearly) isolated photons is the
interaction of the
hard scattered, fast quark with the (thermalized) QGP medium
({\it jet-photon  conversion}~\cite{Fries:2002kt}).  Setting the parton
masses to zero in Eqs.~\ref{eq:wongcompton} and~\ref{eq:wongannihilation} 
we get for annihilation 

\begin{equation}
d\sigma/dt=8\pi\alpha_{em}\alpha_s e^{2}_q(u/t+t/u)/9s^2
\end{equation}

\noindent
and the contribution is largest when $t\rightarrow 0$ ($p_q \approx p_{\gamma}$)
or $u\rightarrow 0$ ($p_{\bar{q}} \approx p_{\gamma}$).   For the
Compton process 

\begin{equation}
d\sigma/dt=-\pi\alpha_{em}\alpha_s e^{2}_q(u/s+s/u)/3s^2
\end{equation}

\noindent
and the largest contribution comes again from
$u\rightarrow 0$ ($p_q \approx p_{\gamma}$).  In all cases the photon
is mostly collinear with the original $q,\bar{q}$ and provides a
direct measurement of the quark momentum.  In~\cite{Fries:2002kt} the
authors predicted that at RHIC energies jet conversion photons will be
the dominant source of photons up to \pt=8\,\gevc, ``outshining'' prompt
photons from initial hard scattering, but this appears to be an
overestimate.  If it were true, there should be a significant
(factor of 2 or more) enhancement of the photon \raa up to 8\,\gevc,
but it is not observed in the data (see Fig.~\ref{fig:ppg139_140}).  
A more detailed calculation that includes realistic parton energy 
loss\footnote{It is hard to over-emphasize the importance of treating
  \piz (hadrons) and photons simultaneously.
}
before jet-photon conversion~\cite{Turbide:2005fk} finds a 30\% 
reduction of the photon \raa at 
8\,\gevc~\cite{Turbide:2007mi}\footnote{Note that this model has been
  challenged in~\cite{Renk:2013kya}, which claims a much reduced
  jet-photon conversion cross section.
}.
Also,
since the probability of jet-photon conversion increases with the
pathlength of the quark in the medium, such photons should exhibit a
quite unique azimuthal asymmetry.  While the initial prompt photons
are produced uniformly in azimuth, jet-photon conversions should be
enhanced in the direction orthogonal to the reaction plane, resulting
in a quite unusual {\it negative} \vtwo of those high \pt photons.
Unfortunately the effect is small~\cite{Turbide:2005bz} and so far not
observed in the 
data~\cite{Adare:2011zr,Acharya:2018bdy,Khachatryan:2018uzc}, 
nor have jet-photon conversion photon yields been measured yet.
More sophisticated analysis
techniques might help in the future\footnote{Jet-conversion photons
  are isolated or nearly isolated (\ie there is no fully evolved
  jet), and have a negative \vtwo.  The only other process producing 
  isolated photons is initial hard scattering with \vtwo=0 and
  calculable rates.  Studying \vtwo of isolated, medium \pt photons
  could in principle reveal the fraction (and thus the spectrum) of
  jet-conversion photons. 
}.

\subsubsection{Photon - hadron/jet correlations: energy loss in the medium}
\label{sec:eloss}

\begin{figure}[htbp]
  \includegraphics[width=0.49\linewidth]{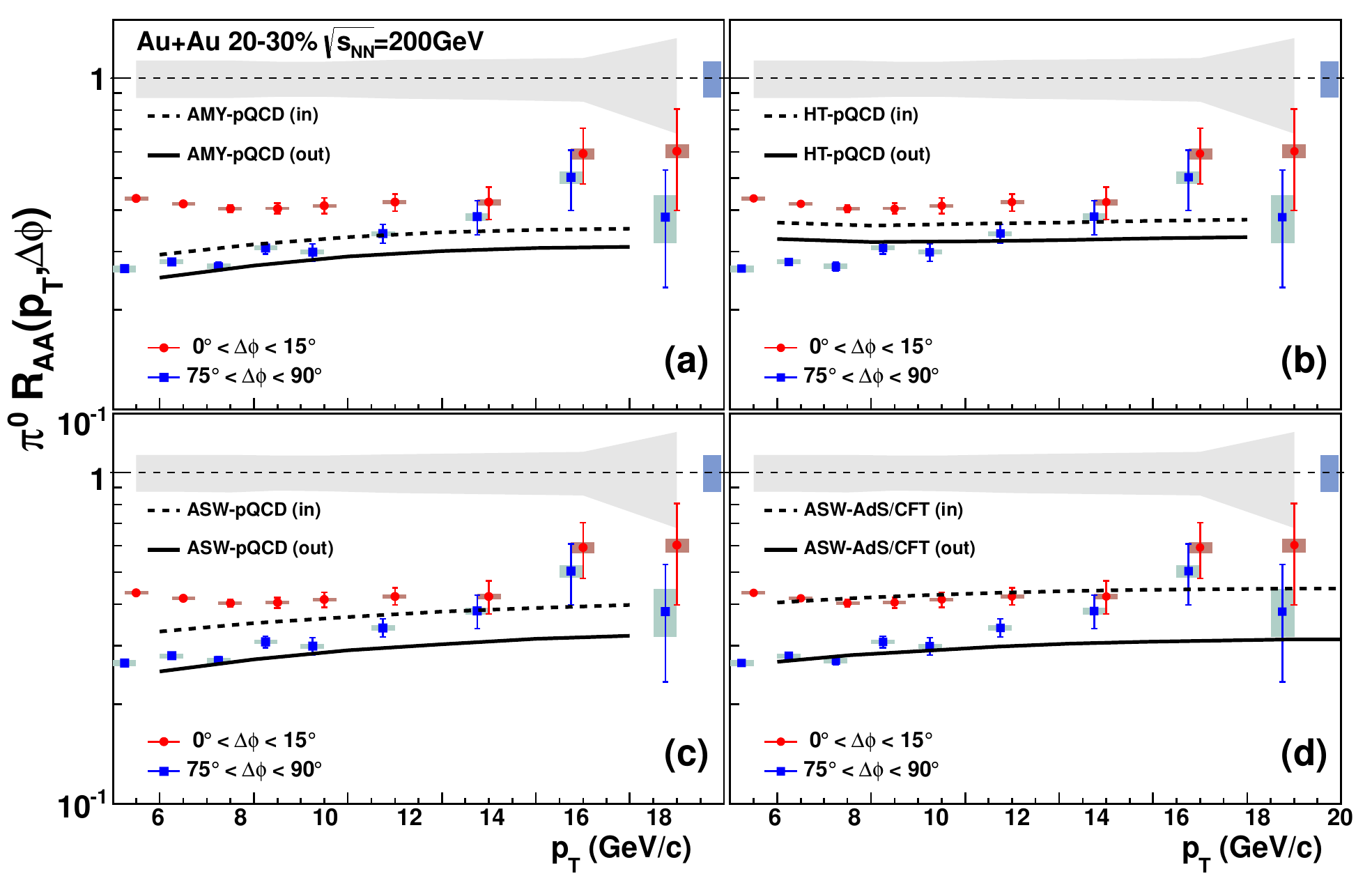}
  \includegraphics[width=0.49\linewidth]{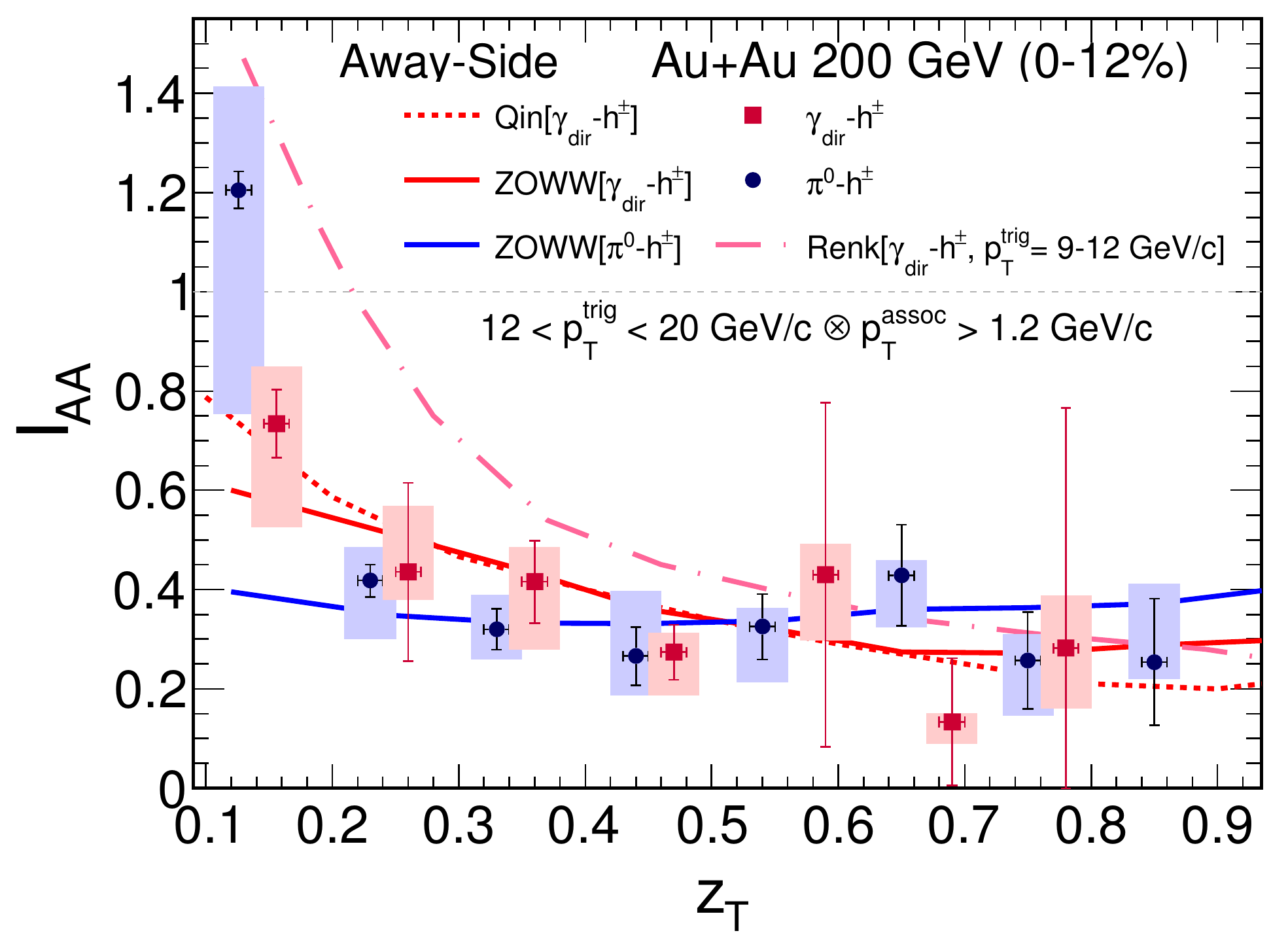}
  \caption{(Left) \raa($\Delta\phi$) as a function of \pt in 20-30\%
    \auau collisions for in-plane (red circles) and out-of-plane (blue
    triangles) compared to four model calculations:
    (a) AMY~\cite{Arnold:2001ba,Arnold:2002ja},
    (b) HT~\cite{Wang:2001ifa},
    (c) ASW~\cite{Salgado:2003gb} and
    ASW-AdS/CFT~\cite{Marquet:2009eq}.
    (Figure taken from~\cite{Adare:2012wg}.)
    (Right) STAR measurements~\cite{STAR:2016jdz} of $I_{AA}^{\gamma_{dir}}$ and
    $I_{AA}^{\pi^{0})}$, the ratio of the per-trigger conditional
    yield in central \auau and \pp collisions, as a function of $z_T$.
    The points for $I_{AA}^{\gamma_{dir}}$ are shifted by 0.03 in
    $z_T$ for visibility.  The curves represent theoretical model
    predictions~\cite{Zhang:2009rn,Renk:2006qg,Qin:2009bk,Chen:2010te}.
    (Figure taken from~\cite{STAR:2016jdz}.) 
}
    \label{fig:STAR_PLB760_IAA}
\end{figure}

Energy loss of hard scattered partons, traversing the medium formed
in heavy-ion collisions, was established early on as the cause of jet
quenching, but the nature of energy loss remained disputed. The
simplest, single-particle observable, the nuclear modification factor
\raa ``integrates'' too many possible effects (surface bias, relative
role of collisional and radiative energy loss, providing little if any
``tomographic'' information~\cite{Renk:2006qg}).  It does not
have good discriminative power among models with widely different
assumptions and mechanisms\footnote{The interested reader can find a
  short but excellent summary of the situation, and many references in
  the introduction of~\cite{Qin:2009bk}.
}.
Measuring \raa \vs the reaction plane~\cite{Afanasiev:2009aa} provides
some more constraint on models,
as do dihadron-correlations~\cite{Adler:2002tq}. 
For instance, in~\cite{Adare:2012wg}
the in-plane and out-of-plane \raa for high \pt \piz is compared
to four model calculations (see Fig.~\ref{fig:STAR_PLB760_IAA}, left plot).  
The first three models
are pQCD-based and exhibit an L$^2$-dependence of the energy loss on
the pathlength in the medium, while the ASW-AdS/CFT model has an
energy loss proportional to L$^3$, which appears to be favored over
the other three scenarios.  Nevertheless, the energy of the parent
parton is still ill-constrained.

This can be remedied by studying back-to-back photon-hadron (or
photon-jet) correlations, since the high \pt trigger photon, being a
penetrating probe, calibrates the initial energy of the recoil parton
before it could lose energy in the medium or start to fragment.  
The direct photon sets the scale of the initial hard scattering.  The
method is illustrated in Fig.~\ref{fig:STAR_PLB760_IAA},
right plot, from~\cite{STAR:2016jdz}, where (associated) 
away-side charged hadron
yields are measured as a function of $z_T=\pt^{assoc}/\pt^{trig}$ 
in \pp and \auau with 
high \pt direct photon and \piz triggers.  While these data are not
definitive, the idea is that
the measured $D(z_T)$ fragmentation functions for photon and \piz 
triggers should be very similar
in \pp, but quite different in \auau, where the trigger \piz now comes
from a parton which already lost energy in the medium.  By forming the
ratio of the conditional yields (having a trigger photon/\piz of a
given \pt)

\begin{equation}
I_{AA} = \frac{D(z_T)^{AuAu}}{D(z_T)^{pp}}
\end{equation}

\noindent
and, in particular, comparing the high- and low-end of the $z_T$
distributions in principle one can discriminate between a
picture where the medium substantially modifies the parton shower
(modified FF) and one where mostly a single parton carries the energy
through the medium and the lost energy shows up only at extremely low
energies and angles~\cite{STAR:2016jdz}.

Analyzing the azimuthal distribution of
the back-to-back photon-jet pair in \pbpb 
CMS concluded~\cite{McGinn:2017opi} that while there is no
indication of in-medium deflection of partons, there is strong parton
energy loss in the medium (jet quenching), that depends on centrality
and photon \pt.  While this observation is not new, photon-jet
correlation data provide strong constraints to in-medium energy loss
models in \AA collisions.  PHENIX and ATLAS also published
photon-hadron and photon-jet measurements (see
Table~\ref{tab:flowplus}).  A recent publication by
ATLAS~\cite{Aaboud:2018anc} on
photon-jet \pt correlations in 5.02\,TeV \pbpb includes many
comparisons to model calculations, none of them being completely
satisfactory.

\subsubsection{Nuclear PDFs and fragmentation function modification}

As pointed out in Sec.~\ref{sec:processes}, the cross-section for hard
scattering processes {\it factorizes} into short distance
(parton-parton scattering) and long distance effects (the incoming
state, \ie the parton distribution functions and the fragmentation
functions of the scattered partons into colorless final particles).
Deep inelastic lepton-nucleus scattering~\cite{Aubert:1983xm,Arneodo:1996ru}
revealed than the PDF of nucleons bound in a nucleus is different from
the ones of free nucleons.  In terms of Bjorken-$x$ the nuclear
effects are~\cite{Eskola:1998iy}: 
\begin{itemize}
\item{depletion at $x<0.1$
(shadowing)} 
\item{excess at $0.1<x<0.3$ (anti-shadowing)} 
\item{depletion at $0.3<x<0.7$ (EMC-effect)} 
\item{excess towards $x\rightarrow 1$ (Fermi motion)}
\end{itemize}
The nuclear PDFs $f_i^{p/A}(x,Q^2)$ for parton species $i$ are defined
relative to the free-proton PDF $f_i^{p}(x,Q^2)$ as

\begin{equation}
f_i^{p/A}(x,Q^2) = R_i^A(x,Q^2)f_i^{p}(x,Q^2)
\end{equation}

\noindent
and usually the modification $R_i^A(x,Q^2)$ is shown.  Ideally, the
free proton and nuclear PDFs should be fitted within the same analysis
to reduce uncertainties~\cite{Eskola:2016oht}, but this hasn't been
done so far.  --  Various sets of nuclear PDFs,
including impact-parameter dependent ones are available and updated
from time to 
time~\cite{Eskola:1998df,Helenius:2012wd,Kovarik:2015cma,Kusina:2016pms}.
The uncertainties are particularly large for the gluon nPDF at low
$x$ (shadowing region)~\cite{Arleo:2011gc}.  As we have seen, photons
are sensitive to the gluon distribution, and the low $x$ region is
experimentally accessible with prompt photon measurements at high
rapidity in $p$A or $d$A collisions\footnote{Selecting a small
  ``projectile'' $p,d$ and a large ``target'' A  and measuring prompt
  photons at high rapidity in the projectile-going direction ensures
  that the parton in the target will have small $x$.   Also, the
  overall multiplicity will be small, making the measurement
  technically feasible, while it would be virtually impossible in an
  AA collision.
}.
While there are ongoing efforts at RHIC and a detector upgrade plan at
LHC~\cite{Cosentino:2017zac},  data on prompt photon production at high 
rapidity are not available yet.  However, inclusive photon
multiplicity measurements at high rapidity, pioneered by
WA93~\cite{Aggarwal:1998bd} at CERN SPS have been also performed by
WA98~\cite{Aggarwal:1999uq},
STAR~\cite{Adamczyk:2014epa} at RHIC and ALICE~\cite{ALICE:2014rma} at
the LHC, which provide some loose constraints on PDFs.

\begin{figure}[htbp]
  \includegraphics[width=0.49\linewidth]{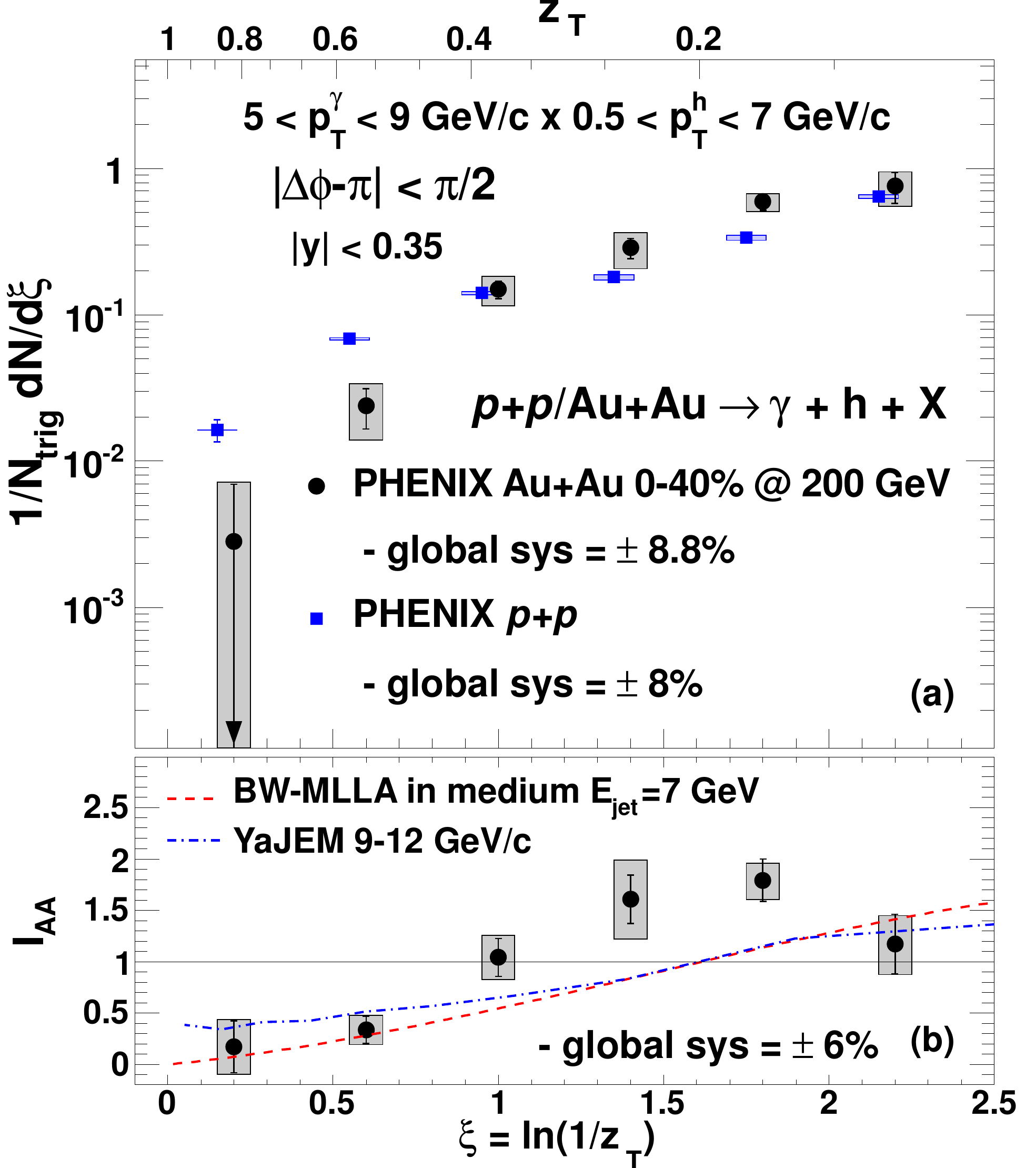}
  \includegraphics[width=0.49\linewidth]{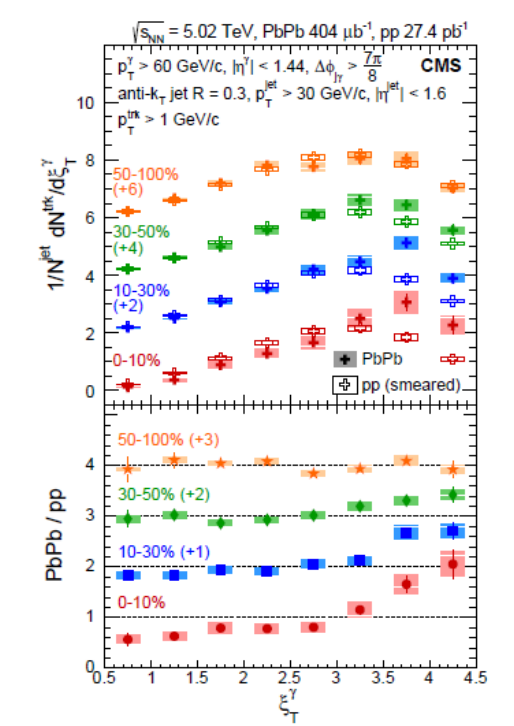}
  \caption{Left: panel (a) shows the charged hadron yield per photon 
    trigger as a function of $\xi = ln(1/z_T)$ for \pp and \auau 
    collisions ($I_{AA}$).  On
    the top the usual $z_T$ scale is also indicated.  Panel (b) shows
    the ratio of the \auau over \pp fragmentation functions.  (Figure
    taken from~\cite{Adare:2012qi}.)  Right top: the centrality
    dependence of $\xi_T^{\gamma}$ for jets associated with an
    isolated photon for \pbpb and \pp collisions.  Right bottom: the
    ratio of \pbpb over \pp distributions, the experimental measure of
    fragmentation function modification.  (Figure taken
    from~\cite{Sirunyan:2018qec}.)
}
    \label{fig:ff}
\end{figure}

Fragmentation functions in hadron-hadron collisions have been
discussed in Sec.~\ref{sec:processes} as the probability of a parton
to produce a particular final state particle at a given momentum
fraction $z$ of the original parton momentum.  
The high $z$ part of the FF characterizes the {\it leading particle}
of the jet.  As we have seen, in
back-to-back isolated photon-hadron correlations the photon
served as the measure of the original parton momentum, setting the jet
energy scale when calculating the (transverse) momentum fraction
$z_T=p^{h}_{T}/p_T^{\gamma}$  of an observed
final state hadron.  This picture implies that fragmentation occurs
fully in the vacuum.  In heavy ion collisions the situation is somewhat
less clear, because the parent parton usually loses energy in the
medium, for instance by radiating gluons or photons.  The opposing
isolated photon still measures the initial momentum of the parton,
but the experimentally observed ``fragmentation function'' is now
depleted at high $z_T$ due to the parton energy loss before
fragmentation.  The low $z_T$ behavior is less trivial: the lost
energy can go into very soft and diffuse modes (no appreciable
enhancement at low $z_T$ in the jet cone), or it can still be
spatially strongly correlated with the jet, in which case it is
interpreted as an in-medium modification of the parton 
shower~\cite{Renk:2009ur}.  It turns out that such enhancement
is manifest both at RHIC~\cite{Adare:2012qi,STAR:2016jdz} and at the
LHC~\cite{Sirunyan:2018qec}, as seen in Fig.~\ref{fig:ff}, where the
results are shown as a function of $\xi = ln(1/z_T)$ in order to
emphasize the soft part (large $\xi$).  As pointed out
in~\cite{Kharzeev:2012re}, the turn-over at the largest $\xi$ values
is due to experimental cutoff, below which particles are not counted
as parts of the jet.  We should also mention that the picture in which
the parton shower is modified, but hadronization still occurs in the
vacuum is not unique: for instance in~\cite{Kopeliovich:2008uy} an
alternate model with (colorless) prehadrons produced both inside and
outside the medium is proposed.

\subsubsection{A collateral benefit: photon-photon scattering}
\label{sec:collateral}

In classical electrodynamics the Maxwell-equations prohibit
photon-photon interactions.  Inspired by Dirac's positron theory
in 1936 Heisenberg published a seminal paper~\cite{Heisenberg:1996bt}
describing the process of photon-photon scattering (``Streuung von
Licht an Licht''), enabled by a fundamentally new property of quantum
electrodynamics (``grunds\"atzlich neuen Z\"ugen der
Quantenelektrodynamik''), the polarization of the vacuum. One of the
related phenomena, elastic scattering of a photon in the Coulomb filed
of a nucleus via virtual $e^+e^-$ scattering (Delbr\"uck scattering)
has actually been observed~\cite{meitner:1933} {\it before} the
Heisenberg paper and turned out to play an important role in the study
of nuclear structure (for a review see~\cite{Milstein:1994zz}), but
observation of other consequences, like photon splitting in a strong
magnetic field~\cite{Adler:1971wn} or $\gamma\gamma\rightarrow l^+l^-$
and $\gamma\gamma\rightarrow \gamma\gamma$ scattering remained elusive
for a long time.

Addressing a completely different problem, excitation of an atom by a
charged particle passing nearby, in 1925 Fermi realized that the
Fourier-transform of the fast moving electric field is equivalent to a
continuous distribution of photons (``Se noi, con un integrale di
Fourier, decomponiamo questo campo nelle sue componenti armoniche
riscontriamo che esso e eguale al campo elettrico che vi sarebbe in
quel punto se esso fosse colpito da della luce con una conveniente
distribuzione continua di frequenze.''), {\it de facto} laying the
foundation of the equivalent photon approximation (EPA) by
Weizs\"acker~\cite{vonWeizsacker:1934nji} and
Williams~\cite{Williams:1934ad}. With the advent of accelerators
intensive photon sources became in principle available, but the much
coveted photon-photon scattering has not been observed for many more
decades.  In 1967 in a feasibility study of such
experiments~\cite{Csonka:1967ua} the author, after enumerating other
methods, even mentions the possibility of two synchronized underground
nuclear explosions producing instantaneous radiation of
$5\times10^{24} \gamma$ for about 70\,ns (with the instantaneous
neutrons arriving only after 1\,$\mu$s, leaving time to collect and 
transmit the data), and the shock waves arriving even later.  Luckily
the author himself cautions wisely in a footnote: ``The practice of
performing this experiment, in conventional laboratories, should be
discouraged because of the detrimental effect it may have on the
personnel, equipment and general appearance of the neighbourhood.''

The pursuit of colliding heavy ions at relativistic energies had little
if anything  to do with the quest for photon-photon scattering 
experiments -- but as a ``collateral benefit'' it made them possible.
The electromagnetic fields of the highly accelerated charges can be
treated in the EPA as photon beams~\cite{Brodsky:1970vk,Brodsky:1971ud} of
small virtuality ($-Q^2<1/R^2$ where $R$ is the charge 
radius), with an $E_{\gamma}^{-1}$ fall-off up to
$\omega_{max} \approx \sqrt{s_{NN}}/(2MR)$ with $M$ being the mass of
the particle or ion~\cite{dEnterria:2013zqi}.  While the collision rate
and the photon spectrum are harder in \pp, the $\gamma\gamma$
luminosities increase as $Z^4$, strongly favoring heavy ion beams.
Study of so-called 
{\it ultraperipheral  collisions}\footnote{Ultraperipheral collisions
  are defined by the impact parameter larger than the sum of radii of
  the two ions.  The interaction can involve either one photon and a
  nucleus, or two photons only.  
} or UPCs quickly became an important topic both at RHIC and LHC, like
photonuclear production of $J/\Psi$ in ultraperipheral \auau
collisions in  PHENIX~\cite{Afanasiev:2009hy}, or \pbpb collisions in
ALICE~\cite{Abelev:2012ba}.  STAR studied early on
$e^+e^-$ production~\cite{Adams:2004rz},
then the photonuclear
production of $\pi^+\pi^-\pi^+\pi^-$ in \auau
collisions~\cite{Abelev:2009aa}. As for pure photon-photon collisions CMS
studied for instance the 
$\gamma\gamma\rightarrow\mu^+\mu^-$~\cite{Chatrchyan:2011ci} and
$\gamma\gamma\rightarrow e^+e^-$~\cite{Chatrchyan:2012tv} as well as 
$\gamma\gamma\rightarrow W^+W^-$~\cite{Khachatryan:2016mud}. Similar
dilepton measurements by ATLAS have been published in~\cite{Aad:2015bwa},
and the $W$ results in~\cite{Aaboud:2016dkv}.  The interested reader can
find recent results and references in the presentations of the Photon
2017 conference~\cite{photon:2017}.  As for the early developments, a
very thorough historic overview has been published
recently~\cite{Scharnhorst:2017wzh} with the first reference dating 
from 1871! 

\section{\bf ``Thermal'' radiation (low \pt region)}
\label{sec:thermal}

\subsection{SPS, FNAL and AGS}
\label{sec:spsags}

The fixed target heavy ion program at CERN SPS started in 1986 and
soon the search for thermal radiation from the system formed in
nucleus-nucleus collisions was underway.
The first results on inclusive photon production and its comparison to
the expected hadron decay yield in \pA and \AA collisions were
published by the HELIOS/NA34 collaboration~\cite{Akesson:1989tk}.  
Protons, $^{16}$O and $^{32}$S beams were used on W and Pt targets,
and photons were measured with the (external) conversion technique
(Sec.~\ref{sec:expconv}).  The ratio of inclusive photons to those
expected from hadron decays remained unity within uncertainties in the
entire measured \pt range ($0.1<p_T<1.5$\,\gevc), indicating no extra
(``thermal'') source.  Independently, the \pt-integrated yields of
inclusive photons were compared to the expected decay photon yields as
a function of \meanet (a proxy for charged multiplicity $N_{ch}$ or 
collision centrality).  The argument was that while decay photons
should be proportional to $N_{ch}$ ($\propto$ number of final state 
particles), thermal radiation may have a quadratic dependence on
$N_{ch}$~\cite{Cerny:1985gt}.  Such excess has not been observed, the inclusive/decay
photon ratio was constant within uncertainties for all colliding
systems and \meanet, moreover, the absolute value of the ratio agreed
remarkably well with the expected value if hadronic decays are the
only source~\cite{Akesson:1989tk}.

Just one year later another CERN SPS experiment, WA80 published an
upper limit for thermal direct photon production using 60 and
200\,A\gev proton and $^{16}$O projectiles on C and Au nuclei.
Technically, this measurement was complementary to HELIOS/NA34, 
because it measured real photons in a calorimeter with a charged
particle veto detector in front of it (rather than measuring
photons converted to a dielectron pair)\footnote{Observing the same
  signal independently and with a completely different technique
  obviously promotes credibility of the result -- a lesson we don't
  always seem to take seriously enough.
}.
Different from HELIOS, \piz spectra have been measured explicitely in
the same setup.  These were then used to estimate 
(via \mt scaling) the yields of $\eta$, $\eta^{'}$, $\omega$
when simulating the expected hadron decay photons.  The results were
presented both as invariant yields and $\gamma/\pi^0$ ratios 
(see Sec.~\ref{sec:waystopresent}).  For \pA collisions the inclusive
photon spectra were completely consistent with the expected decay
photon spectra (no direct photons); for $^{16}$O+A collisions there
was a hint but no clear excess either within stated uncertainties: the
publication claimed a 15\% upper limit on the signal for all systems.

The WA80 experiment went through a significant upgrade meant to
greatly reduce systematic uncertainties, and took data with
$^{32}$S beam at 200\,AGeV in 1990.  The double ratios
(see middle panel in Fig.~\ref{fig:gammapinch}) indicated no clear
excess within the (reduced) uncertainties, and made it possible to set
a 15\% upper limit at 90\% confidence level on the invariant excess
photon yield for the most central collisions~\cite{Albrecht:1995fs},
providing a useful constraint on theoretical models.    

Using the same 200\,AGeV $^{32}$S beam on a (segmented) Au target the
CERES/NA45 experiment at CERN SPS found almost exactly the same upper
limit (14\%) for the emission of direct photons in central S+Au
collisions~\cite{Baur:1995gt}.  
CERES measured photons via external conversion, using
two ring imaging Cherenkov detectors (RICH) to identify electrons with
high efficiency.  The shape and the absolute yield of inclusive photon
\pt distributions for minimum bias collisions was well described
within systematic uncertainties by known hadronic sources.  
Collision centrality dependence was also studied (see right panel in
Fig.~\ref{fig:gammapinch}).  The ratio of inclusive photons to charged
hadrons, as a function of charged particle multiplicity, has been
fitted with $C(1+\alpha dN_{ch}/d\eta)$.  Within systematic
uncertainties the slope $\alpha$ was consistent with zero 
($\alpha = (+0.5 \pm 1.4 stat.  ^{+4.5}_{-1.5} syst.) 10^{-4}$), 
meaning
that the number of inclusive photons is proportional to $N_{ch}$ at
all centralities, as expected if hadron decays are the sole source of
photons.  That means that for beams as heavy as $^{32}$S none of the
CERN SPS experiments found direct photons (``thermal'' radiation), the
more surprising because at the same time the same experiments found
clear evidence of enhanced dilepton production\footnote{Triggering
the question by the author ``if virtual photons are really there, 
how come real photons are virtually absent''?
} 
($e^{+}e^{-}$ or $\mu^{+}\mu^{-}$) in the same
collisions.  The issue is directly addressed and a plausible
explanation is given in~\cite{Tserruya:1995mh}, namely, that the
dielectron pair mass cut (200\,MeV/$c^2$), by virtue of eliminating
the largest hadronic sources,  increases the sensitivity of
the virtual photon measurement with respect to real photons by 2-3
orders of magnitude.

In contrast to the vigorous photon/dilepton program at CERN SPS only
one experiment was dedicated to (low \pt) photon search at BNL AGS,
where E855 was looking for soft photons with an 18\,\gevc $p$ beam
hitting Be and W targets~\cite{Tincknell:1996ks}.  The data were taken in
1990, and photons were detected in two small, movable electromagnetic
calorimeters,  both consisting of 19 hexagonal BaF$_2$ scintillating
crystals, each 9.5$X_0$ long.  In their various data taking positions
the two  detectors ultimately covered a wide rapidity range,
$-2.4<y_{NN}<+0.5$.  The inclusive photon spectra were measured up to
1\,\gevc in \pt, in eight bins of rapidity, and compared to the photon
spectra expected from hadron decay.  While the fitted slopes changed
significantly with rapidity, no significant excess was found at any of
the rapidities.  --  Due to the high energy resolution of the
BaF$_2$ array, E855 could also study the very low \pt limit
($p_T<100$\,MeV) and found no excess beyond the expected decay yields
and Bremsstrahlung from charged hadrons~\cite{Tincknell:1996ks}.
The result is consistent with what the HELIOS collaboration found in
450\,\gevc $p$+Be collisions~\cite{Antos:1993wv}.

\begin{figure}[htbp]
  \includegraphics[width=0.48\linewidth]{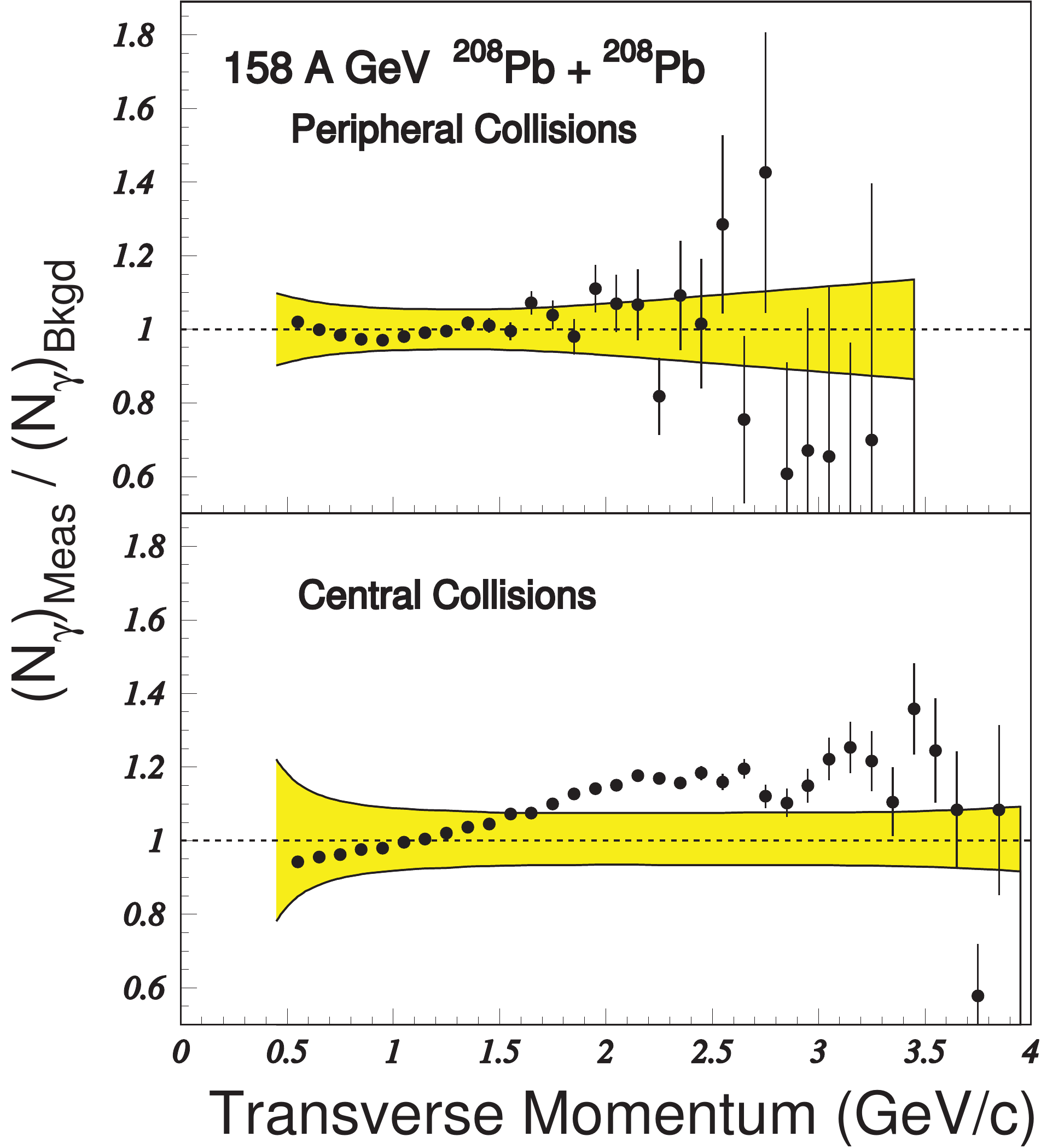}
  \includegraphics[width=0.48\linewidth]{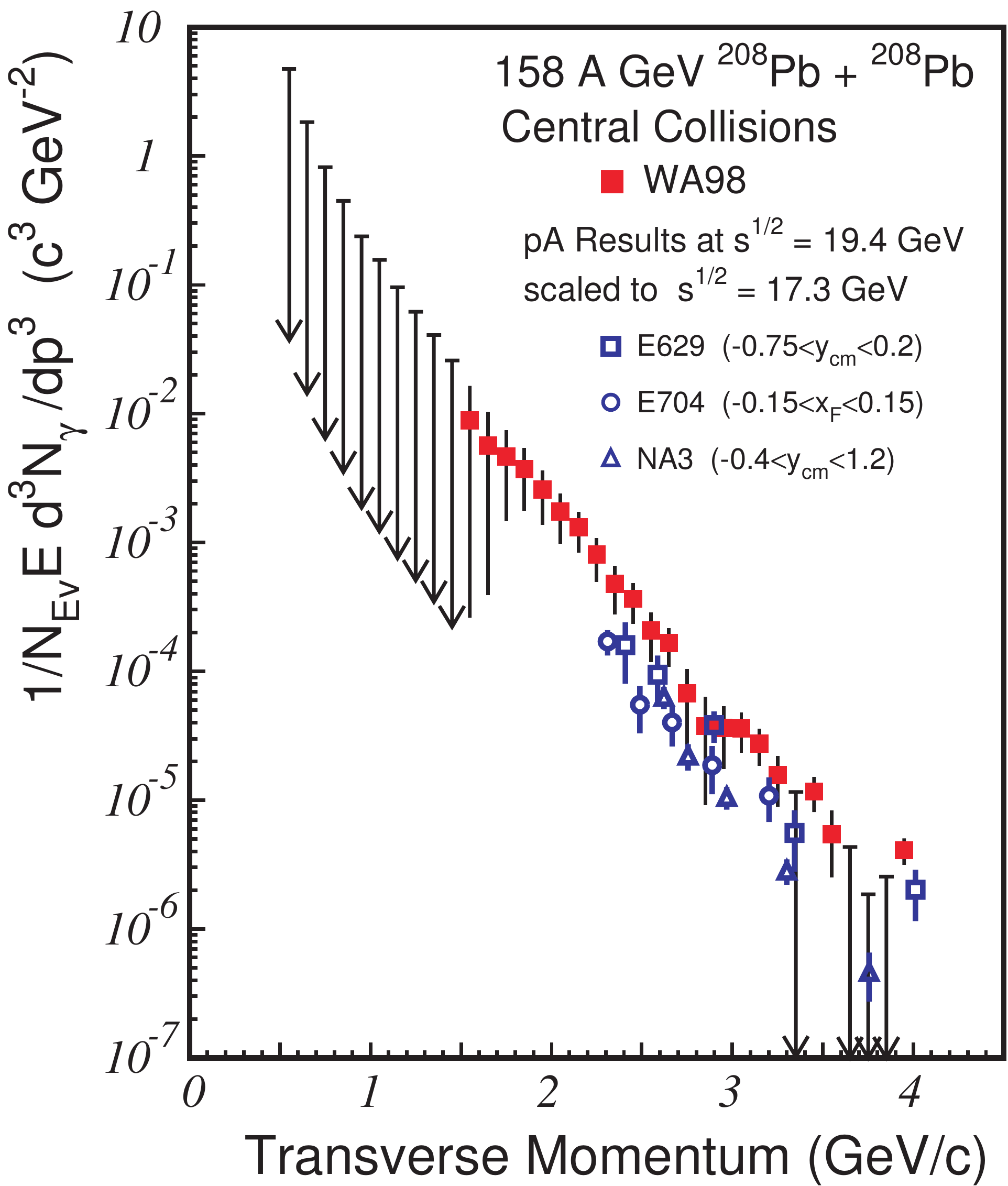}
  \caption{Direct photon result from the WA98 experiment,
    158\,AGeV $^{208}$Pb + $^{208}$Pb collisions~\cite{Aggarwal:2000th}.
    (Left panel) 
    The ratio of measured inclusive photons to calculated decay
  photons as a function of \pt for peripheral (a) and central (b)
  collisions.  Error bars on the
  data are statistical only, the \pt-dependent systematic
  uncertainties are shown as shaded bands.  
  (Right panel) The invariant direct photon yield for central
  collisions.  Error bars indicate
  combined statistical and systematic uncertainties.  Data points
  with downward arrows indicate 90\% C.L. upper limits 
  ($\gamma_{excess} + 1.28\sigma_{upper}$).  The data are compared to
  expected, \Ncoll-scaled \pp yields from three earlier experiments
  (see explanation in the text).
  (Figures taken  from~\cite{Aggarwal:2000th}.) 
  }
    \label{fig:wa98_PbPb_ratio_yield}
\end{figure}

The first positive observation of direct photons in ultrarelativistic
heavy ion collisions was reported by the CERN WA98 experiment 
using 158\,AGeV $^{208}$Pb beams on a Pb
target~\cite{Aggarwal:2000th}, and in the February 10, 2000
announcement at CERN of the discovery of a new state of matter it
served as an ``indication'' (but not the strongest
evidence presented)~\cite{Heinz:2000bk}. 
Photons were measured in the LEDA PbGl calorimeter
supplemented with a charged particle veto in front of it.  In addition
to inclusive photons, WA98 measured simultaneously \piz and $\eta$,
which reduces systematic uncertainties from decay photon subtraction.
The \pt range covered was the largest so far ($0.3<p_T<4.0$\,\gevc,
almost 6 orders of magnitude in invariant yield).
It was found that the ratio of inclusive photons to the expected decay
background exceeded unity at the level of $\approx 2\sigma$ in the
$2.0<p_T<3.5$\,\gevc range in the 10\% most central collisions, while  
no such excess was seen in the 20\% most peripheral collisions (see
Fig.~\ref{fig:wa98_PbPb_ratio_yield}, left panel).  The excess ratio
was used to derive the invariant direct photon yield in central
collisions (see Fig.~\ref{fig:wa98_PbPb_ratio_yield}, right panel).  
The data were compared to the scaled \pp yields from three earlier
experiments\footnote{In essence this was the first attempt to
  establish the nuclear modification factor \raa for photons
  (see Sec.~\ref{sec:aahigh}).
},
FNAL E629~\cite{McLaughlin:1983ba}, using 200\,\gevc $p$ beam on a C
target, FNAL E704~\cite{Adams:1995gg}, using 200\,\gevc $p$ beam on a
$p$ target, and CERN NA3~\cite{Badier:1985wg} also using 200\,\gevc $p$
beam on a C target.  The \pp, \pA data were divided by the \pp
inelastic cross  section (30\,mb) and the mass number of the target to
get the direct photon yield per nucleon-nucleon collisions, then
rescaled for the difference in \sqs, finally multiplied by the
calculated  average number of nucleon-nucleon collisions in central 
Pb+Pb events (660).  WA98 concluded that the shape of the direct
photon spectra in Pb+Pb is similar to that expected from
proton-induced reactions, but the yield is
enhanced~\cite{Aggarwal:2000th}. 

The same 158A\,GeV $^{208}$Pb + $^{208}$Pb data were re-analyzed by
WA98 for two particle correlations of direct photons in central
collisions~\cite{Aggarwal:2003zy}.  In the 0.2-0.3\,\gevc \pt region the
direct photon interferometric radii were quite similar to the pion
radii, indicating that in this \pt region photons are emitted in the
late stage of the collision.  A lower limit on direct photon
production in the same \pt region has been given using the method
described in Sec.~\ref{sec:othertools}; the yield exceeded
predicted yields from the hadron gas.

\begin{table}[h]
  \begin{tabular}{|l|c|c|c|c|c|c|} \hline
   Experiment   &  $\sqrt{s_{NN}}$  &  method  &   $\eta$   &   $p_T$   &
   Publications  & Comment \\ \hline \hline
   UA1 / CERN SPS &  \pbarp 546, 630\,GeV  &  calor.  & $|\eta|<3.0$  
   & 12-30\,\gevc    &   \cite{Albajar:1988im} (1988) &  di-\gam, \kt \\ 
   WA70 / CERN SPS &  $\pi^{-}p$ 280\,GeV  &  calor.  &  
   & 0-3.5\,\gevc    &   \cite{Bonvin:1990kx} (1990) &  di-\gam, \kt \\ 
   CDF / FNAL &  \pbarp 1.8\,TeV  &  calor.  &   $|\eta|<0.9$
   & 10-35\,\gevc    &   \cite{Abe:1992cy} (1993) &  di-\gam, \kt \\ 
   WA93 / CERN SPS &  S+Au 200\,AGeV  &  PMD  &   $2.8<\eta<5.2$
   &     &   \cite{Aggarwal:1998bd} (1998) &  $\gamma$ mult. \\ 
   WA98 / CERN SPS &  Pb+Ni,Nb,Pb 158\,AGeV  &  PMD  &   $2.9<\eta<4.2$
   &     &   \cite{Aggarwal:1999uq} (1999) &  $\gamma$ mult. \\ 
   WA98 / CERN SPS &  \pbpb 158\,AGeV  &  PMD  &   $3.25<\eta<3.75$
   &     &   \cite{Aggarwal:2004zh} (2005) &  $\gamma$ \vtwo \\ 
   PHENIX / RHIC &  \pp 200\,GeV  &  calor.  &   $|\eta|<0.35$
   &   5-15\,GeV/$c$   &   \cite{Adare:2010yw} (2010) &  $\gamma$-jet \kt \\ 
   PHENIX / RHIC &  \auau 200\,GeV  &  calor.  &   $|\eta|<0.35$
   &   1-12\,GeV/$c$   &   \cite{Adare:2011zr} (2012) &  $\gamma$ \vtwo \\ 
   ALICE / LHC &  \pbpb 2.76\,TeV  &  conv.  &   $|\eta|<0.8$
   &   1-5\,GeV/$c$   &   \cite{Lohner:2012ct} (2012) &  $\gamma$ \vtwo \\ 
   ATLAS / CERN LHC &  \pp 7\,TeV  &  calor.  &   $|\eta|<2.37$
   &   0-200\,GeV/$c$   &   \cite{Aad:2011mh} (2012) &  
   di-\gam \\ 
   PHENIX / RHIC &  \auau 200\,\gev  &  calor.  &   $|\eta|<0.35$
   &   5-9\,GeV/$c$   &   \cite{Adare:2012qi} (2013) &  
   $\gamma$ -- hadron \\ 
   ATLAS / CERN LHC &  \pp 7\,TeV  &  calor.  &   $|\eta|<2.37$
   &   50-300\,GeV/$c$   &   \cite{Aad:2013gaa} (2013) &  
   $\gamma$   jet \\ 
   CMS / LHC &  \pp 7\,TeV  &  calor.  &   $|\eta|<2.5$
   &   40-300\,GeV/$c$   &   \cite{Chatrchyan:2013mwa} (2014) &  
   $\gamma$ -- jet  \\ 
   ALICE / LHC &  \pp 0.9-7\,TeV  &  PMD  &   $2.3<\eta<3.9$
   &      &   \cite{ALICE:2014rma} (2015) &  $\gamma$ mult \\ 
   STAR / RHIC &  \auau 200\,GeV  &  comb.  &   $-3.7<\eta<-2.8$
   &      &   \cite{Adamczyk:2014epa} (2015) &  $\gamma$ mult \\ 
   STAR / RHIC &  \pp, \auau 200\,GeV  &  comb.  &   $|\eta^{\gamma}|<0.9$
   &   8-20\,GeV/$c$   &   \cite{STAR:2016jdz} (2016) &  
   $\gamma$ -- jet \\ 
   PHENIX / RHIC &  \auau 200\,GeV  &  comb.  &   $|\eta|<0.35$
   &   1-4\,GeV/$c$   &   \cite{Adare:2015lcd} (2016) &  
   $\gamma$   \vtwo \vthr  \\ 
   PHENIX / RHIC &  \pp 510\,GeV  &  calor.  &   $|\eta|<0.35$
   &   7-15\,GeV/$c$   &   \cite{Adare:2016bug} (2017) &  
   $\gamma$-h $\Delta\Phi$ \\ 
   ATLAS / CERN LHC &  \pp 13\,TeV  &  calor.  &   $|\eta|<3.2$
   &   125-1500\,GeV/$c$   &   \cite{Aaboud:2017kff} (2018) &  
   $\gamma$ -- jet \\ 
   CMS / CERN LHC &  \pbpb 5.02\,TeV  &  calor.  &   $|\eta^{\gamma}|<1.44$
   &   $>$60\,GeV/$c$   &   \cite{Sirunyan:2018qec} (2018) &  
   $\gamma$ -- jet \\ 
   PHENIX / RHIC &  AA 39-2760\,GeV  &    &  
   &   1-15\,GeV/$c$   &   \cite{Adare:2018wgc} (2018) &  
   $\gamma$ scaling  \\ 
   ALICE / CERN LHC &  \pbpb 2.76\,TeV  &  mixed  &   $|\eta^{\gamma}|<0.9$
   &   0.9-6.2\,GeV/$c$   &   \cite{Acharya:2018bdy} (2018) &  
   $\gamma$ \vtwo \\ 
   ATLAS / CERN LHC & \pp \pbpb 5.02\,TeV  &  calor.  &   $|\eta^{\gamma}|<2.37$
   &   63-200\,GeV/$c$   &   \cite{Aaboud:2018anc} (2018) &  
   $\gamma$ -- jet \\ 
   ATLAS / CERN LHC  &  $p+Pb$ 8.16\,TeV  &  calor.  &   $-2.83<\eta<1.90$
   &   20-550\,GeV/$c$   &   \cite{Aaboud:2019tab} (2019) & $R_{pPb}$  \\ 
   \hline
   \end{tabular}
  \vspace{0.3cm}
  \caption{Selected measurements of photon-related observables (other
    than yields)
}
  \label{tab:flowplus}
 \end{table}

\subsection{RHIC and LHC}
\label{sec:rhiclhc}

\begin{figure}[htbp]
  \includegraphics[width=0.49\linewidth]{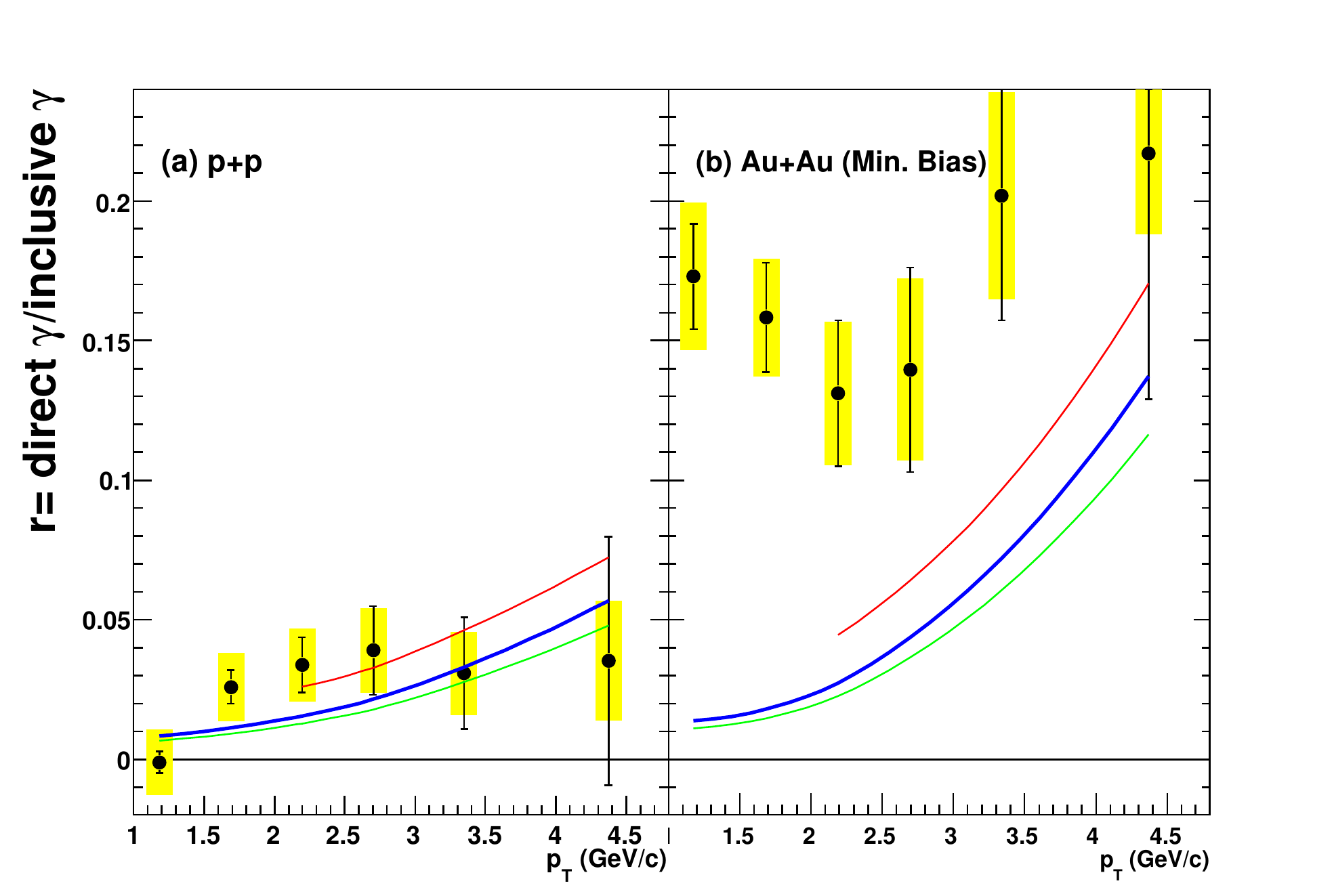}
  \includegraphics[width=0.49\linewidth]{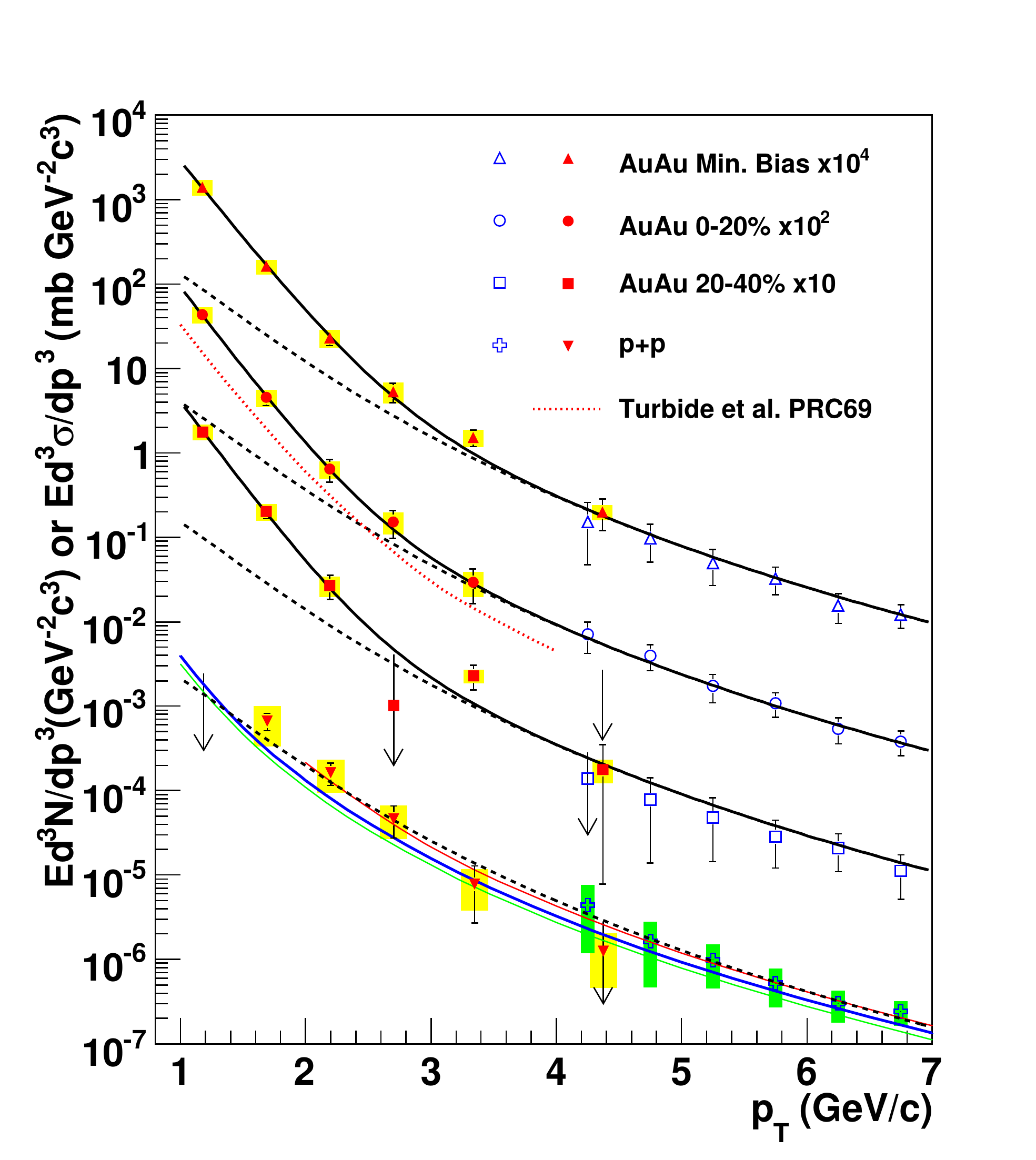}
  \caption{(Left) The fraction $r$ of the direct photon component over
    inclusive photons as a function of \pt in \pp and minimum bias
    \auau collisions.  The curves are NLO pQCD
    calculations~\cite{Gordon:1993qc}. 
    (Right) Invariant cross-section (\pp) and invariant yield (\auau)
    of direct photons as a function of \pt.  The solid markers are
    from~\cite{Adare:2008ab} while the open points
    from~\cite{Adler:2005ig,Adler:2006yt}.  The three curves on the \pp data
    represent NLO pQCD calculations; the dashed curves show a modified
    power-law fit to the \pp data, and the same scaled by \taa.  The
    solid black curves are the sum of an exponential plus the \taa
    scaled \pp fit.  The red dotted curve is a prediction
    from~\cite{Turbide:2003si}. 
    (Figure taken from~\cite{Adare:2008ab}).
  }
    \label{fig:ppg086_fig4}
\end{figure}

The Relativistic Heavy Ion Collider (RHIC) became operational in
2000 colliding \auau at \sqsn=130\,GeV.  Almost immediately two
fundamental observations have been made and published that 
ultimately\footnote{After being confirmed many times by measurements
  at different energies, systems, and ever high statistics.
}
became cornerstones to the claim of discovering the Quark-Gluon
Plasma: the suppression of high \pt hadrons~\cite{Adcox:2001jp} and
elliptic flow of charged hadrons~\cite{Ackermann:2000tr}.  
In contrast results on photons, particularly low \pt, ``thermal''
photons, while shown at conferences as preliminaries starting 2005
and motivating many model calculations, were first published
in a peer-reviewed journal in 2010\footnote{The influential
  2005 paper~\cite{Adler:2005ig} has shown the high \pt direct photon
  yields in \AA collisions and their consistency with \Ncoll
  independent nucleon-nucleon collisions (see Sec.~\ref{sec:aahigh}),
  but it provided only upper limits in the ``thermal'' region due to
  the limitations of the calorimetric measurement.
}~\cite{Adare:2008ab,Adare:2009qk}.  

\subsubsection{``Thermal'' photon yields and
  effective temperatures}
\label{sec:thermalyields}

The first published results~\cite{Adare:2008ab,Adare:2009qk} were obtained with
the internal conversion method (see Sec~\ref{sec:expconv}) by the
PHENIX experiment at RHIC, and are reproduced in
Fig.~\ref{fig:ppg086_fig4}.  Notably, the paper presented results for
200\,GeV \pp and \auau collisions simultaneously.
In \pp the direct photon fraction $r_{\gamma}=r_{direct}/r_{inclusive}$ 
(see Sec.~\ref{sec:waystopresent}) was
consistent with the NLO pQCD calculation\footnote{Note that below
  2\,\gevc the validity of NLO pQCD is questionable, primarily due to
  the poorly constrained yield of fragmentation photons, although they
  are predicted to exceed the prompt photon yield by a factor of
  2-3~\cite{Klasen:2014xfa}. 
}, but for minimum bias \auau a clear excess over primordial photons
has been found.  The absolute yields for various centralities are
shown in the right panel of Fig.~\ref{fig:ppg086_fig4}, with higher
\pt points added from~\cite{Adler:2005ig}. A fit to the \pp spectrum of the
form $A_{pp}(1+p_T^2/b)^{-n}$ and the same fit scaled by the nuclear
overlap function \taa of the respective \auau centrality bin is
superimposed on the data. For \pt$>$4\,\gevc the yields are consistent
with \taa-scaled \pp, allowing little if any additional photon
sources due to the medium.  On the other hand, below \pt$<$4\,\gevc
there is a clear excess, and the shape of the distributions also
changes from power-law ($p_T^{-n}$, characteristic to high \pt, hard
scattering sources) to exponential.  The \auau yields in the entire
\pt-range are reproduced within uncertainties by the two-component fit

\begin{equation}
Ae^{-p_T/T} + T_{AA} A_{pp}(1+p_T^2/b)^{-n}
\end{equation}

\noindent
where the only free parameters are $A$ and the inverse slope $T$ of
the exponential term, which is $T=221 \pm 19^{stat} \pm 19^{sys}$\,MeV
for the most central (0-20\%) collisions~\cite{Adare:2008ab}.
However, the simple picture suggested by the formula above is quite
misleading: there is no reason to equate $T$ with some well-defined
``temperature'' of the system.  

\begin{figure}[htbp]
  \includegraphics[width=0.45\linewidth]{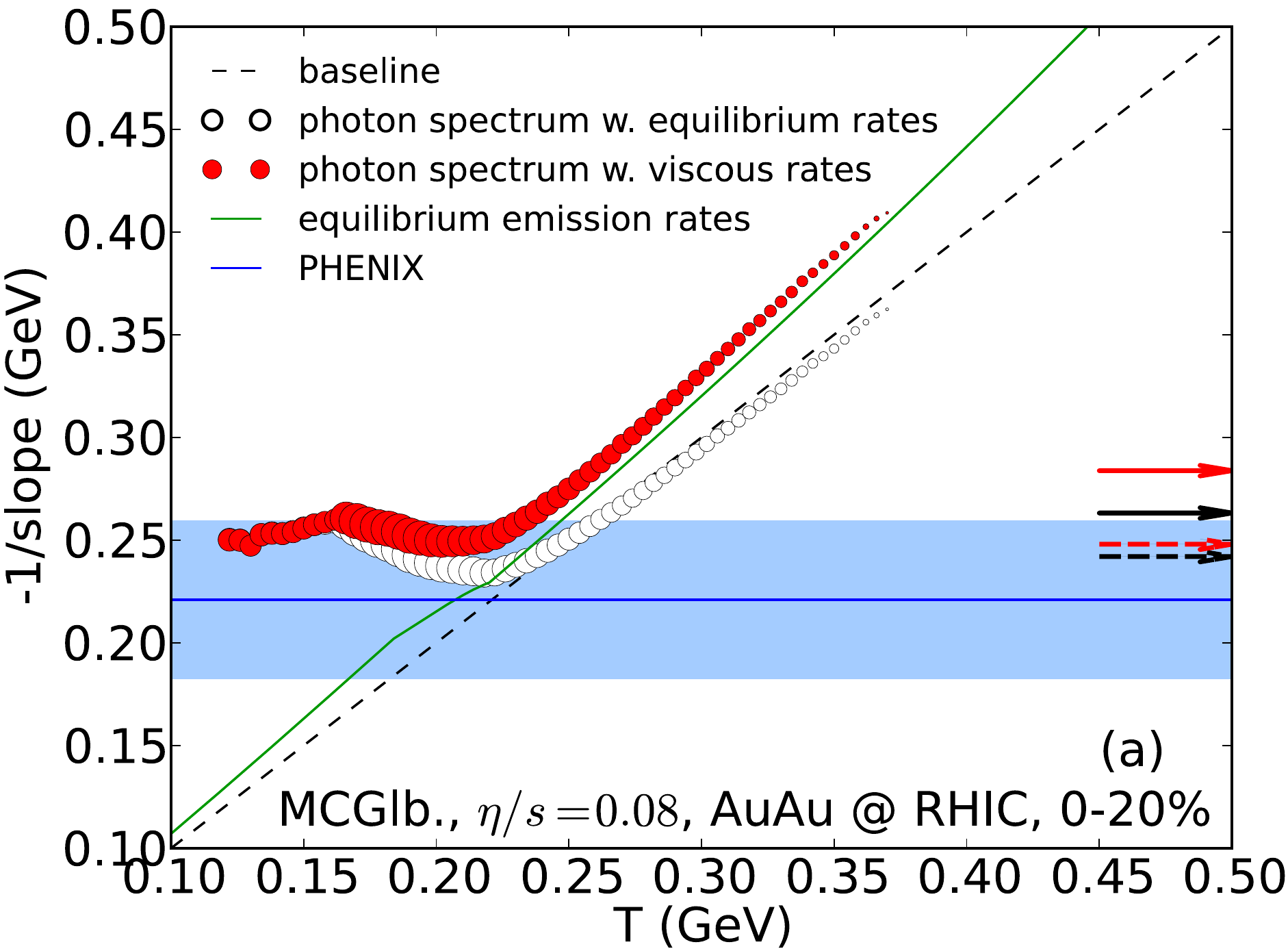}
  \includegraphics[width=0.45\linewidth]{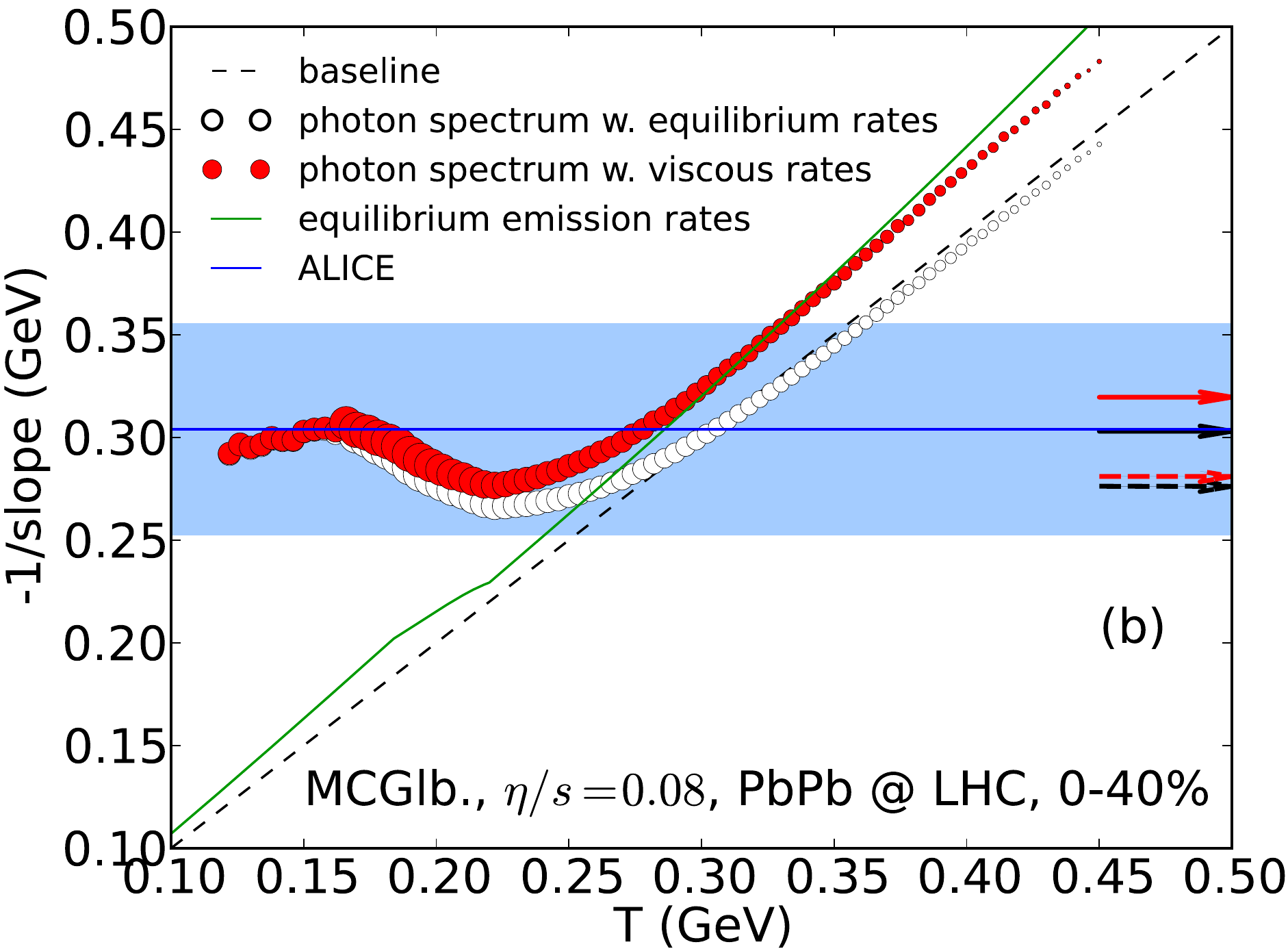}
  \caption{Inverse photon slope parameter $T_{eff}=$-1/slope as a
    function of the local fluid cell temperature, from the equilibrium
    thermal emission rates (solid green lines) and from hydrodynamic
    simulations (open and solid circles)~\cite{Shen:2013vja}, 
    compared to the
    experimental values for (a) \auau collisions at RHIC
    (PHENIX~\cite{Adare:2008ab}) and (b) Pb+Pb collisions at the LHC
    (ALICE~\cite{Wilde:2012wc}.  As explained in~\cite{Shen:2013vja},
    $T_{eff}$ at equilibrium amission rates (green line) is slightly
    higher than the true temperature due to phase-space factors
    associated with the radiation process.
    (Figure taken from~\cite{Shen:2013vja}).
  }
    \label{fig:shen_prc_89_tempevol}
\end{figure}

Even if we assume that the dominant source of excess photons 
in the $1<\pt<4$\,\gevc range is
thermal production from the QGP and/or the hadronic gas\footnote{An
  assumption seriously questioned by some recent models, 
  e.g.~\cite{Cassing:2009vt,Oliva:2017pri,Greif:2016jeb}.
},
due to the ``penetrating probe''
nature of photons the resulting spectrum is a convolution of the {\it entire}
space-time history of the collision and the respective instantaneous
rates. In other words, $T$ reflects some average of radiations from different
temperatures, subject in addition to varying Doppler-shift from the
(time-dependent) radial boost of the system.  The problem is well 
demonstrated in
Fig.~\ref{fig:shen_prc_89_tempevol} in the framework of one particular
hydrodynamic model calculation~\cite{Shen:2013vja}, where the horizontal
axes are the true instantaneous temperatures, the vertical axes are
the observable inverse slopes and the size of the markers reflects the
magnitude of the instantaneous rate.  Different models give different
correlations between the inverse slope and the true instantaneous $T$
(see for instance Fig. 6 in~\cite{Linnyk:2013hta}), but it is clear that
the experimentally measured single inverse slope (often called
``effective'' temperature or $T_{eff}$) in itself provides little
constraint  on the evolution of the system.

\begin{figure}[htbp]
  \includegraphics[width=0.49\linewidth]{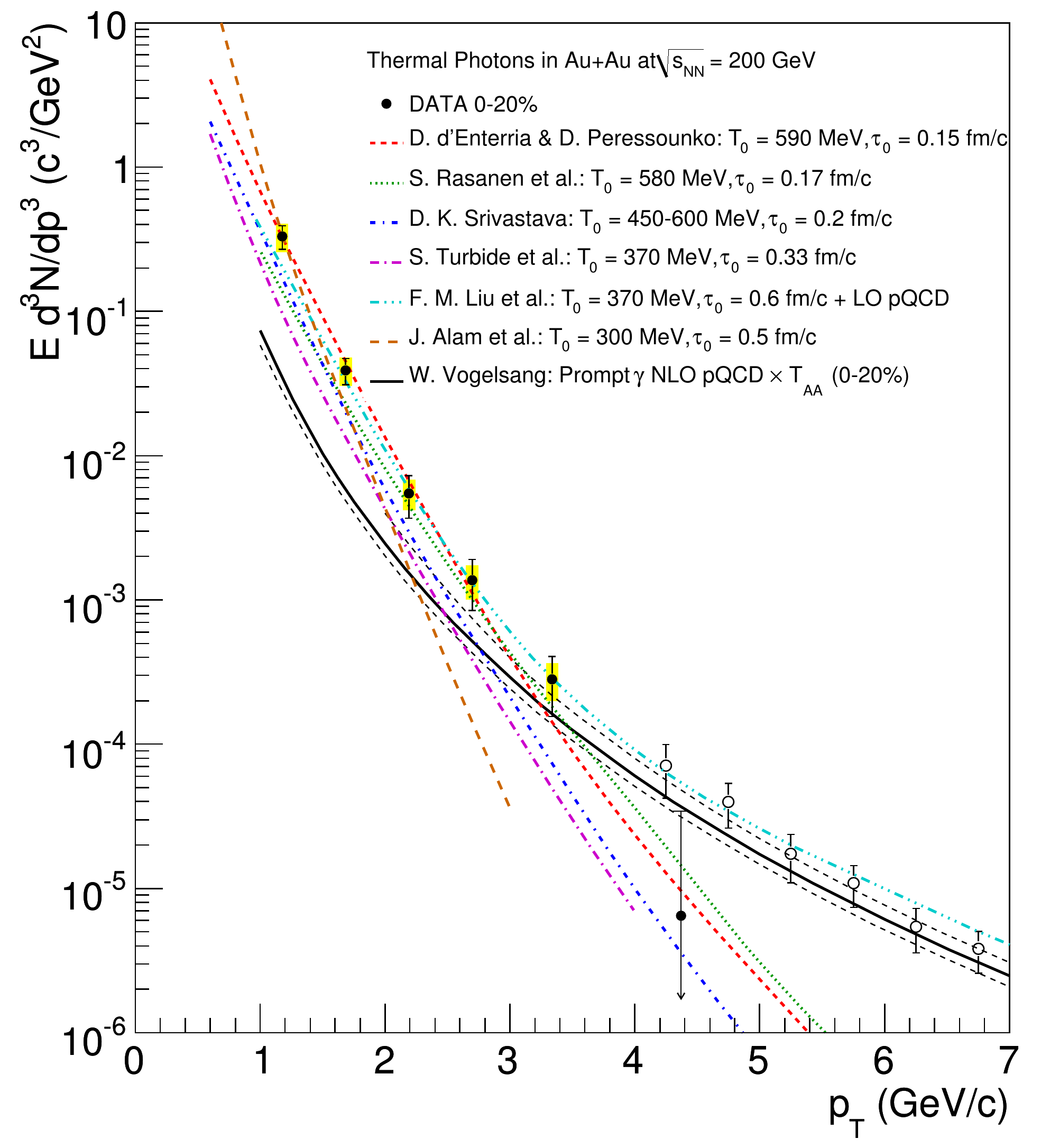}
  \includegraphics[width=0.49\linewidth]{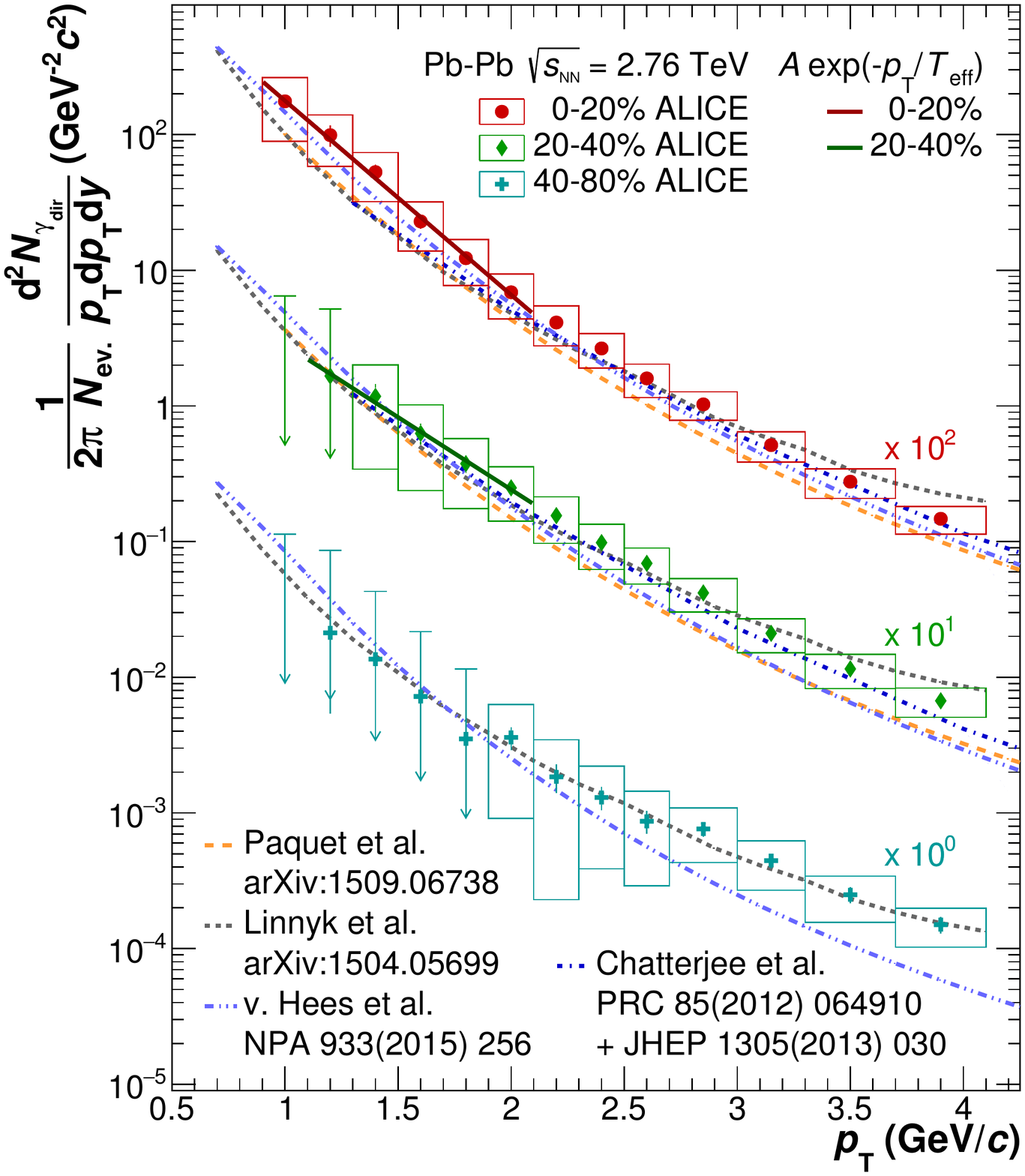}
  \caption{(Left) Theoretical calculations of thermal photon emission are
    compared to direct photon data in central 0-20\% \auau collisions
    by d'Enterria and Peressounko~\cite{dEnterria:2005jvq}, Rasanen 
    {\it et al} (based on~\cite{Huovinen:2001wx}), Srivastava and Sinha 
    (based on~\cite{Srivastava:2000pv}), 
    Turbide {\it et al}~\cite{Turbide:2003si}, 
    Liu {\it et al}~\cite{Liu:2008eh} (this calculation includes pQCD
    contributions), and Alam {\it et al}~\cite{Alam:2000bu}.
    The solid and dotted black lines are pQCD
    calculations~\cite{Gordon:1993qc} varying $\mu$ from 0.5\pt to 2\pt,
    and scaled by \taa.
    (Figure taken from~\cite{Adare:2009qk}.)
    (Right) Direct photon spectra in \pbpb collisons at \sqsn=2.76\,TeV
    measured by the ALICE experiment.  Invariant yields for three
    different centrality classes are shown and compared to model
    calculations~\cite{Paquet:2015lta,vanHees:2014ida,Chatterjee:2012dn,
    Linnyk:2015tha}.
    (Figure taken from~\cite{Adam:2015lda}.)
}
    \label{fig:phenix_alice_gdirect_theories}
\end{figure}

This point is well illustrated in
Fig.~\ref{fig:phenix_alice_gdirect_theories} (left panel)
where various hydro model calculations are overlayed on the 0-20\% 
centrality direct photon data by PHENIX~\cite{Adare:2009qk}.
By varying the
``initial time'' $\tau_0$ (the time when the QGP is formed and system 
can be considered locally thermalized) and the initial temperature
$T_0$ at $\tau_0$ most calculations come reasonably close to the data.  
Note that $T_0$ and $\tau_0$ are anti-correlated: a conservative, late
formation time ($\tau_0$=0.6\,fm/$c$) implies $T_0$=300\,MeV, while
the extreme fast formation ($\tau_0$=0.15\,fm/$c$) would imply
$T_0$=600\,MeV.  While the data could not rule out any of these
scenarios, in~\cite{Adare:2009qk} it has been argued that even the
lowest $T_0$ is well above the $T_c \approx$170\,MeV cross-over transition
temperature from the hadronic phase to the QGP, predicted by lattice
QCD calculations~\cite{Borsanyi:2010bp,Borsanyi:2013bia,Bazavov:2017dsy}.

The first direct photon measurement at the LHC was performed by the
ALICE experiment in \sqsn=2.76\,TeV \pbpb collisions
and first shown as preliminary for the 1$<p_T<$14\,\gevc range in
the 0-40\% centrality bin in 2012.
This analysis has been done with the external conversion technique
(see Sec.~\ref{sec:expconv}), and a clear, exponential excess over 
the NLO pQCD expectation was seen below 4\,\gevc \pt, with an inverse 
slope of 304$\pm$51\,MeV.  For the final publication~\cite{Adam:2015lda}
the analysis has been extended to three centrality bins 
(0-20\%, 20-40\% and 40-80\%), and, more important, a second,
independent measurement with the high resolution, high granularity
PHOS calorimeter has been added.  The agreement of the inclusive
photon spectra between the two methods (external conversion and
calorimetry) was within 1.2 standard deviations, while the double
ratios agreed within 0.4 standard deviation.  The published results
are the error-weighted average of the two independent measurements.

The final ALICE direct photon spectra at low \pt are shown in
Fig.~\ref{fig:phenix_alice_gdirect_theories} (right panel)
and compared to various model
calculations.  In the most central collisions there is a signal down
to \pt =1\,\gevc. The low \pt region 0.9$<$\pt$<$2.1  is fitted with 
an exponential two different ways.  First, the pQCD photons, as
calculated in~\cite{Paquet:2015lta} are subtracted.  In that case
the inverse slope for 0-20\% centrality is 
$T_{eff}$=297$\pm$12$^{stat}\pm$41$^{sys}$), 
close to the preliminary value obtained for 0-40\%, but in the next
centrality bin (20-40\%) $T_{eff}$ is much higher, albeit with very large
uncertainties ($T_{eff}$=410$\pm$84$^{stat}\pm$140$^{sys}$).  
The role of pQCD photons is negligible up to a few \gevc: the inverse
slopes without subtraction are
$T_{eff}$=304$\pm$11$^{stat}\pm$40$^{sys}$
and $T_{eff}$=407$\pm$61$^{stat}\pm$96$^{sys}$ for 0-20\% and 20-40\%,
respectively. Finally, in peripheral collisions (40-80\%) only upper 
limits could be established up to 2\,\gevc.  Also, the spectra in 
Fig.~\ref{fig:phenix_alice_gdirect_theories} (right panel)
are clearly ill-described by a
single exponential in the 0.9$<$\pt$<$4.0\,\gevc range.


Although radically different, the listed models\footnote{For a more
  detailed discussion of the models see Sec.~\ref{sec:puzzle}.
} 
all describe the data within uncertainties.  
Van Hees {\it et al.}~\cite{vanHees:2014ida} uses an updated version of
their thermal fireball model~\cite{vanHees:2011vb} based on ideal
hydrodynamics with initial flow (prior to thermalization).  The
initial time is 
$\tau_0$=0.2\,fm/$c$ with $T_0$=682\,MeV and $T_0$=641\,MeV for the
0-20\% and 20-40\% event classes, respectively.
Chatterjee {\it et al.}~\cite{Chatterjee:2012dn,Chatterjee:2013naa} 
use and event-by-event
(2+1D) longitudinally boost invariant ideal hydrodynamic model with
fluctuating initial conditions.  The initial time is
$\tau_0$=0.14\,fm/$c$ with $T_0$=740\,MeV and $T_0$=680\,MeV for the
0-20\% and 20-40\% event classes, respectively.
Paquet {\it et al.}~\cite{Paquet:2015lta} use
(2+1D) longitudinally boost-invariant viscous
hydrodynamics~\cite{Ryu:2015vwa} with fluctuating, impact-parameter
dependent glasma (IP-Glasma) initial
conditions~\cite{Schenke:2012wb}. The hydro evolution starts at
$\tau_0$=0.4\,fm/$c$ with $T_0$=385\,MeV and $T_0$=350\,MeV for the
0-20\% and 20-40\% event classes, respectively.
For a given $\tau_0$ the initial temperatures are somewhat higher than
for the RHIC data.  In the PHSD model by
Linnyk {\it et al.}~\cite{Linnyk:2015tha} the full evolution of the
system is described microscopically in an off-shell transport
approach, rather than by hydrodynamics.
Given the experimental uncertainties, the data in
Fig.~\ref{fig:phenix_alice_gdirect_theories} (right panel)
can not discriminate
between the models.  This is a common problem so far in the low \pt 
direct photon measurements.

\begin{figure}[htbp]
  \includegraphics[width=0.62\linewidth]{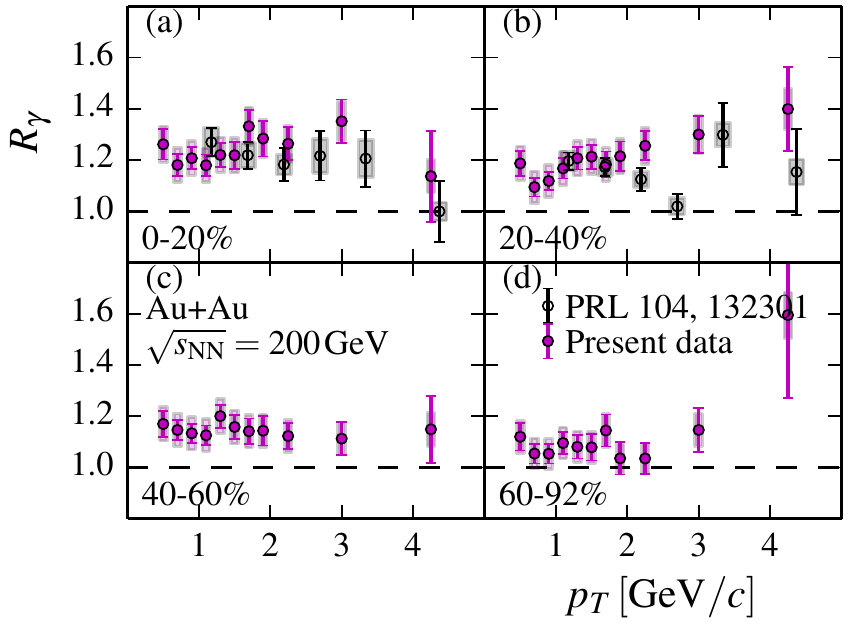}
  \includegraphics[width=0.36\linewidth]{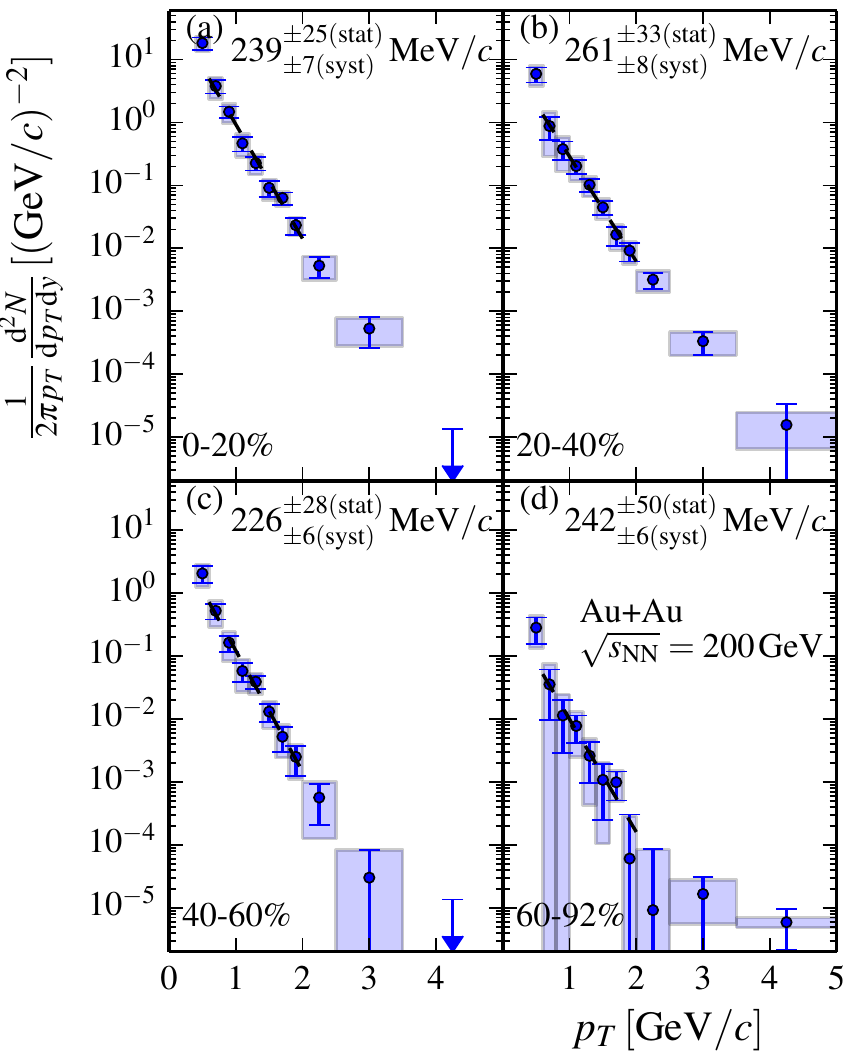}
  \caption{(Left panels) Inclusive over decay photon ratio \rgam in the low
    \pt region for all centralities from the 2007 and 2010
    \sqsn=200\,GeV \auau datasets, analyzed with the external
    conversion method (``Present data'').  For comparison, the same
    quantity obtained earlier with the internal conversion
    method~\cite{Adare:2008ab} is also shown.
    (Right panels) Direct photon \pt spectra after subtraction of the
    \Ncoll-scaled \pp contribution for all centralities, along with an
    exponential fit in the 0.6$<$\pt$<$2.0\,\gevc region.
    (PHENIX data, figure taken from~\cite{Adare:2014fwh}.)
  }
  \label{fig:ppg162_fig6}
\end{figure}

About the same time the PHENIX experiment repeated the measurement of 
low \pt direct photons on a much larger \sqsn=200\,GeV \auau dataset 
(taken 2007 and 2010) with the external conversion 
technique~\cite{Adare:2014fwh}.
The results for all centralities are shown in 
Fig.~\ref{fig:ppg162_fig6}.  The \rgam ratios (inclusive over decay
photons) in the  two most central bins are compared to the
corresponding ratios obtained earlier with the internal conversion
method~\cite{Adare:2008ab}, and found to be consistent within stated
uncertainties, although a $\sim$15\% difference, as predicted
in~\cite{Dusling:2009ej} could not be excluded.  The direct photon yield
can then be obtained from the hadron decay photon yield using

\begin{equation}
\gamma^{direct} = (R_{\gamma} - 1)\gamma^{hadron}.
\end{equation}

Finally, the \Ncoll-scaled \pp contributions\footnote{Measured at and
  above \pt=1.5\,\gevc and extrapolated below.
}
are subtracted, and the resulting spectra shown in
Fig.~\ref{fig:ppg162_fig6} (right panel) along with exponential fits in the
0.6$<$\pt$<$2.0\,\gevc region.  Remarkably, all inverse slopes are
consistent with $\sim$240\,MeV, independent of centrality.  The PHSD
model\footnote{Which is a transport code so it does not have temperature
evolution {\it per se}, but of course an exponential can always be
fitted to the spectra.
}
actually predicts such behavior~\cite{Linnyk:2013wma} and the effective
temperature in the same 0.6$<$\pt$<$2.0\,\gevc region is
$T_{eff}\sim 260\pm 20$\,MeV, very close to the experimental result.
On the other hand, a viscous hydrodynamic calculation~\cite{Shen:2013vja} 
using the VISH2+1 code~\cite{Song:2007ux} for system evolution
predicts a weakly centrality-dependent $T_{eff}$, which is 267\,MeV 
for 0-20\%, dropping monotonically to 225\,MeV for 60-92\% centrality.

\begin{figure}[htbp]
  \includegraphics[width=0.6\linewidth]{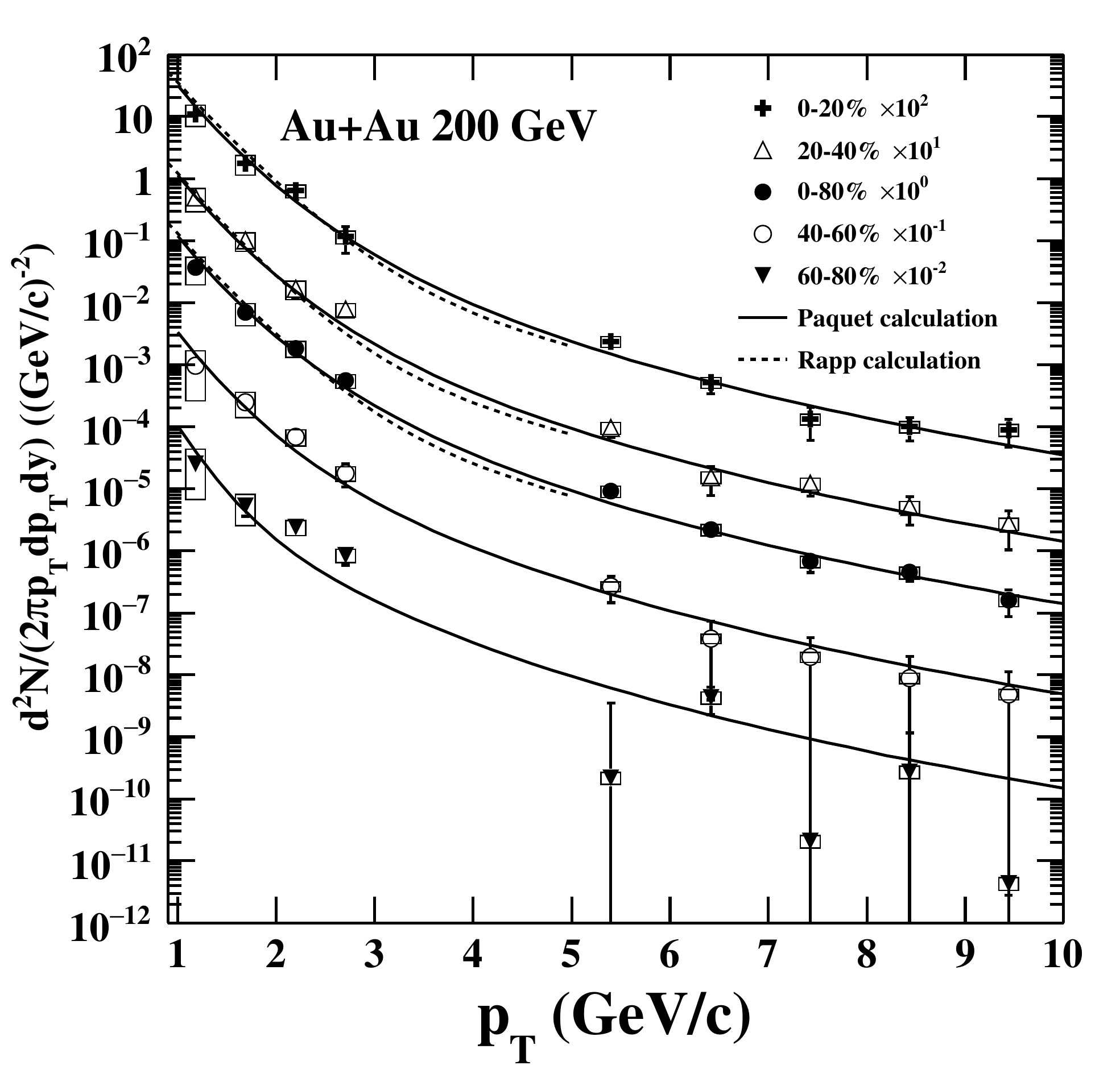}
  \includegraphics[width=0.38\linewidth]{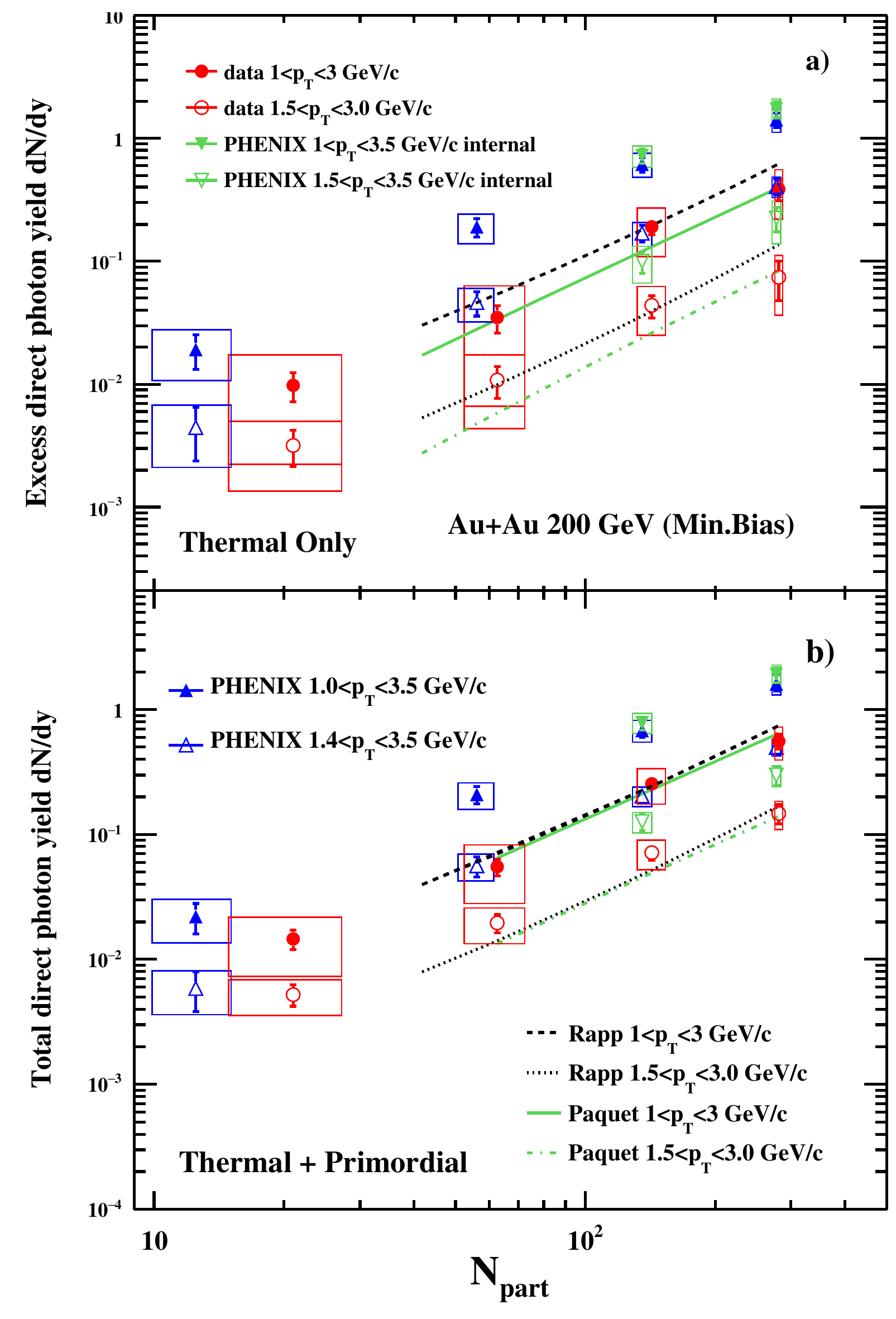}
  \caption{(Left) Direct photon invariant yields as a function of \pt
    in \auau collisions at \sqsntwo by
    STAR~\cite{STAR:2016use}. Results are compared to model
    calculations by van Hees et al.~\cite{vanHees:2011vb,vanHees:2014ida} 
    and Paquet et al~\cite{Paquet:2015lta}.
    (Right) The excess [panel (a)] and total [panel (b)] direct photon
    yields in different \pt ranges as a function of \Npart from STAR
    (circles) and PHENIX (triangles) in \auau collisions at
    \sqsntwo.  The down-pointing triangles represent results
    from the internal conversion method~\cite{Adare:2008ab} while the
    up-pointing triangles represent the results from~\cite{Adare:2014fwh}.
    Model predictions from 
    Rapp {\it et al.}~\cite{vanHees:2011vb,vanHees:2014ida} and
    Paquet {\it et al.}~\cite{Paquet:2015lta} are also shown for the
    excess (a) and total (b) direct photon yields.  Statistical and
    systematic uncertainties are shown by bars and boxes, respectively.
    (Figures taken from~\cite{STAR:2016use}.)
  }
  \label{fig:star_intyields}
\end{figure}

In 2017 the STAR Collaboration at RHIC published a new measurement of
direct virtual photon production in \sqsntwo \auau
collisions~\cite{STAR:2016use} and the derived direct (real) photon
invariant yields as a function of centrality in the
1$<$\pt$<$10\,\gevc range, with a gap between 3 and 5\,\gevc (see
Fig.~\ref{fig:star_intyields}, left plot).  The
basic technique used (internal conversion) is identical to the one in  
the PHENIX publications~\cite{Adare:2008ab,Adare:2009qk}, but the detectors, and
consequently details of the analyses are rather different.
Unfortunately the low \pt results obtained by STAR and PHENIX are 
incompatible, as shown in the right plot of
Fig.~\ref{fig:star_intyields}, where the integrated yields in the low
\pt region are compared.  The small
difference in the regions of integration does not explain the
discrepancies (other possibilities will be discussed in
Sec.~\ref{sec:devil}).  
Interestingly, the centrality dependence is similar,
\ie the slope $\alpha$ of the \dngamdeta \,\vs\, \dnchdeta fit to the STAR
data is (within uncertainties) compatible with the PHENIX
slopes (see the flatness of the STAR points in
Fig.~\ref{fig:intyield_fits_ratios}, 
also~\cite{Lijuan:2017})\footnote{It should  be noted that
  the high \pt part in STAR agrees with the \taa-scaled pQCD
  calculation, just as in PHENIX~\cite{Afanasiev:2012dg} -- see the argument
  about and significance of the finding that at high \pt the direct
  photon \raa is unity in Sec.~\ref{sec:highpt}.  On the other hand
  the high \pt measurement in PHENIX was done with a different technique
  (calorimetry), while in STAR both the low and high \pt regions were
  measured with the same method (internal conversion).
}.  
Extrapolation of
the $\eta$ spectrum to low \pt (where it is not measured but is an
important source of background) is done differently in STAR and
PHENIX.  Unfortunately, if STAR adopts the PHENIX extrapolation
(everything else unchanged) the discrepancy {\it increases}.  Despite
a joint effort by the two experiments {\it the issue of the
  discrepancy is so far unresolved}.  Unwelcome as they are, such
situations happen, as they did in the past (and with time found a
resolution).  Ongoing analysis of the 2014 200\,\gev \auau dataset by
PHENIX, larger than all previous datasets combined, and with yet
another method (external conversion) might help to put an end to the
controversy.

The PHENIX experiment also measured low \pt direct photons with the
internal conversion technique in \sqsn=200\,GeV \cucu 
collisions~\cite{Adare:2018jsz}, and
via external conversion in \sqsn=62.4\,GeV and \sqsn=39\,GeV \auau
collisions~\cite{Adare:2018wgc}.  
Preliminary results have been shown at the Quark Matter
2017 conference and reported in~\cite{Sharma:2017tht}.  The inverse
slope for minimum bias \cucu data was 
$T_{eff}= 288 \pm 49(stat) \pm 50(syst)$\,MeV (note the large
uncertainties), while for the 62.4 and 39\,GeV minimum bias data
$T_{eff}= 211 \pm 24(stat) \pm 44(syst)$\,MeV and
$T_{eff}= 177 \pm 31(stat) \pm 68(syst)$\,MeV, respectively.
It should be noted that for the 62.4 and 39\,GeV \auau data the \pp
contribution has not been subtracted.  With the inclusion of the
\sqsn=2.76\,TeV ALICE data large colliding systems from \cucu to 
\pbpb and almost two orders of
magnitude in collision energy are now covered.  

\begin{figure}[htbp]
  \includegraphics[width=0.9\linewidth]{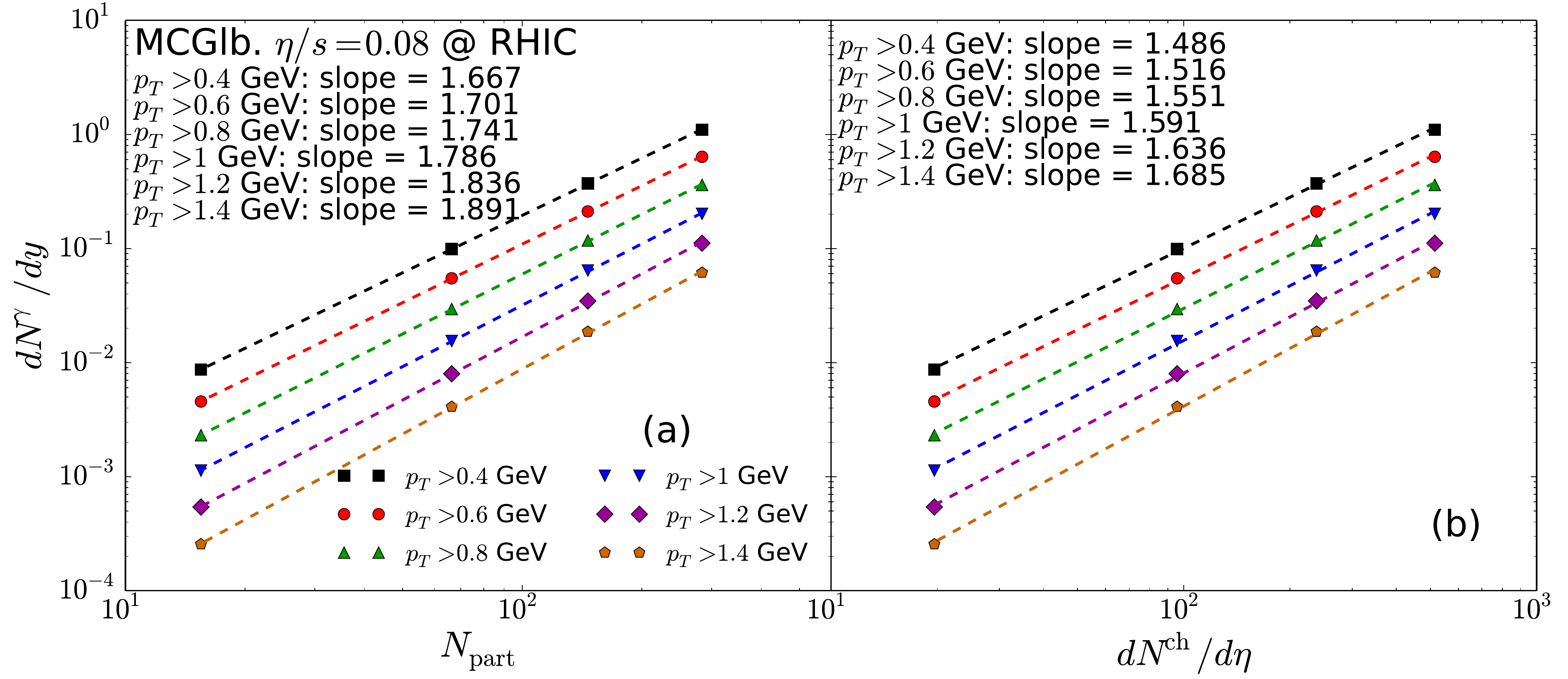}
  \caption{Model calculation (viscous hydrodynamics~\cite{Shen:2013vja})
    of the centrality dependence of the photon yield for
    \sqsn=200\,GeV \auau collisions at RHIC.  Centrality is expressed
    both in terms of (a) \Npart and (b) $dN^{ch}/d\eta$.
    (Figure taken from~\cite{Shen:2013vja}.)
  }
  \label{fig:shen_2014_npartnch}
\end{figure}

While the inverse slopes in Fig~\ref{fig:ppg162_fig6} do not show a
clear centrality-dependence, the yields certainly do.  
As discussed in Sec.~\ref{sec:sources}, the emission of hadrons,
originating  from a medium (QGP or hadron gas) and characterized by
$dN_{ch}/d\eta$ should roughly be proportional to the
number of constituents\footnote{The appropriate degree of freedom can be
\Npart participating nucleons or $N_{qp}$ participating constituent
quarks, as pointed out in~\cite{Adler:2013aqf}.
},
or a small power thereof, due to
rescattering~\cite{Kajantie:1986dh,Srivastava:1999jz}, 
while photons, coming from binary collisions of the constituents,
should be produced at a higher power $AN_{part}^{\alpha}$.  Naively
one would expect $\alpha \sim 2$, but the power is substantially
decreased by the rapid cooling and expansion/dilution  of the system.
Alternately, instead of comparing the (integrated) photon yields to
\Npart or $N_{qp}$ (not directly measured) one can compare them to the
observed charged particle density $dN^{ch}/d\eta$, essentially the
number of final state charged hadrons.  In the viscous hydro model
cited above~\cite{Shen:2013vja} the dependence of the integrated photon
yields on both \Npart and $dN^{ch}/d\eta$ are calculated; also, for
both sets the lower integration limit is varied from \pt=0.4\,\gevc to
1.4\,\gevc.  The results, including the non-negligible variation of
the slopes with the lower integration limit, are shown in
Fig.~\ref{fig:shen_2014_npartnch}, and reflect the fact that the ratio
of yields coming from the QGP and the hadron gas (HG) changes with
\pt.  In fact, the authors point out that taking only the HG yields
for \pt$>$0.4\,\gevc they would scale as a function of \Npart with
power 1.46 and as a function of $dN^{ch}/d\eta$ with power 1.23; the
corresponding powers for the QGP are much larger, 2.05 and 1.83.

\begin{figure}[htbp]
  \includegraphics[width=0.75\linewidth]{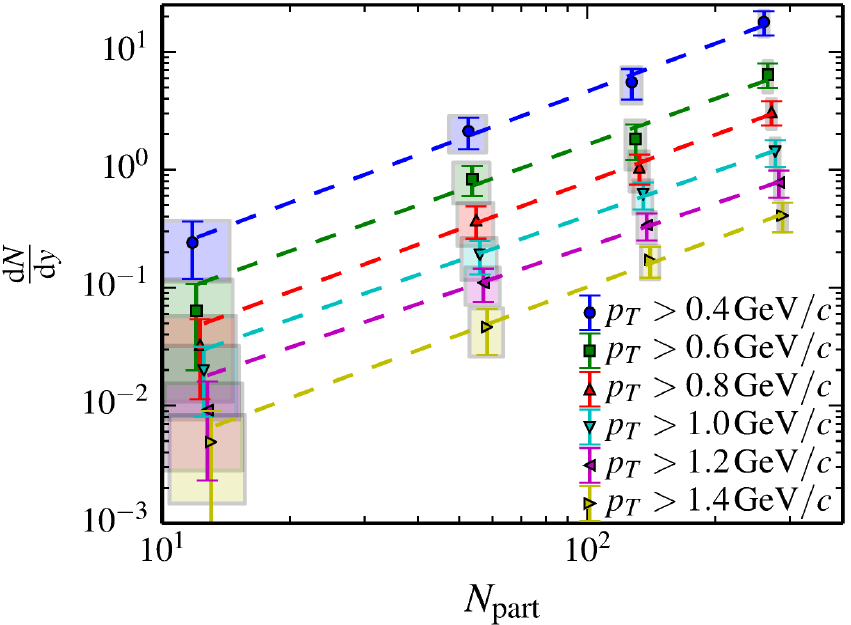}
  \caption{Integrated ``thermal'' photon yields in \sqsn=200\,GeV
    \auau collisions measured by the PHENIX experiment as a function of
    \Npart for different lower \pt integration limits.  The points at
    a given \Npart are slightly shifted for bette visibility.  The dashed
    lines are independent fits to a power law $N_{part}^{\alpha}$.
    (Figure taken from~\cite{Adare:2014fwh}.)
  }
  \label{fig:ppg162_fig10}
\end{figure}

The first experimental results are shown in
Fig.~\ref{fig:ppg162_fig10}. Contrary to the prediction
in~\cite{Shen:2013vja}, the slopes $\alpha$ in 
$dN_{\gamma}/dy = AN_{part}^{\alpha}$ did not change significantly with
the \pt integration limit, and the average value is
$\alpha = 1.38 \pm 0.03(stat) \pm 0.07(syst)$, even somewhat lower
than the purely HG slope in~\cite{Shen:2013vja}.  This constancy of the
slope was quite unexpected.  The spectra are steeply falling, so the
integral is mostly determined by how the spectrum looks like near the
lower integration limit.  If one neglects non-conventional sources and
thinks only in terms of the ``radiation comes either from the QGP or
the HG'' dichotomy, yields with the lowest \pt integration limit would
clearly be overwhelmed by HG production, while yields integrated from
the highest, \pt=1.4\,\gevc limit should have some (substantial?) QGP
contribution, which in turn should have a steeper centrality (\Npart)
dependence.  This is reflected in other models, too. In the PHSD transport
model~\cite{Linnyk:2013wma}  $\alpha \sim 1.5$; the scaling power from
the QGP contribution alone would be higher ($\alpha \sim 1.75$), but
the ratio of photons from the QGP and the hadron phase is quite small,
only about 10\% in the most peripheral, and 30\% in the most central
collisions. Also, it is strongly dependent on \pt in the low \pt
region (up to \pt=1\,\gevc, see Fig. 6 in~\cite{Linnyk:2013wma}).  --  
A simple (and admittedly incomplete) model of photon
production in the Glasma~\cite{Chiu:2012ij} puts the scaling power in the
range 1.47$\leq\alpha\leq$2.2, but the shape and the centrality
dependence of the spectra are well described, once the normalization
is fixed for one centrality and ``geometric scaling'' is applied.  --
The model claiming enhanced photon emission due to strong initial
magnetic fields~\cite{Basar:2012bp} predicts {\it less} enhancement in
central collisions, since the magnetic field decreases with decreasing
impact parameter, disfavored by the data. -- It is interesting to note
that recent models tend to
{\it deprecate} QGP radiation: they assume that the bulk of the
``thermal'' yield is produced either before QGP is formed, or at the
transition from QGP to HG and later.  We will 
discuss these models and their implications in more detail in the
context of the ``direct photon puzzle'', see Sec~\ref{sec:puzzle}.

\begin{figure}[htbp]
  \includegraphics[width=0.99\linewidth]{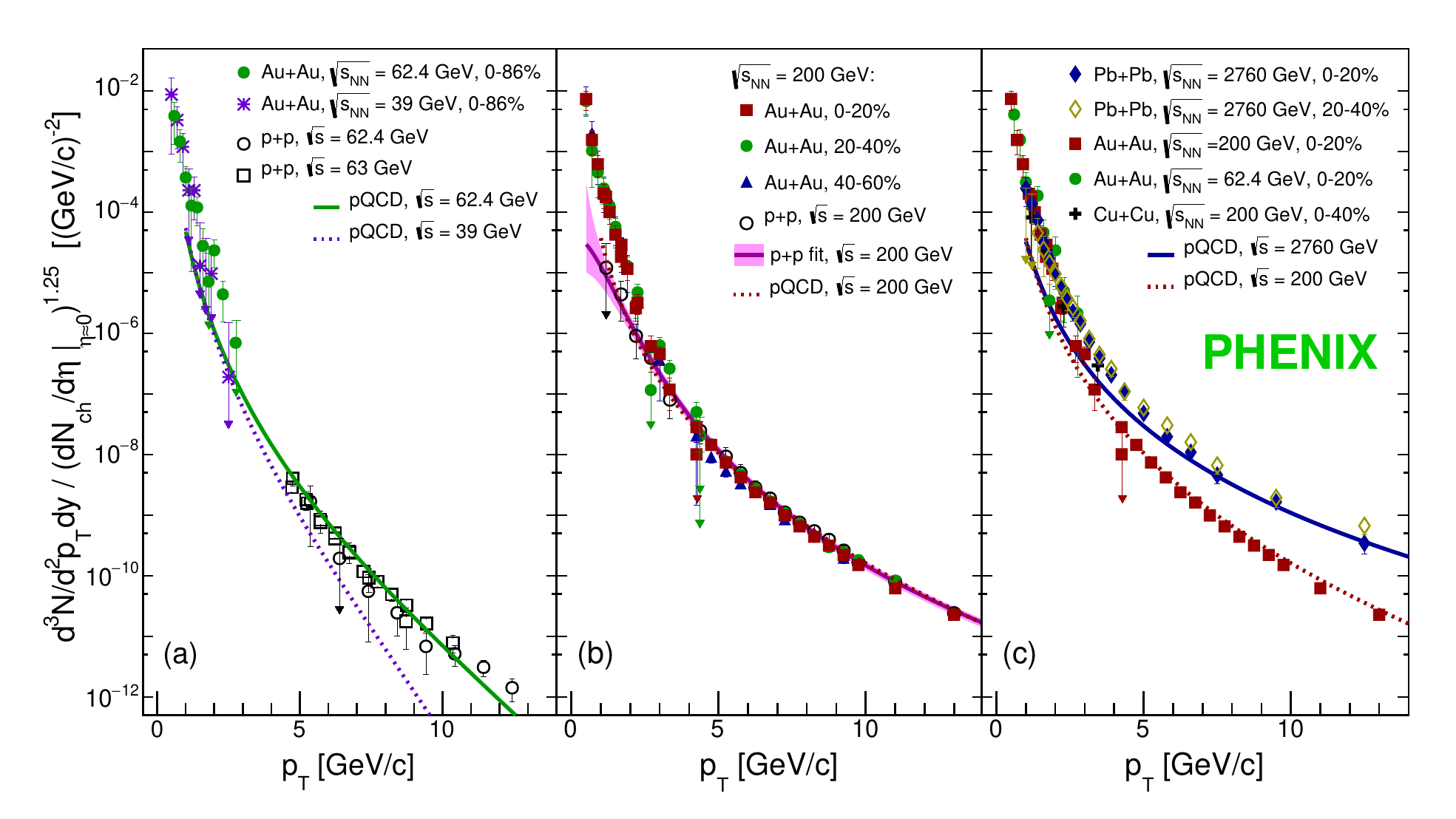}
  \caption{Direct photon spectra normalized by (\dnchdeta)$^{1.25}$
    for \auau at \snn 39 and 62.4\,GeV (panel (a)) and 200\,GeV 
    (panel (b)).  Panel (c) compares spectra for different $A+A$
    systems at different \snn.  (Figure taken from~\cite{Adare:2018wgc}.)
  }
  \label{fig:ppg212_Fig1_Final}
\end{figure}

\begin{figure}[htbp]
  \includegraphics[width=0.7\linewidth]{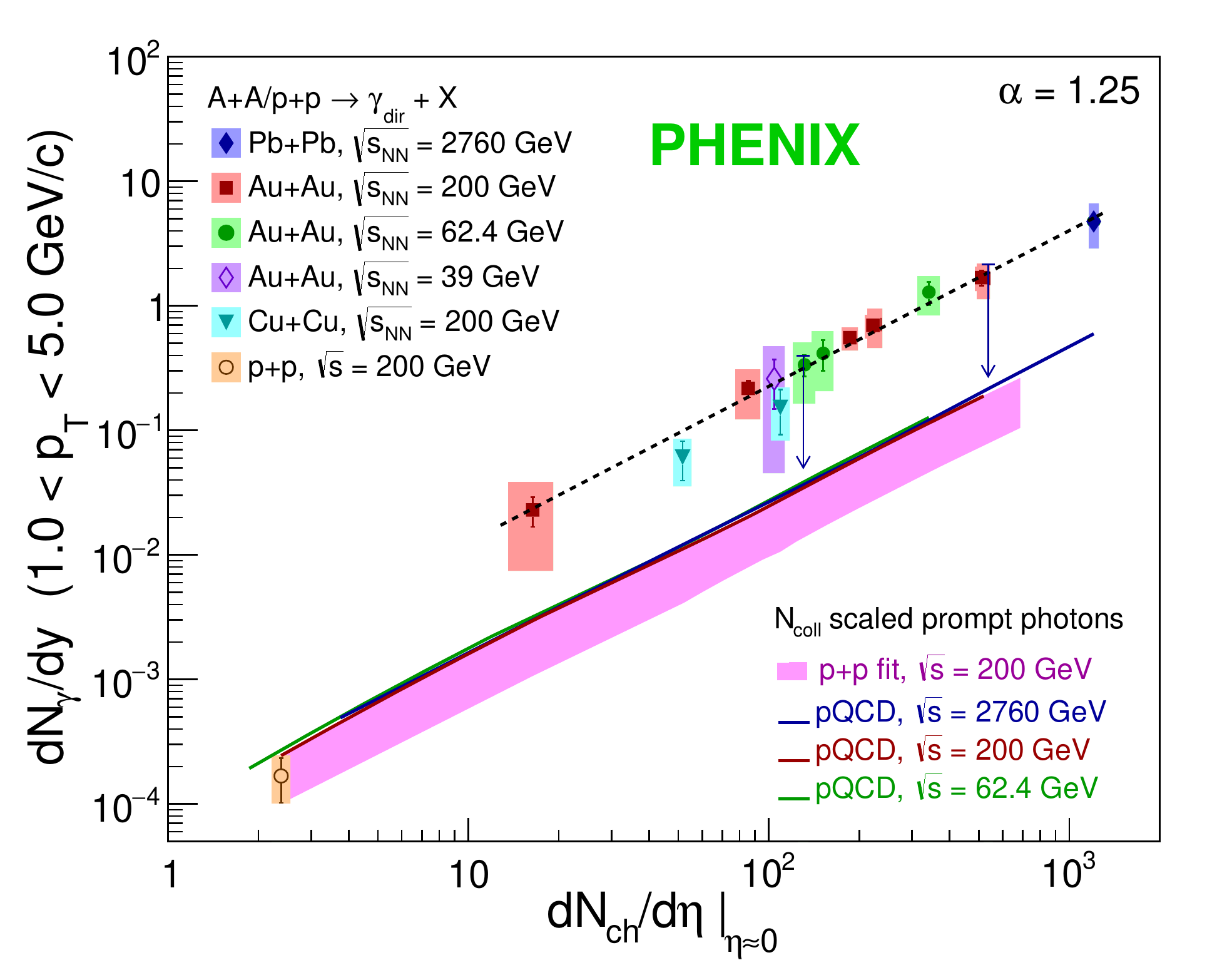}
  \caption{Direct photon yields integrated for \pt$>$1.0\,\gevc vs
  \dnchdeta for various colliding systems and energies.  The \pbpb
  data are from ALICE, all other data are from PHENIX.  (Figure taken
  from~\cite{Adare:2018wgc}.)
  }
  \label{fig:ppg212_Fig3_Final}
\end{figure}

\subsubsection{Scaling of direct photon yields with \dnchdeta}
\label{sec:scaling}

The integrated yield \vs \Npart provides good insight in the centrality
dependence in a particular \AA system at a specific \sqsn collision
energy, but \Npart is ill-suited for comparisons between systems and
energies (also, it is not a direct experimental observable).  When
investigating the system size and energy dependence of photon
production, the (pseudo)rapidity density of produced, final state
charged particles, \dnchdeta is a more natural choice -- and it is 
a measured, model-independent quantity.

In a recent publication~\cite{Adare:2018wgc} by PHENIX it has been
pointed out that if the direct photon yields at mid-rapidity are
normalized by the corresponding (\dnchdeta)$^{1.25}$ then the
``thermal'' photon yields are similar for a wide range of colliding
nuclei\footnote{As long as the nuclei are both sufficiently large;
  as of now it is still unclear what happens in very asymmetric
  collisions, like \pau or \dau.
}, collision centralities and collision energies 
(see Fig.~\ref{fig:ppg212_Fig1_Final}).  Alternately, the photon
yields integrated above $\pt>1.0$\,\gevc as a function of \dnchdeta
scale with a power $\alpha=1.25$ for the same wide range of systems
and energies and centralities (see Fig.~\ref{fig:ppg212_Fig3_Final}).
It is important to note that
$\Ncoll\sim(\dnchdeta)^{1.25}\sim\Nqp^{1.25}$ where \Nqp is the number
of quark participants~\cite{Adler:2013aqf}\footnote{In other words,
  direct photon production apparently scales with \Ncoll, which is
  fully expected   at high \pt, but quite surprising in the
  ``thermal'' region. 
}.
The origins of that scaling are unclear, but, at least qualitatively,
it would fit a picture where most photons are produced in space-time 
near the QGP$\rightarrow$HG transition, which in turn would be largely
independent of the initial conditions (as long as QGP is formed).
While highly speculative at the moment, it is an avenue worth
exploring, among others by filling the \dnchdeta gap between the \pp
and the most peripheral \AA points in Fig.~\ref{fig:ppg212_Fig3_Final}.
This can be done two different ways.  The cleaner one is to continue
exploring 
large-on-large ion collisions (where the scaling has been observed)
but decrease \sqsn further, and also to analyze very peripheral data
in smaller and smaller centrality bins,
which in light of the increasing \AA datasets should be
possible.  The other possibility is to look at \pA and \dau data --
an ongoing effort -- and maybe at very high multiplicity \pp events.
However, one has to be careful with the conclusions.  As recent
results from other observables indicate, kinematics and initial state 
effects~\cite{Alvioli:2013vk,Kordell:2016njg} play
much bigger role in very asymmetric collisions (\pA, \dau) than in
\AA, and comparing very high multiplicity, \ie extreme \pp
events to average \AA collisions may also be misleading.
Unfortunately telling apart new physics from experimental bias is not
always trivial.

\begin{figure}[htbp]
  \includegraphics[width=0.45\linewidth]{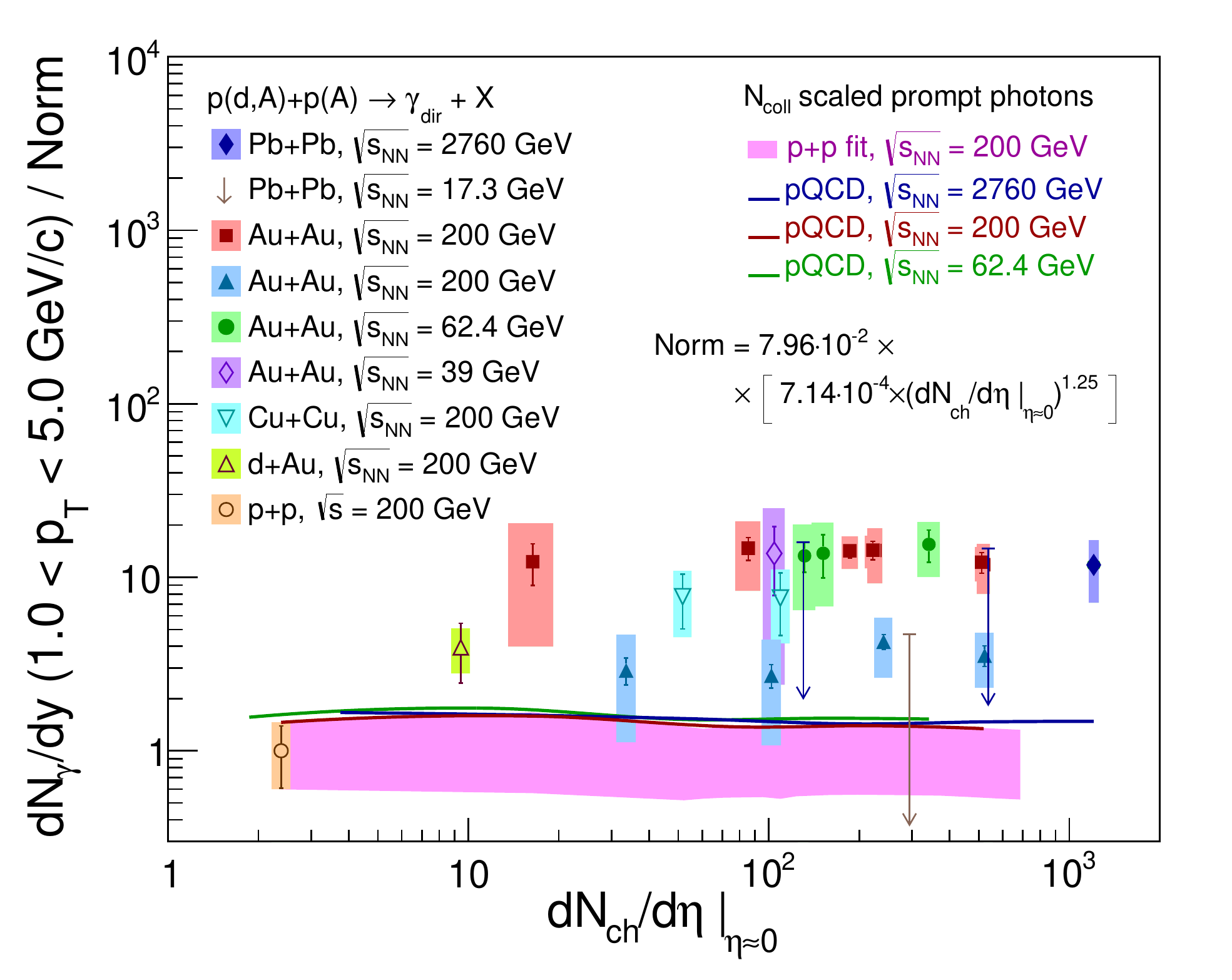}
  \includegraphics[width=0.45\linewidth]{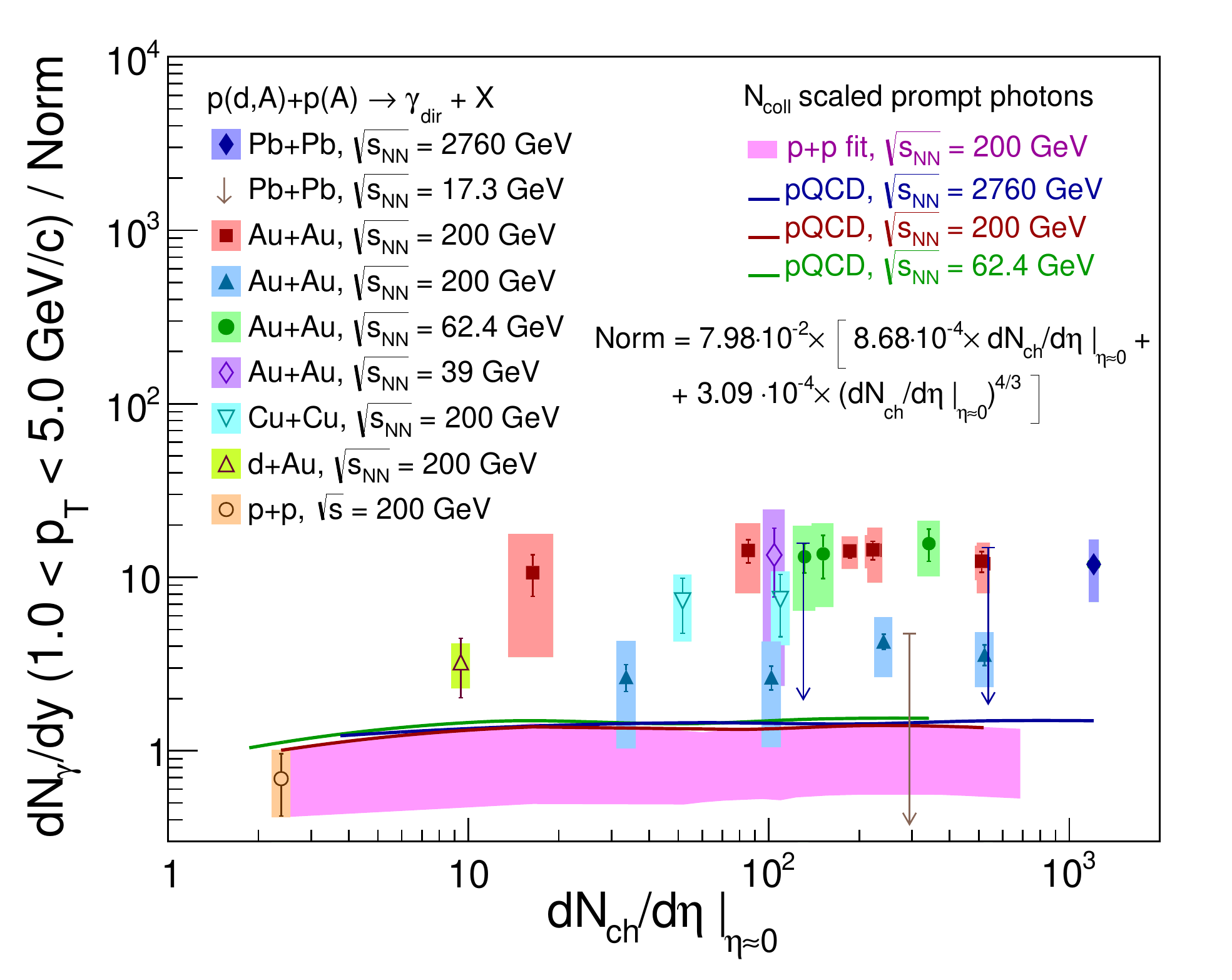}
  \caption{Low \pt integrated direct photon yield for various
    colliding systems and energies, divided by fits with two different
    functions, $A(\dnchdeta)^{\alpha}$ and 
    $A(\dnchdeta) + B(\dnchdeta)^{4/3}$.  The fits are normalized to
    the 200\,\gev \pp point.  In addition to the data shown in
    Fig.~\ref{fig:ppg212_Fig3_Final} the PHENIX \dau
    data~\cite{Adare:2012vn} and the STAR virtual photon
    data~\cite{STAR:2016use} are presented, too (upward pointing blue
    solid triangles, well separated from the PHENIX and ALICE points),
    and upper limits calculated from WA98 data~\cite{Aggarwal:2000th} 
    are also included (arrows).    Both fit functions
    have only two free parameters, and the data/fit is virtually
    indistinguishable. 
    (Figure courtesy of Vladimir Khachatryan.)
  }
  \label{fig:intyield_fits_ratios}
\end{figure}

It should be pointed out (and in fact it is mentioned 
in~\cite{Adare:2018wgc}),
that the scaling function $A(\dnchdeta)^{\alpha}$ used in
Fig.~\ref{fig:ppg212_Fig3_Final} is not unique; the
integrated yields can be equally well fitted for instance with 

\begin{equation}
\frac{dN_{\gamma}}{dy} =
A\frac{dN_{ch}}{d\eta} + B\Big(\frac{dN_{ch}}{d\eta}\Big)^{4/3}
\end{equation}

\noindent
another function with just two free parameters, but completely
different form.  In order to show that
the two functions provide equally good fits,  
in Fig.~\ref{fig:intyield_fits_ratios} the low \pt integrated yields
are divided by the two different fit functions discussed above.  The
functions are normalized to the \pp point.
Fig.~\ref{fig:intyield_fits_ratios}  has a few more data points than 
Fig.~\ref{fig:ppg212_Fig3_Final}, including the \dau results from
PHENIX~\cite{Adare:2012vn}, upper limits calculated from the WA98
data~\cite{Aggarwal:2000th} and the STAR virtual photon
data~\cite{STAR:2016use} that are incompatible with the PHENIX results.
Obviously the two fits describe the data equally well, although the
functional form is quite different, suggesting different physics.
Remarkably, the power $4/3$ is exactly that of the second, nonlinear
term in Feinberg's 1976
paper~\cite{Feinberg:1976ua} where he calculates the number of photons
produced in hadron-hadron collisions if ``an intermediate stage of
hadronic matter'' exists (see Sec.~\ref{sec:intro}).  Very suggestive,
but one has to be careful before drawing any conclusions.    
Nevertheless, it appears that the ``excitation function'' of low \pt
direct photon production over a wide range of colliding systems,
centralities and energies can be described with just two free
parameters, a tantalizing hint that there might be some 
{\it fundamental commonality in the underlying physics}.

\section{\bf Direct photon flow -- the era of the ``direct photon puzzle''}
\label{sec:puzzle}

\subsection{First results on \vtwo and \vthr}
\label{sec:v2v3}

In a 2008 review of the electromagnetic probes~\cite{David:2006sr} the 
authors pointed out that a {\it coherent and quantitative} description 
of the sQGP -- including direct photon observations -- 
in heavy ion collisions is still missing.  
The measured large photon yields in the ``thermal'' region~\cite{Adare:2008ab}
could be explained in a hydrodynamic framework with inverse slopes 
$T_{eff}$ ranging from 370 to 660 MeV (varying the initial
thermalization time $\tau_i$, see~\cite{dEnterria:2005jvq} and references
therein).  Independently, large elliptic flow \vtwo, scaling with the
number of quarks  was found for hadrons.  The usual interpretation was
(and still is) that hadrons inherit the final momentum anisotropies of the
sQGP which in turn build up gradually from the initial pressure
anisotropies, starting at $\tau_i$ and lasting until the time of
chemical freeze-out.  Photons are produced predominantly early
(highest temperatures), when the pressure gradients, \ie acceleration 
is highest, but velocities are still small, while hadrons are
imprinted by the large velocities at the end of sQGP expansion.  
Accordingly, direct photons from the QGP were expected to have very
small \vtwo compared to photons from the hadron
phase~\cite{Chatterjee:2005de,Turbide:2005bz,Chatterjee:2008tp}, 
which itself is still smaller than \vtwo of the final state hadrons, 
and, per extension, \vtwo of the \piz decay photons.
Interestingly, the first
report on direct photon flow appeared to confirm this
expectation of small photon \vtwo~\cite{Adler:2005rg}, albeit with 
important caveats.

All this changed radically at the Quark Matter 2011 conference where
PHENIX presented preliminary results on direct photon \vtwo in the
1-12\,\gevc \pt region and found that ``{\it for \pt$>$4\,\gevc the
  anisotropy for direct photons is consistent with zero, which is as
  expected if the dominant source of direct photons is initial hard
  scattering.  However, in the \pt$<$4\,\gevc region dominated by
  thermal photons, we find a substantial direct-photon \vtwo
  comparable to that of hadrons, whereas model calculations for
  thermal photons in this kinematic region underpredict the observed
  \vtwo.}''~\cite{Adare:2011zr}. 

\begin{figure}[htbp]
  \includegraphics[width=0.45\linewidth]{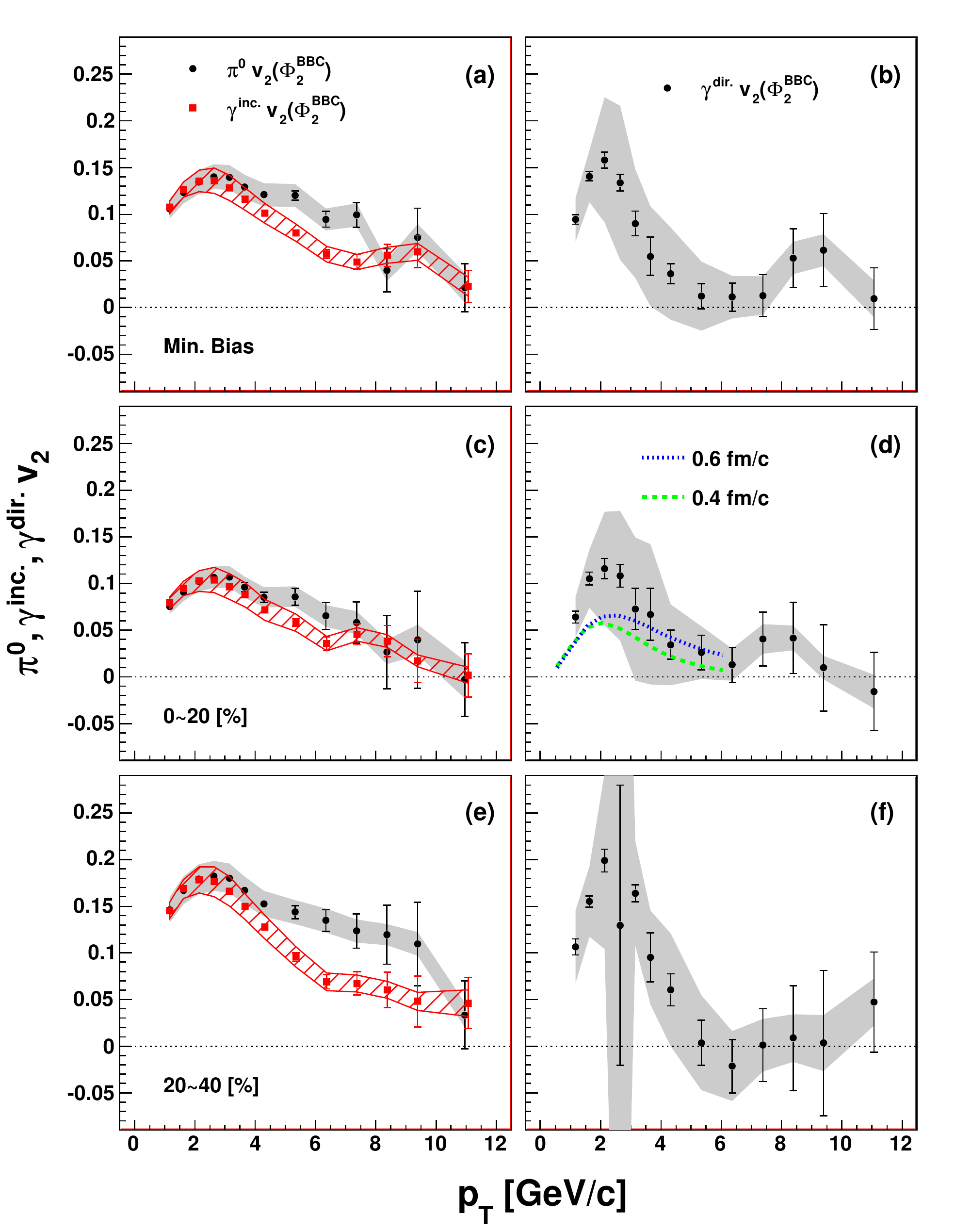}
  \caption{(a),(c),(e) Centrality dependence of \vtwo for \piz and
     inclusive photons measured for minimum bias, 0-20\% and 20-40\%
     centrality Au+Au collisions at \sqsn=200\,\gev.  (b),(d),(f)
     shows the direct photon \vtwo, vertical bars showing statistical
     while  shaded areas showing systematic uncertainties.
     Superimposed on panel (d) are calculations from
     ~\cite{Chatterjee:2008tp} using two different initial times.
    (Figure taken from~\cite{Adare:2011zr}.)
  }
    \label{fig:ppg126_fig2}
\end{figure}

The results, reproduced in Fig~\ref{fig:ppg126_fig2}, were
received with significant scepticism because they meant that ``thermal''
photons have \vtwo just as large as final state hadrons.  This was simply
incompatible with the old paradigm.  The observed \vtwo is a
convolution of the rates (largest early on, when $T$ is highest) and 
the anisotropic boost from the expansion (largest at late times, when 
$T$ is in turn smallest).  Hadron \vtwo encodes the final, maximum velocities,
but photons, being penetrating probes, are boosted only by the
velocities experienced at the moment of their creation, which
initially are close to zero.  No theory
predicted or could readily accomodate the simultaneous observation of
large yields and large \vtwo for ``thermal'' photons, and the issue
quickly became dubbed the ``{\it direct photon puzzle}'', spawning
workshops~\cite{tpd:2011,trw:2012,ect:2013,gsi:2014,tpd:2014,ect:2015,ect:2018},
 impromptu collaborations of experimentalists and theorists,
and a remarkable number of papers.

\begin{figure}[htbp]
  \includegraphics[width=0.9\linewidth]{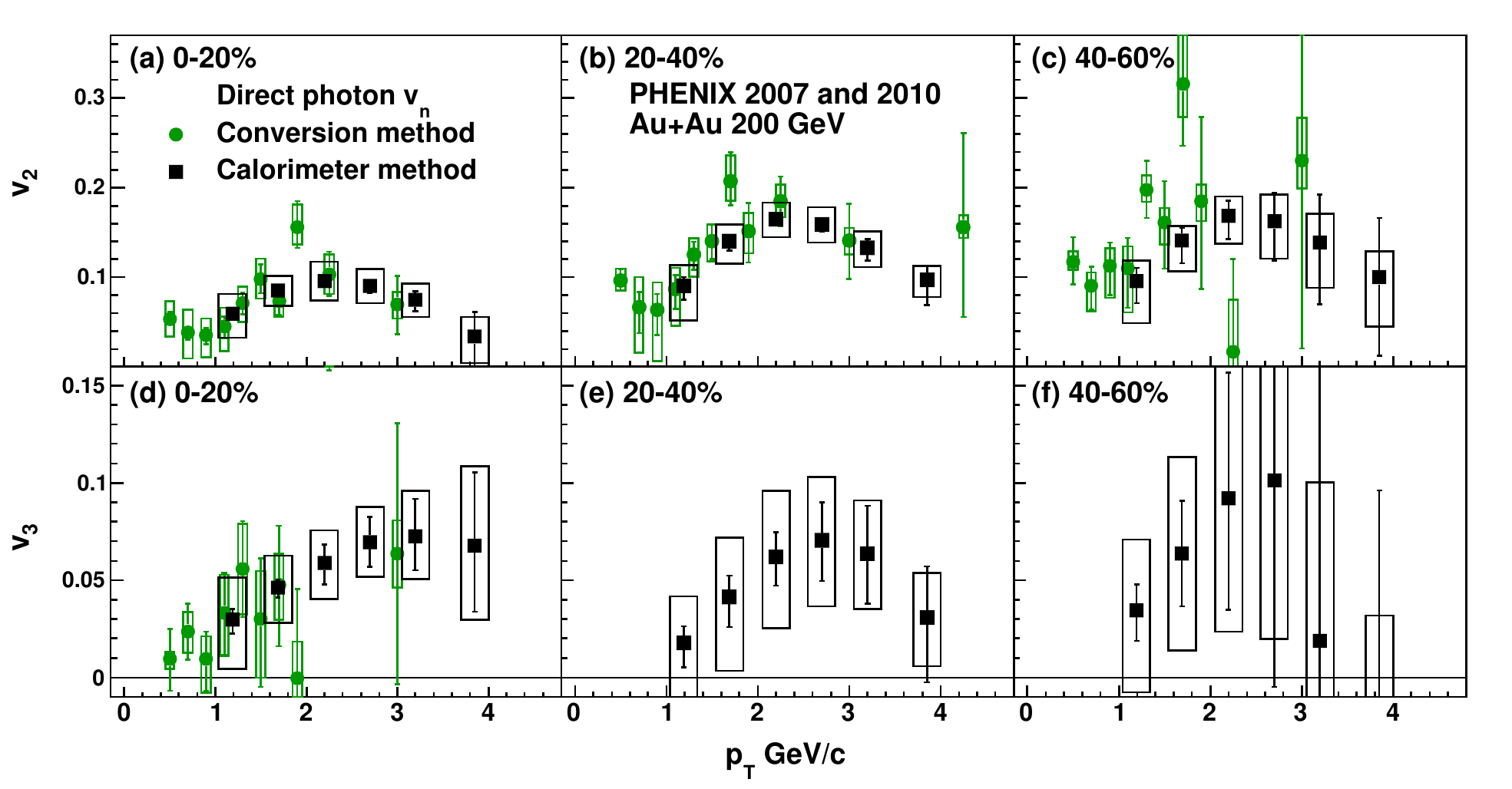}
  \caption{Direct photon \vtwo and \vthr at midrapidity
    ($|\eta|<0.35$), for different centralities, measured by PHENIX with the
    conversion method (solid circles, green) and the calorimeter
    method (solid squares, black).  The event plane was determined by
    the reaction plane detector ($1<|\eta|<2.8$).  The error bars
    (boxes) around the data points are statistical (systematic)
    uncertainties. 
    (Figure taken from~\cite{Adare:2015lcd}.)
  }
    \label{fig:ppg176_fig6}
\end{figure}

\begin{figure}[htbp]
\includegraphics[width=0.9\linewidth]{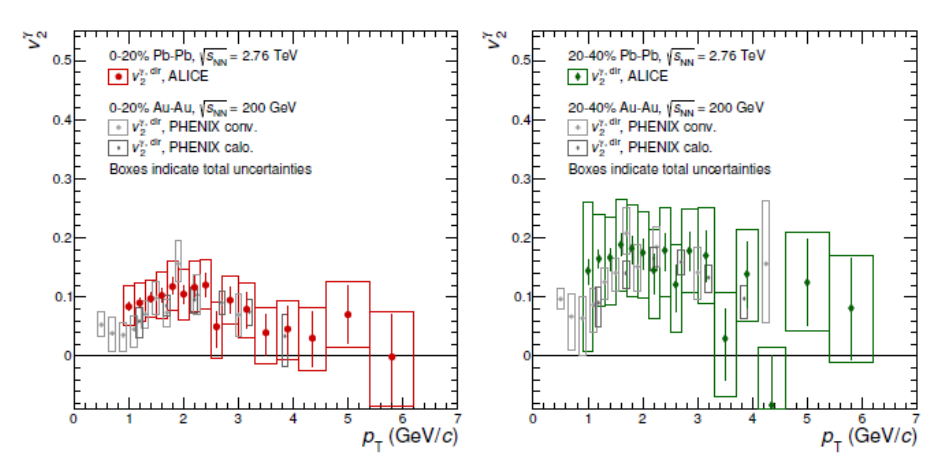}
  \caption{
    Elliptic flow of direct photons in 2.76\,TeV \pbpb collisions
    (ALICE) compared to PHENIX results~\cite{Adare:2015lcd} in 200\,\gev
    \auau.  Vertical bars are statistical, boxes are the total
    uncertainty.     (Figure taken from~\cite{Acharya:2018bdy}.)
  }
    \label{fig:alice_newflow}
\end{figure}

\begin{figure}[htbp]
\includegraphics[width=0.45\linewidth]{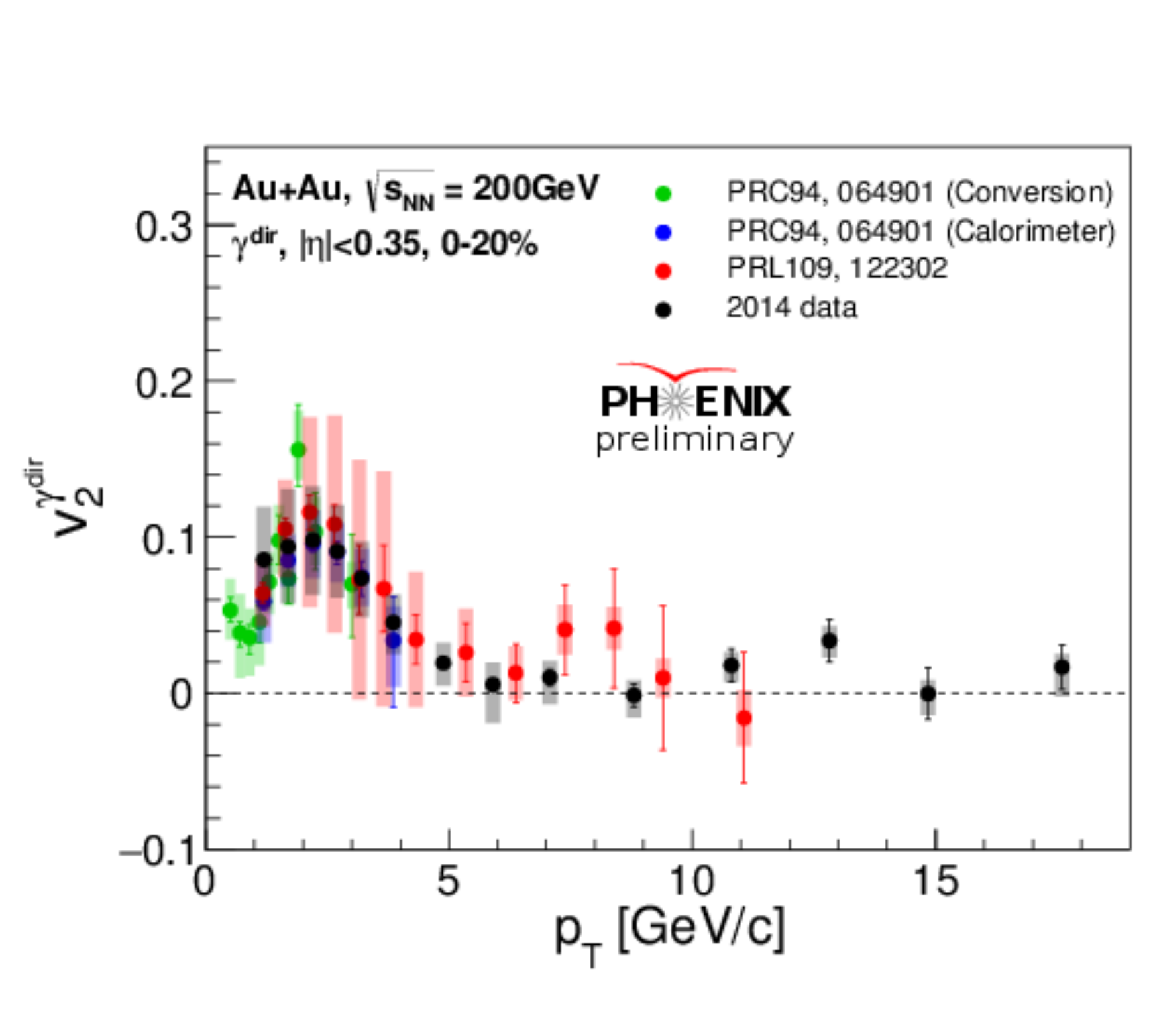}
\includegraphics[width=0.45\linewidth]{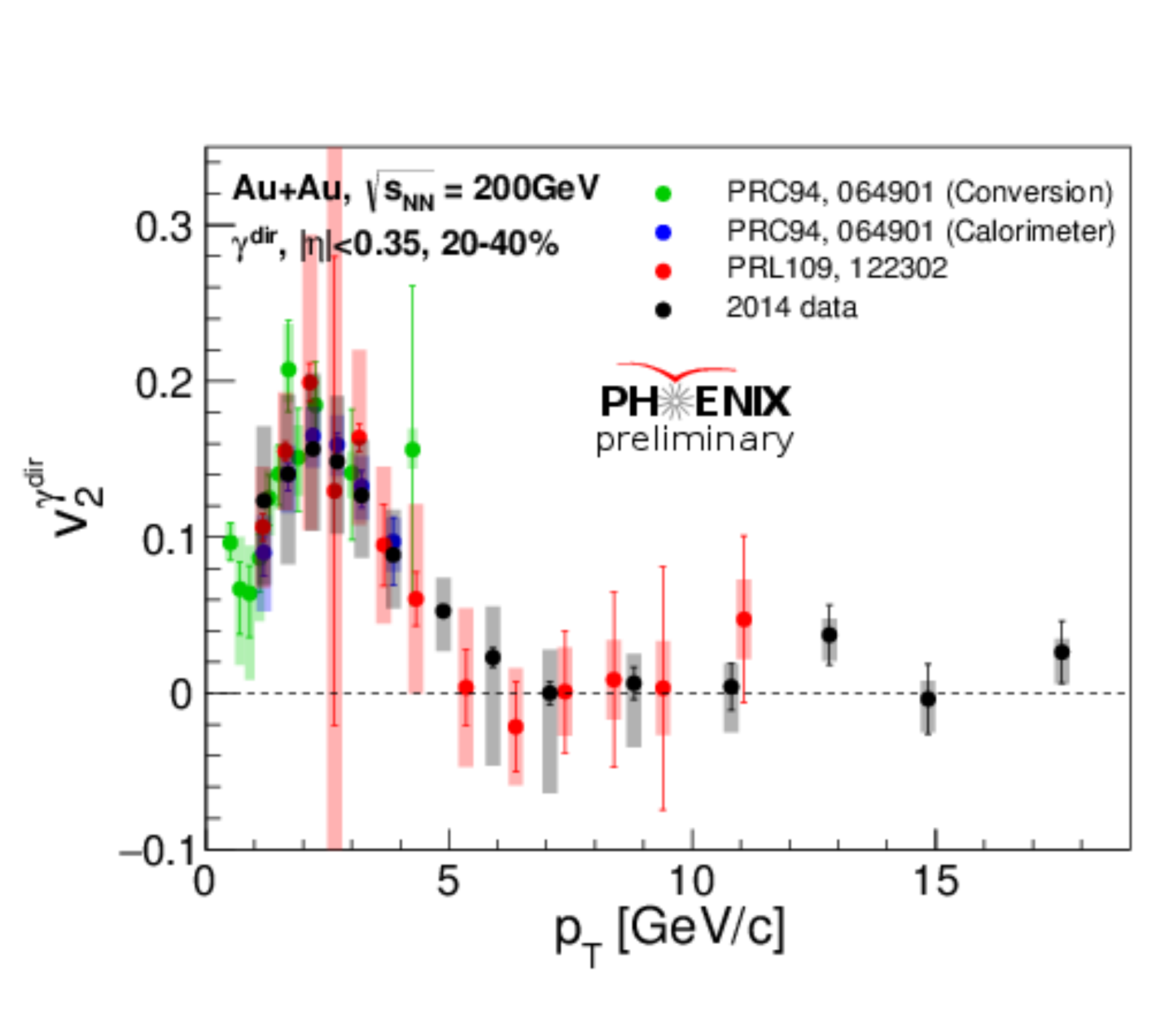}
  \caption{Direct photon \vtwo in 200\,\gev \auau collisions, measured
    by PHENIX with three different techniques.  The green, published
    data are from conversions at a known radius (HBD backplane, the
    outer shell of the Hadron Blind Detector, 60\,$cm$ from the beam
    crossing), while
    the black, preliminary ``2014 data'' are obtained with the method
    that does not assume {\it a priori} the conversion radius (they
    can happen at different layers of the VTX detector).  For
    more details see Sec.~\ref{sec:expconv}.
  }
    \label{fig:wenqing_newflow}
\end{figure}

In 2012 the ALICE Collaboration made a similar observation at LHC
energies, and has shown it as preliminary result at the Quark Matter
2012 conference~\cite{Lohner:2012ct}, but in the subsequent years
there was some doubt whether the observed \vtwo is really significant
or still consistent with zero 
(no flow at all)~\cite{tpd:2014,gsi:2014,ect:2018}.
In 2016 the PHENIX Collaboration published another
paper on direct photon \vtwo and \vthr~\cite{Adare:2015lcd}, concentrating
only on the \pt$<$4\,\gevc (``thermal'') region, measuring photons two
different ways (calorimeter and conversion, see
Sec.~\ref{sec:experimental}), and confirmed the previous findings: the
direct photon \vtwo is large, comparable to the hadron \vtwo.  The
new results are shown in Fig.~\ref{fig:ppg176_fig6}.  
The final results by ALICE on photon \vtwo in 2.76\,TeV \pbpb
collisions~\cite{Acharya:2018bdy,Sas:2018oxv}  are just
being published (2019) and the authors
conclude that ``{\it A comparison to RHIC data shows a similar
  magnitude of the measured direct-photon elliptic flow.  Hydrodynamic and
  transport model calculations are systematically lower than the data,
  but are found to be compatible.}''~\cite{Acharya:2018bdy}  
Different from the 2012 (preliminary)
measurement this time (2019) ALICE applied two independent methods,
conversion and calorimetry.  The final data are shown in
Fig.~\ref{fig:alice_newflow}.  Remarkably, the results are consistent
with the PHENIX \vtwo presented in Fig.~\ref{fig:ppg176_fig6}.
There is an ongoing
effort in PHENIX to repeat the analysis with a {\it third} 
method~\cite{Khachatryan:2018uzc} and 
on a dataset that is an order of magnitude larger than earlier ones;
preliminary results are shown in Fig.~\ref{fig:wenqing_newflow}.  
The low \pt part once again confirms earlier findings.
The high \pt part should be compared to the first (2011) measurement,
shown in Fig.~\ref{fig:ppg126_fig2}.  The \pt range is extended while
the uncertainties are considerably smaller, and the clear message is
that at high \pt the direct photon \vtwo is consistent with zero, as
expected, if the bulk (or all) of those photons are produced in
initial hard scattering.

\begin{figure}[htbp]
\includegraphics[width=0.45\linewidth]{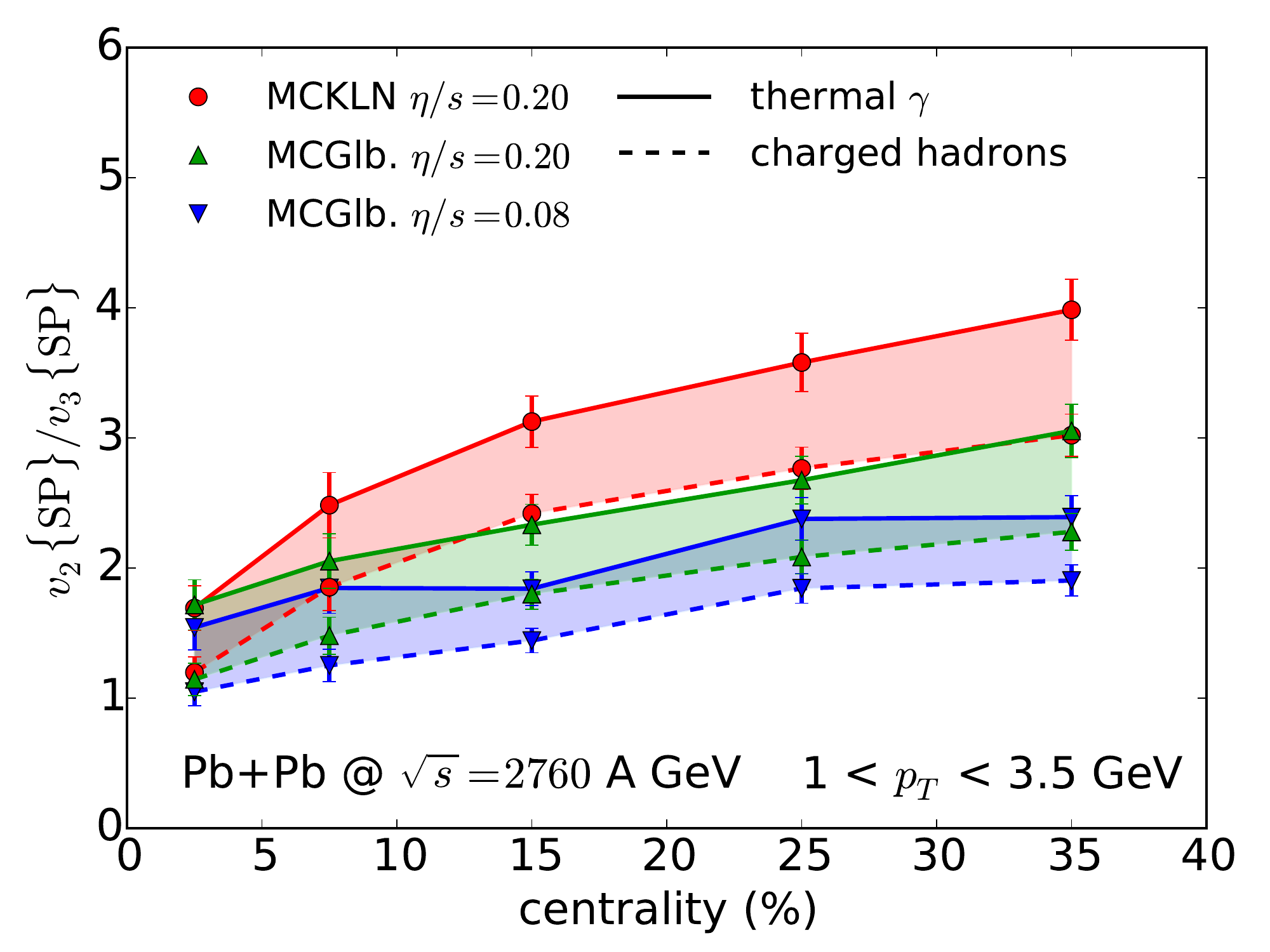}
  \caption{The ratio of the integrated \vtwo to the integrated \vthr
    for the LHC, for different initial conditions and $\eta/s$.  The
    ratio is shown as a function of centrality.
    Solid and dashed lines show the ratio for photons
    and charged hadrons, respectively.
    (Figure taken from~\cite{Shen:2013cca}).
  }
    \label{fig:v2v3ratio}
\end{figure}

There is only one published direct photon \vthr measurement so
far~\cite{Adare:2015lcd}, and it has large uncertainties 
(see Fig.~\ref{fig:ppg176_fig6}).  As pointed out
in~\cite{Shen:2013cca,Chatterjee:2014dqa}, 
\vthr is purely driven by initial density fluctuations, therefore, it
carries information on pre-equilibrium photons, including what, if any
role the initial magnetic field plays.  Also, the ratio of photon
\vtwo/\vthr serves as a ``viscometer''.  It has been argued
already in~\cite{Dusling:2009bc} that shear viscosity ($\eta$) effects
both the photon \vtwo and \Teff extracted at low \pt.  The idea was
expanded in~\cite{Shen:2013cca,Shen:2014cga} by studying higher order
$v_n$ and suggesting that \vtwo/\vthr of ``thermal'' photons is
different from that of hadrons, because it is weighted toward earlier
times, and viscous effects are largest at early times when the
expansion rate is largest.  The expected ratio of integrated \vtwo and
integrated \vthr for photons and charged hadrons is shown in
Fig.~\ref{fig:v2v3ratio} for two types of initial conditions 
(Monte-Carlo Glauber and Monte-Carlo Kharzeev-Levin-Nardi)
and two values of specific shear viscosity $\eta/s$.
In all settings the differences between photons and hadrons 
(shaded regions) are substantial, as are the predictions for the 
different settings (different colors), making \vtwo/\vthr a powerful
observable to support or rule out models.  Unfortunately current
experimental uncertanties (see Fig.~\ref{fig:ppg176_fig6}) are still
too large to allow this, but hopefully the data will improve soon,
since large existing datasets have not been analyzed so far (or the
results are not public yet).

\subsection{The Devil's advocate}
\label{sec:devil}

\begin{figure}[htbp]
  \includegraphics[width=0.6\linewidth]{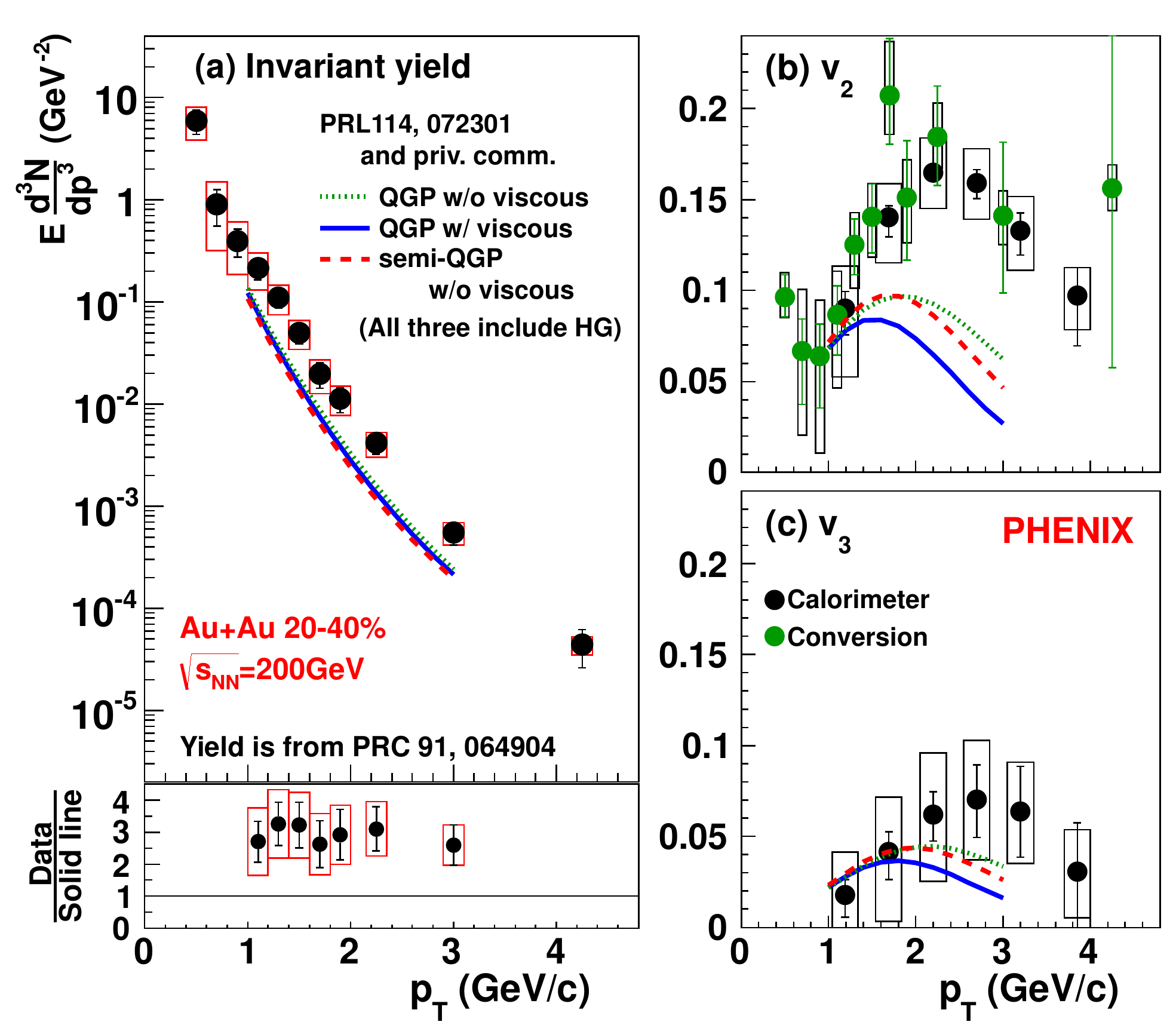}
  \caption{Comparison of the direct photons yields and \vtwo, \vthr
    with a hydrodynamical model~\cite{Paquet:2015lta} calculated under 
    three different assumptions, including the ``semi-QGP'' 
    scenario~\cite{Gale:2014dfa}.  For more comparisons see
    Sec.~\ref{sec:attempts}. 
    (Figure taken from~\cite{Adare:2015lcd}.)
  }
    \label{fig:ppg176_fig8}
\end{figure}

In Sec.~\ref{sec:attempts} we will review the current theories, but it
is fair to say that none of them is able to simultaneously describe
both the observed large yields and the large \vtwo.
This is demonstrated for instance in Fig.~\ref{fig:ppg176_fig8}, with
more comparisons in Sec.~\ref{sec:attempts}.
Moreover, those models that are at least partially
successful tend to downplay the role of radiation from the QGP: they
either emphasize pre-equilibration sources with some built-in
anisotropy, or production in the hadron phase, where large \vtwo comes
naturally.  But a {\it partonic medium almost without radiation} would
be a major paradigm change, and of course it should be proven that all
other, non-electromagnetic phenomena are also consistent with it.
Such unorthodox ideas would not be entirely new
(see for instance~\cite{Shuryak:2014zxa} and references therein),
but before making such leap, 
scientific rigor requires to pose the uncomfortable and provocative
question: what if the {\it data} are wrong\footnote{After all,
  ``{\it paranoia is the experimentalist's best friend.}''  Direct
  photon measurements are extremely difficult, and in the past decades
  it happened more than once that a questionable result stirred up the
  field for quite some time.  The author, an experimentalist himself,
  believes such ``sacrilegious'' questions to us are best asked by
  ourselves, rather than insisting on the line ``the data are what they
  are; if you cannot explain them, it's your problem''.
}?

Although the substantial direct photon \vtwo is now confirmed both by
(semi)-independent analysis techniques and different experiments,
at RHIC and the LHC\footnote{Inclusive photon \vtwo has already been
  measured at the SPS (see for instance~\cite{Aggarwal:2004zh,Aggarwal:2004ub}),
  but no attempt has been made to extract the direct photon \vtwo.
},
we shouldn't forget that at the moment (early 2019) there are still
unresolved discrepancies between the measured photon yields between
PHENIX and STAR (see Fig.~\ref{fig:star_intyields})\footnote{It would
  be interesting to see a direct photon \vtwo measurement from STAR
  based on their lower yields!
},
and that the yields, more precisely \rgam plays a decisive role in
measuring direct photon \vtwo.  Moreover, a recent statistical
analysis of the ALICE \vtwo data (Reygers~\cite{ect:2018}) indicates
that the significance of the non-zero \vtwo might only be
$O(1\sigma)$.  Since the ``direct photon puzzle'' remains an
unresolved issue with potentially far-reaching consequences, utmost
caution is warranted, and we believe it is useful to briefly discuss
the potential pitfalls of the measurements.

\begin{figure}[htbp]
  \includegraphics[width=0.52\linewidth]{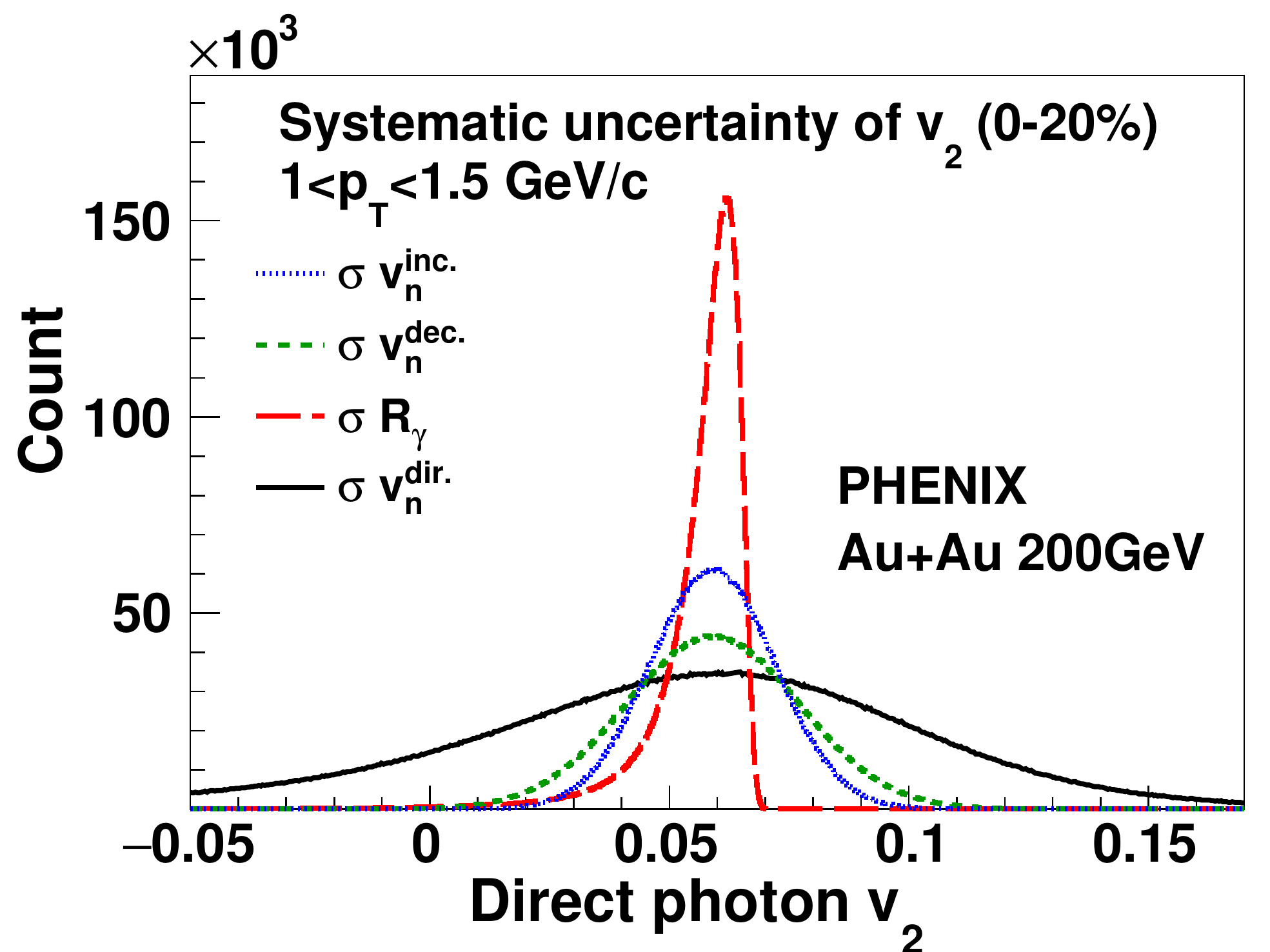}
  \includegraphics[width=0.42\linewidth]{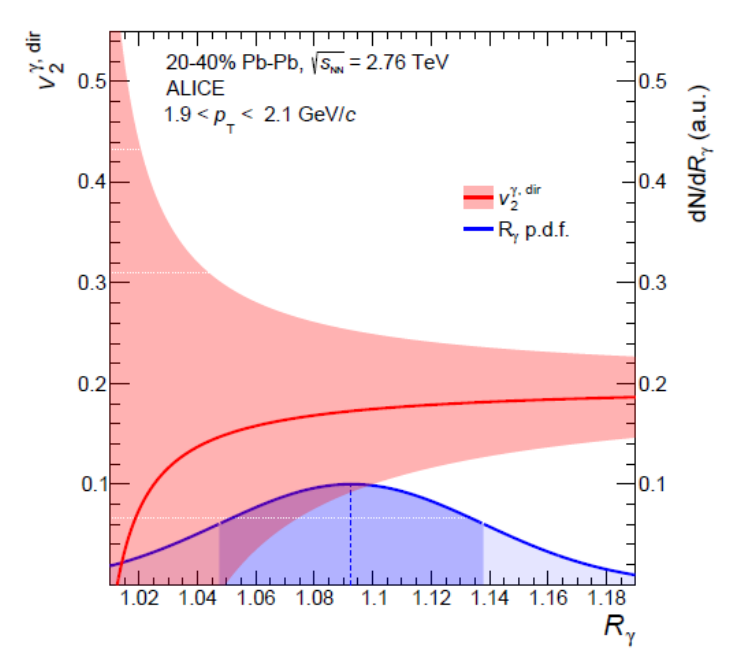}
  \caption{(Left) An example of the direct photon $v_2^{dir}$ measured with
    the calorimeter in PHENIX.  Each of the various dashed curves
    indicate the probability distribution of $v_2^{dir}$ result due to
    the variation of a single term in Eq.~\ref{eq:vndir}.  While
    varying $v_2^{inv}$ and $v_2^{dec}$ leaves the uncertainty
    Gaussian, varying $R_{\gamma}$ results in strongly asymmetric
    shapes.  The black solid curve shows the result when all
    uncertainties are taken into account simultaneously.
    (Figure taken from~\cite{Adare:2015lcd}.)
    (Right) Central value (solid red line) and uncertainty of the
    direct photon \vtwo for a selected \pt interval.  The upper and
    lower edges of the red shaded area correspond to the total
    uncertainty of the direct photon \vtwo as obtained from linear
    Gaussian propagation of the uncertainties of the inclusive and
    decay photon \vtwo.  The blue Gaussian reflects the $\pm 1\sigma$
    uncertainty of the measured \rgam in this \pt interval.
    (Figure taken from~\cite{Acharya:2018bdy}.)
  }
    \label{fig:ppg176_fig5}
\end{figure}

Looking at the formula from which $n$-th order direct photon flow is
calculated 
\begin{equation}
v_n^{\gamma,dir} = 
\frac{R_{\gamma}v_n^{\gamma,inc} -  v_n^{\gamma,dec}}{R_{\gamma}-1}
\label{eq:vndir}
\end{equation}

\noindent
where \rgam is the excess photon ratio, 
$v_n^{\gamma,dir},v_n^{\gamma,inc},v_n^{\gamma,dec}$ are the $n$-th
order Fourier-coefficients for direct, inclusive and decay photons,
respectively, the difficulties are obvious.  Since most inclusive
photons come from hadron decay,
$v_n^{\gamma,inc} \simeq v_n^{\gamma,dec}$ almost per definition.
Moreover, \rgam, the direct photon excess is usually close to
unity, often less than $2\sigma$ above it.  Therefore, both the
numerator and denominator in Eq.~\ref{eq:vndir} are small and within
uncertainties might even change sign.  As pointed out originally by
ALICE and discussed in detail in~\cite{Adare:2015lcd}, Gaussian error
propagation cannot be used\footnote{Except in the case when
$v_n^{\gamma,inc} \equiv v_n^{\gamma,dec}$, meaning that either there
  are no ``thermal'' photons at all (\rgam =1), or their
$v_n^{\gamma,dir} \equiv v_n^{\gamma,dec}$.
}.
Instead, the probability distribution for the possible values of
$v_n^{\gamma,dir}$ has to be modeled using Eq.~\ref{eq:vndir} and
randomizing its components with their (individually Gaussian) errors.
The procedure is illustrated in Fig.~\ref{fig:ppg176_fig5}; there the
resulting uncertainty of $v_2^{\gamma,dir}$ is clearly asymmetric,
and while the central values are essentially unchanged, the
probability distribution would even allow $v_n^{\gamma,dir}$ to change
sign.  It is also obvious that the asymmetry comes from the nonlinear
dependence of $v_n^{\gamma,dir}$ on \rgam, as seen in
Eq.~\ref{eq:vndir}, so one key to better $v_n^{\gamma,dir}$ in the
future is significant improvement on \rgam (or direct photon 
yield) measurements, which in turn starts with the cleanest possible
inclusive photon sample, \ie the best photon/electron identification 
(hadron rejection) achievable.

A recent study of the effects of hadron contamination in
conversion photon ($e^{+}e^{-}$) measurements of
\vtwo~\cite{Bock:2016jus}, now the most common method
in the ``thermal'' region, 
pointed out how important it is to derive $v_2^{\gamma,inc}$ from a 
completely clean sample.  Already 1\% $\pi^{\pm}$ contamination may
cause a 10-20\% change in $v_2^{\gamma,inc}$, and the deviation also
has a very strong \pt dependence\footnote{Note that in the low \pt
  region $e/\pi$ separation is usually cleaner in experiments using
  Cherenkov detectors than in TPC's if they rely only on $dE/dx$
  information, but no time-of-flight. 
}.

Another serious issue is insufficient information on yields and
spectra of higher mass neutral mesons, decaying in part into photons 
(anything above the \piz).  Out of these $\eta$ is by far the most
important, but for instance in \sqsn=200\,\gev \auau there is no
measurement of the $\eta$ yields or \vtwo at low
\pt~\cite{Adler:2006bv,Adare:2013wop}.  While it is firmly established that at
high \pt the $\eta/\pi^0$ ratio is constant over a large range of
colliding systems and energies~\cite{Adler:2006bv}, there is no universally
accepted method to extrapolate the $\eta$ spectrum to low \pt.
In a recent publication of the direct virtual photon
yields~\cite{STAR:2016use} the STAR Collaboration has shown that using
two different -- equally justifiable -- assumptions\footnote{A Tsallis
  blast-wave model, with the freeze-out parameters obtained by fitting
  other hadrons simultaneously (the standard procedure at STAR), and
  $m_T$ scaling the measured \piz spectrum (the PHENIX method), in both
  cases normalized to the known $\eta/\pi^0$ ratio at high \pt.
}
the resulting direct/inclusive photon ratio can change up to 43\% in
minimum bias Au+Au collisions.  Note that in~\cite{STAR:2016use} the STAR
Collaboration found that ``{\it ... the excess and total yields are
  systematically lower than the PHENIX results in 0-20\%, 20-40\% and
  40-60\% centrality bins.}''  In other words, \rgam, that plays a
crucial role in the \vtwo measurement, is much smaller than in PHENIX.
A  direct photon $v_n$ measurement by STAR with 
the same apparatus and on the same  dataset would be very helpful step to
resolve the ``direct photon puzzle''.

\subsection{Methodology: $v_n$ with respect to what and how?}
\label{sec:whichplane}

\subsubsection{Event plane method}
\label{sec:epmethod}

The traditional definition of $v_n$ comes from the Fourier-expansion
of the event-by-event azimuthal distribution of the emitted
particles with respect to a symmetry plane $\Psi$
characterizing the specific event.  If for each order $n$ of the
expansion a separate plane $\Psi_n$ is defined, the expansion with
respect to the azimuth $\phi$ reduces to 

\begin{equation}
\frac{dN(...)}{d\phi} = \Big\langle \frac{dN(...)}{d\phi}\Big\rangle
\Big( 1 + \sum_{n} 2v_n\cos[n(\phi-\Psi_n)] \Big)
\label{eq:fourier}
\end{equation}

\noindent
where $N(...)$ means the number of particles (total or some subset in
bins of \pt, $\eta$, etc.), $\Psi_n$ is the azimuth of the $n$-th
order symmetry plane in an absolute coordinate system and $v_n$ is the
amplitude of the $n$-th order term.  A key question is how -- and to
what accuracy -- $\Psi_n$ is defined\footnote{Risking being pedantic
  we should point out that {\it any} azimuthal distribution can be
  expanded in Fourier-series, irrespective of the physical origins of
  the asymmetries (anisotropy) in it.  An obvious example of
  azimuthally asymmetric events are events with jets, which will have
  non-zero $v_n$ coefficients.  Regrettably, $v_n$-s are often
  referred to as ``$n$-th order flow'', strongly suggesting
  collective, hydrodynamic behavior where there might be none at all.
}.

In each heavy ion collision for a theorist there is always a clearly
defined ``reaction plane'' (RP) spanned by the impact parameter
$\vec{b}$ and beam direction $\vec{z}$.  For smooth initial geometry
and spherical nuclei this is expected to be a mirror symmetry plane
for the overlap area, against which azimuthal distributions (\vtwo) 
are calculated.  If, in contrast, the continuous density distribution 
of nuclei is replaced by discreet, randomly placed constituents 
(nucleons or even partons), another symmetry plane called
``participant plane'' (PP) can be calculated from the constituents
that actually interact~\cite{Chatterjee:2014dqa}.  The distinction is
important: since $v_n(PP)$ is {\it defined} by the de facto
interacting particles, usually it is larger than $v_n(RP)$, defined by
an average of all, interacting and spectator particles (centers of
gravity of the two nuclei).  Note that RP and PP live both only in
theory, they are not directly accessible in an experiment.

Instead, experiments reconstruct an ``event plane'' (EP) from the
azimuthal distributions of the measured final state particles -- 
usually charged hadrons\footnote{In fixed target experiments in
  principle the event plane can be deduced the complementary way,
  from the distribution on the non-interacting ``spectator'' nucleons,
  but at colliders this is practically impossible.
}. 
This can lead to biases, most obviously from
jets\footnote{One single parton in a particular $\varphi$ direction
  produces large final state multiplicity in a relatively narrow cone.
}.  The jet bias can be mitigated (but not fully eliminated) if the EP
is determined from particles at a large (pseudo)rapidity gap from the
region where the actual measurement takes place, but note that this
method tacitly assumes that the the $n$-th order symmetry 
angle (plane) $\Psi_n$ is independent of pseudorapidity.  There are 
other sources of bias, too, like resonance decays, Coulomb-effects,
etc~\cite{Adler:2002pu,Luzum:2012da}.  Such correlations, unrelated to
the event plane, are usually called {\it non-flow correlations}.  Finally, 
it is concievable that the EP for direct photons is not always exactly 
the same as for final state hadrons -- from which EP is usually 
measured~\cite{paquet:2014}.  If so, this is a serious problem with no
simple solution in sight: while one could in principle measure the event
plane using (identified) photons instead of charged hadrons, this EP
would still be essentially identical to the hadron EP, since the
majority of photons come from FS hadron decays and inherit {\it their}
event plane, rather than the -- possibly decorrelated -- direct photon
EP.

After all these physics caveats we should mention an important
practical problem with the event plane method.  EP can only be
measured with some finite resolution $\sigma_{EP}$\footnote{The
  dispersion between the experimentally reconstructed $\Psi_n$ and the
  true $\Phi_n$ of the underlying distribution~\cite{Poskanzer:1998yz}. 
}, which depends on
multiplicity, is a strongly non-monotonic function of collision
centrality, 
and connects the ``raw'' $v_n^{raw}$ to the ``true'' $v_n^{true}$ via

\begin{equation}
v_n^{true} = \frac{v_n^{raw}}{\sigma_{EP}}
\end{equation}

\begin{figure}[htbp]
  \includegraphics[width=0.45\linewidth]{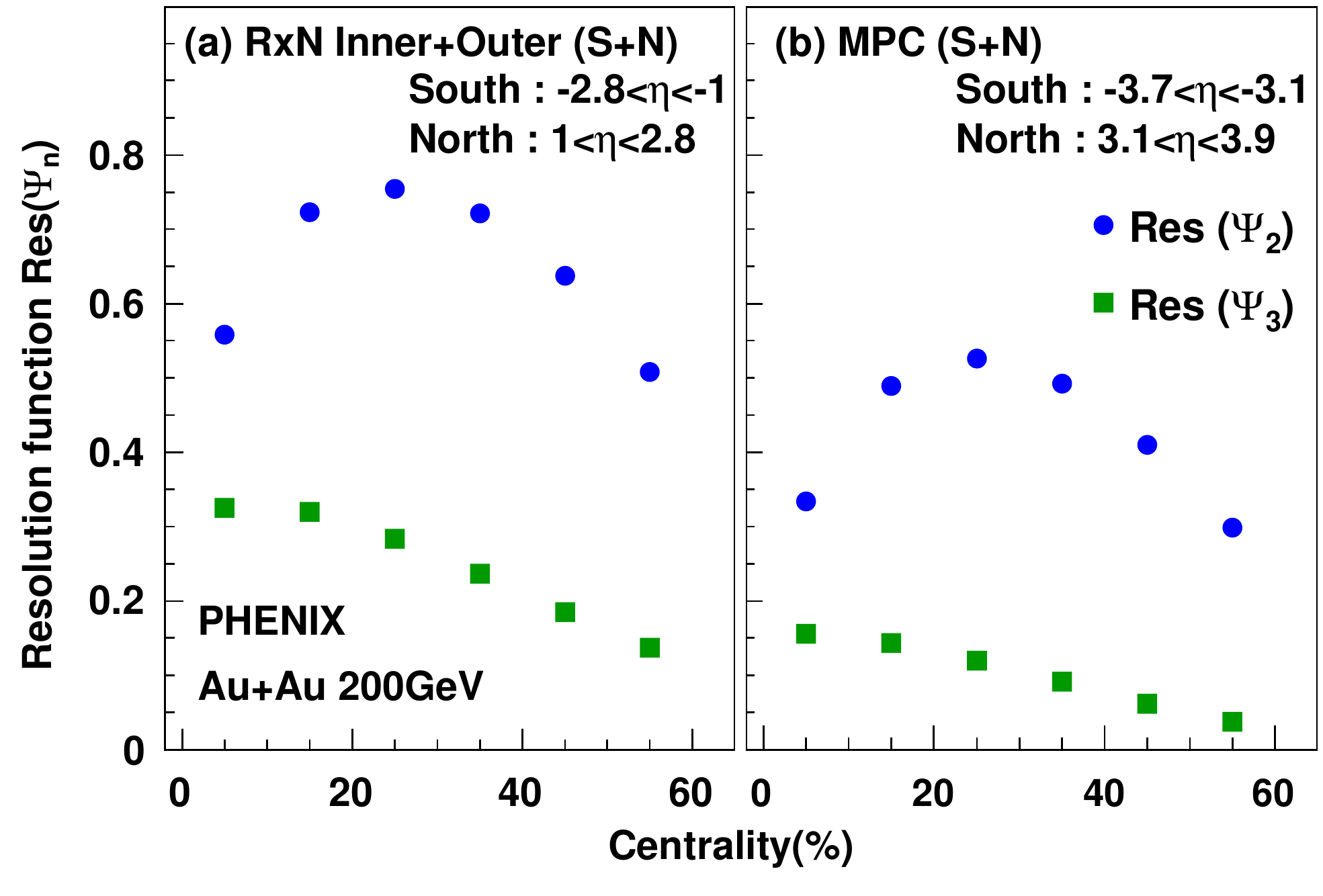}
  \caption{Event plane resolution as a function of collision
    centrality for the second and third order $v_n$ in PHENIX.  The
    so-called Reaction Plane Detector (RxN) and the Muon Piston
    Calorimeter (MPC) have different granularities and cover different
    $\eta$ ranges.
    (Figure taken from~\cite{Adare:2015lcd}.)
  }
    \label{fig:ppg176_fig1}
\end{figure}

Typical event plane resolutions are shown in
Fig.~\ref{fig:ppg176_fig1}.  Note that the $1/\sigma_{EP}$ corrections
are relatively minor for \vtwo, but depend very strongly on
centrality, while they can be a factor of 5 or higher for \vthr. 
That means that the actual modulations $v_n^{raw}$ measured in the
experiments are sometimes quite small and often similar across
centralities\footnote{Of course from a purely mathematical point of
  view the procedure is correct, but the critical reader of
  experimental papers should be aware that the final results are small
  measured numbers subject to large corrections.
}.  Using two or more EP detectors with different resolutions (and
still getting consistent results) alleviates some of these concerns.
Typical second- and third-order event plane resolutions are shown in
Fig.~\ref{fig:ppg176_fig1}. 

Another serious issue comes from the fact that $v_n$ can rarely be 
measured in a single event, instead, it is calculated from a large 
ensemble of events (each having its own set of $\Psi_n$ event planes 
in Eq.~\ref{eq:fourier}).  If event-by-event fluctuations
of $v_n$ were negligible, this would not be a problem, but apparently
they are not.  Relative fluctuations can reach 
30-50\%~\cite{Alver:2006wh,Alver:2007rm}, so the measured $v_n$ is
the mean $\langle v_n \rangle$ at best -- when the event-plane
resolution is high.  However, if it is low, the event plane
measurement yields the root-mean-square value
$\sqrt{\langle v_n^2 \rangle}$, and in general the result lies
somewhere between those two values~\cite{Luzum:2012da}.  This
ambiguity makes comparisons between experimental results (and theory)
difficult.

\subsubsection{Scalar product method}
\label{sec:spmethod}

Less biased methods exist
to measure azimuthal asymmetries of particle distributions, notably
the cumulant expansion of multiparticle
correlations~\cite{Borghini:2000sa,Borghini:2001vi} and a variant of the
event plane method called ``scalar product'' method~\cite{Adler:2002pu}.
These ``make for a superior measurement because they consistently
yield the rms value of $v_n$, while introducing no disadvanage
compared to the traditional event-plane
measurements''\cite{Luzum:2012da}. 
The scalar product method is based on the $n$-th order flow
vector of $N$ particles defined as

\begin{equation}
Q_n = |Q_n|e^{in\Psi_n}= \frac{1}{N}\Sigma_j e^{in\phi_j}.
\end{equation}

The flow vector $Q_n$ of the particle in question (e.g. a photon) is
then related to the flow vectors $Q_{nA},Q_{nB}$ of reference particles
in two ``subevents'' $A,B$ (in practice particles in some reaction
plane detectors, separated in $\eta$ from the region where \vtwo,
\vthr are measured) and the scalar product
$v_n$[SP] is then calculated as

\begin{equation}
v_n[SP] = \frac{\langle Q_nQ_{nA}^* \rangle}
{\sqrt{\langle Q_{nA}Q_{nB}^* \rangle }}.  
\end{equation}

Note that independent of multiplicity, the scalar product method
always yields the root mean square 
$v_n$ ($\sqrt{\langle v_n^2  \rangle}$), 
just as the low resolution
limit in the event plane method~\cite{Luzum:2012da,Paquet:2015lta}

\begin{equation}
v_n[EP_{low res}] = v_n[SP] = 
\frac{\langle v_n^{\gamma}v_n^h
  \cos(n(\Psi_n^{\gamma}-\Psi_n^h))\rangle}
{\sqrt{\langle (v_n^h)^2\rangle}}
\end{equation}

\noindent
where the upper index $^h$ refers to the reference particles (usually
hadrons). 

While earlier photon flow measurements used the event plane method, in 
the recent direct photon elliptic flow publication by
ALICE~\cite{Acharya:2018bdy} the scalar product method was applied.

\subsection{Attempts to resolve the direct photon puzzle}
\label{sec:attempts}

In the following the PHENIX results will be compared to various model
calculations\footnote{ALICE now also has yields and \vtwo from the
  same dataset (albeit no \vthr), but since the PHENIX data triggered
  the ``direct photon puzzle'' in 2011, a wider range of models have
  been compared to them.  Note that both the PHENIX and ALICE data
  have been obtained by more than one analysis method involving
  different sets of subdetectors, which means an (almost) independent
  confirmation of the results within the experiments.
}.

As discussed earlier, the essence of the direct photon puzzle is the
simultaneous presence of large yields and large azimuthal asymmetries
-- comparable to that of hadrons -- for a signal that is penetrating, 
produced continuously, thus integrates the entire time history of 
instantaneous rates and expansion velocities.  The conventional line
of thought was that the overwhelming part of low \pt photons is of
thermal origin from the QGP and the hadron gas. It was expected that
 the instantaneous
rates are highest early on, when temperature is highest but production 
is still isotropic, while large anisotropies come from
photons produced late, thus inherit the maximum velocity 
anisotropy of their parent partons/hadrons.  Plausible and not much
contested {\it before} the observation of a photon \vtwo similar to
that of hadrons, but {\it after} the results 
in~\cite{Adare:2011zr,Adare:2015lcd}
such scenario appears to be incomplete at best, completely false at
worst.

Predictably, many theorists took up the challenge.  Large
amount of work was invested, 
new models, new insights were published or presented at
conferences and at dedicated workshops~\cite{tpd:2011,trw:2012,ect:2013,
gsi:2014,tpd:2014,ect:2015,ect:2018}.  
New ideas ranged from tuning the traditional ``thermal production /
hydrodynamic evolution'' picture, through suggestions of
new mechanisms that would give large asymmetries to the earliest
photons,
to the quite radical concept of suppressing
photon production from the QGP phase due to slow chemical
equilibration of  quarks.  Hydrodynamic
models were upgraded from ideal to viscous, included
event-by-event fluctuations of the initial geometry.
Transport models (circumventing hydro) re-emerged and proved to
be quite successful, as did those that re-defined the
initial state based on the Color Glass Condensate (CGC)
picture or the process of thermalization.  
Below we will discuss a few models in some detail.
So far none of them provided a completely satisfactory description of
the experimental observations, but they gave us a huge influx of
creative new ideas at a time when they are really 
needed\footnote{A personal remark.  A few years ago one could often
  hear that ``we are close to being able to formulate the Standard
  Model of heavy ion collisions''.  With the advent of surprising
  observations like apparent strong collectivity even in \pp
  collisions, perplexing observations in very asymmetric collisions,
  losening connection between event activity and collision
  geometry, hints of the critical point at very different collision
  energies depending on the observable, etc.  -- with all these new
  developments a comprehensive and coherent description of heavy ion
  collisions is not around the corner yet.  Direct photons didn't
  fully unveil the history encoded in them so far -- some day they
  will, I hope.  But they already did something equally important --
  {\it they denied us intellectual complacency}.
}.
The relevant literature is quite extensive, and reviewing it in its
entirety would far exceed the scope of this paper.  While we will
present several interesting ideas that address only part of the
``puzzle'', our selection gives precedence to models that attempt to 
explain yield and flow simultaneously.

\subsubsection{Hydrodynamical models}
\label{sec:hydro}

\vspace{0.1in}
{\it Expanding elliptic fireball model and its extensions} \\
\vspace{0.1in}
The first attempt to simultaneously explain the observed high
``thermal'' photons yields~\cite{Adare:2008ab} and \vtwo~\cite{Adare:2011zr} has
been published 2011 in~\cite{vanHees:2011vb}.  The expansion dynamics has 
been modeled by an elliptic blast-wave source, with parameters
adjusted to measured spectra and \vtwo of light and multi-strange
hadrons.  The QGP radiation is a parametrization of the complete
leading order in $\alpha_s$ rate as given in~\cite{Arnold:2001ms}, 
the emission in hadronic matter is based on~\cite{Turbide:2003si}.  In
order to compare to data, ``primordial'' (prompt) photons are added
two different ways: from an NLO pQCD calculation and from a fit to
the measured PHENIX \pp data -- both scaled by the number of $NN$
collisions.  An important addition is the use of effective meson and
baryon chemical potentials (rather than chemical equilibrium
throughout the hadronic phase); this enhances photon production in
the later hadronic stages, increasing \vtwo.  When the transverse
acceleration of the fireball is increased from earlier $a_T$=0.053$c^2$/fm
to $a_T$=0.12$c^2$/fm, the photon spectrum is well described while
\vtwo is at the lower end of the systematic uncertainty band of the
data\footnote{An interesting consequence was the ``disappearence of
the QGP window''.  In earlier models each of the
three most important contributors to photon production -- prompt
radiation, thermal yields from the QGP and from the hadron phase
have a characteristic \pt range where they dominate over other
sources.  Depending on the model, this ``window'' was somewhere in
the 2-5\,\gevc  \pt range, but in this model at maximum $a_T$ (needed
to reproduce the invariant yields) the QGP window virtually disappears.
}.

The model has been updated in~\cite{vanHees:2014ida} by implementing a
lattice-QCD based EOS, employing an ideal hydrodynamic model with
non-vanishing  initial  flow, and introducing a ``pseudo-critical''
enhancement of the QGP and hadronic rates around
$T_{pc}\approx$170\,MeV.  Meson-meson Bremsstrahlung is also
included.  The results are compared to the PHENIX \sqsn=200\,GeV
Au+Au ``thermal'' photon spectra~\cite{Adare:2008ab} and the updated \vtwo
and \vthr results~\cite{Adare:2015lcd} in Fig.~\ref{fig:ppg176_fig7_9}.
For both observables the calculations underpredict the data.

\begin{figure}[htbp]
  \includegraphics[width=0.2\linewidth,angle=-90,origin=c]{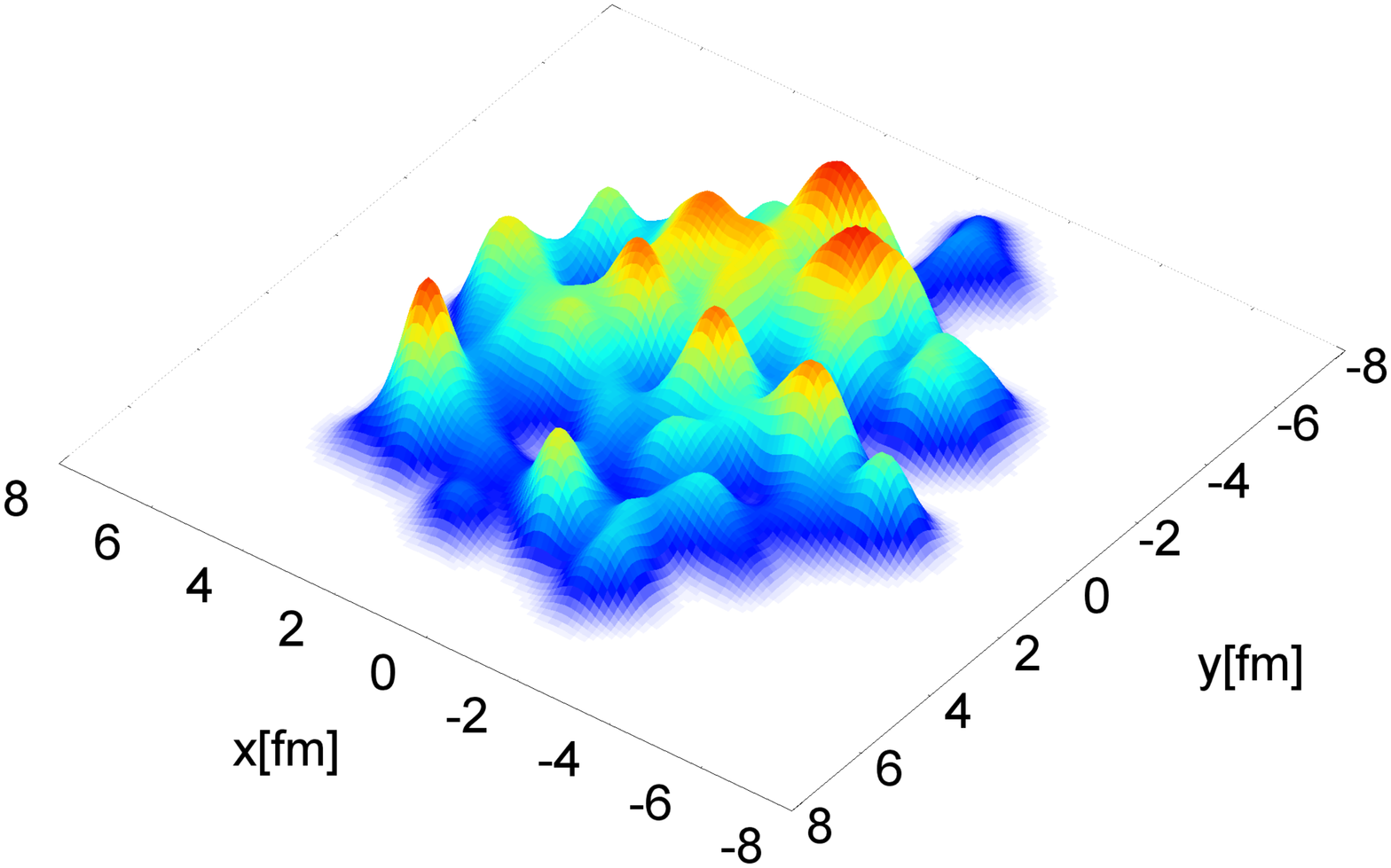}
  \includegraphics[width=0.2\linewidth,angle=-90,origin=c]{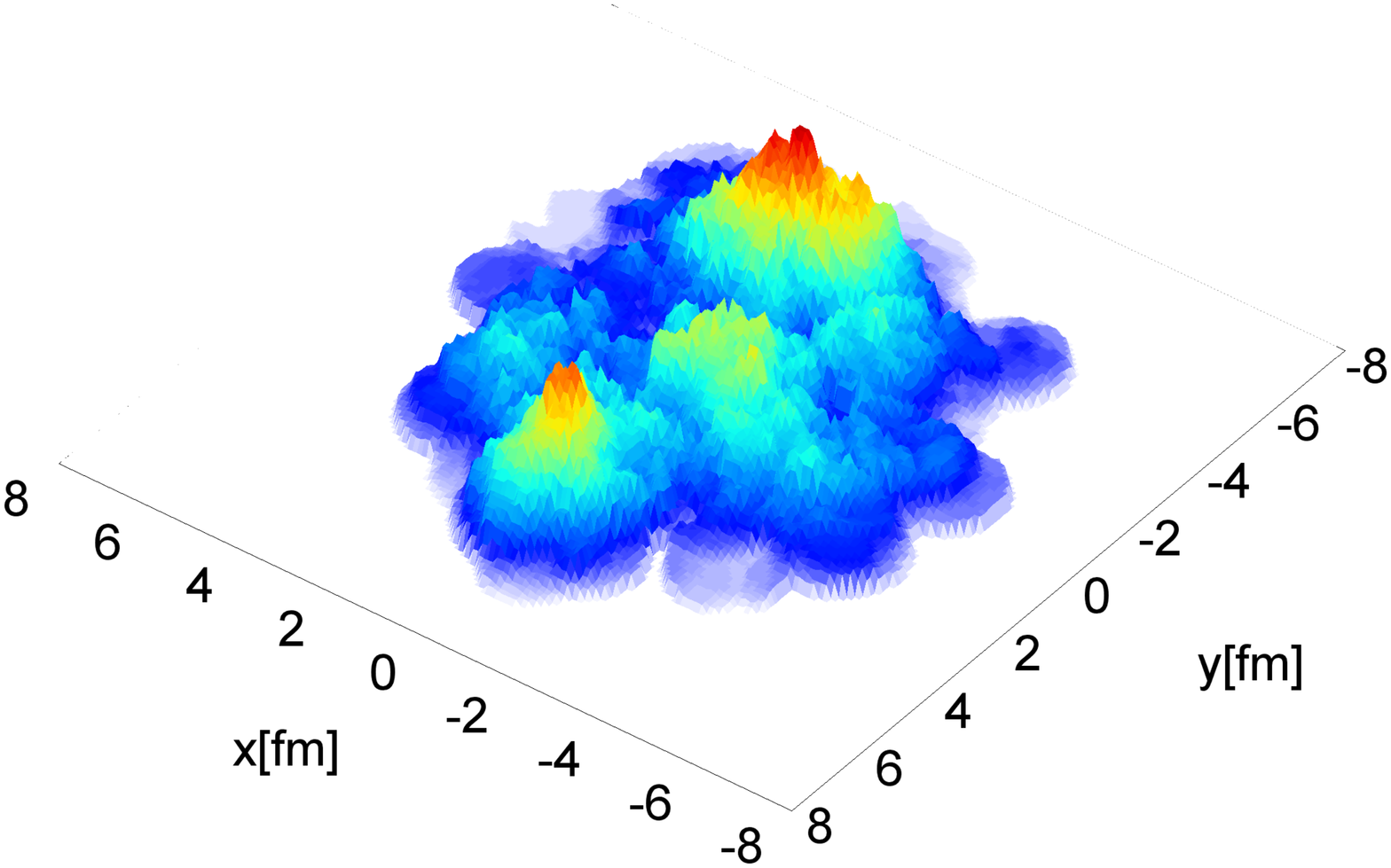}
  \includegraphics[width=0.2\linewidth,angle=-90,origin=c]{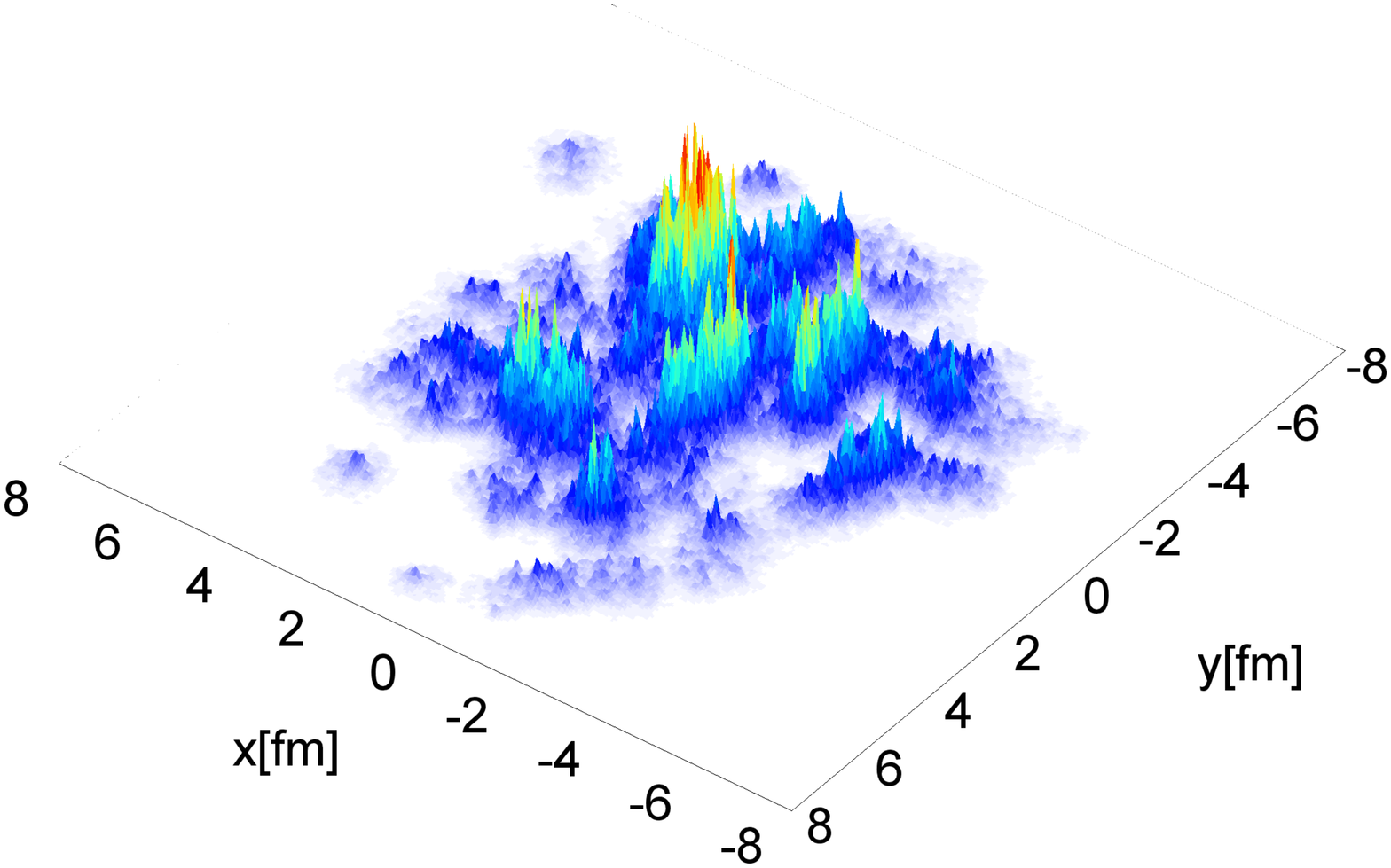}
  \caption{Initial energy density (arbitrary units) in the transverse
    plane for a heavy ion collision using three different models.
    Left: MC Glauber.  Middle: MC-KLN.  Right: IP-Glasma.
    (Figure taken from~\cite{Schenke:2012wb}.)
  }
    \label{fig:prl108.252301}
\end{figure}

\vspace{0.1in}
{\it Ideal and viscous hydrodynamics} \\
\vspace{0.1in}
Elliptic flow of ``thermal'' photons using relativistic hydrodynamics
has first been calculated in~\cite{Chatterjee:2005de} assuming
0.2\,fm/$c$ thermalization time and 520\,MeV initial temperature.
Using the same (2+1)D hydrodynamic code in~\cite{Turbide:2007mi}
the \pt range was extended by including hard processes
(also in~\cite{Turbide:2005bz} it has been pointed out that at high 
\pt {\it negative} \vtwo is expected for photons coming from
jet-plasma interaction).  Since no meaningful measurements of low \pt
photon \vtwo existed at the time, these can be considered predictions.
Both calculations predicted the trend of photon \vtwo first rising,
then falling with \pt and essentially vanishing at high \pt, but
grossly underestimated its magnitude, claiming that it will be
significantly lower than the hadron \vtwo.

Once it became obvious that from hadronic data estimates of $\eta/s$
(shear viscosity to entropy density) ratio can be made, viscous
relativistic hydro calculations became the rule.  The space-time
evolution of the system (and thus \vtwo) are obviously affected, but
so are the rates, too.  The effects of both
shear and bulk viscosity on photon production has been studied for
instance in~\cite{Shen:2013cca,Shen:2014cga,Paquet:2015lta}.  Viscous
corrections to the rates are small, particularly in the ``thermal''
region relevant here (they increase with \pt), but photon \vtwo at low
\pt is actually {\it suppressed} by 20-30\%, making the ``puzzle''
even bigger.

\vspace{0.1in}
{\it Event-by-event hydrodynamics} \\
\vspace{0.1in}
Traditionally hydrodynamic calculations started with assuming
smooth initial conditions, including geometry, a good starting point as
long as the colliding ions are large, only the first few Fourier
components are considered in azimuth {\it and} both nuclei are
relatively large\footnote{Even then, \vtwo in the most central
  collisions is underpredicted if smooth initial conditions are chosen.
}.  
Beyond that -- and particularly for very asymmetric
collisions, like $p/d$+A -- the fluctuating inner structure of the 
nucleus has to be taken into account on an event-by-event basis; the
resulting initial density profile serves as an input to the hydro
calculation.  The initial configuration (density profile) can be
obtained for instance with a Glauber Monte-Carlo~\cite{Miller:2007ri}.
CGC-inspired models include the early 
MC-KLN~\cite{Kharzeev:2001gp,Kharzeev:2004if}, where no pre-thermal
evolution of the gluons was considered, 
or the more recent impact-parameter dependent Glasma
flux tube picture~\cite{Schenke:2012wb} 
(see Fig.~\ref{fig:prl108.252301}).  
The EKRT model~\cite{Eskola:1999fc} combines the idea of gluon
saturation with the dominance of few GeV jets (``minijets'') as
principal sources of particle production~\cite{Niemi:2015qia}, 
and also results in a
``lumpy'' initial state for the hydro evolution.

Assuming that thermal radiation is exponential in temperature and
linear in radiating volume, initial ``hotspots'' from fluctuating
initial conditions enhance photons at higher \pt, while the low \pt
part of  the spectrum from the plasma is less effected, because it
comes mostly from the volume-dominated, later plasma
stage~\cite{Chatterjee:2011dw}.

Applying event-by-event fluctuating initial conditions to \cuau
collisions in~\cite{Dasgupta:2019whr} the authors argue that $v_1$
is more sensitive to the initial formation time of the plasma compared
to \vtwo, \vthr, and simultaneous measurement of $v_1$, \vtwo, \vthr 
in \cuau would be very helpful in clarifying the direct photon
puzzle\footnote{The prediction is that $v_1$ is negative at low \pt,
  and changes sign around \pt=2.5\,\gevc.
}.

\subsubsection{Initial state, (fast) thermalization}
\label{sec:initialstate}

\vspace{0.1in}
{\it Glasma, slow chemical equilibration of quarks} \\
\vspace{0.1in}
Glasma is a conjectured transient state of matter between the initial
state Color Glass Condensate or CGC~\cite{McLerran:1993ni} and the thermally
equilibrated Quark-Gluon Plasma.  In CGC most of the energy of the
colliding nuclei is carried by gluons with a flat distribution
up to relatively high momenta (the saturation momentum
$Q_s\approx$2\,\gevc).  Gluon splitting then provides the mechanism
for early (gluon) thermalization~\cite{Monnai:2014xya}, and
recombination populates the high \pt gluon spectrum above
$Q_s\approx$2\,\gevc.  But to leading order quarks are needed to produce
photons.   Quarks are created via pair production from gluons and
equilibrate later than gluons (typical times 0.8\,fm/$c$ and
2\,fm/$c$)~\cite{Monnai:2014xya,Monnai:2014kqa}. 
Photon emission from the Glasma has first been studied
in~\cite{Chiu:2012ij} and shown that Glasma photon production can
describe the centrality dependence of low \pt photon production
(geometric scaling), and since quarks become substantial only at later
stages of the Glasma, the photon \vtwo also becomes
higher~\cite{Monnai:2014kqa}, a step in the right direction, but the
yields and \vtwo are not adequately reproduced yet.

\vspace{0.1in}
{\it Bottom-up thermalization} \\
\vspace{0.1in}
There is little argument that the medium formed in \AA collisions
becomes (locally) thermalized relatively fast\footnote{Somewhere in
  the range of 0.15-1\,fm/$c$.
}, 
but little is known about {\it how} this happens.  In a 2001
paper~\cite{Baier:2000sb} the ``bottom-up'' thermalization scenario was
put forward to explain it.  Assuming that the saturation momentum 
is high ($Q_s \gg \Lambda_{QCD}$) the basic idea was that early on
($\tau \approx \alpha^{-5/2}Q_{s}^{-1}$) emission
of soft gluons dominate which quickly equilibrate and form a thermal
bath, in which hard gluons lose their energy (and heat it further up)
until about $\tau \approx \alpha^{-13/5}Q_{s}^{-1}$.  Thermalization
begins with the soft momentum modes.  A very recent
study~\cite{Berges:2017eom} found that the ``bottom-up'' scenario
(called BMSS) captures the correct physics of the glasma.  Based on
this it provides a parametric estimate of the pre-equilibrium photon
production, claiming that at RHIC energies the glasma contribution is
even {\it larger} than the thermal contribution from the QGP,
particularly for more peripheral collisions (at LHC energies this is
no  longer true).  This scenario might explain the large yields and
the surprising observation that the effective temperatures barely
change with collision centrality, but its effect on $v_n$ is not
studied yet.

\subsubsection{Transport calculations}
\label{sec:transport}

\begin{figure}[htbp]
  \includegraphics[width=0.49\linewidth]{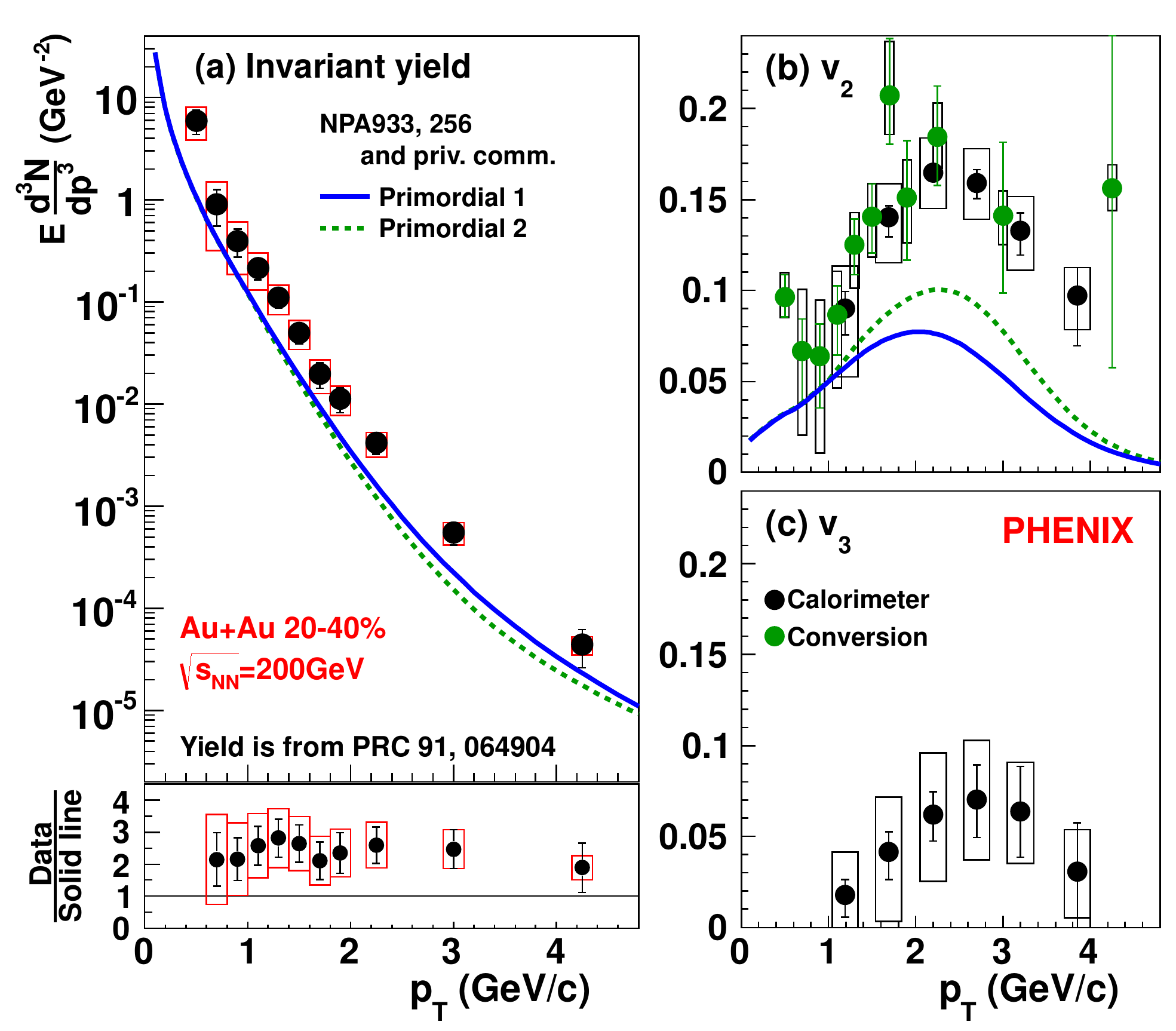}
  \includegraphics[width=0.49\linewidth]{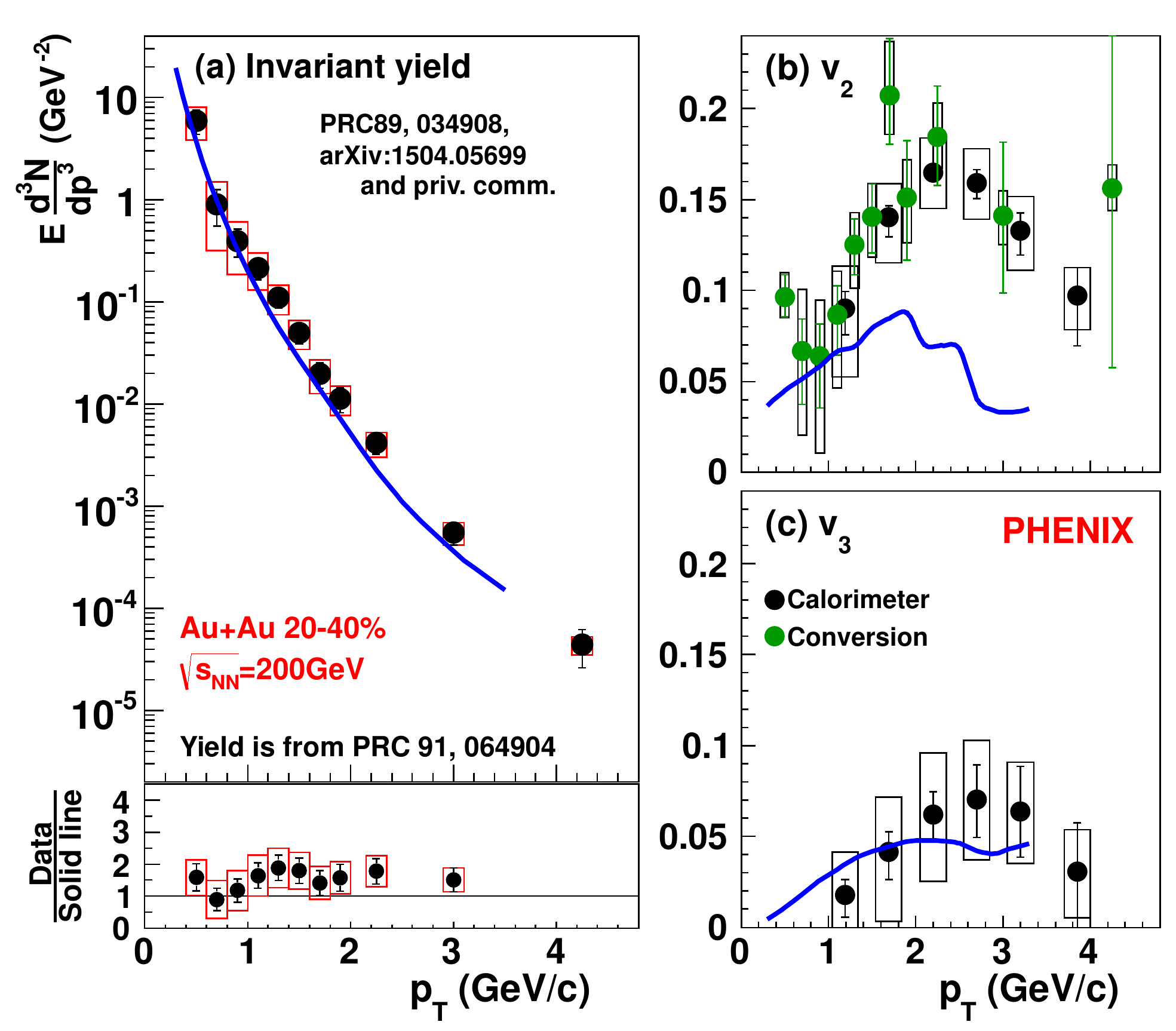}
  \caption{(Left)
    Comparison of the direct photon yields~\cite{Adare:2014fwh} and
    the updated \vtwo measurement~\cite{Adare:2015lcd} 
    with the updated fireball model implementing pseudo-critical
    enhancement of thermal yields~\cite{vanHees:2011vb,vanHees:2014ida}.
    The two curves for the yield and \vtwo correspond to two different
    parametrizations of the prompt photon component as described in
    the text.  
    (Figure taken from~\cite{Adare:2015lcd}.)
    (Right) Comparison of the direct photon yields 
    and \vtwo, \vthr    with the PHSD model~\cite{Linnyk:2013wma,Linnyk:2015tha}.
    (Figure taken from~\cite{Adare:2015lcd}.)
  }
    \label{fig:ppg176_fig7_9}
\end{figure}

\vspace{0.1in}
{\it Parton-hadron-string dynamics (PHSD)} \\
\vspace{0.1in}
The parton-hadron-string dynamics (PHSD) model~\cite{Cassing:2009vt} 
``is an off-shell transport approach
that consistently describes the full evolution of a relativistic
heavy-ion collision from the initial hard scatterings and string
formation through the dynamical deconfinement phase transition to the
QGP as well as hadronization and the subsequent interactions in the
hadronic phase.''~\cite{Linnyk:2015tha}.  
Correspondingly, it is a skillful combination and extension of
previous models describing various phases of the collision.  Prompt
photons (initial hard scattering) are calculated with standard pQCD.
Description of the strongly interacting system out-of-equilibrium
is based on the Kadanoff-Baym 
theory~\cite{Baym:1961zz,Baym:1962sx,Vanderheyden:1998ph},
transport description of equilibrated 
quarks and gluons is based on the dynamical
quasiparticle model (DQPM)~\cite{Cassing:2007nb} that has nonvanishing
widths of the partonic spectral functions\footnote{One of the
  consequences is that the evolution of the system closely resembles
  to that found in hydrodynamics with the same $\eta/s$ and initial
  spatial eccentricity~\cite{Cassing:2008sv}.
}
and is tuned to match lQCD results for the QGP.  For the hadronic
sector the hadron-string dynamics (HSD)
approach~\cite{Ehehalt:1996uq,Bratkovskaya:1996qe} is used.  Processes
incorporated include hadron decays 
($\pi^0 \rightarrow \gamma\gamma$, $\eta \rightarrow \gamma\gamma$,
$\omega \rightarrow \pi^0 + \gamma$, 
$\eta^{'} \rightarrow \pi^0 + \gamma$,
$\phi \rightarrow \eta + \gamma$,
$a_1 \rightarrow \pi^0 + \gamma$)\footnote{The model includes 
  {\it all} hadron decay photons, including those that experiments
  subtract, like \piz and $\eta$; for comparisons with measured data
  these obviously have to be subtracted.  An advantage of calculating
  all photon contributions in the same framework is that when
  comparing to data, the exact same contributions can be discarded
  that have been subtracted by the particular experiment.  This is
  true for both the yields and $v_n$.
},
their interactions ($\pi \pi \rightarrow \rho \gamma,
\rho \pi \rightarrow \pi \gamma$), meson-meson and meson-baryon
Bremsstrahlung ($m+m \rightarrow m+m + \gamma$, 
$m+B \rightarrow m+B + \gamma$),
vector meson - nucleon interactions 
(like $V+p \rightarrow \gamma + p/n$), and the 
$\Delta \rightarrow N\gamma$ resonance decay.
The Landau-Pomeranchuk-Migdal (LPM) effect~\cite{Knoll:1993ic},
suppressing radiative photon production in dense systems due to
coherence, is studied and found to influence the total yields from the
QGP below 0.4\,\gev, but negligible for photons from the hadronic
phase.   --  Similarly, \vtwo ($v_n$) is the weighted sum of the
``partial flow'' from the individual photon sources

\begin{equation}
v_n(\gamma) = \sum_i v_n(\gamma^i) w_i(p_T)
\end{equation}

\noindent
where the summation goes over the various processes $i$ and $w_i(p_T)$
is the relative contribution of process $i$ at a given \pt 
($N_i(p_T)/\sum_i N_i(p_T)$).
-- Quantitatively, almost half of the total direct photon yield in
PHSD is produced in the QGP~\cite{Linnyk:2015tha}.  The corresponding
partial \vtwo is small, and not fully compensated by the large \vtwo
of photons from the hadronic phase.  On the other hand \vthr 
(``triangular flow'') is manifestly non-zero and consistent with the
data at low \pt.  Comparisons to PHENIX data in mid-central \auau
collisions are shown in Fig.~\ref{fig:ppg176_fig7_9}.  The yield is well
reproduced down to the lowest \pt, but \vtwo is underpredicted except
at the lowest \pt where radiation from hadrons and Bremsstrahlung is
expected to dominate.

\vspace{0.1in}
{\it Hybrid approaches} \\
\vspace{0.1in}
Hybrid approaches are an excellent, pragmatic solution to the problem
that various stages of the collision, like the initial stage and
formation of a (thermal) medium, then the space-time evolution of this
medium, and the late stages before complete kinetic freeze-out can
each be most conveniently described in a different framework.
The centerpiece is typically a relativistic viscous hydrodynamics code
describing the evolution of the medium (for instance
MUSIC~\cite{music:2010,Schenke:2010nt}), sufficiently modular to
accept various inputs for the initial (and pre-thermal) state,
instantaneous rate models, and ``afterburners'' for photon production
by hadrons in the post-hydro era.  One typical incarnation is the
calculation in~\cite{Ryu:2015vwa} where the initial state is
determined with the IP-Glasma model, and the afterburner for hadronic 
rescattering is the ultrarelativistic quantum molecular dynamics
(UrQMD) code.  A similar scheme is applied for (``thermal'') photons 
in~\cite{Paquet:2015lta}; the photon emission rates used are
from~\cite{Turbide:2003si,Heffernan:2014mla,Holt:2015cda}.
The calculations include corrections for both bulk and shear
viscosity, and underestimate both the spectrum and \vtwo (viscosity
softens the spectra somewhat, and decreases \vtwo significantly).
Late stage emission (from UrQMD) on the other hand increases direct
photon \vtwo~\cite{Paquet:2015lta}.

Coarse graining is a method to
overcome the difficulties that electromagnetic emissivities are
typically calculated near thermal equilibrium, but transport
formulations usually don't provide local
temperatures~\cite{Paquet:2015lta}.   Instead, the
transport final states are divided into cells on a space-time grid 
(coarse grained) and local temperatures assigned using the equation of
state~\cite{Huovinen:2002im,Endres:2015fna}.

\vspace{0.1in}
{\it Boltzmann approach to multiparton scattering (BAMPS)} \\
\vspace{0.1in}
The Boltzmann approach to multiparton scattering (BAMPS) has been
introduced in~\cite{Xu:2004mz} as a 3+1 dimensional Monte Carlo cascade
for on-shell partons obeying the relativistic Boltzmann equations.  In 
a recent study~\cite{Greif:2016jeb} of nonequilibrium photon production
with BAMPS finds that it is larger than that of the QGP, the spectra
are harder.  This is mainly due to scatterings of energetic jet-like
partons with the medium\footnote{Similar to ``jet-photon conversion'',
  discussed in Sec.~\ref{sec:isospin}.
}.
So far only the QGP has been studied in~\cite{Greif:2016jeb}, and with a
special set of initial conditions, where gluons dominate and quarks
are produced by inelastic scattering, delaying their appearance.
Therefore, the yields are far below what
traditional hydro codes or a full transport simulation (PHSD) would
give.  Nevertheless there are three important partial lessons to be
learned for nonequilibrium processes in the QGP proper.  
First, nonequilibrum spectra
are harder than thermal ones.  Second, $2\rightarrow 3$ processes are
important, almost on par with $2\rightarrow 2$ processes at
2-3\,\gevc.  Third, using running $\alpha_s$ rather than
$\alpha_s=0.3$ fixed, increases the yield by almost a factor of 2.
Finally, the \vtwo of these ``BAMPS photons'' is actually 
{\it  negative}, aggrevating, instead of alleviating the \vtwo
problem.

\vspace{0.1in}
{\it ``No dark age'' -- Abelian flux tube model}\\
\vspace{0.1in}
The role of photons from the early stages of the collision
(pre-equilibrium) is examined in~\cite{Oliva:2017pri} using the Abelian
flux tube model (AFTm) to define the initial gluon fields and their
evolution into the QGP.  The fast decay of the fields results in
quarks and gluons, which in turn scatter and produce photons very
efficiently early on, before the QGP is formed.  The model is somewhat
similar to the ``bottom-up thermalization'', but it is embedded into a
relativistic transport code and follows the dynamical evolution of the
system up to freeze-out.  The processes implemented in the collision
integral are the basic $2\rightarrow 2$ processes as shown in
Fig.~\ref{fig:partongraph} with cross-sections as in
Eqs.~\ref{eq:wongcompton} and~\ref{eq:wongannihilation} with $m=0$, 
modified by a
temperature-dependent overall factor $\Phi(T)$ to account for higher
order (radiative) processes.  $\Phi(T)$ is chosen such that the overall
AMY production rate~\cite{Arnold:2001ms} is reproduced when the system
is in the equilibrated QGP phase.  The final number of quarks and
gluons with this AFTm initialization is the same as with a traditional
Glauber initialization for hydro (equilibrium assumed at 
$t_0\approx 0.6$\,fm/$c$), but due to early appearance of quarks the
photon radiation with AFTm is about 30\% higher at RHIC energies than
that obtained with the Glauber model.  At \pt$\approx$2\,\gevc the 
contribution of early stage photons is comparable to the yields from 
the fully formed QGP, in other words, there is ``no dark age'', the
early stage is quite bright.  The calculated photon spectrum from the
QGP at RHIC energies is compared to other transport calculations.
PHSD~\cite{Linnyk:2015tha} is consistent with AFTm at higher \pt 
(2\,\gevc and above), but exceeds AFTm below that.
BAMPS~\cite{Greif:2016jeb}, in contrast, falls below AFTm in the entire
$0.5<p_T<3$\,\gevc range due to the delayed appearance of quarks and
thus reduced emission of photons.  Note that currently AFTm does not
address the question of direct photon \vtwo, but due to the early
production it would presumably underpredict the data.

\subsubsection{Other ideas}
\label{sec:otherideas}

\vspace{0.1in}
{\it ``Semi-QGP''}\\
\vspace{0.1in}
The transition from QGP to hadron gas, now widely believed to be a
cross-over\footnote{At top RHIC and LHC energies, it is less clear
  what happens at lower RHIC energies and at SPS, but new
  medium-energy facilities in construction might help to clarify the
  situation.}, 
happens around a critical temperature $T_c$.  High above $T_c$
perturbative methods can be used, at lower temperatures hadronic
models are valid, but near $T_c$ both approaches break down.  In order
to understand the transition to confinement the degree of ionization
of the color charge, characterized by the expectation value of the
Polyakov loop is used~\cite{Hidaka:2008dr,Fukushima:2017csk}.  The Polyakov 
loop is unity (full ionization of color charge) at $T\gg T_c$, but 
decreases as $T\rightarrow T_{c}^{+}$, because color fields gradually 
evaporate and are replaced by color singlet excitations (hadrons).  
This intermediate state is called ``semi-QGP'' and its effects on
``thermal'' photon production and \vtwo are studied in~\cite{Gale:2014dfa}.
The conclusion is that the net yield from the QGP phase decreases, but
at the same time the total \vtwo is enhanced, since relatively more
photons come from the hadronic phase, where the instantaneous \vtwo is
maximum.  Results for yield, \vtwo and \vthr calculated with this
model are shown as dashed red curves in Fig.~\ref{fig:ppg176_fig8}.
The yield is still low, as is \vtwo, and, more significantly, the
maximum of \vtwo is at much lower \pt than in the data.  The authors
themselves point out that photons from parton fragmentation should be
included\footnote{This would enhance the yields, but not necessarily
  \vtwo, since the jet axis does not necessarily have to coincide with
  the \vtwo plane.
}
and that the semi-QGP addresses only the $T\rightarrow T_{c}^{+}$
region, but not the possibly enhanced production in the hadronic
phase, claimed for instance in~\cite{Lee:2014pwa}.

\begin{figure}[htbp]
  \includegraphics[width=0.6\linewidth]{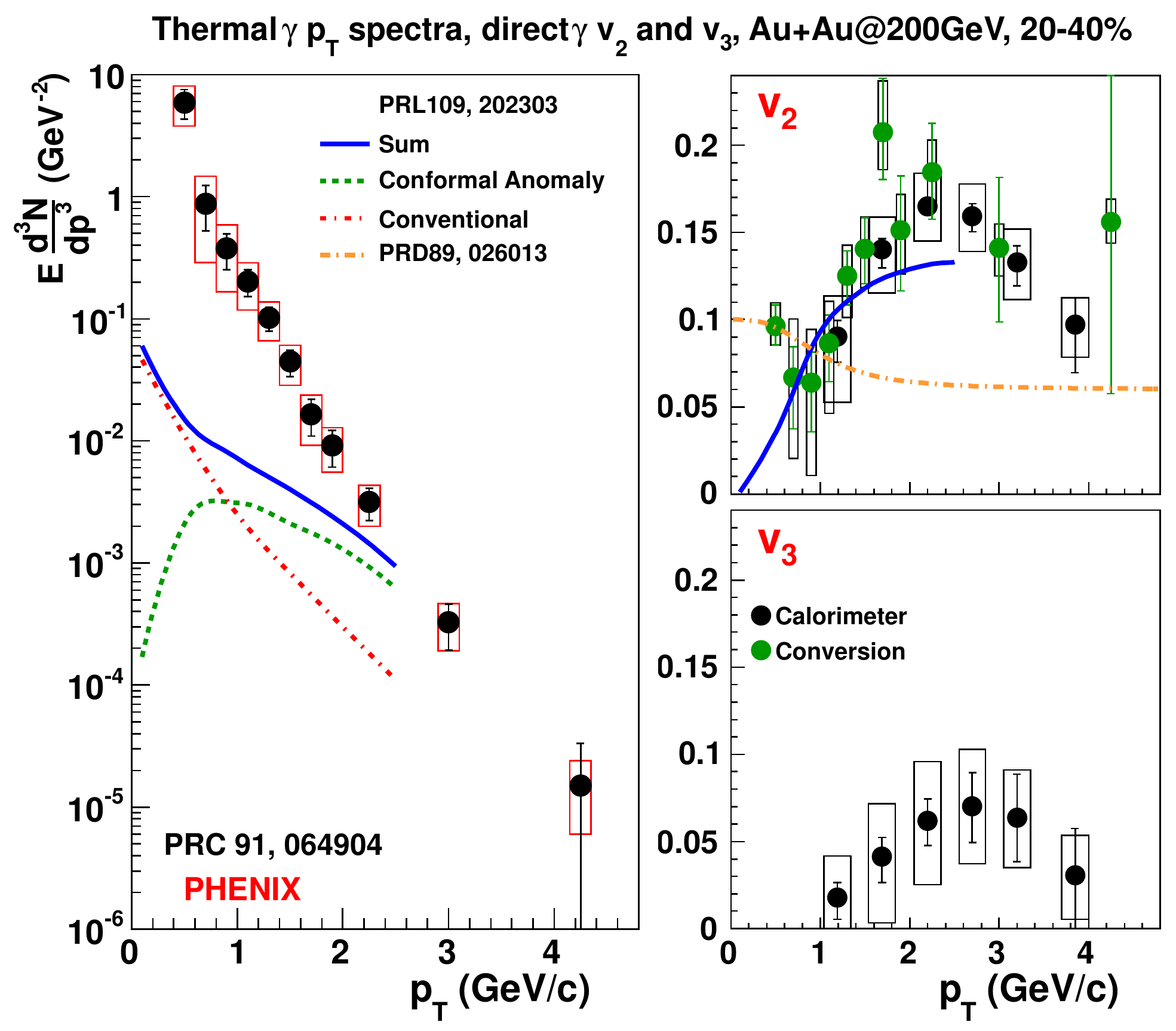}
  \caption{
    Yield and anisotropy in models that include
    strong initial magnetic field.  Calculations of the yield
    (conformal, conventional and sum as solid blue line) are 
    based on~\cite{Basar:2012bp} where the respective \vtwo is also
    calculated.  The yellow dash-dotted curve for \vtwo is
    from~\cite{Muller:2013ila} where yield estimates are not given.  Note
    that for models where \vtwo is due to magnetic field effects,
    \vthr is zero.  (Figure courtesy of Takao Sakaguchi.)
  }
    \label{fig:ppg176_magnetic}
\end{figure}

\vspace{0.1in}
{\it Early \vtwo from photons in presence of a magnetic field.} \\
\vspace{0.1in}
An interesting scenario has been put forward in~\cite{Muller:2013ila} to
circumvent the problem that momentum anisotropies need time to build
up, therefore, early ``thermal'' photons will have little \vtwo.
The existence of a strong magnetic field in non-central heavy ion
collisions has long been conjectured and its effect on both photon
production and anisotropy studied for weak coupling
in~\cite{Basar:2012bp} (see Fig.~\ref{fig:ppg176_magnetic}).  
For a strongly coupled plasma
the AdS/CFT correspondence is used in~\cite{Muller:2013ila}. The
  authors find that -- due to the magnetic field -- photons with
  in-plane or out-of-plane polarization acquire different \vtwo.
  Actually, photons with out-of-plane polarization photons will have
  very large positive \vtwo as \pt$\rightarrow 0$, making this an
  unusual scenario (but also proposed in~\cite{Ayala:2019jey}) where 
  the photon \vtwo does not vanish at low \pt.
  However, the authors themselves warn that ``the \vtwo obtained in
  our model should be regarded as the upper bound generated solely by
  a magnetic field in the strongly coupled scenario'' and for the full
  observed \vtwo contributions from viscous hydrodynamics should be added.
Moreover, if the strong initial magnetic field were really the
principal source of large photon \vtwo, then \vthr should be zero in
first approximation~\cite{Bzdak:2012fr}, in contrast to currently
available data (see Fig.~\ref{fig:ppg176_fig6}).  --  A recent
study~\cite{Ayala:2017vex} calculates photon production at early times
from nonequilibrium gluon fusion induced by magnetic field\footnote{A
  process otherwise forbidden due to charge conjugation invariance,
  and made possible only by the presence of the magnetic field.
}
and finds that both the low \pt yields and \vtwo are enhanced,
mostly below 1\,\gevc, a trend not incompatible with the data in
Fig.~\ref{fig:ppg176_fig6}.  However, the uncertainties on the data
points are too large  --  a problem that ongoing analyses of much
larger datasets will hopefully solve.

\newpage

\section{\bf Concluding remarks}
\label{sec:concluding}

{\it Disclaimer} \\
\vspace{0.1in}
So far the author tried to provide a comprehensive, balanced review of the
field, as free of any personal bias or preference as possible.
In contrast, this last section is strictly personal; it will not
necessarily reflect the consensus or even majority opinion of the
field, nor is it intended to do so.

\vspace{0.2in}
{\it Direct photons in heavy ion collisions: promises kept, broken --
  and open} \\
\vspace{0.1in}
Let us recapitulate some major ``promises'' of direct photons
in studying the physics of heavy ion collisions in the past decades --
which expectations were fulfilled, which were not, and which are the
ones where ``the jury is still out''.

\vspace{0.1in}
Validating the use of the Glauber-model in large-on-large ion
collisions, proof of sanity of the concept of \Ncoll and the way it is
calculated is a major success: at large \pt the direct photon \raa 
is indeed around unity (see Sec~\ref{sec:standardcandle}).  
Based on this there is a strong hope
that photons (or other electromagnetic probes) will solve the problem
of ``centrality'' biases is small-on-large collisions, and separate
genuine new high \pt physics from experimental artifacts.
(measuring $d+d$ collisions would be of additional help).
In summary,
{\it the promise to be a reliable reference for the initial geometry
  was kept.}

\vspace{0.1in}
Using photons in back-to-back photon-jet measurements to set the
initial (unquenched) parton energy scale is ongoing (see
Sec.~\ref{sec:eloss}), there are some early successes and no
indication of problems with the method so far.
{\it The promise to be the ultimate parton energy calibration tool was
  kept.} 

\vspace{0.1in}
In both cases above the probes involved are high \pt photons.  For low
\pt the situation is somewhat less clear.

\vspace{0.1in}
Establishing the temperature at initial thermalization time $\tau_0$
remained elusive; the reasons are described in
Sec.~\ref{sec:thermalyields} and illustrated in
Figs.~\ref{fig:shen_prc_89_tempevol} 
and~\ref{fig:phenix_alice_gdirect_theories}, left panel.
The space-time integral of strongly varying rates smears out all
information on the initial state.  While dileptons may provide 
somewhat more differential information\footnote{Inverse \pt slopes \vs
  pair mass can meaningfully differentiate between early and late
  times~\cite{David:2006sr}. 
},
{\it the promise to measure initial temperature using real photons is
  broken}. 

\vspace{0.1in}
Thermal radiation from the QGP was thought to be the dominant low \pt
photon source up to the early 90's, and even after that most models
indicated that there is a ``QGP window'', a range in the \pt spectrum
where the bulk of the photons come from the QGP, \ie reflect its
properties.  As we discussed in Sec.~\ref{sec:attempts}, with the
observation of large photon \vtwo radiation from the QGP (at least
from its ``classic'' form) became more and more deprecated, so it is
fair to say that
{\it the promise to study the QGP via its directly observed thermal
  radiaton is broken}.

\vspace{0.1in}
In~\cite{Shen:2013cca,Shen:2014cga} it has been pointed out that the
direct photon \vtwo/\vthr ratio can serve as a ``viscometer'' of the
QGP (see Sec.~\ref{sec:v2v3}), particularly when compared to the
corresponding ratio for charged hadrons.  Current experimental
uncertainties do not allow to draw conclusions yet, but significant
improvements are expected in the near future, so
{\it the promise to put constraints on the early values (and maybe the
  time evolution) of $\eta/s$ remains open.}

\vspace{0.1in}
The role of a large, albeit short-lived, magnetic field created in
non-central collisions is actively investigated both theoretically and
experimentally.  As discussed in Sec.~\ref{sec:otherideas}, it may
help to enhance both the photon yield and \vtwo at early times,
alleviating the ``direct photon puzzle''.  One of the scenarios even
predicts large positive \vtwo as $\pt \rightarrow 0$, a tantalizing
possibility disfavored, but not conclusively excluded by the available
data.  At this point
{\it the promise to provide independent information on the initial
  magnetic field is still open}.

\vspace{0.1in}
The fine structure of the initial energy density distribution
plays a decisive role 
in any hydrodynamic calculation.  Even for collisions in the same
centrality class the positions of the nucleons fluctuate
event-by-event, giving rise to odd harmonics in the $v_n$ expansion.
The energy density attributed to each binary collision fluctuates,
too, it is model-dependent (see Fig.~\ref{fig:prl108.252301}), and
its actual shape affect both the final spectra and
$v_n$~\cite{Schenke:2012wb}.  This topic will be even more important
as the study of very asymmetric collisions, as well as \pp, \AA 
collisions with extremely high or low multiplicity intensifies.
Uncertainties of current published
results do not allow yet to differentiate between scenarios, but
{\it the promise to provide information on the initial state is
  still open}. 

\vspace{0.1in}
HBT-correlations to measure the size and shape of the system
(``femtoscopy'') were immensely successful for hadronic observables,
but those reflect (mostly) the status at freeze-out time.  Since high
\pt photons are created very early, they could provide unique
information on the geometry at the initial time.  This potential of
high \pt photon HBT has already been pointed out
in~\cite{Bass:2004de}, but so far not attempted in any of the
experiments, not the least because it requires huge statistics.
Nonetheless, there are indications that at LHC it might be
feasible~\cite{ect:2018}, so
{\it the promise to provide a direct view of the geometry at earliest
  times is still open}.

\vspace{0.1in}
As stated repeatedly, by their penetrating nature direct photons are
``historians'' of the entire collision, including the dynamics of the
expansion.  However, decyphering their {\it space-time integrated}
message is very model-dependent so far.  As discussed in
Sec.~\ref{sec:puzzle}, none of the models describes simultaneously all
real photon observables, nor embeds them into an
all-encompassing ``standard model'' of heavy ion collisions.  There is
substantial progress both in microscopic transport and hydrodynamic
models, as is in understanding of the role of the initial state,
thanks to the copious $p/d$+A data, but a uniform picture is still
elusive.  Both theorists and experimentalists have to overcome huge
challenges before such picture can be drawn.  Eventually we will
succeed.  Until then,
{\it the journey is just as exciting and fascinating as the arrival
  will be.}

\vspace{0.2in}
{\it Quo vadis, direct photon?} \\
\vspace{0.1in}

There is a wide consensus in the field that direct photons, being
penetrating probes, are unique tools to report on the entire history
of the space-time evolution in heavy ion collisions, but also, for the
very same reason they are singularly hard to interpret.  In light of
this, it is quite unfortunate that so far there was not a single
experiment fully optimized for real photon measurements\footnote{
  Dileptons fared somewhat better, and it almost immediately paid off:
  NA60, the current standard in precision measurements quickly decided
  the decades-long argument on the fate of the $\rho$ in the
  medium~\cite{Arnaldi:2006jq}. 
}.
While almost all current experiments have real photon capabilities,
photons are only part of a much larger program, not the main (or single)
focus, thus none of the detectors are without compromises from the
point of view of direct photon measurements.\footnote{This is true
  even if one takes into account the benefits of measuring
  simultaneously the hadron spectra, a huge and  irreducible source of
  background for direct photon measurements.
}  
One consequence is larger
systematic uncertainties than state-of-the-art technology would
allow to achieve, \ie less power to reject some models.  A very recent 
proposal~\cite{pbm:2018}
to install a ``near-massless'' detector in IP2 of the LHC during Long
Shutdown 4 is an important step, hopefully successful, to remedy the
situation. 

Another issue is that independent confirmation of certain results (by
a different experiment) is often impossible.  This is particularly
true for low \pt (``thermal'') photons, which are the richest in
information, but hardest to interpret theoretically.  
Also, the external (and internal) drive to constantly move to uncharted
territories and discover new phenomena sometimes overshadows the need
to thoroughly explore and exploit previous discoveries to their full
depth and potential.  Science lives on new ideas but may die on
insufficiently tested old ones.  {\it Sapienti sat.}

\section{\bf Acknowledgements}

The author greatfully acknowledges decades of fruitful cooperation,
enlightening discussions and unwavering friendship of
Peter Braun-Munzinger, 
Axel Drees,
Charles Gale,
Abhijit Majumder,
Norbert Novitzky,
agdm,
Klaus Reygers,
Takao Sakaguchi,
Johanna Stachel,
Michael J. Tannenbaum
and the late Gerry Brown.

\clearpage

\bibliography{ropp_v4}

\end{document}